\documentclass[10pt]{book}
\usepackage[sectionbib]{chapterbib}
\usepackage{float}
\usepackage[utf8]{inputenc} 
\usepackage{geometry}
\usepackage{times}
\usepackage[dvipdfmx]{graphicx}
\usepackage{amssymb,amsmath}
\usepackage{bigints}
\usepackage{hyperref}
\hypersetup{
setpagesize=false,
 bookmarksnumbered=true,%
 bookmarksopen=true,%
 colorlinks=true,%
 linkcolor=blue,
 citecolor=green,
}
\usepackage[square,numbers,sort&compress]{natbib}
  
\newif\iffigure
\figuretrue

\newcommand{\Nconf}{ M} 

\newcommand{\Slash}[1]{\ooalign{\hfil/\hfil\crcr$#1$}}
\newcommand{\be}{\begin{equation}}
\newcommand{\ee}{\end{equation}\noindent}
\newcommand{\bea}{\begin{eqnarray}}
\newcommand{\eea}{\end{eqnarray}}
\newcommand{\nn}{\nonumber}
\newcommand{\tr}{{\rm tr}}

\newcommand{\1}{{\bf 1}}

\newcommand{\maprightb}[1]{\smash{\mathop{
\hbox to 1cm{\rightarrowfill}}\limits_{#1}}}

\def\calS{{\cal S}}
\newcommand{\del}{\partial} 
\newcommand{\bc}{\begin{center}}
\newcommand{\ec}{\end{center}}

\newcommand{\calC}{{\cal C}}
\newcommand{\calD}{{\cal D}}

\newcommand{\calL}{{\cal L}}
\newcommand{\calO}{{\cal O}}

\newcommand{\hsm}{\hspace{-1mm}}
\newcommand{\matTwo}{\left(\begin{array}{rr}}
\newcommand{\matThree}{\left(\begin{array}{rrr}}
\newcommand{\emat}{\end{array}\right )}
\newcommand{\detTwo}{\left|\begin{array}{rr}}
\newcommand{\detThree}{\left|\begin{array}{rrr}}
\newcommand{\edet}{\end{array}\right |}
\newcommand{\Det}{{\rm Det}}
\newcommand{\bra}{\langle}
\newcommand{\ket}{\rangle}
\newcommand{\Nred}{N_{\rm red}}

\newcommand{\eqspc}{{\,}}
\newcommand{\eps}{{\epsilon}}
\newcommand{\bibun}[2]{ \frac{\partial #1}{\partial #2}}
\newcommand{\nsave}[1]{\left\langle #1 \right\rangle_\eta}
\newcommand{\tileta}{\tilde{\eta}}
\newcommand{\delx}{\partial_x}
\newcommand{\HFP}{H_{\rm FP}}
\newcommand{\half}{1/2}
\newcommand{\abst}{\rm s.\, t.}

\newcommand{\normns}{N_\eta}
\def\inteta{\int {\cal D} \eta}
\def\Gdistns{e^{-\frac{1}{4} \int d\tau (\eta_R^2 + \eta_I^2)}}
\newcommand{\etaR}{ {\eta^{(R)}}}

\newcommand{\llangle}{\left\langle}
\newcommand{\rrangle}{\right\rangle}
\newcommand{\barz}{\bar{z}}
\newcommand{\delz}{\partial_z}
\newcommand{\dely}{\partial_y}

\newcommand{\link}[2]{U_{#2}(#1)}
\newcommand{\calU}{{\cal U}}

\newcommand{\calN}{{\cal N}}
\newcommand{\SUN}[1]{ {\rm SU}(#1)}
\newcommand{\SLNC}[1]{{\rm SL}(#1, \mathbb{C})}

\newcommand{\re}{{\rm Re}}
\renewcommand{\Re}{{\rm Re}}
\renewcommand{\Im}{{\rm Im}}
\newcommand{\ds}{\displaystyle}

\usepackage{color}

\begin{document}

\title{Finite-density lattice QCD and sign problem:\\ 
current status and open problems}
\author{Keitaro Nagata}
\date{}
\maketitle

This article was translated from Japanese to English by Masanori Hanada and Etsuko Itou. The original Japanese version can be found at \url{http://www2.yukawa.kyoto-u.ac.jp/~soken.editorial/sokendenshi/vol31/sokendenshi_2020_31_1.html}.

Keitaro Nagata passed away on November 21, 2019. MH and EI fixed a few typos, removed Japanese references, and corrected several obvious errors. While requests for major changes and citations cannot be accepted, suggestions for correcting typos or mistranslations are welcomed. 

MH and EI thank Sinya Aoki, Atsushi Nakamura, and Jun Nishimura for carefully reading the manuscript and providing useful comments and advice regarding the context of the article.

\vspace{4cm}
\begin{figure}[htbp] 
\begin{center}
\scalebox{0.5}{\includegraphics{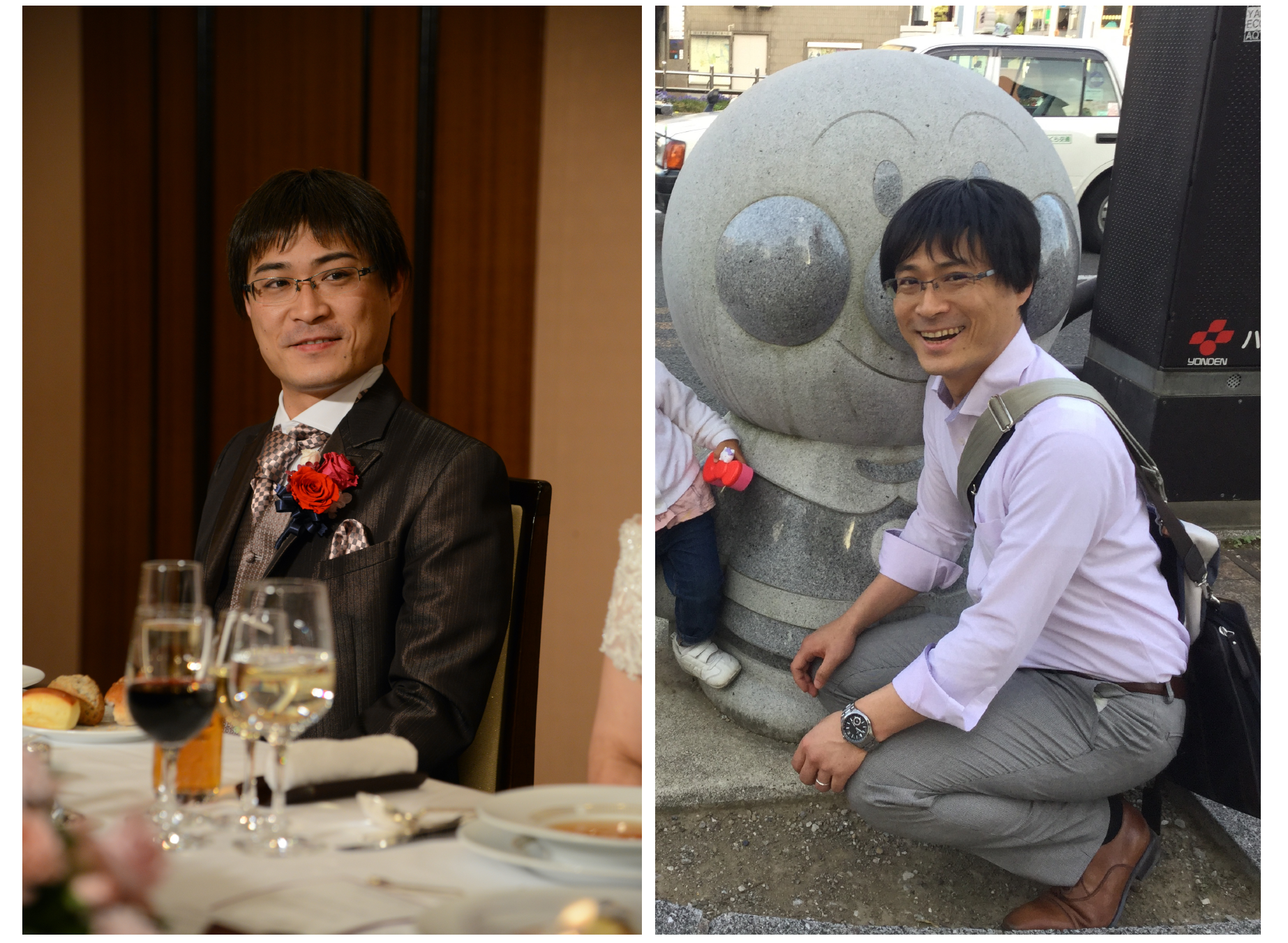}}
\end{center}
\end{figure}

\newpage
\section*{Preface by Sinya Aoki}
\noindent
This article, entitled ``Finite-density lattice QCD and sign problem: current status and open problems", is the English translation of the last manuscript by Dr. Keitaro Nagata, who passed away on November 21, 2019 at the age of 40. 

I received the Japanese version of the manuscript on September 2, 2019 from Keitaro. He asked me to read his review of lattice QCD at finite density, which was written as a summary of his research career when he decided to leave the field. As I immediately recognized it was an excellent review that critically summarized the status of the sign problem in lattice QCD at finite density, I recommended that he publish it in ``Soryushiron Kenkyu", a Japanese researchers' community journal on elementary particle theory, and I sent him some comments regarding the draft.
Unfortunately, he was in critical condition when he sent me the manuscript and he passed away before finishing the revisions I recommended. After his death, Dr. Etsuko Itou, Keitaro's dear wife, found Keitaro's incomplete manuscript, which he had struggled to finish with the help of voice input equipment until the end of his life. By speculating on his intentions, Etsuko and I supplemented several words and sentences at places where his voice input was unclear. In doing so, we were able to complete his manuscript and submit it to the journal.

The sign problem in lattice QCD at finite density is one of the most difficult problems in theoretical physics and is so challenging that even quantum computers may not solve it. As studies on this subject are ongoing, it is difficult to find good reviews, particularly in Japanese, which give readers a wide and comprehensive overview of the problem. While reviews of hot topics being studied tend to stress the reviewers' own achievements, Keitaro introduced various methods objectively and evaluated their pros and cons critically in this article. I believe the article will become one of the best introductory resources for young students and researcher to learn ``the sign problem in lattice QCD at finite density" and start their own work on this subject.      

I always enjoyed discussions with Keitaro, which were stimulating and made me feel the joy of physics research. Even now I often remember that Keitaro gave sincere but critical feedback on others' work with a gentle smile on his face. In Keitaro's memory, I would be very happy if his last work motivates many young people to take on this challenging problem.
\\
\\
July 2, 2021\\
Sinya Aoki\\
Professor/Director\\
Yukawa Institute for Theoretical Physics, Kyoto University

\newpage 
\section*{Preface by Keitaro Nagata}
\noindent
\noindent

{\small 

Finite-density lattice QCD aims for the first-principle study of QCD at finite density, which describes the system consisting of many quarks. The main targets are systems such as quark-gluon plasma, nuclei, and neutron stars. Explaining macroscopic physics from the microscopic theory is a natural path in the development of physics. 
To understand the strong interaction completely, we have to solve finite-density QCD. Each of the systems mentioned above has open problems which cannot easily be accessed by experiment or observation, 
so it is important to make progress in finite-density lattice QCD. 

The main theme of this article is the sign problem, which is a very serious obstacle in the study of QCD at finite density. With the sign problem, the Markov Chain Monte Carlo method fails. In short, this is a difficulty associated with the numerical integration of multi-variable functions with bad behavior. It can happen in any many-body system, but empirically, it often happens in physically important situations such as phase transition. Presumably, this is not a coincidence: due to the Lee-Yang zero-point theorem, the phase factor of the Boltzmann factor tends to fluctuate more near the phase transition. The solution of the sign problem is required for detailed studies of various theories, and hence, it is one of the most prominent tasks in modern physics. 

In the past, major progress in physics was accompanied by the invention of computational techniques, such as classical mechanics and differential/integral calculus, general relativity and Riemannian geometry, quantum mechanics and linear algebra, quantum field theory and renormalization, and particle theory and group theory. 
Given the importance of numerical methods to analytically un-tractable problems, it is natural to add computational science to this list. QCD urged the use of computers in particle physics, and more recently, numerical relativity provided important predictions for the detection of the gravitational wave.

Only a small number of experts studied finite-density QCD until around 2000, but more groups work recently because powerful computers are widely available. In the 1980s and 1990s, several methods to circumvent the sign problem, such as the reweighting method and the canonical method, were proposed. In the 2000s, these methods are applied to, and generalized to, various cases. By the mid-2010s, there was progress in the study of the high-temperature low-density region of the QCD phase diagram. Still, the High-density region has been a very difficult target. Recently, the sampling methods applicable to the complex actions, such as the complex Langevin method and the Lefschetz thimble method, have been invented and some parameter regions previously out of reach are under investigation.

In this article, we summarize the past development and current status of the field of finite-density lattice QCD. The difficulty in the study of theories with the sign problem is that the numerical methods which are correct in principle do not necessarily work in practice and it is hard to know when it fails. We will introduce various approaches in this article, but all of them have pitfalls, which lead to unphysical results unless we study carefully.  
We will explain what kinds of studies were done in the past, to what extent they succeeded, and what kinds of obstacles they encountered, 
and why the approaches correct in principle can lead to wrong answers. 
In this way, we would like to provide lessons from the past for ambitious researchers who plan to work on the finite-density lattice QCD.    

}
\newpage
\tableofcontents

\chapter{Motivations for finite-density QCD}

In this section, we give the motivations to study finite-density QCD. 
We explain the kind of physics is considered, the open problems, and things understood so far. 
Then we explain why the first-principle analysis based on QCD is important.

\section[Open problems in finite-density QCD]{Open problems in the study of baryonic matter and the necessity of finite-density lattice QCD}

\subsubsection{The origin of matter}

What are the matters around us made of, and how were they created? 
The mysteries associated with the origin of matter have been the driving force for the development of natural science. The achievements in the 20th century such as the standard model of particle physics and big bang cosmology revolutionized our understandings of the origin of matter.

We can answer this question as follows. The universe has begun with inflation. After the big bang, it cooled down as it expands. The mass of the particles has been created associated with the phase transition of the vacuum. Quarks formed protons and neutrons. Protons and electrons formed hydrogen atoms. Hydrogen atoms formed stars. Light elements are formed in stars, and the supernovae created heavy elements. Of course, this scenario is not established completely. In the history of science, it happened many times that the mainstream idea turned out to be wrong. It is important to check our understandings carefully.

\subsubsection{Quark, Gluon, QCD}

Matters made of quarks, gluons or hadrons, such as nuclei or quark-gluon plasma (QGP), are called baryonic matters.\footnote{
In this article, we use the terminology `baryonic matter' for quark matter as well, unless otherwise stated.
}
About 5\% of the total mass in the universe comes from baryonic matter. 
Although it is a small portion, the majority of `matter' familiar to us is nuclei, and hence, baryonic matter. Quantum Chromodynamics (QCD) gives the microscopic description of baryonic matter.
However, it is not easy to explain the macroscopic hadronic phenomena from QCD. 
There is steady progress, but it has not been completed because the analysis of QCD is very hard.

QCD is a non-abelian gauge theory describing quark (matter) and gluon (gauge field). Quantum Electrodynamics (QED) is a well-known example of gauge theory. Because QED and QCD have different gauge groups, 
they have very different properties. For example, the gauge field in QED (photon) does not have self-interaction, unlike gluons. 
The coupling constant in QCD is larger (smaller) at low energy (high energy), unlike QED. Among the properties of QCD, asymptotic freedom, color confinement and spontaneous breaking of chiral symmetry (SBCS) are particularly important. The asymptotic freedom means that the effective coupling constant becomes weaker at a short distance. 
It has been found experimentally in the $ep$ deep inelastic scattering experiment at SLAC, and later Gross, Wilczek, and Politzer pointed out that non-Abelian gauge theory can explain this property. The color confinement is the property that particles with color charge cannot be observed individually. The proof of the color confinement remains a theoretical challenge. Phenomenologically, the color confinement is characterized by the property that the color charge cannot be observed individually and that the inter-quark potential calculated in lattice gauge theory diverges at long-distance. Because of the color confinement, the experimentally observable dynamical degrees of freedom are colorless hadrons such as protons and neutrons.

The pion is the Nambu-Goldstone (NG) boson associated with the SBCS. 
Because the chiral symmetry of QCD is not exact due to the small quark mass, the pion has a small but nonzero mass ($m_\pi \sim 140$ MeV). 
It gives rise to the typical distance scale for the nuclear force $m_\pi^{-1}$, and the separation of the distance scale for the electromagnetism and strong interaction. 
It is also known that SBCS characterize the hadronic interactions at low-energy (the PCAC hypothesis). Atomic nuclei are formed by nucleons bounded by the nuclear force arising from the interactions such as the exchange of pions.  

\subsubsection{Phase structure of QCD and nuclei, quark-gluon plasma, neutron stars
}

Water --- the many-body system of H$_2$O --- can be in various phases depending on temperature or pressure. 
In the same manner, the many-body system of quarks can be in various phases 
depending on temperature or chemical potential.

Usually, QCD is in the hadronic phase (or the confinement phase), where chiral symmetry is spontaneously broken and colors are confined. At high temperatures, the chiral symmetry is restored, the confining potential between quarks disappears, and the phase transition to the quark-gluon plasma (QGP) phase (or the deconfinement phase) takes place. Originally, the existence of this phase transition was conjectured because the asymptotic freedom can set in at high temperatures. The nonperturbative analysis via lattice QCD simulation confirmed this conjecture. A QGP-like phase is expected at high density as well because the mean distance between quarks becomes small and asymptotic freedom sets in. However, the details are not known yet because the study of the high-density region is difficult both theoretically and experimentally.

The phase structure of QCD is not just of academic interest. It is related to the origin of matter. It is believed that the baryons were in the QGP phase in the early, hot universe, and the phase transition to the confinement phase took place and quarks and gluons are converted to hadrons when the universe expanded and cooled down sufficiently. If we understand the confinement/deconfinement phase transition, we can understand the universe before the recombination.

After the transition from the QGP phase to the hadronic phase, hydrogen and light elements are formed, light elements accumulate to form stars, and light nuclei are formed in the stars. If a star is sufficiently heavy, a supernova explosion takes place, heavy elements are formed, and the neutron star is left. A neutron star consists mainly of neutrons; this is, in some sense, a big nucleus. However, unlike usual nuclei, its inner structure is not understood well. The density of usual atomic nuclei does not become high because of the repulsive nature of the nuclear force at a short distance (repulsive core). However, the neutron star is compressed to a very high density by gravitational force, and the density at the core is estimated to be a few to ten times larger than usual atomic nuclei. If the many-body system of nuclei is compressed to such high density, the wave functions of neighboring nuclei overlap, and it is impossible to tell which quark belongs to which nucleus. Then, intuitively, the transition from the system of nuclei to the system of quarks is expected. However, the phase realized in a neutron star is currently unknown. If we understand the equation of state of QCD at high density, we can understand the phase realized inside a neutron star.

\subsection{Open problems}

In this section, we discuss a few important open problems in the study of baryonic matter. 

\subsubsection{The evidence of the formation of QGP, heavy-ion collision, and QCD critical point 
}

It is important to understand the phase transition between the QGP phase and the hadron phase, which should have happened in the early universe. To understand this phase transition, relativistic heavy-ion collision experiments have been performed.
Whether the phase transition can be observed in the experiments depend on the order of the transition and transition temperature. It is expected that the signal could be detected if the transition were of first order. 
However the `transition' is likely to be a cross-over around 150 MeV, according to recent lattice QCD simulations~\cite{Aoki:2006we},
and there seems to be no discontinuity at a vanishing chemical potential~\footnote{
For this reason, this transition temperature is sometimes called pseudo critical temperature.}.
At the Relativistic Heavy Ion Collider (RHIC) at Brookhaven National Laboratory, the temperature at the early stage of the collision has been estimated from the temperature distribution of the lepton pairs created by the collision. This temperature is sufficiently higher than the transition temperature obtained by lattice QCD simulation, and hence, almost certainly the RHIC created QGP. Recently, the heavy-ion collision experiment is performed also at the Large Hadron Collider (LHC) at CERN, and an even higher temperature has been realized. However, because the QCD transition is cross-over, it is difficult to see a direct signal of the transition.

Beam Energy Scan (BES)~\cite{Aggarwal:2010wy,Adamczyk:2013dal,Adamczyk:2014fia} is one of the attempts to directly observe the phase transition between the QGP phase and hadron phase. It is expected that, if the system goes through the QCD critical point while the fireball created by the heavy-ion collision expands and cools down, the information of the early stage of the collision can be obtained via the critical phenomena. However, the QCD critical point has not been discovered so far. Although there are many phenomenological studies regarding the QCD critical point, the location of the critical point on the phase diagram is not known. Even the very existence is not established. The first-principle study based on QCD is needed to establish the existence or non-existence of the QCD critical point, and if it does exist, to determine the critical temperature and density.

\subsubsection{The effect of the nuclear medium on the property of hadrons, and the phase realized inside a neutron star
}

Another open problem is the measurement of the hadron mass inside nuclei or at finite baryon density. If we regard the nuclei as baryonic matter, the chiral condensate may decrease inside nuclei. Because the hadron mass and couplings are associated with SBCS, their values may change as the chiral condensate decreases. This is called a partial restoration of chiral symmetry and was studied as a way to detect SBCS directly. While the change of hadron mass inside nuclei has been suggested in one of the experiments, there seems to be no independent cross-check. Although there are many papers based on phenomenological approaches, however, it is impossible to get reliable results because of the ambiguity associated with the lack of experimental data for the hadron mass, coupling constant, and the interaction length,
which changes at finite density. To understand the properties of hadrons inside the nuclear medium quantitatively, it is mandatory to use QCD.

Yet another open problem is the phase inside a neutron star. Because the density at the center of neutron stars is much higher than the usual nuclear density, it would be natural to expect that the deconfined phase is realized. However, the detail is not known.
\subsection{Why QCD is needed, and what are the challenges}

\iffigure
\begin{figure}[htbp] 
\centering
\includegraphics[width=8cm]{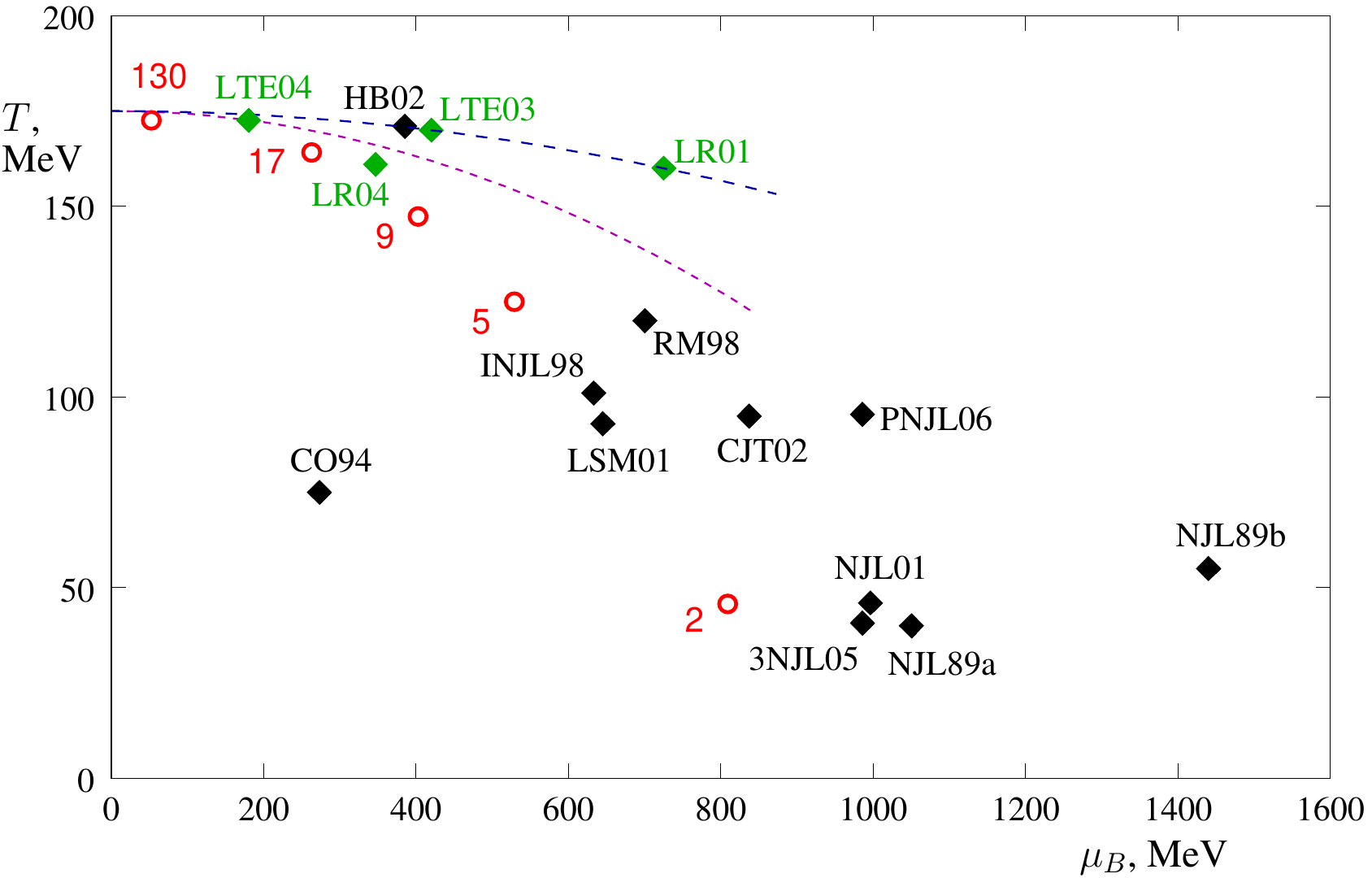}
\begin{minipage}{0.9\linewidth}
\caption{\small 
The location of the QCD critical point claimed in various references. 
This plot appeared originally in Ref.~\cite{Stephanov:2007fk}. 
}\label{Stephanov2007fkfig}
\end{minipage}
\end{figure} 
\fi

The open problems we have discussed above have been studied over a quarter of a century. 
It goes without saying that the experiments are the most unambiguous way to reach the answer. 
However, the range of temperature and density which can be achieved experimentally is limited,
and not all physical quantities are measurable.

In the past, most studies were based on phenomenological approaches via effective models. Effective models can be handled relatively easily, but there are various uncertainties --- e.g., uncertainties associated with the choice of the effective Lagrangian, quantum corrections, or parameters --- which make the conclusion less objective. Many quantities of interest in the study of the QCD phase diagram are not universal, and hence, they depend on the details of the models. Effective Lagrangians are constructed by focusing on a part of the properties of QCD, such as the chiral symmetry or confinement, but there are many possible choices of the effective Lagrangians. Even worse, there are various ways to incorporate quantum corrections into the analyses. Often the higher-order loops are truncated, but the validity of the truncation is not apparent. 
The parameters in the Lagrangians are determined by using experimental results as input (e.g., the mass of nuclei or pion, the pion decay constant $f_\pi$); they are obtained at zero-temperature and zero-density, which may or may not be used at finite density. Because of these uncertainties, the outcomes of the phenomenological studies are different depending on the references. For example, in a review by Stephanov, a large uncertainty regarding the location of the QCD critical point is reported~\cite{Stephanov:2007fk}(Fig.~\ref{Stephanov2007fkfig}).  
Another example that demonstrated the limits of the phenomenological approach was the discovery of a neutron star whose mass is twice the Solar mass~\cite{Demorest:2010bx}. Such a neutron star is heavier than the theoretical upper limit based on the phenomenological approaches, and hence, the validity of such approaches was questioned. 
Although attempts to improve phenomenological models continue, it is very difficult to guarantee the validity of such models.
Those models contributed to the qualitative understanding of the QCD, but they cannot give quantitative predictions without relying on experimental data.

At this moment, there is no successful approach to the study of finite-density baryonic matter. To understand the origin of baryonic matter, a breakthrough is needed. Because baryonic matters consist of quarks, by taking the statistical approach of the quantum field theory, all baryonic matters should be described by introducing temperature and chemical potential to QCD. Even when experiment or observation is difficult, a quantitative study should be possible by solving QCD.

\section{History of finite-density lattice QCD}

To study the properties of baryonic matters quantitatively without uncertainty, we need to solve the QCD with quark chemical potential, 
which is often called `finite-density QCD'. Very unfortunately, there is no established way to solve finite-density QCD. In lattice QCD simulations, usually, the importance sampling method is used for the path integral. 
When the quark chemical potential is nonzero, the Euclidean action becomes complex, and the importance sampling is not applicable. 
The problem that the importance sampling does not apply to the theories with complex action or non-Hermitian Hamiltonian is called the `sign problem' or `complex action problem'. The sign problem is one of the challenges in modern physics, which shows up in various important problems such as the determination of the phase transition. In the past, many studies are performed to beat the sign problem. In this article, we will explain some of them. Because the main part of this article does not follow the historical order, let us recall the history of finite-density QCD briefly here.

Lattice QCD was proposed by K.~G.~Wilson in the mid-1970s, as a nonperturbative approach to QCD. The application of lattice QCD to finite-density systems began in the mid-1980s. In lattice QCD, the Monte Carlo method is used to perform the path integral. Even with the Monte Carlo method, the numerical path integral requires a large computational cost~\footnote{
In computational science, the `computational cost' means the CPU-time and/or the memory size needed for the computation. If either of them is too much, the computation is very difficult.}
Especially, the simulation with fermions requires the determinant of a large matrix, so the first attempts for the finite-density simulation targeted the two-color QCD~\cite{Nakamura:1984uz}, which is computationally less demanding. After the study of the two-color QCD, it turned out that the three-color QCD with quark chemical potential has a complex action and the Monte Carlo method is not applicable. Without using the Monte Carlo method, it is very hard to perform the path integral. 

The study of finite-density QCD is inseparable from the invention of new methods to avoid the sign problem. In the past, there were two turning points at around 2000 and 2010. The 1980s and 1990s were the dawning age; many of the basic ideas used in modern simulations were proposed in this period. Popular topics during this time were on the low-temperature finite-density region related to the nuclei and neutron star. 
However, the lattice QCD simulations were not successful due to a certain difficulty. The study of the low-temperature finite-density region is still challenging even today, as we will repeatedly mention in this article. 
In the lattice QCD community, old papers tend to be ignored because the computer power steadily grow and the simulations become more precise; 
however, it is worthy of special attention that many ideas are proposed and actual simulations are performed in such an early period, without relying on powerful computers. 

In the 2000s, the study of the high-temperature low-density region got attention because of the RHIC experiment. The search of the QCD critical point by Fodor and Katz~\cite{Fodor:2001pe} triggered a stream of studies on the QCD critical point and the hadron-QGP phase transition. The number of researchers in the field increased probably because high-performance computers became widely available. Multiple independent groups worked on the simulations and more precise results were obtained thanks to the increasing computer power, which led to the comparison of the simulation results and substantial progress in the field. Kratochvila and de Forcrand summarized the multiple results regarding the cross-over line between the hadron phase and QGP phase obtained by independent collaborations via different methods and showed that consistent results are obtained in the low-density high-temperature region while there is no consensus regarding the high-density region~\cite{Kratochvila:2005mk}.  
Given that a consensus is obtained among multiple collaborations using different methods, the result can be trusted. We can safely say that the research on finite-density QCD reached a certain level of maturity. 
As we will explain in detail, in the low-density high-temperature region, the sign problem is milder than in the high-density low-temperature region. Namely, the low-density high-temperature region is an easier target, and substantial progress in the 2000s partly relies on this fact. 

Another turning point was around 2010. Firstly, as progress was made by using methods like reweighting, the limitation of those methods was recognized, and it was widely realized that completely new approaches are needed for the high-density region. Since then, there are two streams in the study of finite-density QCD. One of them is to simulate the low-density region more and more precisely by improving the existing methods and seek for applications to the RHIC experiment, such as the fluctuation of the baryon number near the critical temperature. The other direction is to invent new methods to tackle the high-density region. 
Many ideas were proposed since 2010, including the complex Langevin method, Lefschetz thimble method, tensor network method, dual variable method, and diagrammatic Monte Carlo method. Among them, the tensor network method is an application of the tensor renormalization group which was developed in the field of quantum information and condensed matter physics. Such interdisciplinary development is worth mentioning. 
A related topic is quantum entanglement and entanglement entropy; the study of quantum entanglement in gauge theory and QCD is an interesting new direction. These new methods are proven to give the right answers to certain theories with known exact solutions. Regarding the application to QCD, the complex Langevin method has been studied, and actual simulation for the high-temperature low-density region has been performed in 2015. There were new developments in the experiments and observations, such as the Beam Energy Search in RHIC~\cite{Aggarwal:2010wy} and the discovery of the neutron star with twice the Solar mass~\cite{Demorest:2010bx}. At the same time, the difficulty of studying the low-temperature high-density region has been recognized; the complex Langevin method does not work there, and the Lefschetz thimble encounters a difficulty associated with the summation of multiple thimbles. Those problems share the cause with the problems recognized in the 1990s and take place at the same parameter region. 
Several papers suggested that these problems are related to the vanishing fermion determinant. These problems seem to be a particularly difficult type of sign problem. Attempts to improve these methods continue, toward the study of the QCD critical point and the parameter regions describing the nuclei and neutron star. 

\section{Structure of this article}

Although finite-density lattice QCD has been making huge progress, it does not seem to be understood well by non-experts. One possible reason is that it is not easy to understand the concept of importance sampling without having experience in the actual simulations. Without a proper understanding of the importance sampling, it is hard to understand what kinds of troubles come from the failure of the importance sampling, and how the proposed cures can work. Another possible reason is that, when the cures to the sign problem are discussed, formally correct arguments can lead to practically wrong results, and it is hard to figure out when such things happen.

The sign problem is a failure of the importance sampling in the Monte Carlo method, which makes it impossible to extract the points contributing to the multiple integral. We can also say that we have to invent an efficient numerical integration method for badly-behaved functions. It is a problem associated with the computational cost, which is not the problem of yes or no such as Fermat's conjecture or Poincare conjecture. As there is no generic method for the integral, probably there is no numerical integration method that works efficiently with arbitrary integrands. As the accuracy of the numerical integration depends on the property of the integrand, the level of difficulty of the sign problem depends on the property of the action. In finite-density lattice QCD, it often happens that a method useful in the high-temperature or low-density region becomes less effective as the temperature goes down or the density goes up and eventually ceases to work. Usually, this is not because the method is wrong in principle, rather the property of the QCD action changes. In Chapter~\ref{chap:latticeQCD}, we give a brief summary of lattice QCD, and then explain how the sign problem appears in lattice QCD. Furthermore, we consider what it means to solve the sign problem, and introduce the results regarding the temperature- and chemical-potential-dependence of the average phase factor, which is a characterization of the hardness of the sign problem.

In Chapter~\ref{sec:oldmethod}, we focus on the reweighting method, the Taylor expansion method, and the analytic continuation from the imaginary chemical potential. We will explain the idea, the validity, and the scope of those methods in detail. All those methods are similar in that they use the importance sampling for the gauge-configuration generations. They are effective at the high-temperature low-density region of the QCD phase diagram. In Sec.~\ref{sec:silverblaze}, we explain the early onset problem which appears in the finite-density region of the hadron phase. It is difficult to solve the early onset problem via the reweighting method or the Taylor expansion method; it is necessary to invent a configuration-generation method/path-integral method that can work for the complex action. Recently, the research in that direction is getting active. Among such attempts, in Chapter~\ref{sec:newmethod} we introduce the complex Langevin method which is already applied to QCD.

If we count the small variations as well, there are a huge number of references in finite-density lattice QCD and the sign problem, and new results are still coming out. In this article, we do not aim to review all results because it is beyond the author's ability. Rather, we put the materials which the author studied at the core, and introduce the key ideas in finite-density lattice QCD. There are several reviews on finite-density lattice QCD. The results from the 1980s and 1990s by the Glasgow group are summarized in Ref.~\cite{Barbour:1997ej}. Ref.~\cite{Muroya:2003qs} is a little bit old, but it is a good review for beginners which contains various methods. More recent reviews are Refs.~\cite{deForcrand:2010ys,Aarts:2015tyj}. These reviews are published in 2010 and 2015; we can see various developments during these 5 years.

Most of the development in finite-density QCD is reported in the annual Lattice Field Theory conference and summarized in the proceedings. By looking at them, it is possible to follow the latest development to some extent.   

\bibliographystyle{utphys}
\bibliography{ref_list}

\newpage
\chapter{Finite-density QCD and sign problem
}
\label{chap:latticeQCD}

In this section, we review the origin of the sign problem in finite-density QCD and the troubles associated with the sign problem. Firstly, we define lattice QCD and explain the role of the importance sampling in the numerical simulation. Then we see that, when the quark chemical potential is introduced, the QCD action becomes complex and the importance sampling ceases to work. Then we discuss what it means to ``solve the sign problem". Because the property of the imaginary part of the action depends on temperature and density, the level of difficulty of the sign problem varies with temperature and density.

Regarding lattice QCD, we explain only the very basic materials needed for this article. To learn more advanced materials, see e.g.~Ref.~\cite{Rothe:book}. A recent textbook \cite{Gattringer:book} contains some discussions regarding finite-density QCD. 

\section{Lattice QCD}
\label{sec:LQCDandSP}

\subsection{Lattice in Euclidean spacetime}
Let us consider QCD on 4d Euclidean spacetime. 
The gauge part and the fermion part of the action $S_g$ and $S_f$ are given by\footnote{
The $\theta$-term $\calL = \epsilon^{\mu \nu \alpha \beta} {\rm tr} F_{\mu\nu} \tilde{F}_{\alpha \beta}$ can be added to the Lagrangian, preserving the gauge invariance and renormalizability. 
In this article, we neglect the $\theta$-term, because it is very small even if it exists.}
\begin{align}
S_g &= \frac{1}{2} \int d^4x \, {\rm tr} F_{\mu\nu} F_{\mu\nu} , \\
S_f &= \int d^4x \, \bar{\psi}(x) ( \gamma_\nu D_\nu  + m + \mu \gamma_4) \psi(x)
\equiv
 \int d^4x \, \bar{\psi}\Delta\psi. 
\label{eq:QCDgaugeaction}
\end{align}
Here, $x$ represents the four-dimensional Euclidean coordinate $(x_1, x_2, x_3, x_4)$.
$D_\nu$ is the covariant derivative, 
$F_{\mu \nu}$ is the field strength, and $\psi$ is the quark field. 
$m$ and $\mu$ represent the mass of the quark and chemical potential, respectively.  
$\gamma_\nu (\nu=1,2,3,4)$ are gamma matrices which satisfy the anticommutation relation
$\{ \gamma_\alpha, \gamma_\beta\} = \delta_{\alpha \beta}$. 
The gamma matrices are Hermitian ($\gamma_\nu^\dagger = \gamma_\nu$), 
while the covariant derivative is anti-Hermitian ($D_\nu^\dagger = - D_\nu$).

We quantize the system via the path integral. 
The quark fields can be integrated out analytically because they appear in the action in the bilinear form. 
We use the formula 
\[
\int \calD \psi \, \calD \bar{\psi} \, e^{- \int \bar{\psi} \Delta \psi } = \det \Delta
\]
which involves the fermion determinant $\det \Delta$.
For simplicity, we consider $N_f$ quarks with degenerate mass. 
Then the partition function is given by  
\begin{align}
Z = \int \calD A_\mu (\det \Delta)^{N_f} e^{-S_g}.
\label{eq:2017Jul03eq1}
\end{align}

In finite-density QCD, we need to study various energy scales ranging from the QGP phase to the hadron phase, and hence the non-perturbative effects are important. Lattice QCD is the standard approach to study the non-perturbative aspects of QCD. In lattice QCD, spacetime is discretized by using a lattice. Let the numbers of lattice points along
$x, y, z$ and $t$ directions be $N_x, N_y, N_z$ and $N_t$, respectively. 
We take the lattice spacing to be $a$. Then the ultraviolet modes (with wavelength $<a$) are removed. We need to take the physical volume of the lattice $a N_x, a N_y, a N_z$ and $a N_t$ to be sufficiently large. 
To perform the simulations with realistic computational resources, we cannot take the lattice size too large. As we will discuss, finite-density QCD requires more resources compared to zero-density QCD, and hence small lattice size is often used. QCD in continuum spacetime is reproduced in the continuum limit ($a\to 0$).

\subsection{Discretization of the QCD action
}
\label{sec:latticeaction}

We introduce the link variable $U_{n \mu}=e^{ i g a A_\mu(n)}$ on the link between two points 
$n=(n_x, n_y, n_z, n_t)$ and $n+\hat{\mu}$ (Fig.~\ref{fig:linkvar}). 
Here $\hat{\mu}$ is a unit vector along the $\mu$-direction ($\mu=1, 2, 3, 4$). 
The link variable $U_{n \mu}$ is an SU(3) matrix
 ($3\times 3$ unitary matrix with determinant $\det U_{n \mu}=1$). 
The partition function $Z$ and the expectation value of the observables $O$ are given by 
\begin{align}
Z &= \int \left(\prod_{n, \mu} dU_{n, \mu} \right) ( \det \Delta)^{N_f} e^{-S_g},
\label{eq:latticeZ}\\
\langle O \rangle &= Z^{-1} \int \left(\prod_{n, \mu} dU_{n, \mu} \right) O \cdot ( \det \Delta)^{N_f} e^{-S_g}.
\label{eq:latticeO}
\end{align}
Here $dU$ is the Haar measure (see Appendix~\ref{sec:haarmeasure}).

\subsubsection{Discretization of the gauge part of the action $S_g$}

\iffigure
\begin{figure}[htbp] 
\centering
\includegraphics[width=6cm]{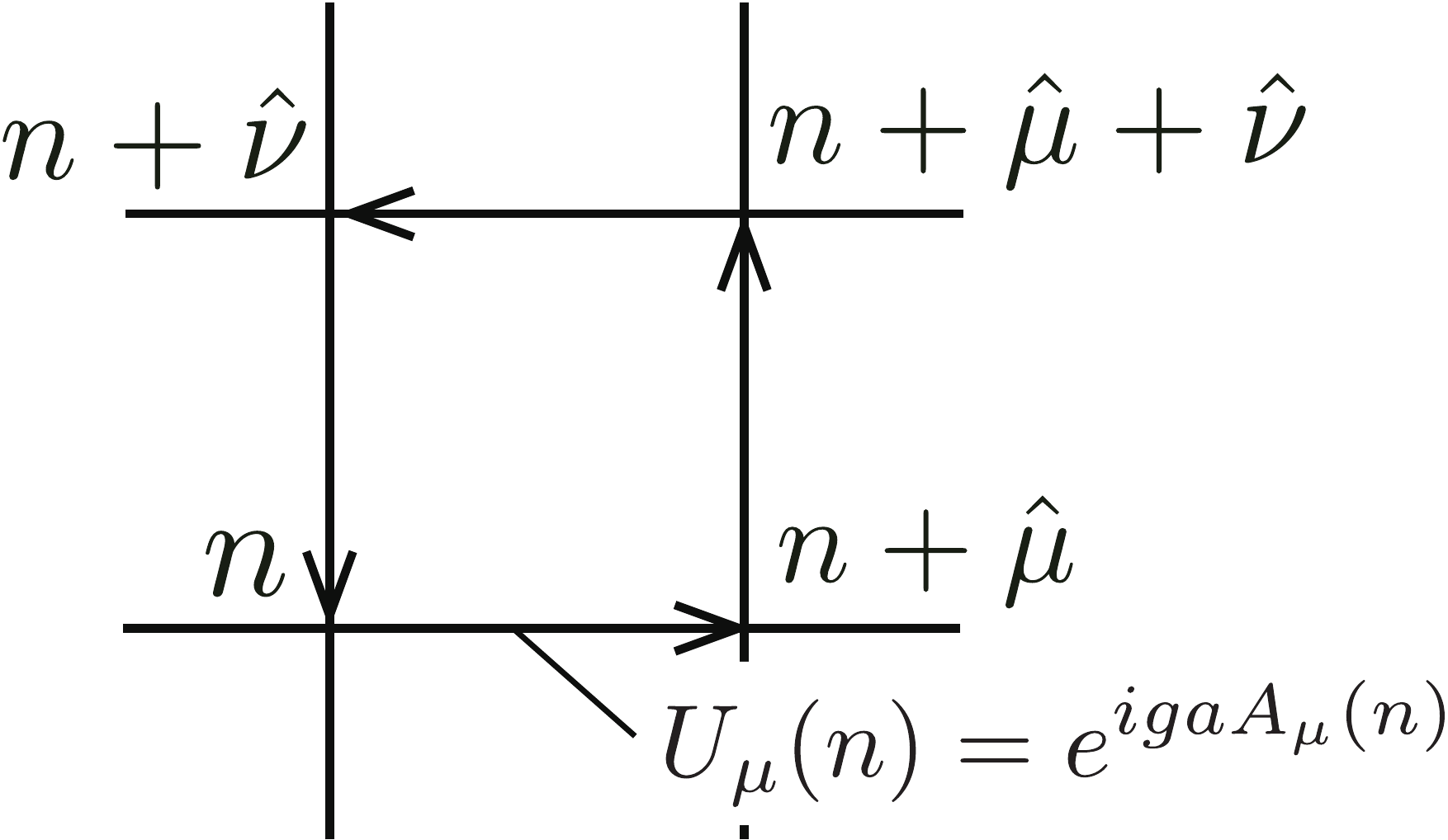}
\begin{minipage}{0.9\linewidth}
\caption{\small 
Plaquette: The link variable $U_{n\mu}$ is defined on the link connecting the lattice points $n$ and $n+\hat{\mu}$. The product of the link variables along the smallest square is called plaquette. 
}
\label{fig:linkvar}
\end{minipage}
\end{figure}
\fi

The QCD action on the lattice has to satisfy two conditions, 
(i) invariance under the gauge transformation on the lattice, and 
(ii) QCD in the continuum spacetime is reproduced in the continuum limit ($a\to 0$).

To build an action satisfying the condition (i), firstly we need to define the gauge transformation on the lattice. The gauge transformation of the link variable is defined by 
\begin{align}
U_{n\mu} \to U'_{n\mu} = g_n U_{n\mu} g_{n+\hat{\mu}}^{-1}. 
\label{eq:2017Jun07eq1}
\end{align}
Here $g_n$ is an element of SU(3) which satisfies $g_{n}^{-1} = g_{n}^\dagger$.
We can see that the closed loops made of link variables are gauge invariant. 
The simplest quantity of that kind is the trace of the plaquette $U_{\mu\nu}(n)$ given by  
\begin{align}
U_{\mu\nu}(n) & = U_{n,\mu} \, U_{n+\hat{\mu},\nu} \, U_{n+\hat{\nu},\mu}^\dagger \, U_{n,\mu}^\dagger \, . 
\label{eq:2017Jun07eq2}
\end{align}
See Fig.~\ref{fig:linkvar}. 
By substituting (\ref{eq:2017Jun07eq2}) to (\ref{eq:2017Jun07eq1}), we obtain  
\[ 
U_{\mu \nu } (n) \to g_n U_{\mu\nu}(n) g_n^\dagger. 
\]
Therefore, ${\rm tr} U_{\mu \nu}(n)$ is gauge-invariant. 
The same transformation law applies to any closed loop, and hence the traces of the loops are gauge invariant.

Next, let us consider the condition (ii). The simplest gauge-invariant quantity which leads to the QCD action is the plaquette. By expanding the plaquette with $a$, we obtain~\footnote{
For each combination of $n, \mu, \nu$, 
$F_{\mu \nu}(n)$ is a $3\times 3$ matrix. 
In \eqref{eq:2017Jun07eq3}, 
$F_{\mu \nu}^2$ means the product of $F_{\mu\nu}$, with out the sum with respect to $\mu$ and $\nu$. In other words, we do {\it not} use the Einstein's summation convention here. }
\begin{align}
U_{\mu\nu}(n) &= e^{i g a^2 F_{\mu\nu}},  \nn \\
 & = \1+ i g a^2 F_{\mu\nu} +  \frac12 ( i g a^2)^2 F_{\mu \nu}^2 + \cdots. 
\label{eq:2017Jun07eq3}
\end{align}
The third term on the right-hand side contains the $\mu, \nu$-component of the gauge part of the action $F_{\mu\nu}^2$. To extract this term, we note that $F_{\mu \nu}$ is Hermitian and ${\rm tr} \, (F_{\mu\nu})^n$ is real. 
Hence, if we take the trace of \eqref{eq:2017Jun07eq3}) and keep the real part, we obtain 
\[ 
 {\rm Re \, tr} \, U_{\mu\nu}(n) =  {\rm tr} \left[ \1 - \frac12 g^2 a^4 F_{\mu \nu}^2 + \cdots\right].  
\]
We can rewrite it as 
\[ 
a^4 {\rm tr} F_{\mu \nu}^2  =  \frac{2}{g^2} {\rm Re \, tr} \left[ \1 -  U_{\mu\nu}(n) \right]
+O(a^6).
\]
Therefore, by defining $S_g$ as 
\begin{align}
S_g & = - \frac{2}{g^2} \sum_{n} \sum_{\mu \neq \nu} {\rm Re} \,{\rm tr} [ U_{\mu \nu}(n)], \\
   & = - \frac{1}{g^2} \sum_{n} \sum_{\mu \neq \nu} \,{\rm tr} [ U_{\mu \nu}(n) + U_{\mu \nu}(n)^\dagger], 
\label{eq:gauge_action}
\end{align}
we can make $S_g$ gauge-invariant on the lattice, and the gauge part of the action \eqref{eq:QCDgaugeaction} is reproduced as $a\to 0$. 
Often, $\beta = 2 N_c/g^2$ is used instead of $g$; see e.g.~~\cite{Rothe:book}.

\subsubsection{Discretization of the fermion part of the action $S_f$}

For the fermion part of the action, we need to impose the chiral symmetry, in addition to the two conditions we discussed above (the gauge invariance and the appropriate continuum limit). 
However, it is not straightforward to realize the chiral symmetry on a lattice.

The staggered fermion, which is also called the Kogut-Susskind (KS) fermion, is one of the popular options for the lattice fermion.
The operator associated with the KS fermion is given by 
\begin{align}
\Delta_{x\, y} = m \,  \delta_{x\, y} + \sum_{\nu} \frac{\eta_\nu(x)}{2a} \left[ e^{\delta_{\nu 4}\mu a} U_{x\nu} \delta_{x+\hat{\nu},y}
- e^{- \delta_{\nu, 4}\mu a} U_{x-\hat{\nu}, \nu}^\dagger \delta_{x-\hat{\nu},y}\right].  
\label{Stfermion}
\end{align}
Here $x$ and $y$ are four-vectors representing the lattice sites, 
$m$ and $\mu$ are the mass and chemical potential of the quark, 
and $\eta_\nu(x)$ is a scalar quantity which is called the Kawamoto-Smit phase. 
The operator $\Delta_{xy}$, which corresponds to a derivative on the continuum space, 
is a matrix whose indices represent the lattice points and the internal degrees of freedom such as colors. 
$\Delta_{x y}$ is the matrix element associated with the lattice points $x$ and $y$; 
it is still a matrix that has color indices. When the chemical potential $\mu$ is zero, the term corresponding to the Dirac operator (the second term on the right-hand side of \eqref{Stfermion}) is anti-Hermitian; see Appendix~\ref{sec:anti_hermite_KS} for proof. 
Furthermore, $\Delta$ itself has the $\gamma_5$-Hermiticity 
\begin{align}
\eps_x \Delta_{x y} \eps_y = \Delta^\dagger_{yx}, 
\end{align}
where $\eps_x = (-1)^{x_1 + x_2 + x_3 + x_4}$ corresponds to $\gamma_5$ in the continuum theory. We discuss the anti-Hermiticity of the Dirac operator in \ref{sec:signproblem}, because it is related to the sign problem. 

The KS fermion is used frequently in actual lattice QCD simulations because it requires less simulation cost compared to other popular formulations of the lattice fermion. The propagator of the KS fermion has $2^4=16$ poles. This is more than the four poles needed for the quark field; the KS fermion contains so-called fermion doublers. Sometimes the KS fermion is regarded as the 4-flavor fermion action, by utilizing those 16 degrees of freedom.\footnote{
These `flavors' are sometimes called `tastes', to distinguish them from physical flavors.}
To describe just one fermion by using the KS fermion, it is necessary to take the fourth root of the fermion determinant. Whether this `fourth root trick' is legitimate is a controversial issue. It is believed that the fourth root trick does not lead to a problem when $\mu=0$, after the continuum limit is taken. On the other hand, when $\mu\neq 0$, the fourth-root trick can lead to the ambiguity of the phase of the fermion determinant \cite{Golterman:2006rw}. As of today, the fourth root trick has not been a serious issue, because the lattice QCD simulations at finite $\mu$ are not yet precise enough. Hence, we do not discuss the validity of the fourth root trick in this review. In the future, when the precision of the simulations is improved, careful checks will be needed.

Another popular option is the Wilson fermion. In the Wilson fermion action, the doublers are removed by the deformation term. Written explicitly, the fermion matrix is modified to the following form: 
\[ 
\Delta = D_\mu \gamma_\mu + m + r a D_\mu D_\mu. 
\]
The third term, which is called the Wilson term, removes the doublers.
The Wilson fermion on the lattice is defined by 
\begin{align}
\Delta_{x,y} &= \delta_{x,y}
 - \kappa \sum_{\nu=1}^{4} \left\{
        e^{+\mu a \delta_{\nu 4}} (r-\gamma_\nu) U_{x\nu} \delta_{y,x+\hat{\nu}}
      + e^{-\mu a \delta_{\nu 4}} (r+\gamma_\nu) U_{y\nu}^{\dagger} \delta_{y,x-\hat{\nu}} \right\} \nn \\
&- \delta_{x, y} C_{SW} \kappa \sum_{\mu \le \nu} \sigma_{\mu\nu} F_{\mu\nu}.  
\label{Wfermion}
\end{align}
The terms involving $r$ are the Wilson terms. $\kappa$ is called the hopping parameter. As one can easily see from the expression in the continuum spacetime, the Wilson term is in the diagonal entries concerning the spinor indices, in the same way as the mass term. 
This is the reason that the fermion doubling can be resolved. However, the price is that the chiral symmetry is explicitly broken. In the continuum limit, the Wilson term does not affect the low-energy modes, and the correct fermion action in the continuum spacetime is reproduced. \eqref{Wfermion} is normalized such that the mass term becomes 1, 
and $\kappa$ is a function of the actual quark mass. This convention is used widely for the Wilson fermion. The last term containing $C_{\rm SW}$, called the clover term,  is introduced to reduce the discretization error~\cite{Sheikholeslami:1985ij}.\footnote{
When the Wilson fermion is adopted for the simulations the clover term is often used. It is useful for removing the unnatural results associated with the naive Wilson fermion (e.g.~ the quark-mass dependence of the order of confinement-QGP phase transition). When the lattice is coarse without the clover term, the discretization effect is severe.}

So far we have introduced two kinds of lattice fermions. The KS fermion \eqref{Stfermion} has the chiral symmetry, but it has unphysical doublers. The Wilson fermion \eqref{Wfermion} does not have doublers but breaks the chiral symmetry. Neither of them is completely satisfactory. Actually, there is a No-Go theorem proven by Nielsen and Ninomiya, which states that the local bilinear lattice fermion action must have doublers when it has the translational invariance, Hermiticity, and chiral symmetry. Therefore, the chiral symmetry and the absence of the doublers cannot be compatible, as long as other conditions are satisfied.
The chiral symmetry can be preserved by using the domain-wall fermion or overlap fermion, but there is no established procedure to introduce the chemical potential to those lattice fermions. For example, the sign function used for the overlap fermion cannot be determined uniquely when the chemical potential is nonzero. While some proposals are made \cite{Bloch:2006cd,Bloch:2007xi}, there are not so many applications to actual lattice simulations.  
In this review, we will consider only the KS fermion and the Wilson fermion.

Lattice QCD involves discretization errors associated with the approximation of the continuum spacetime by a lattice and the finite-volume effect associated with the finite lattice size. However, because QCD is asymptotically free, the continuum limit ($a\to 0$) can be taken, 
and in the continuum and large-volume limit the actual physics --- QCD in the continuum, infinite-volume spacetime --- is reproduced. In this sense, lattice QCD simulation is called the first-principle calculation of the strong dynamics.  

\subsubsection{Temperature in Euclidean path integral}
In the Euclidean path integral formulation, based on the correspondence between quantum statistical mechanics and the partition function, the temperature is defined by 
\[
T = \frac{1}{aN_t}. 
\]
The partition function in the Minkowski spacetime is given by $Z=\langle \Phi_{\rm final} | e^{ i H  t} | \Phi_{\rm initial} \rangle $. Via the analytic continuation of the imaginary time $t \to i t$ and the compactification of the imaginary time with period $aN_t$, the partition function changes to $Z={ \rm tr} \, e^{ - aN_t H}$. In quantum statistical mechanics, the partition function is given by $Z = { \rm tr} \, e^{ - H/T}$. By identifying them, we obtain the expression regarding the temperature.

Therefore, in lattice QCD, the temperature of the system depends on the lattice spacing $a$. By sending $a$ to zero fixing $T=(a N_t)^{-1}$, actual, physical QCD at temperature $T$ can be studied. To study the temperature dependence of the observables, ideally one should take continuum limit $a\to 0$ for various values of $T$. While such a continuum limit is achieved in recent finite-temperature QCD simulations at $\mu=0$, it is not easy to study the continuum limit for $\mu\neq 0$ because of the large computational cost. To vary temperature $T$ without taking the continuum limit, one can (a) fix $N_t$ and vary $a$, or (b) fix $a$ and change $N_t$. Option (a) is convenient in that $T$ can be varied continuously, however different ultraviolet cutoff $a^{-1}$ gives different discretization error, which makes the analysis complicated. Option (b) does not suffer from this problem, but the temperature can be varied only discretely.

\begin{table}[tbh]
\begin{center}
\caption{\small The qualitative relationship between the lattice spacing $a$, coupling constant $g$, $\beta \propto 1/g^2$, 
and temperature $T= 1/ (a N_t)$. 
Quantitative details depend on the choice of the lattice action. } 
\label{tab:avsT}
\label{tab:agbetaT}
\begin{tabular}{c|cc}
\hline $a$ & small (fine lattice) & large (coarse lattice) \\
$g$ & small  & large \\
$\beta$ & large & small \\
$T$ & high temperature & low temperature  \\
\hline
\end{tabular}
\end{center}
\end{table}

The lattice QCD action does not depend on the lattice spacing $a$ explicitly, and hence, the value of $a$ cannot be manipulated directly. 
Instead, the lattice coupling constant $\beta$ (or $g$) can be dialed, 
and $a$ is determined as a function of $\beta$ via renormalization: $a=a(\beta)$. While different lattice actions or renormalization schemes can lead to different $\beta$-dependence of $a$, qualitative features can be understood from the asymptotic freedom and non-perturbative effects at long distances. Larger $a$ means smaller ultraviolet cutoff and hence larger $g$. $\beta$ increases as $g$ becomes large, and hence, as $a$ becomes large. Temperature $T=1/(aN_t)$ goes down as $a$ becomes large, if $N_t$ is fixed. These relations are summarized in Table~\ref{tab:agbetaT}. 
\subsubsection{Diagrammatic interpretation of the effect of the chemical potential
}

\iffigure
\begin{figure}[htbp] 
\centering
\includegraphics[width=12cm]{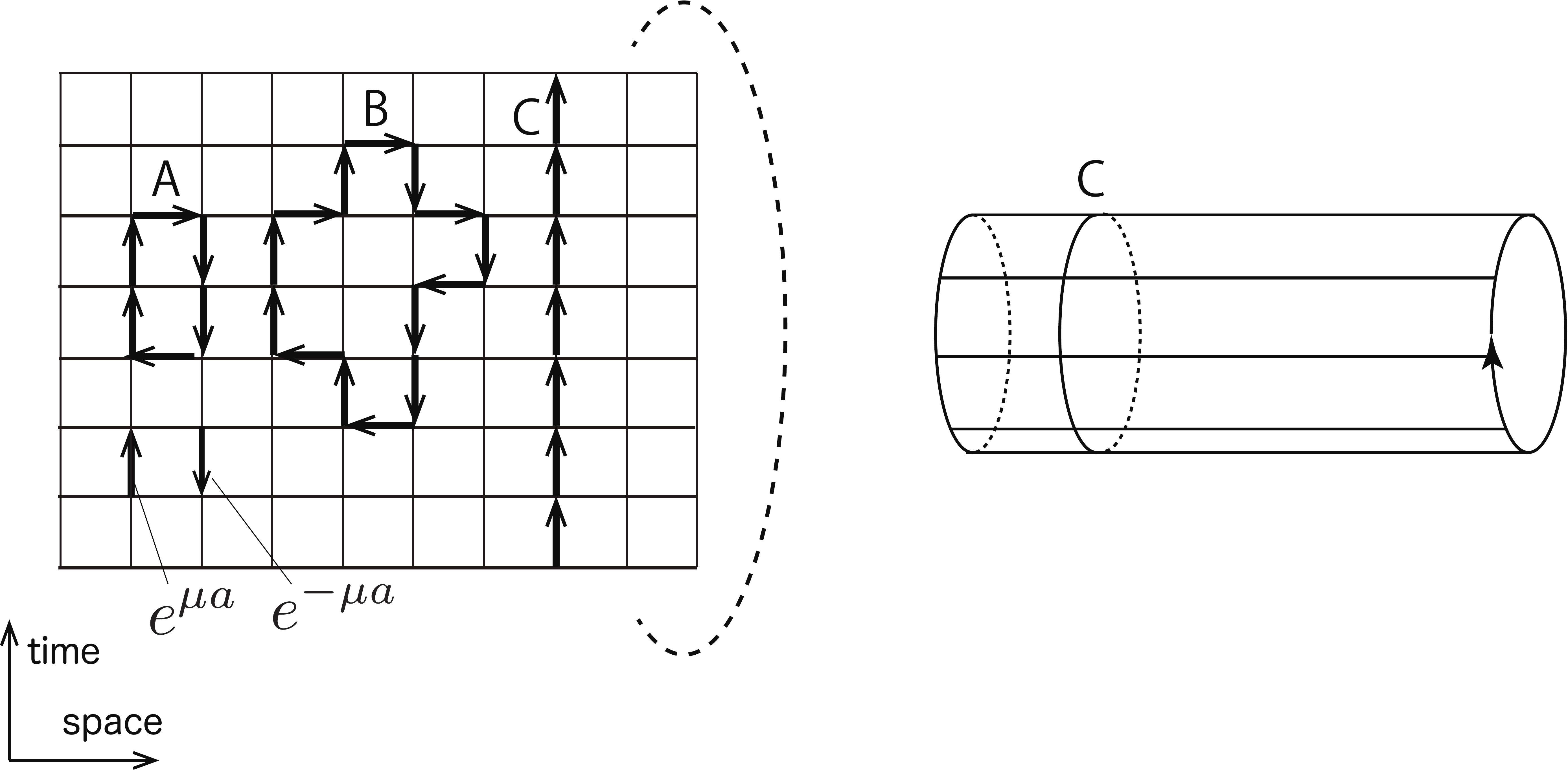}
\begin{minipage}{0.9\linewidth}
\caption{\small 
Closed loops from the fermion determinant. 
A, B: Loops which are not winding on the time direction. Because such loops contain the same number of upward and downward quark lines, $e^{\mu a}$ and $e^{-\mu a}$ cancel with each other, and there is no $\mu$-dependence. 
C: This loop wraps on the time direction. There are only upward quark lines, which leads to a $\mu$-dependence. The closed loop with the opposite direction can be constructed as well. 
}
\label{fig:chemipote}
\end{minipage}
\end{figure}
\fi

In the continuum theory, the chemical potential is introduced by adding $\mu \bar{\psi} \gamma_4 \psi$ to the Lagrangian. On the lattice, it can be done by multiplying $e^{\pm \mu a}$ to the link variables along the temporal direction ($\nu = 4$) in the Dirac operator. Formally, even on the lattice, a linear term $(\sim \mu U)$ can be used, but then, a strong divergence arises as $a\to 0$ and the correct continuum limit is not realized~\cite{Hasenfratz:1983ba}. The multiplication of $e^{\pm \mu a}$ to the link variables is physically natural because it corresponds to adding $i\mu$ to $A_4$. To see this, note that this multiplication gives a term coupled to the fourth component of the current ($\mu\bar{\psi}\gamma_4\psi$) in the continuum limit, as one can check directly.

The chemical potential introduced in this way makes it easier to understand the $\mu$-dependence of the fermion action diagrammatically. Let us consider a term $\bar{\psi}_x \Delta_{xy} \psi_y$ in the Lagrangian. When the fermion is integrated out, this term can be interpreted as if a quark is created at $x$ and annihilated at $y$.
Note that $\Delta_{xy}$ contains a term proportional to $\delta_{y,x\pm\hat{\mu}}$, that describes the creation and annihilation of a quark at $x$ and $x\pm\hat{\mu}$. This can be interpreted as a quark hopping to the neighboring site, and hence this term is often called the hopping term. The hopping parameter $\kappa$ controls the frequency of the hopping to the neighboring site. 

The fermion determinant can be expressed by using the products of $\Delta$'s. Pictorially, $\Delta_{xy}$ can be expressed by using the arrow connecting $x$ and $y$, as in Fig.~\ref{fig:chemipote}. Those arrows have to be arranged such that all of them belong to some closed loop. 
When the quark moves upward or downward, the factor $e^{\mu a}$ or $e^{-\mu a}$ appears, respectively. Among such loops, only the ones winding on the temporal circle have the dependence on the chemical potential (loop C in Fig.~\ref{fig:chemipote}). In the loops not winding on the temporal circle, the same number of $e^{\mu a}$ and $e^{-\mu a}$ appear and cancel with each other (loop A and loop B in Fig.~\ref{fig:chemipote}). The loops winding once along the temporal direction involve the factor $(e^{\pm \mu a})^{N_t}=e^{\pm \mu/T}$. The ones winding on the temporal circle $n$ times involve $[(e^{\pm \mu a})^{N_t}]^{n}=e^{\pm n\mu/T}$. The factor $e^{\mu/T}$ is called fugacity. In Sec.~\ref{sec:canonical}, we prove that $\det \Delta$ can be expanded in powers of $e^{\mu/T}$. Such expansion corresponds to the expansion with respect to the winding number along the temporal direction.

\subsection{Basic quantities}
\label{sec:observables}

Below, we introduce several basic quantities which are frequently used in the study of the QCD phase diagram. 
\subsubsection{Polyakov loop}

The Polyakov loop is often used to characterize the confinement/deconfinement phase transition. It is defined by 
\begin{align}
L(\vec{x})= \frac13\,\tr\, {\rm P} \biggl\{ \exp\biggl(i g \oint_0^{1/T} A_4(\vec{x},\tau) d \tau \biggr)\biggr\}, 
\label{eq:Ploop_cont}
\end{align}
where ${\rm P}$ stands for the path ordering, 
$\vec{x}$ is the spatial coordinate, $\tau$ is the imaginary time, $g$
is the QCD coupling constant, and $A_4$ is the $\mu=4$ component of the gauge field $A_\mu$. $A_\mu$ ($\mu=1,2,3,4$) are su(3) matrices ($3\times 3$ traceless Hermitian matrices) with the color indices, and the trace is taken over the color indices. On the lattice, the Polyakov loop is defined as 
\begin{align}
L(\vec{x}) = \frac{1}{3} {\rm tr} \prod_{i=1}^{N_t} U_4(t_i) = \frac{1}{3} {\rm tr} U_4(\vec{x},t_1) U_4(\vec{x},t_2) \cdots U_4(\vec{x},t_{N_t}).  
\label{eq:Ploop}
\end{align}
The expectation value of the Polyakov loop is related to the free energy of the probe quark in the heavy-mass limit as 
\[ 
\langle L \rangle \propto e^{ - F/T}. 
\]
Let us consider the $\mathbb{Z}_{N_c}$-transformation which sends all links at a time slice $t_i$ 
\[ 
U_4 (\vec{x}, t_i ) \to \omega U_4 (\vec{x}, t_i). 
\]
Here $\omega$ is an element of $\mathbb{Z}_{N_c}$. The Yang-Mills action (QCD in the quench approximation) is invariant under this $\mathbb{Z}_{N_c}$-transformation. This is called center symmetry. 
The Polyakov loop is not invariant under this $\mathbb{Z}_{N_c}$-transformation. The hadron phase is the ordered phase with the unbroken center symmetry, where the expectation value of the Polyakov loop is zero. This corresponds to $F\to\infty$, and hence infinite free energy is needed to extract a quark. In the QGP phase, the center symmetry is broken spontaneously, the expectation value of the Polyakov loop is nonzero, and $F$ is finite. In this way, the Polyakov loop serves as the order parameter for the confinement. 
When there are dynamical quarks (i.e.~when we do not take the heavy-quark limit), the center symmetry is explicitly broken in the fermion part of the action. Then, strictly speaking, the Polyakov loop is not the order parameter. Still, however, even with dynamical quarks the Polyakov loop is often used as a rough characterization of the phases because it is small in the hadron phase and large in the QGP phase.

\subsubsection{Thermodynamic quantities}

When the volume of the system is large, the pressure in the grand canonical ensemble is given by
\begin{align}
p(\mu, T) = \frac{T}{V} \log Z(\mu, T)
\label{eq:2017Mar27eq1}. 
\end{align}
The quark number density $n_q$ is~\footnote{
Let the number operator for quark be $\hat{N}$.
By using the spatial volume $V$, 
\begin{align}
n_q & = \langle \hat{N} \rangle / V, \nn \\ 
   & = \frac{1}{Z}  {\rm tr} \left[ \frac{\hat{N}}{V} e^{ - ( \hat{H} - \mu \hat{N}) /T} \right].
\end{align}
Combining it with 
$Z={\rm tr} e^{ - (\hat{H} - \mu \hat{N})/T}$, we obtain this relation. 
} 
\begin{align}
n_q 
   & = \frac{1}{V} \frac{\partial}{\partial \mu/T} \log Z.  
\end{align}
By substituting \eqref{eq:latticeZ} to this, we obtain the quark number density on a lattice: 
\begin{align}
n_q & = \frac{1}{V} \left\langle N_f {\rm tr} \left[ \Delta^{-1} \frac{\partial }{\partial \mu/T} \Delta \right] \right\rangle. 
\label{eq:2017Jul03eq2}
\end{align}
$\Delta^{-1}$ and $\partial \Delta/\partial(\mu/T)$ are the propagator and the vertex function of the quark. Note that the propagator is a function of the link variables, and via the link variables, it depends on the gluons; this is not the propagator of the free quark, rather it is the full propagator which takes into account the exchange of gluons with the vacuum. In the continuum limit $a\to 0$, the vertex function becomes $\gamma_4$. This is the same form as in the continuum theory.

The expression \eqref{eq:2017Jul03eq2} is the quark number density in the actual physical unit. To determine the actual value, the value of the lattice spacing $a$ is needed. It is determined, for example by using the experimental value of the mass of pion as an input. In the finite density QCD, due to the large simulation cost, it is often difficult to study a sufficiently large lattice in which the pion mass can be determined. In such cases, dimensionless quantities such as $p/T^4$ and $n_q/T^3$ are used. The latter is determined by 
\begin{align}
\frac{n_q}{T^3} & = N_f \left(\frac{N_t}{N_s}\right)^3 \left\langle  {\rm Tr} \left[ \Delta^{-1}(\mu) \frac{\del\Delta(\mu)}{\del(\mu/T)} \right]\right\rangle. 
\label{eq:2017Mar29_ndens}
\end{align}

The quark number susceptibility $\chi_q = \partial n_q/\partial \mu$ is used for the determination of the phase transition at finite density. The dimensionless version $\frac{\chi_q}{T^2}$ is given by 
\begin{align}
\frac{\chi_q}{T^2} &= \frac{1}{VT^3} \left(\frac{ \partial}{\partial (\mu/T)}\right)^2 \log Z, \nn \\
                   &= \frac{1}{VT^3} \left\{ \frac{1}{Z} \left(\frac{ \partial}{\partial (\mu/T)}\right)^2  Z 
                      - \left( \frac{1}{Z}\frac{ \partial}{\partial (\mu/T)} Z\right)^2  \right\}.  
\label{eq:2017Mar29eq2}
\end{align}
Let us write down the lattice version of this quantity. For simplicity, we use the notation $\partial \Delta / \partial (\mu/T) = \Delta'$. Then, 
\begin{align}
\frac{\del^2 [\det \Delta(\mu)]^{N_f}}{\del(\mu/T)^2}  = &
\left\{  -  N_f {\rm Tr} \left[ \Delta^{-1} \Delta'  \Delta^{-1} \Delta' \right] 
+ N_f{\rm Tr} \left[ \Delta^{-1} \Delta'' \right] + N_f^2 \left( {\rm Tr} \left[ \Delta^{-1} \Delta'  \right] \right)^2 \right\} [\det \Delta(\mu)]^{N_f}. \nn
\end{align} 
By substituting this to \eqref{eq:2017Mar29eq2}, we obtain 
\begin{align}
\frac{\chi_q}{T^2}  = & \left(\frac{N_t}{N_s}\right)^3 \left\{
-  N_f \left\langle {\rm Tr} \left[ \Delta^{-1}  \Delta'  \Delta^{-1} \Delta' \right] 
\right \rangle
+ N_f \left\langle {\rm Tr} \left[ \Delta^{-1} \Delta'' \right] \right\rangle 
 + N_f^2 \left\langle \left( {\rm Tr} \left[ \Delta^{-1} \Delta' \right] \right)^2 \right\rangle
\right\}_. 
\end{align}
We can calculate the higher-order derivatives, but it is hard to determine them numerically because the number of terms increases quickly, 
and also because larger and larger statistics are needed. The higher-order derivatives are related to the fluctuations in the BES experiment.
We will discuss related topics  in Sec.~\ref{sec:Taylor} and Sec.~\ref{sec:pt_canonical}.

\subsection{Configuration generation via the importance sampling}
\label{sec:MonteCarlo}

We have seen the formal definition of lattice QCD. However, to calculate the values of the observables, we have to perform the path integral. 
\eqref{eq:latticeZ} and \eqref{eq:latticeO} are the integrals over (number of links)$\times (N_c^2-1)$ variables. Unless we perform this integral, the formal definition of lattice QCD does not tell us about actual physics. The analytic methods such as the perturbative expansion and strong coupling expansion can work only in a limited parameter region. The practical and quantitative method is the Monte Carlo simulation based on the importance sampling.

\subsubsection{Trapezoidal rule}

Let us first recall the trapezoidal rule for a one-dimensional integral $S=\int_a^b f(x) dx$. 
By representing the interval of integration $[a,b]$ with finite number of points 
$x_i\in[a,b], (i=1,2,\cdots N)$, the integral $S$ is approximated by 
\[
S = h \sum_{i=1}^N f(x_i), \, h= \frac{b-a}{N}. 
\] 
The number of points $N$ needed for the calculation depends on 
the required accuracy and the property of the integrand. Typically, in the equal-interval-division method such as the Simpson method and the Gaussian quadrature such as the Gauss-Legendre formula, $N=O(10)$ or $O(100)$ gives a sufficiently accurate estimate. The number of points needed in the integration methods of this kind increases very rapidly with the dimensions, e.g,~$N^2$ for two-dimensional integral, $N^3$ for three-dimensional integral, and $N^M$ for $M$-dimensional integral. 
In lattice QCD, $M$ is proportional to the number of lattice sites: $M\sim 10^4$ and $10^5$ when the number of sites along each dimension is 10 and 20, respectively. Needless to say, such calculation is practically impossible. 
\subsubsection{Importance sampling}

In the Monte Carlo method, the points $x_i$ are chosen randomly. While naive Monte Carlo methods are not particularly efficient compared to other methods such as the trapezoidal rule, the importance sampling, which is an improved version of the Monte Carlo simulation, is powerful. In the Simpson method and the Gaussian quadrature, the points $x_i$ are chosen from the entire integration region, regardless of the integrand. This is often called uniform sampling. On the other hand, in the importance sampling more (resp.~less) points are sampled in the region where $f(x)$ is large (resp.~small), namely the points which have larger contributions to the integral are sampled more frequently. 

To understand the key idea of the importance sampling, let us take the statistical physics of the system of many particles as an example. The microscopic state is specified by the location and velocity of the particles $\{x_1, v_1, x_2, v_2, \cdots, x_N, v_N\}$, or equivalently by specifying a point in the $2N$-dimensional phase space. Such a set of numbers, or equivalently a point in the phase space, is called the `configuration'. We use $\calC$ to denote a configuration: 
\[ \calC = \{x_1, v_1, x_2, v_2, \cdots, x_N, v_N\}. \] 
In the Euclidean path integral, the partition function of the system and the expectation value of the observable $O$ are given by 
\begin{align*}
Z & = \int \prod_{i=1}^N dx_i dv_i e^{-S}, \\
\langle O \rangle & = Z^{-1} \int \prod_{i=1}^N O dx_i dv_i e^{-S}.  
\end{align*}
By choosing $\Nconf$ points in the phase space $\{ \calC_1, \calC_2, \cdots, \calC_\Nconf\}$, the expectation value of $O$ can be approximated as 
\begin{align}
\langle O \rangle & = \frac{1}{\Nconf} \sum_{i=1}^\Nconf O( \calC_i). 
\end{align}
In many integrations such as the Simpson methods, the phase space is divided into a mesh, but among the configurations considered there are unrealistic ones, e.g.~a state in which many particles are localized in a tiny region, which makes the calculation inefficient. In thermal equilibrium, particles should uniformly spread in the space, and the velocities should be distributed around the mean value, and such configurations should reproduce the expectation value of the observables. In the Euclidean path integral, configurations with small values of $S$ have a large contribution because the integrand is $e^{-S}$ is large then. By sampling such states more frequently, the path integral can be performed efficiently. This is the importance sampling.

The link variables are the integration variables in lattice QCD. 
The configuration is specified by $\calC = \{U_l | l=1, 2, \cdots N_{\rm link} \}$, where $N_{\rm link}$ is the number of the link variables, and called the `gauge configuration'. For brevity, the gauge configuration is sometimes written as $\{ U \}$, or even more simply, $U$. In the analogy to the particles used above, each link variable corresponds to the location and momentum of a particle, and $\calC$ corresponds to the set of the locations and momenta of all particles. Physical quantities such as the plaquette and Polyakov loop are functions of the gauge configuration, $O=O(\calC)$. When $\Nconf$ gauge configurations $\{ \calC_1, \calC_2, \cdots, \calC_\Nconf\}$ are chosen from the phase space, the expectation value of the observable $O$ is approximated by 
\begin{align}
\langle O \rangle = \frac{1}{\Nconf}\sum_{i=1}^{\Nconf} O (\calC_i). 
\label{eq:2016Sep1eq1}
\end{align}
In the expression \eqref{eq:2016Sep1eq1}, the choice of $\calC$ has not been specified yet. To determine $\langle O \rangle$ we need to specify the way how the gauge configurations are picked up. 

In the importance sampling, configurations with larger $e^{-S}$ are picked up more frequently. This is achieved by the Metropolis step which compares the values of $e^{-S}$. Suppose there are two configurations $\calC_1$ and $\calC_2$. Then by comparing $S_1 = S(\calC_1)$ and $S_2 = S(\calC_2)$, smaller one should be chosen.

The algorithms for the generation of the gauge configuration have been studied intensively. As the standard method, the Hybrid Monte Carlo (HMC) algorithm, which is one of the Markov Chain Monte Carlo methods, is currently used widely.~\footnote{
See Chapter 16 of Ref.~\cite{Rothe:book} for the details of the HMC algorithm. }
The number of gauge configurations needed for the calculation of the observables depends on the parameters (quark mass, temperature, lattice spacing, lattice size, etc), the observable, and the precision required for specific purposes. In many lattice QCD simulations, $\Nconf$ is as small as 100. This is much smaller than the number of points required in the uniform sampling. One might wonder whether such a calculation is valid, but the fact that recent lattice QCD calculations reproduced low-lying hadron spectra rather precisely serves as a good sanity check. The phase transition from the hadron phase to the QGP phase has also been calculated with high accuracy, and it has been concluded that this transition is a cross-over around $T\sim 150$ MeV. Because $T_c$ has not yet been determined experimentally, the prediction from lattice QCD is used as an important input for the study of QCD at high temperatures.

\section{Sign problem}
\label{sec:signproblem}

\subsection{Importance sampling does not work for complex actions}
While the importance sampling is the standard approach to the Euclidean path integral, it does not apply to arbitrary theories. This is because of the Metropolis step which compares $e^{-S}$. Let the values of the action corresponding to two configuration $\calC_1, \calC_2$ be $S_1 = S( \calC_1), S_2 = S(\calC_2)$. If $S\in \mathbb{R}$, then one can compare
\[
e^{-S_1}, e^{-S_2}
\]
and decide which is bigger, but it is impossible if $S\in \mathbb{C}$. 
For example, when $S_2 = S_1 + i\pi$, we have  
\[
e^{-S_1} = a, e^{-S_2}  = -a,  (a \in\mathbb{R}).
\]
Then which of $\calC_1$ and $\calC_2$ contribute more to the path integral? Because they cancel with each other, one cannot ignore one of them. Furthermore, after they cancel with each other, the configurations with a smaller value of $|e^{-S}|$ may give the important contribution for the integral, and hence, it may not be appropriate to ignore configurations with small $|e^{-S}|$. Therefore, when the action is complex, the path integral cannot be approximated just by collecting the configurations with `small values of $S$', and hence, the importance sampling does not work. 
This is called the sign problem or complex-action problem. 

\subsection{Complex action in QCD}

The gauge part of the action $S_g$ defined by \eqref{eq:gauge_action} is real. On the other hand, the fermion determinant $\det \Delta$ is real positive when $\mu=0$ but it can be complex when $\mu\neq 0$. Let us see the details of how it happens.

In the Euclidean spacetime, the covariant derivative $D_\nu$ is anti-Hermitian ($(D_\nu)^\dagger = - D_\nu$) and the gamma matrices are Hermitian $\gamma_\nu^\dagger = \gamma_\nu$. Therefore, the Dirac operator $D \equiv D_\nu \gamma_\nu$ is anti-Hermitian, and its eigenvalues are pure imaginary. Because $D_\nu \gamma_\nu$ and $\gamma_5$ anti-commute with each other, the eigenvalues form pairs $\pm i x (x\in \mathbb{R})$~\footnote{
Let $\lambda$ and $\psi$ be an eigenvalue and eigenvector of $D$: 
$D \psi = \lambda \psi$. Due to the Hermiticity of $D$, $\lambda$ is pure imaginary. Then
\begin{align}
\gamma_5 D |\psi \rangle = \lambda (\gamma_5 | \psi \rangle).  
\end{align}
By using $\gamma_5 D = - D \gamma_5$, we obtain 
\begin{align}
\gamma_ 5 D | \psi \rangle = - D \gamma_5 | \psi \rangle. 
\end{align}
Therefore, 
$\gamma_5 |\psi \rangle \equiv |\phi\rangle$ satisfies 
\begin{align}
D | \phi \rangle = - \lambda | \phi \rangle. 
\end{align}
Hence $|\phi\rangle$ is an eigenvector of $D$ with the eigenvalue $-\lambda$. Therefore, the nonzero eigenvalues form pairs $\pm \lambda$. 
}.
The fermion determinant can be expressed by using the eigenvalues $\pm i x_n$ as 
\begin{align}
\det ( D + m) & = \prod_{n=1}^N ( i x_n + m) ( - i x_n + m), \nn \\
            & = \prod_{n=1}^N ( x_n^2 + m^2).
\end{align}
Therefore, the fermion determinant is real. The same proof applies to the lattice staggered fermion if the Dirac operator is anti-Hermitian. (See Appendix~\ref{sec:anti_hermite_KS}).

The proof given above does not apply to the Wilson fermion because 
the Wilson term $( D_\nu D_\nu)$ spoils the anti-Hermiticity. Still, the $\gamma_5$-Hermiticity holds, namely $\Delta = D_\nu \gamma_\nu + m + r' D_\nu D_\nu, (r'=r a) $ satisfies the following relation: 
\begin{align}
\gamma_5 \Delta^\dagger \gamma_5 & = \gamma_5 \left( D_\nu \gamma_\nu + m + r' D_\nu D_\nu\right)^\dagger \gamma_5 \nn \\
                                 & = \gamma_5 \left((D_\nu)^\dagger (\gamma_\nu)^\dagger + m + r' (D_\nu D_\nu)^\dagger \right)\gamma_5\nn \\
                                 & = \gamma_5( - D_\nu \gamma_\nu + m  + r' D_\nu D_\nu )\gamma_5 \nn \\
                                 & = D_\nu \gamma_\nu + m + r' D_\nu D_\nu =  \Delta. 
\label{eq:g5ahrelation1}
\end{align}
By using $\det AB = \det A \cdot \det B$, $(\det \Delta)^* = \det \Delta$ follows from \eqref{eq:g5ahrelation1}, and hence the fermion determinant is real. Next, let us show the positivity. For any complex number $\alpha$, the following relation holds: 
\begin{align}
\det ( \Delta - \alpha ) & = \det \gamma_5 ( \Delta - \alpha ) \gamma_5 , \nn \\
& = \det (\Delta^\dagger - \alpha) , \nn \\
& = \det (\Delta - \alpha^*)^\dagger , \nn \\
& = (\det (\Delta - \alpha^*))^* 
\label{eq:g5hermite}
\end{align}
If $\alpha=\lambda$ is equal to an eigenvalue of $\Delta$, the left-hand side is zero. Then the right-hand side must be zero as well, and hence, 
$\alpha^*= \lambda^*$ has to be the eigenvalue of $\Delta$ too.  
Hence the eigenvalues of $\Delta$ form pairs $\lambda, \lambda^*$. 
Therefore, $\det \Delta$ is real positive, and the effective action for the fermion part $S_f = - \log \det \Delta$ is real.

When $\mu \neq 0$, $D \equiv D_\nu \gamma_\nu + \mu \gamma_4$ is not anti-Hermitian. Furthermore, 
\begin{align}
\gamma_5 \Delta^\dagger(\mu) \gamma_5 & = \gamma_5 ( D_\nu \gamma_\nu + m + \mu \gamma_4)^\dagger \gamma_5, \nn \\
                                      & = \gamma_5 ( - D_\nu \gamma_\nu + m + \mu \gamma_4) \gamma_5, \nn \\
                                      & = D_\nu \gamma_\nu + m - \mu \gamma_4 , \nn \\
                                      & = \Delta(-\mu),  
\label{eq:g5ahrelation2}
\end{align}
and hence the $\gamma_5$-Hermiticity is also lost. The fermion determinant is not real in general, $\det \Delta (\mu) \in \mathbb{C}$
and $S_f = - \log \det \Delta$ is complex. The same holds for the Wilson fermion as well. Therefore, when the chemical potential is introduced, 
the configurations cannot be generated via the importance sampling.  
\subsection{
What does it mean to `solve the sign problem'?}
\label{sec:solve_signproblem}

How can we perform the multiple integral, when the importance sampling cannot be used? As of today, no numerical method applicable to generic path integral with the complex action is know. Therefore, in the study of finite-density lattice QCD, we need to find a solution to the sign problem. 
In the following sections, we will discuss various approaches. Before that, let us discuss a more basic conceptual issue: when we say `solve the sign problem', what exactly do we mean? Along the way, we will mention the notion of the computational cost, which is very important when we discuss the sign problem. By understanding this point clearly, the reader can read the following sections smoothly.

`To solve the sign problem' can be rephrased as `to perform the path integral for the theory with complex action'. The crucial requirement is that it has to be practically doable. For example, while elementary integration methods such as the Simpson method are formally correct, 
they cannot be applied to multiple integral with hundreds of thousands of variables, at least with currently available computational resources.  
Even if a given algorithm is correct in principle, it is not a `solution' unless it can be carried out and lead to a concrete number. In this sense, it is reasonable to define the problems as `to perform the path integral for the theory with complex action by using realistic computational cost'.

However, then what is `realistic computational cost'? The computational cost would mean the number of operations needed in the algorithm. It is not easy to give a sharp definition of `realistic' because it can depend on the spec of the computer, the number of people working on the calculation, or the time frame of the project. In computational science, whether the method is realistic is usually judged by looking at the scaling of the cost concerning the typical scale of the problem, such as the system size. 
If the cost increases exponentially ($\sim e^N$) with respect to the number of variables $N$, even with very powerful computers it is practically impossible to study a large system size. Even when the scaling is power ($\sim N^x$), if the power $x$ is large, the simulation is practically hard.

Even when the action is complex, from the physical requirement the partition function has to be positive. Necessary simulation costs can depend on the parameters. Simple toy examples to understand the sign problem include $\int dx e^{i n x}$ and $\int dx \cos n x$. Although a good accuracy can be achieved with a small number of points when the oscillation is mild ($n\sim 1$), however, more and more points are needed when the fluctuation becomes more rapid ($n \gg 1$). If the integrand has only one peak (e.g.,~the Gaussian distribution), one only has to pick up the points around the peak. For the oscillating functions, however, there are multiple peaks, and one has to sample the points around each of them, and hence, more and more points are needed as the number of peaks becomes larger. In the path integral, the oscillation comes from the imaginary part of the action. Because the action is the integral of the Lagrangian over the spacetime volume, the size of the action is proportional to the volume of the system: $S \propto V$. Therefore, the imaginary part of the action is also proportional to the volume: ${\rm Im} S \propto V$.
It is proportional to the chemical potential as well because the origin of the phase is the density term $S \propto \mu \int d^4x \bar{\psi} \gamma_4 \psi$, and by using the average density $\bar{n}$
it can be written as $S \propto \mu V \bar{n}$. In this way, the imaginary part of the action is proportional to the chemical potential and volume, as $\mu$ and $V$ increase, the oscillation of the integrand becomes more violent, and the simulation cost increases. The fact that the simulation cost depends on the parameter means the level of difficulty depends on the parameter. In the study of finite density QCD, many methods effective at small $\mu$ and $V$ become less reliable at large $\mu$ or $V$ due to the rapid increase of the computational cost. Often it is not easy to estimate the necessary simulation cost, and hence, it is not easy to judge the validity of the simulation results, which makes the sign problem even more challenging. 

\subsection{Phase fluctuation in QCD and the difficulty of the sign problem}
\label{sec:phaseflucuation_nQCD}

Let us see how the phase of the fermion determinant depends on temperature and density. For that purpose, the calculation at $\mu \neq 0$ is needed, and one has to  handle the sign problem. Here we use the reweighting method; we only show the results, postponing the details of the method to the next section.

We separate the fermion determinant to the absolute value and the phase as 
\begin{align}
(\det \Delta (\mu ))^{N_f} = | \det \Delta(\mu) |^{N_f} e^{ i \theta },  
\end{align}
where $N_f$ is the number of flavors. 
The argument $\theta$ corresponds to the imaginary part of the action, ${\rm Im} S$. 
Because the expectation value of $\theta$ is zero,\footnote{
Because the partition function $Z$ is real and positive, 
\[
 \int \calD U | \det \Delta(\mu) |^{N_f} \sin \theta \, e^{-S_g}   = 0  
\]
holds. 
$Z$ is the even function of $\theta$, and the expectation value of $\theta$ is zero. 
}
to describe the behavior of $\theta$ the variance is often considered. 
In general, the variance of $X$ is given by $\langle (X - \langle X \rangle)^2\rangle$, and by taking into account $\langle \theta \rangle = 0$ one obtains
\begin{align}
\langle (\theta - \langle \theta \rangle )^2 \rangle = \langle \theta^2 \rangle. 
\end{align}
The average phase factor $\langle e^{i \theta} \rangle $ is commonly used as well. By using the Euler's formula, it is written as 
\begin{align}
\langle e^{i \theta} \rangle = \left\langle \left( 1- \frac{1}{2}\theta^2 + \cdots \right) \right\rangle.  
\end{align}
Note that the odd powers of $\theta$ vanish. When the phase fluctuation is violent, the average phase factor approaches zero.

\iffigure
\begin{figure}[htbp] 
\centering
\includegraphics[width=9cm]{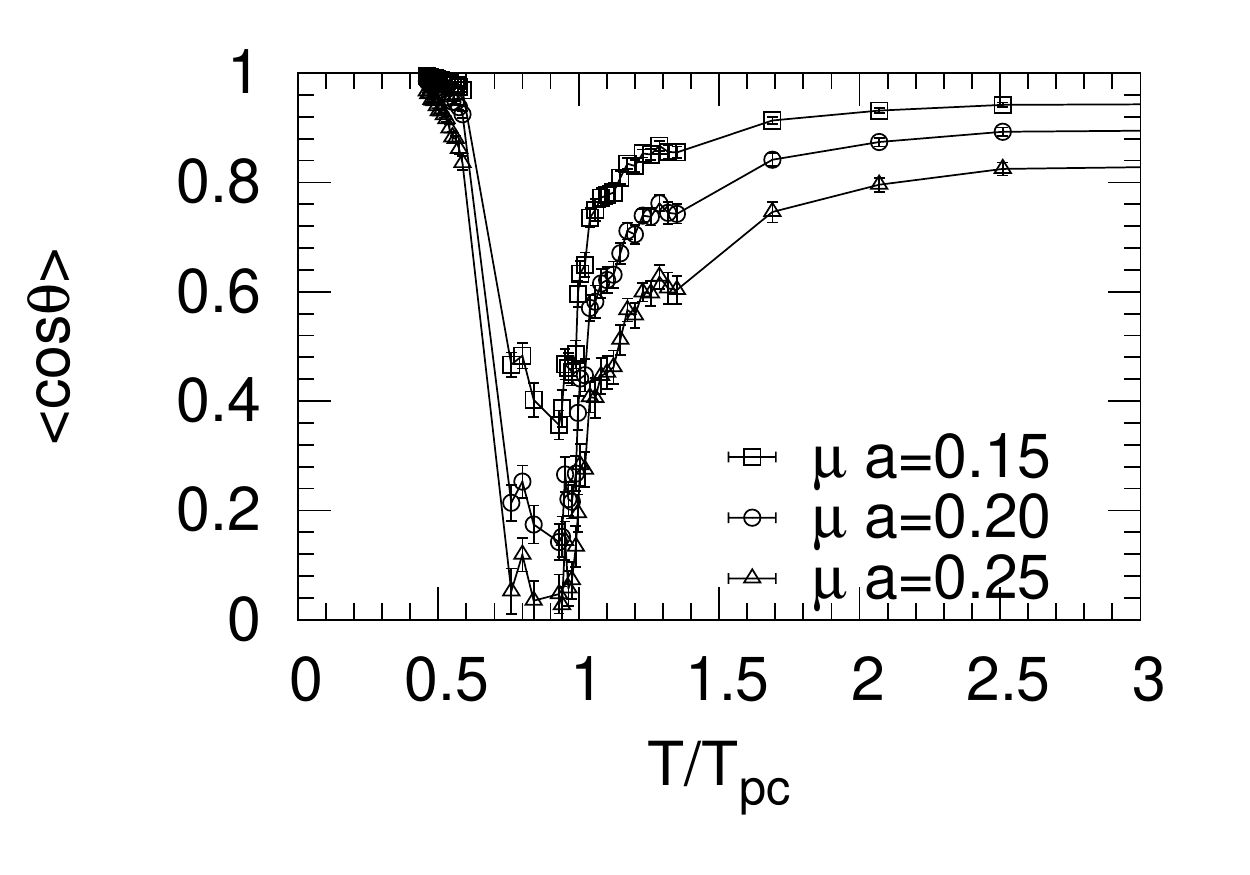}
\begin{minipage}{0.9\linewidth}
\caption{\small 
The $T$- and $\mu$-dependence of the average phase factor $\langle e^{i\theta}\rangle_0=\langle\cos\theta\rangle_0$. When the value of $\langle\cos\theta\rangle_0$ is smaller, the sign problem is severer. Near $T_{pc}$ and at large $\mu$, the average phase factor becomes small. This result is for the two-flavor QCD at lattice volume $8^3\times 4$. The average phase factor becomes smaller also as the volume increases. 
}
\label{fig:averagephase}
\end{minipage}
\end{figure}
\fi

In Fig.~\ref{fig:averagephase}, the average phase factor $\langle e^{i \theta} \rangle_0$ is shown as a function of temperature. The lower index $0$ means that the reweighting method, which will be introduced in the next section, is used. $T_{\rm pc}$ is the transition temperature between the hadron phase and the QGP phase at $\mu=0$.  

When $\mu$ is fixed and $T$ is varied, the average phase factor is close to 1 at low temperature, decreases quickly as $T$ approaches $T_c$, and it increases again at $T>T_c$. In other words, the variance of $\theta$ is small at low and high-temperature regions, and large near $T_c$. We can see that the sign problem is severe slightly below $T_c$ $(T= (0.7\sim 1)T_c$. At low temperature $(T< 0.5 T_c)$, the phase fluctuation is small, and the $\mu$-dependence is almost negligible. This behavior comes from an important property of the Fermi distribution: at low temperature, until $\mu$ exceeds a threshold value the $\mu$-dependence does not set in. The values of $\mu$ in Figure~\ref{fig:averagephase} are below the threshold. When $\mu$ is increased further, beyond a certain value $\langle e^{i \theta} \rangle_0$ becomes small very quickly. The challenge in the low-temperature region is explained in later sections.

When $T$ is fixed and $\mu$ is varied, $\langle \cos \theta \rangle_0$ decreases and the variance of $\theta$ increases as $\mu$ increases. The same tendency has been found in the analyses based on the chiral perturbation theory~\cite{Splittorff:2007zh}. This is because the imaginary part of the action is proportional to $\mu$. While the volume dependence of the average phase factor is not shown in Figure~\ref{fig:averagephase}, it would be easy to imagine that the average phase factor quickly approaches zero as the volume is increased because $\theta$ is proportional to the volume as well.

In summary, the level of difficulty of the sign problem in finite-density QCD depends on the parameters such as temperature $T$, the chemical potential $\mu$, and the volume $V$. The phase fluctuates more violently as $\mu$ and $V$ becomes larger. When the temperature is varied, a large phase fluctuation is seen slightly below $T_c$. In the low-temperature regime, the sign problem suddenly sets in beyond a threshold value of the chemical potential. 

\bibliographystyle{utphys}
\bibliography{ref_list}

\newpage
\chapter{Solution to sign problem 1: Methods based on the importance sampling}
\label{sec:oldmethod}

There are two types of strategies to deal with the sign problem. One is to develop an algorithm for the configuration generation which can be applied to a complex-valued action as well. If such an algorithm could be found, then it would be a solution to the sign problem, but it is not easy.
The other direction is to generate the configurations by using a sign-problem-free action and use those configurations to calculate the physical quantities in the original system with the sign problem. It is a kind of approximation method which does not directly solve the sign problem, so it is often called the method for circumventing the sign problem. In the latter case, various methods and technical improvements have been proposed. In this chapter, as the methods for circumventing the sign problem, we introduce the reweighting method, the Taylor expansion method, the imaginary chemical potential method, and the canonical approach. For each method, we explain basic ideas, application examples, and the advantages. We also mention the limitations common to those methods at the end of this chapter.

\section{Reweighting method}
\label{sec:reweight}

In the importance sampling method, the configurations should be generated at each point in the parameter space of the theory (e.~g., temperature, mass, chemical potential), because the properties of the integrand $e^{-S}$ such as the location and width of the peak can change depending on the parameters and the configurations contributing to the path integral change as well. However, if the variation of the parameters is sufficiently small, then the variation of the action should also be small, and more or less the same configurations contribute to the path integral. For instance, if we consider $e^{-S}=e^{-(x-\alpha)^2}$, as long as $\alpha$ is small the configurations near $x=0$ should give us a good approximation.

Ferrenberg and Swendsen proposed the reweighting method, in which we can obtain a physical quantity with a set of lattice parameters by using the configurations generated with a different set of parameters~\cite{Ferrenberg:1988yz}. This method enables us to obtain the information of several sets of parameters from the configurations generated at one point of the parameter space. Originally, the reweighting method was introduced to reduce the cost of configuration generation. It applies to various situations, whether the integral suffers from the sign problem or not. The basic idea of the reweighting method is a shift of parameters that can be applied to avoid the sign problem. The reweighting method is a key idea to all methods explained in this chapter: the lattice QCD action at finite density is modified such that the importance sampling method can be applied to the modified action, and then the difference from the original theory is corrected by the reweighting method. The crucial issue is when the reweighting method can be practically useful. It cannot be applied when the fluctuation of the phase is large, so it is useful when the phase fluctuation is mild. We will explain this point in detail because it is important when we try to understand various other methods for avoiding the sign problem based on the Monte Carlo method.

Below, we use an example without sign problem to explain the basic idea and some important remarks regarding the actual calculations. Then, we will see the application of the reweighting method to the sign problem. We will also introduce the multi-parameter reweighting method, which is an improved version of the reweighting method. In the last part of this section, we will investigate the cause of the increase in the simulation cost in the finite-density QCD.

\subsection{Basic idea of the reweighting method}

Here, we introduce the reweighting method for the partition function of the quantum statistical system. The same argument goes through for the path integral formulation of QCD. Let us consider the quantum system described by the Hamiltonian $\hat{H}$. The canonical partition function and the expectation value of physical quantity $O$ at temperature $T$ are given by
\begin{align}
Z(T) & = {\rm tr} e^{- \hat{H}/T}, \\
\langle O \rangle(T) & = Z(T)^{-1} {\rm tr} \left[ O e^{- \hat{H}/T} \right].
\end{align}
For simplicity, we assume that $\hat{H}$ does not explicitly depend on the temperature. The expectation value of $O$ at $T$ can be expressed by 
\begin{align}
\langle O \rangle(T) & = \frac{1}{\Nconf}\sum_{i=1}^{\Nconf} O(\calC_i),
\label{eq:2017Jun28eq2}
\end{align}
if we generate $\Nconf$ configurations $\{ \calC_i | i= 1, 2, \cdots \Nconf \}$ via the importance sampling. Now, we transform $Z(T)$ as follows
\begin{align}
Z(T) &= {\rm tr} e^{ - \hat{H} (1/T - 1/T_0) - \hat{H}/T_0}, \nn \\
  &= {\rm tr} \biggl[ R \,  e^{ - \hat{H}/T_0} \biggr], \,\,\,R\,  \equiv e^{ - \hat{H} (1/T - 1/T_0)}.
\label{eq:reweight1}
\end{align}
This is just an identity, because $e^{ - \hat{H}/T_0} \times R=e^{ - \hat{H}/T}$. We interpret $R$ as an observable, then obtain 
\begin{align}
Z(T)   &= Z(T_0) \times \langle R \rangle_0. 
\label{eq:20190809eq1}
\end{align}

Here, $\langle \cdot \rangle_0$ stands for the average over the configuration generated at the temperature $T_0$,
\begin{align}
\langle O  \rangle_0 = \frac{ {\rm tr} \, O \, e^{- \hat{H}/T_0 }}{{\rm tr} \, e^{- \hat{H}/T_0}}.
\end{align}
$Z(T)$ on the left hand side in (\ref{eq:20190809eq1}) is the partition function at the temperature $T$. On the right hand side  $Z(T_0)$ and $\langle R \rangle_0$ are calculated by using the configuration generated at $T_0$.
Thus, in (\ref{eq:20190809eq1}), a quantity at $T$ ($Z(T)$) is expressed in terms of the quantities at $T_0$ ($Z(T_0)$ and $\langle R \rangle_0$).
By using the configuration $\{ C'_i | i=1, 2, \cdots, \Nconf' \}$ generated at $T_0$, 
\begin{align}
Z(T) & = Z(T_0) \times \frac{1}{\Nconf'} \sum_{i=1}^{\Nconf'} R(C_i').
\end{align}
Similarly, the expectation value of physical observable $O$ at temperature $T$ can be expressed by using the quantities evaluated at temperature $T_0$: 
\begin{align}
\langle \hat{O} \rangle(T) &= \frac{{\rm tr} \, \hat{O} \, e^{ - \hat{H}/T}}{Z(T)} , \nn  \\
  &= \frac{{\rm tr} \left[ O \, R \, e^{ - \hat{H}/T_0}\right] }{{\rm tr} \left[ R \, e^{ - \hat{H}/T_0} \right] }.
\label{eq:reweight2}
\end{align}
This is also just a identity, and using the configuration generated at $T_0$, (\ref{eq:reweight2}) is expressed as
\begin{align}
\langle \hat{O} \rangle(T)   &= \frac{ \langle O R \rangle_0  }{ \langle R  \rangle_0 }, \nn \\
 & = \frac{\sum_i  O(C_i') R(C_i')}{\sum_i  R(C_i')}.
\label{eq:reweight3}
\end{align}
Equation~(\ref{eq:reweight3}) gives the formula to evaluate the physical observables at temperature $T$ by using the configurations generated at a different temperature $T_0$. In the ideal gas simulation, for instance, the pressure at $310$K can be calculated by the configuration generated at $300$K. Although the shifted parameter above the example is the temperature, we can shift any parameter of the system, e.~g., an external magnetic field in the Ising model or a quark mass in lattice QCD. The same method can be applied to discrete parameters as well.

This method is called the reweighting method. More generally, similar methods which use the configurations generated at different parameter are also called the reweighting methods. The reason for this name is obvious from the meaning of the equations. By using the `reweighting factor'
\begin{align}
r(\calC_i') = \frac{ R(\calC_i')}{ \ds{\sum_{i=1}^{\Nconf'} R(\calC_i')}}, 
\label{eq:20190810eq1}
\end{align}
\eqref{eq:reweight3} can be written as 
\begin{align}
\langle \hat{O} \rangle = \sum_{i=1}^{\Nconf'} O(\calC_i') r(\calC_i').
\label{eq:20190830eq1}
\end{align}
Thus, $\langle \hat{O} \rangle$ is obtained by averaging the observable multiplied with the weight factor $r(\calC_i')$ in \eqref{eq:reweight3}.
The ensemble average is defined by using the configurations at $T$ in the original definition \eqref{eq:2017Jun28eq2}, while the ones in \eqref{eq:reweight3} are given by using the configurations at $T_0$.
In \eqref{eq:20190830eq1}, the weight factor $r$ corrects the difference between two types of configurations, hence the physical observable at $T$ can be calculated by using the configurations at $T_0$.

The reason why we can calculate the physical quantities by using the configuration at different temperature is that $e^{- \hat{H}/T}$ contains all the information of the temperature dependence. Roughly speaking, we are only saying that the extrapolation is doable if the form of the function is known. In this way, the idea of the reweighting method is simple.

\subsection{Application of reweighting method and overlap problem}
\label{sec:overlap_problem}

\subsubsection{Application of reweighting method to QCD}
\iffigure
\begin{figure}[htbp] 
\centering
\includegraphics[width=7cm]{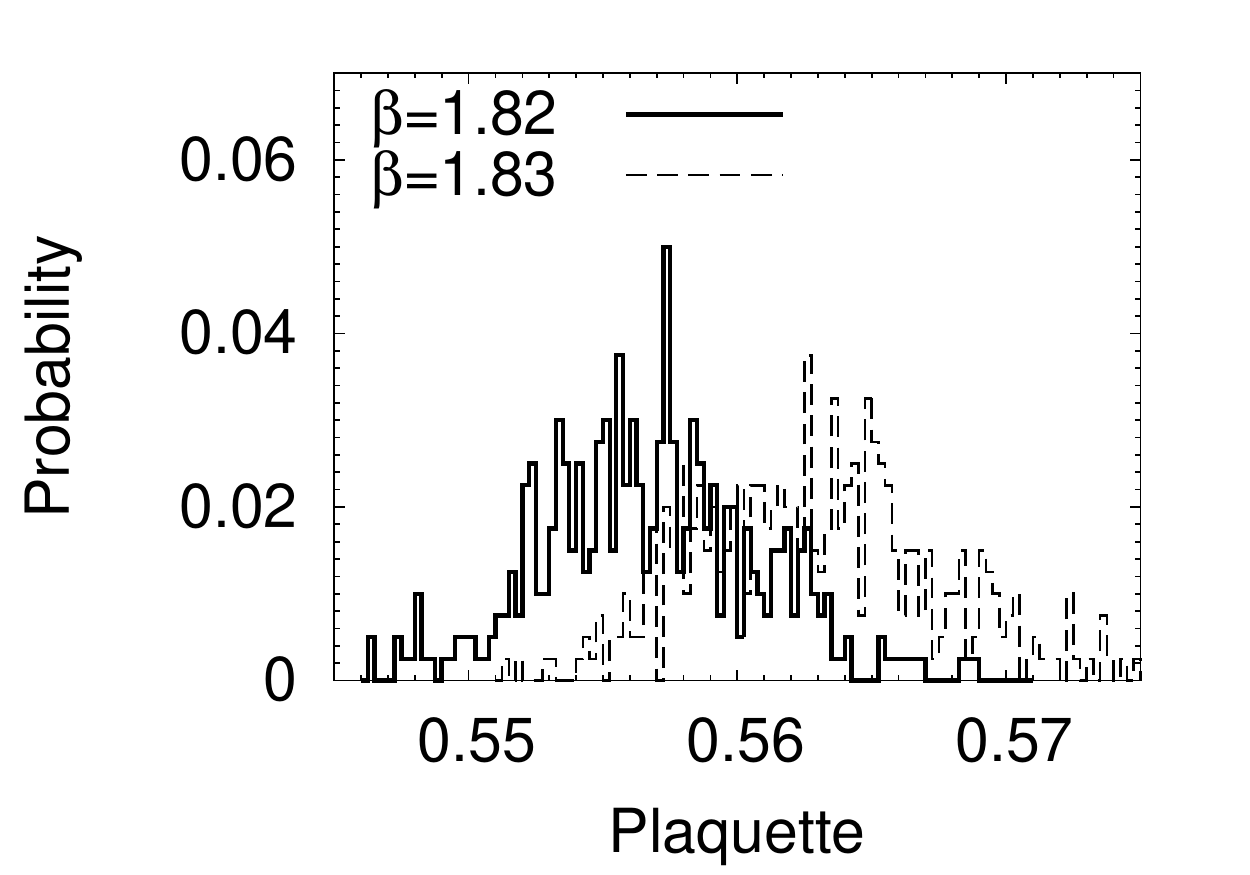}
\includegraphics[width=7cm]{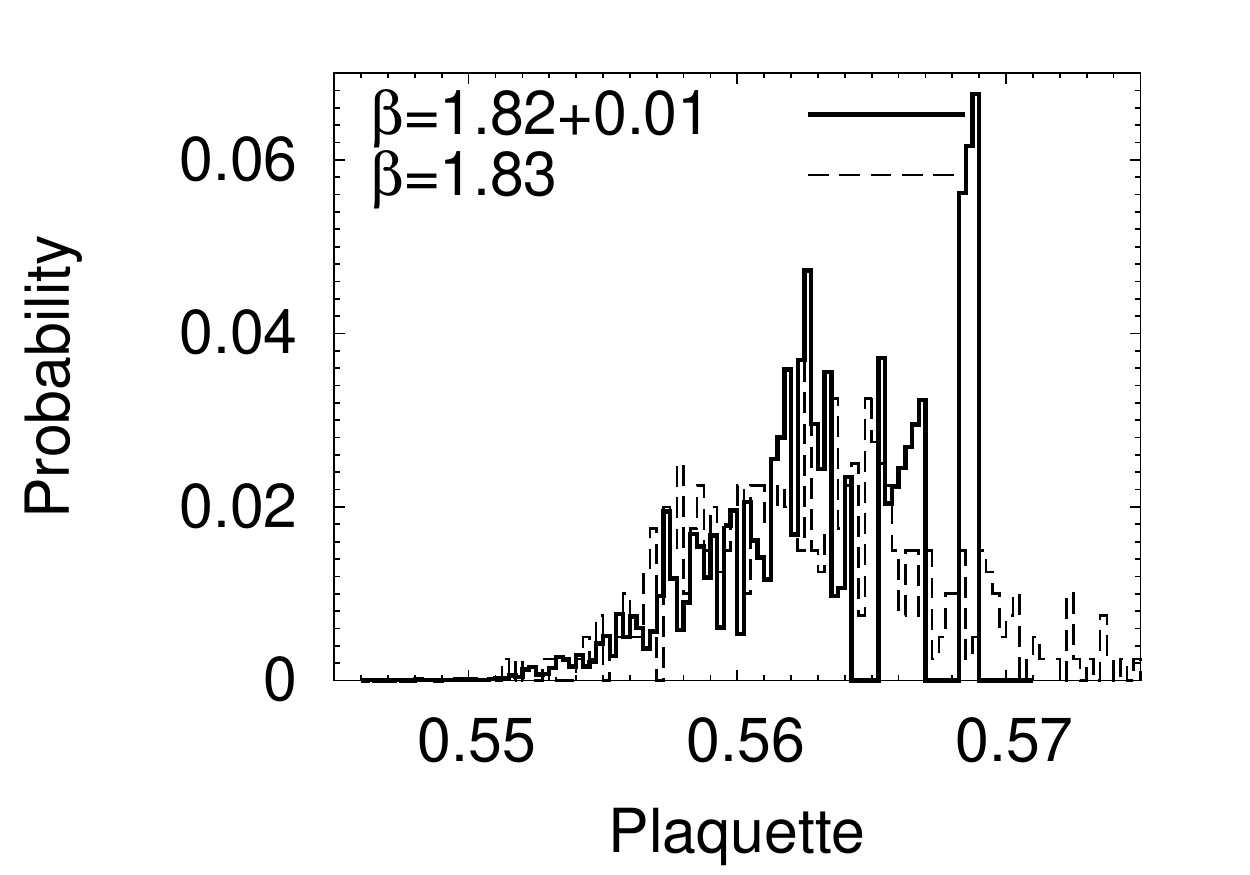}
\begin{minipage}{0.9\linewidth}
\caption{\small 
Application of the reweighting method. [Left] The distribution of the values of the plaquette, generated at $\beta=1.82$ and $\beta=1.83$. 
[Right] Comparison of the plaquette distributions at $\beta = 1.83$, obtained by applying the reweighting method to the configuration generated at $\beta = 1.82 $ (solid line) and the plaquette distribution obtained directly at $\beta = 1.83$ (dashed line). If the reweighting works properly, these two distributions should be consistent with each other, but there is a discrepancy in the large plaquette regime. This discrepancy of the distributions comes from the overlap problem. This numerical simulation is carried using the Iwasaki gauge and the clover-improved Wilson fermion on $8^3 \times 4$. The overlap problem becomes severer as the volume increases.
}
\label{fig:reweight}
\end{minipage}
\end{figure}
\fi

Although the reweighting method can be applied to various theories,
it is also known to have a side effect called the overlap problem.
In this section, to understand the characteristics features of the reweighting method and the overlap problem, we apply the reweighting method to the lattice coupling constant $ \beta $ of QCD when there is no sign problem.

Here, it is convenient to define $S_g = \beta S_G$, since we use the reweighting method to the parameter $\beta$. The $\beta$-independent part $ S_G $ corresponds to the gauge action $ S_g $ with the lattice coupling constant set to 1. The partition function can be rewritten as 
\begin{align}
Z(\beta) & = \int {\cal D}U (\det \Delta)^{N_f} e^{- \beta S_G}, \nn \\
         & = \int {\cal D}U  e^{ - (\beta-\beta_0) S_G} (\det \Delta)^{N_f} e^{- \beta_0 S_G}. \nn
\end{align}
 $(\det \Delta)^{N_f} e^{- \beta_0 S_G}$ and $R=e^{ -(\beta-\beta_0) S_G}$ can be interpreted as the weight of the importance sampling and a part of the physical observable, respectively. Then, we obtain 
 \begin{align}
Z(\beta)  & = Z(\beta_0) \langle e^{ - (\beta-\beta_0) S_G} \rangle_0.
\end{align}

Although the reweighting method is formally an identity, the overlap of configurations is important for actual simulations. The configurations generated by the simulation cover the region dominating the integral. The regions covered by the sets of configurations generated at $\beta$ and $\beta_0$ are different. The ``overlap" means the overlap between these sets of configurations. Let us estimate the overlap by using actual simulation results. We generated the $200$ configurations at $\beta_0 = 1.82$ and $\beta=1.83$. The values of these $\beta$ correspond to temperatures slightly lower than $ T_c $, namely, in the hadron phase.

We denote the gauge configurations generated at $\beta_0=1.83$ and $\beta=1.82$ as $ \{C_i \} $ and $ \{C'_i \} $, respectively. In principle, the overlap between the sets of configurations can be seen by plotting the region covered by the two types of configurations in the phase space.
However, in practice, it is difficult to see it visually, since the dimension of the phase space is high. As a method of observing the similarity of configurations, here we compare the distribution of the physical quantities which strongly correlate with the parameter to which the reweighting is applied. In the case of reweighting of $ \beta $, the plaquette is one of such physical quantities. By using \eqref{eq:2017Jun07eq2}, we defined a quantity $P$ as
\[P = \frac {1} {N_c N_ {\rm plaq}} \sum_ {n, \mu, \nu} {\rm tr} \, U_ {\mu \nu} (n). \] 
Here, $ N_ {\rm plaq} $ denotes the number of plaquette. By using \eqref{eq:20190830eq1}, the expectation value calculated via the reweighting method is expressed as 
\[
\langle P \rangle_0 = \sum_i P(C'_i) r(C'_i).
\]
In the left panel of Fig.~\ref{fig:reweight}, the histograms of $P$ at $ \beta = 1.82 $ and $ 1.83 $ are shown. The larger $ \beta $ corresponds to the higher temperature (see table~\ref{tab:agbetaT}). The value of the plaquette corresponds to the energy density so that the right side of the horizontal axis corresponds to the higher energy states. It is physically natural that the distribution shifts toward the high energy side as $ \beta $ increases. The histogram is not smooth because the number of samples is small. If the number of samples is increased, the distribution approaches a smooth bell shape. Here, to emphasize the important remarks regarding the reweighting method, the number of configurations is intentionally taken to be small. Note that the number of gauge configurations here, namely $200$, is a standard number of samples in lattice QCD research, although it depends on the physical quantity of interest.

Let $ w (P; \beta) $ be the probability distribution of $P$. The solid line in the right panel of Fig.~\ref{fig:reweight}  is the probability distribution at $ \beta_0 = 1.82 $, $ w_0 = w(P; 1.82) $, multiplied by the reweighting factor $ e ^ {-(\ beta- \ beta_0) S_G} $, namely 
\[ w (P r; 1.82).\] 
When the reweighting method functions properly, this distribution should be consistent with the distribution $w (P; 1.83)$ obtained directly at $\beta=1.83$. At $0.55 \lesssim P\lesssim 0.56$, these two distributions show a good agreement. Thus, by applying the reweighting method to the plaquette distribution generated at $\beta_0$, we can obtain the plaquette distribution at $\beta$. On the other hand, as the value of the plaquette increases, the distributions deviate. In the region of $P>0.57$, the probability distribution of the solid line is zero, and the two distributions do not match at all. Even with reweighting, which is formally a mathematical identity, sometimes the correct result is not being reproduced. This problem is caused by the mismatch of the distributions, namely the overlap problem.

It is important to note that the overlap problem leads to systematic errors.~\footnote{
There are two types of errors; systematic error (bias) and statistical error (variation). It is necessary to control both errors to measure physical quantities correctly. The statistical error is the variation that occurs accidentally. For example, when measuring the length of a rod of $1$ meter, numbers such as $1.005$ meters or $0.995$ meters can be obtained for each measurement. If the measurement method is correct, the measured values will fluctuate around the true value, $1$ meter, and the average value will approach the true value as more and more data are accumulated. On the other hand, if the measurement is not performed correctly due to abnormalities in the measuring equipment or an incorrect/inaccurate measurement method, the measured values fluctuate around a wrong value, and the true value will not be obtained even if the data are accumulated. Such error is called a bias. In statistics, these are also called precision and accuracy. An accurate result means that there is no bias in the measured values, and a precise result means that the number of measurements is sufficient. If the measurement is not accurate, the true value cannot be obtained even if the precision is improved.}. In the right panel of Fig.~\ref{fig:reweight}, the distribution generated by the reweighting method (solid line) does not contain the large-$P$ region, and hence, the expectation value of the plaquette is underestimated compared to the actual value. On the other hand, if $ \beta $ is shifted to a smaller value, the left tail of the actual distribution will be truncated, then the expectation value of the plaquette is overestimated. In general, the overlap problem has the effect of cutting off one of the left or right tails, resulting in a systematic error.

\subsubsection{Causes of overlap problems}

The overlap problem takes place because the probability distribution $w_0$ obtained by simulation is different from the actual probability distribution. Let $\tilde {w} _0$ be the actual probability distribution of the gauge configuration at $\beta_0$. The distribution $ w_0 $ obtained by the simulation is approximately the same with $\tilde {w} _0$. However the configurations with very small values of $ \tilde {w} _0 $ do not appear in the finite-time simulation, and then $\tilde{w}_0$ is approximated to be zero. The point of reweighting is to reproduce the probability distribution of the configurations by multiplying the reweighting factor \eqref{eq:20190830eq1}; but if the probability distribution is approximated by zero the correct probability cannot be extracted by multiplying the reweighting factor. The approximate distribution $ w_0 $ does not precisely describe the low-probability configurations at the tail of the distribution, therefore the distribution obtained by the reweighting also deviates from the actual distribution. 
The importance sampling is a method that focuses on the configurations giving a large contribution to the integral, but in other words, a method of discarding configurations that do not contribute to the integral value.
The tail regions of the distribution have a small contribution to the integral, and it is unavoidable that these regions are cut off by the importance sampling. In this sense, the overlap problem is caused by the characteristic feature of the importance sampling method.

The precision of $ w_0 $ is roughly determined by the importance of the configurations and the number of samples (number of configurations obtained in the simulation). For example, let us throw ten dices simultaneously. The probability that all ten dices return $1$ is $(1/6)^{10}$. If the number of trials is small, say $10$ trials, such a rare event may not occur, but it will be observed if the number of trials is large enough.

Let us make a rough estimate for lattice QCD. Since the plaquette represents the energy per unit volume, the action $ S $ is approximated as (plaquette) $ \times $ (volume of the system). (Strictly speaking, there are contributions from the number of states and fermions, but here we consider only the gauge action because we are interested only in a qualitative overview of the overlap problem.) The appearance probability of a configuration with plaquette $ P $ is given by $e^{-\beta_0 S_G} \sim e^{-\beta_0 V P}$. In the above calculation, the lattice volume is $ 8 ^ 3 \times4 $. At $\beta_0=1.82$, the ratio of the appearance probability at $P=0.555$, which is around the peak of distributions,  and $P=0.57$, which is a tail,  is
\[ \frac {e^{-0.57 \beta_0 V}} {e^{-0.555 \beta_0 V}} = e^{-(0.57-0.555) \beta_0 V} \sim e^{-56}. \]
Therefore, it is negligibly small compared to the number of configurations, $\Nconf=200$.

\subsubsection{Remarks on the overlap problem and phase transition}

When we use the reweighting method, we should carefully check if the results are affected by the overlap problem. If the simulation at the target point is doable as in the above example, the effect of the overlap problem can be studied by comparing the histograms. For the comparison, t is better to choose the physical quantities which are strongly correlated to the parameters to be reweighted, e.g., the plaquette for the $\beta$-reweighting and the quark number density for the $\ mu$-reweighting. If the simulation at the target point is not doable (which is usually the case when the reweighting method is used to circumvent the sign problem), the failure of the reweighting method can be detected to some extent by looking at the histogram of the reweighting factors.
If there is a peak around the edge of the $R$ histogram, a small number of gauge configurations near the peak dominate the integral. In such cases, the reweighting method is likely to fail. Another simple test is to confirm the stability of the expectation values when the number of configurations is varied. Since the overlap problem is often accompanied by systematic errors, the expectation value changes when the number of samples is increased.

The overlap problem can create a fake signal similar to a phase transition, even if no phase transition exists. Therefore, we should pay extra attention to the discussion of phase transitions. In the right panel of Fig.~\ref{fig:reweight}, there is a peak around $ P \sim 0.57 $.
This comes from the truncation of the right side of the distribution, which does not exist in the actual distribution. As the difference between $\ beta $ and $\beta_0$ increases, this peak becomes sharper and the configurations near the peak become dominant in the reweighted calculations. Then, the effective number of statistics is decreasing since the configuration away from the peak does not contribute to the expectation value. The decrease of the effective statistics makes the error of the physical quantity large because the error $e$ and the statistics $n$ have a relationship of $e \sim 1 / \sqrt{n}$. This increase in error is sometimes misidentified with the increase of the fluctuation near the phase transition point. Note also that the reweighted distribution in Fig.~\ref{fig:reweight} (solid line) has an asymmetric shape because one of the two sides of the distribution is truncated. These sharp peaks and asymmetries of the distribution are unphysical features caused by the overlap problem, which are absent in the actual distribution (dashed line). If such asymmetric distribution is obtained by the reweighting method, it is necessary to check the overlap problem.

\subsection{Avoiding of the sign problem using the reweighting method}

\subsubsection{Formulation of the chemical potential reweighting}

As explained in \S.~\ref{sec:signproblem}, when the chemical potential is introduced the action takes complex values and the importance sampling cannot be used. However, this problem can be avoided by using the reweighting method.

We separate the integrand into $\mu$-dependent and $\mu$-independent parts, 
\begin{align}
\langle O \rangle(\mu)  & = \frac{ \bigintss \calD U \, O \, (\det \Delta(\mu))^{N_f} e^{-S_g} }{Z}, \nn \\
 & = \frac{ \bigintss \calD U \, O \, \left(\ds{\frac{\det \Delta(\mu)}{\det \Delta(0)}}\right)^{N_f} (\det \Delta(0))^{N_f} e^{-S_g}  } 
{\bigintss \calD U \, \left(\ds{\frac{\det \Delta(\mu)}{\det \Delta(0)}}\right)^{N_f} (\det \Delta(0))^{N_f} e^{-S_g}}.  \label{eq:reweightingQCD1}
\end{align}
Since $(\det \Delta (0))^{N_f} e^{-S_g}$ is real, we interpret this part as the weight of the importance sampling. The remaining factor
\begin{align}
R = \bigg(\frac{\det \Delta(\mu)}{\det \Delta(0)}\biggr)^{N_f}
\label{eq:reweightfact1}
\end{align}
is regarded as a part of the measured quantities. Thus, Eq.~\eqref{eq:reweightingQCD1} is expressed as
\begin{align}
\langle O \rangle(\mu)  = \frac{\langle R O \rangle_0}{\langle R \rangle_0}.
\label{eq:reweightingQCD2}
\end{align}
Here $\langle \cdot \rangle_0$ denotes the expectation value using the configuration generated at $\mu = 0$. If $ \Nconf $ gauge configurations $\{ \calC_1, \calC_2, \cdots, \calC_\Nconf \}$ are generated by the simulation at $ \mu = 0 $, then
\begin {equation}
\langle O \rangle (\mu) = \frac {\displaystyle {\sum_ {i = 1} ^ {\Nconf} R (\mu, \calC_i) O (\mu, \calC_i)}} {\displaystyle {\sum_ {i = 1} ^ { \Nconf} R (\mu, \calC_i)}}.
\label{eq:reweightingQCD3}
\end {equation}
Thus, we can obtain the expectation value of the physical quantity at $\mu\neq 0$ by using the gauge configuration generated at $\mu=0$.

\subsubsection{Important remarks on the reweighting method applied to the sign problem}

The overlap problem explained in Sec.~\ref{sec:overlap_problem} also occurs in \eqref{eq:reweightingQCD3}. When the reweighting method is applied to the sign problem, it is necessary to pay attention to the phase fluctuation of the reweighting factor.

We regard the fermion determinant $\det \Delta (\mu)$ as a part of the reweighting factor $ R $ and interpret it as part of the physical quantity since $\det \Delta (\mu)$ takes complex values and the importance sampling cannot be applied directly. Therefore, the reweighting factor $R$ is also complex. If the complex phase of $R$ fluctuates significantly, $\langle R \rangle_0$ can be very close to zero. The partition function $Z(\mu)$ is expressed by
\begin{align}
Z(\mu ) \propto \langle R (\mu) \rangle_0 
\label{eq:2017Jul18eq1}
\end{align}
by using the reweighting method, so $Z(\mu)$ cannot be distinguished from zero if $\langle R \rangle_0 \sim 0$. The partition function should be real and positive, however, if the phase fluctuation is extremely large and the accuracy of the simulation is not good enough, then $Z(\mu)$ is zero within the statistical uncertainty. If it happens, the denominator in the right-hand-side of \eqref{eq:reweightingQCD3} is zero, so this relation ceases to make sense. 

\iffigure
\begin{figure}[htbp]
\centering
\includegraphics[width=7.5cm]{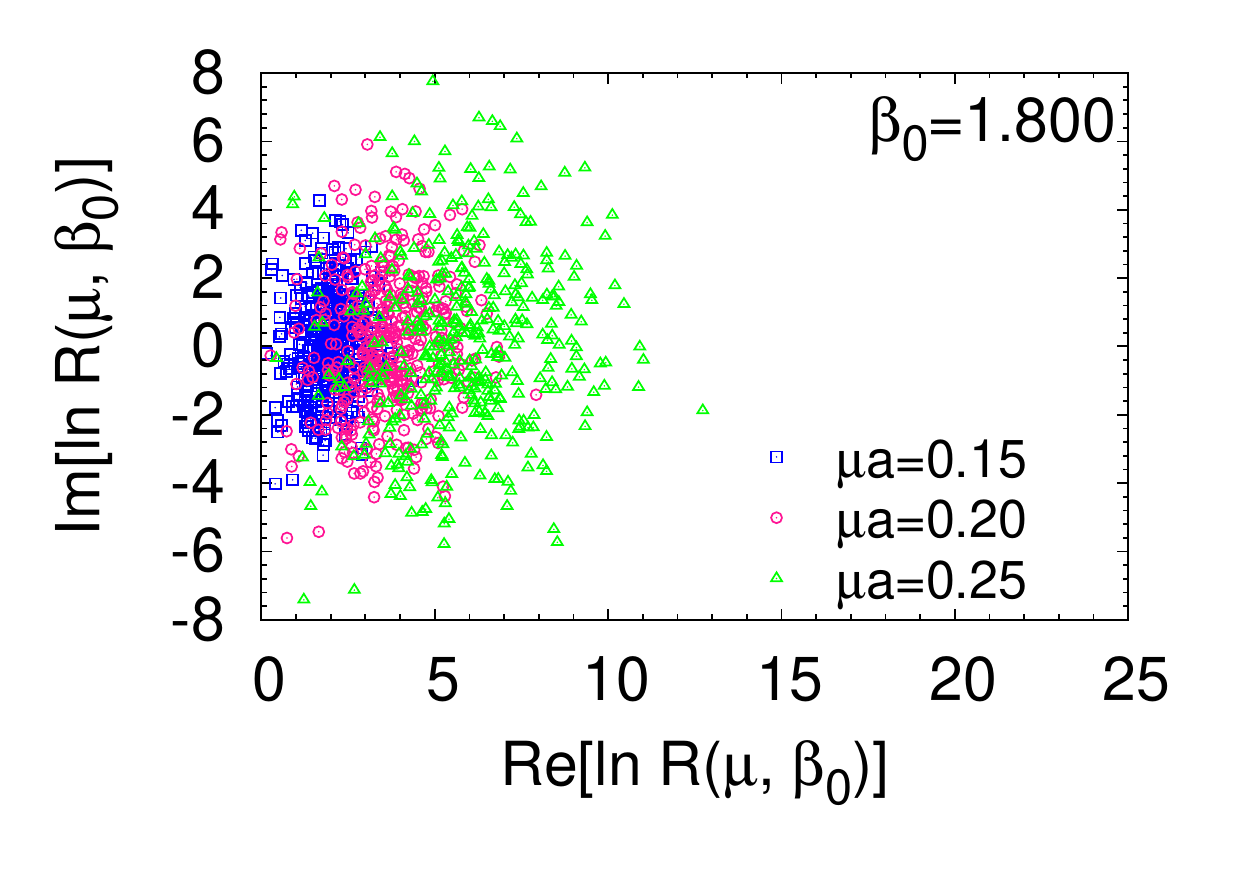}
\includegraphics[width=7.5cm]{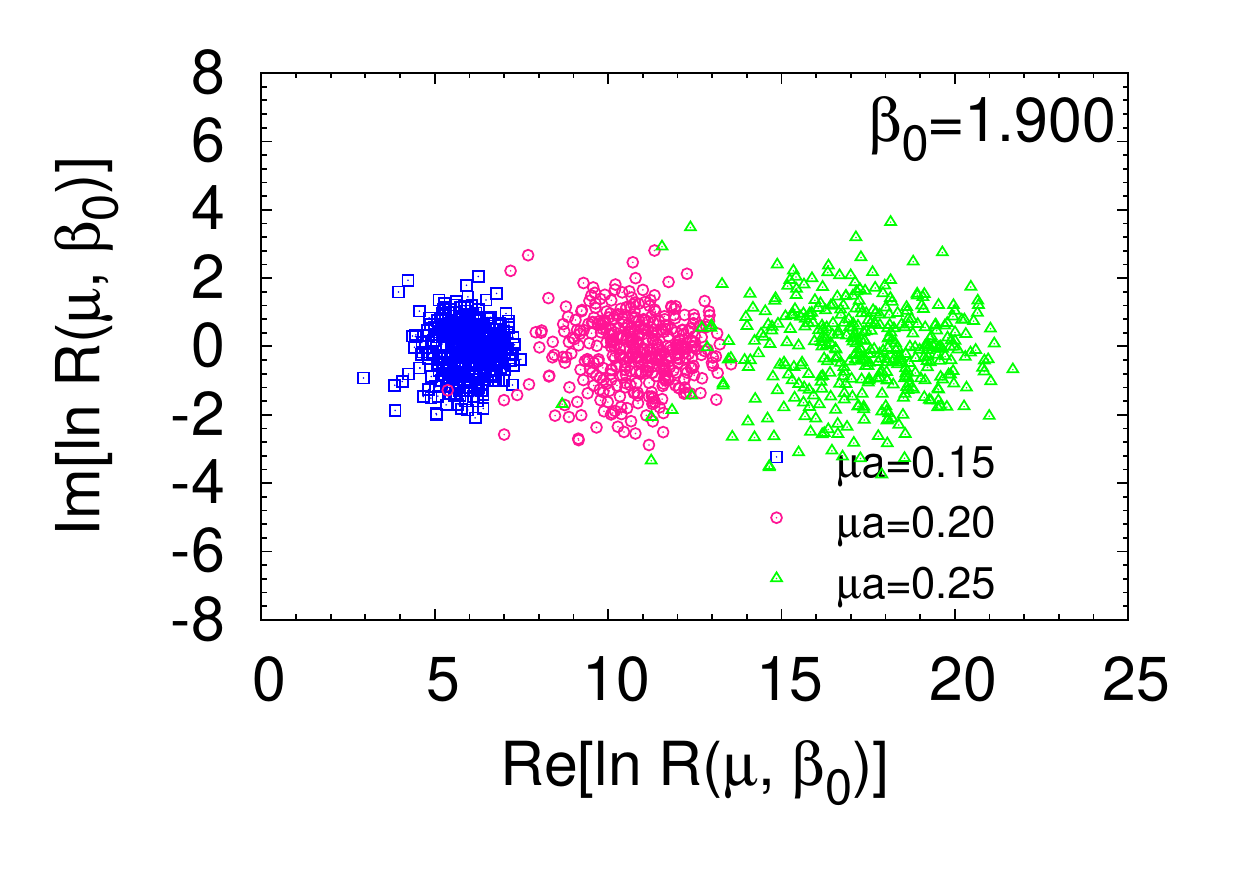}
\begin{minipage}{0.9\linewidth}
\caption{\small 
The fluctuation of the fermion determinant for $8^3\times 4$ lattice~\cite{Nagata:2012pc}. 
$\beta_0$ is the value of $ \beta $ where the gauge configuration are generated. $\beta_0=1.8$ is slightly below $ T_c $ and $1.9$ is slightly above $T_c$. The almost circular shape of the distribution means that the correlation between the magnitude and the phase part of $R$ is weak.
}
\label{Oct142011Fig1}
\end{minipage}
\end{figure}
\fi

Let us see how $R$ depends on $ \mu$. Fig.~\ref{Oct142011Fig1} shows the distribution of $R(\mu, \calC_i)$ in the complex plane. The gauge configurations $\calC_i$ are generated at $\mu=0$ and lattice coupling $\beta_0$. The left and right panels correspond to the hadronic phase (slightly lower temperature than $T_c$) and the QGP phase (slightly higher temperature than $T_c$), respectively. The horizontal axis of the figure, ${\rm Re} \ln R$, represents the absolute value of $R$ on a logarithmic scale, and the vertical axis, $ {\rm Im} \ln R $, is the complex phase of $R$. At $\mu=0$, $R (\mu = 0) = 1$ for all gauge configurations by definition, which is sitting at the origin of the plot.
In both hadronic and QGP phases, the distribution of $R$ expands as $\mu$ increases. The size of the distribution along the horizontal axis describes the difference of the absolute values of $R$ depending on the configuration, from which we can see the severeness of the overlap problem. In the right panel, at $ \mu a = 0.25 $, ${\rm Re} \ln R$ spreads from below $15$ to above $20$. It can be seen that the large and small $|R|$ differ by about 5 digits. The configuration average of the reweighting factors is
\[ \langle R \rangle_0 = \frac {1} {\Nconf} \sum_ {i} R (\calC_i). \]
However, if one of the $R (\calC_i)$'s (say $ R (\calC) $) in the sum becomes extremely large,
\[ \langle R \rangle_0 \simeq \frac {1}{\Nconf} R (\calC). \]
Since the expectation value is determined by one configuration, the effective statistic is $ 1 $ instead of $ \Nconf $.
Such a decrease in the effective statistics corresponds to the overlap problem mentioned above.

Next, we consider the phase fluctuation of the complex phase of $R$. If the phase of $R$ exceeds $|\pi/2|$, then the real part of $R$ becomes negative. When the phase is $|\pi|$, $R$ reaches on the negative real axis of the complex plane. If the distribution in Fig.~\ref{Oct142011Fig1} spreads out in the vertical direction, the positive and negative contributions cancel with each other in the configuration average of $R$, which makes $\langle R\rangle_0$ very close to $0$. In Fig.~\ref{Oct142011Fig1}, the phase fluctuation is severe slightly below $T_c$. It is due to this property of $ R $ that $ \langle \cos \theta \rangle_0 $ shown in Fig.~\ref{fig:averagephase} approaches $0$ at a temperature slightly lower than $ T_c $. To avoid the sign problem using the reweighting method, it is necessary to pay attention to these two problems.

\subsection{Improvement of the reweighting method: Multi-parameter reweighting}
\label{sec:mpr}
\newcommand{\RRW}{R} 
\newcommand{\ketSP}{\rangle_0}
\newcommand{\Rbar}{\bra R\ket_0}
\newcommand{\Csw}{C_{\rm SW}}
\def\rwfac{ \left(\frac{\det \Delta(\mu) }{ \det \Delta(0)}\right)^{N_f} e^{-(\beta-\beta_0) S_G} }

Simple solutions to the two problems mentioned in the previous section are to keep the parameter shift in the reweighting small enough or to increase the number of configurations exponentially when the parameters are shifted more. As a slightly more sophisticated method, Ferrenberg and Swendsen~\cite{Ferrenberg:1989ui} proposed a method that uses multiple ensembles generated at multiple points in the parameter space (multi-ensemble method). They also proposed another method that applies the reweighting to multiple parameters (the multi-parameter reweighing (MPR) method). The multi-ensemble reweighting method improves the overlap by performing configuration-generation at multiple points instead of at one point in the parameter space. On the other hand, multi-parameter reweighting is a method of shifting parameters in multi-dimensional parameter space. Fodor and Katz used the MPR method by performing the reweighting to temperature and chemical potential~\cite{Fodor:2001au}, and showed that the accuracy of the simulation can be improved compared to the one-parameter reweighting~(Fig.\ref{0104001phase2}). The key points in MPR are how to select the configuration-generation points and how to move to the target parameters. When the configuration is generated near the critical point, information on both sides of the phase transition can be captured, so a wider parameter region can be studied via the reweighting of such points. Furthermore, there is a way to approach the target parameters suppressing the overlap problem~\cite{Ejiri:2004yw, Csikor:2004ik, Nagata:2012pc}. Fodor and Katz performed the MPR to $\mu$ and $T$ on the configuration-generation point near the pseudo-critical temperature of QCD and determined the QCD critical point~\cite{Fodor:2001pe, Fodor:2004nz}. In this section, we explain the MPR method.

\subsubsection{Formulation of multi-parameter reweighting}
\iffigure
\begin{figure}
\centering
\includegraphics[width=7.5cm]{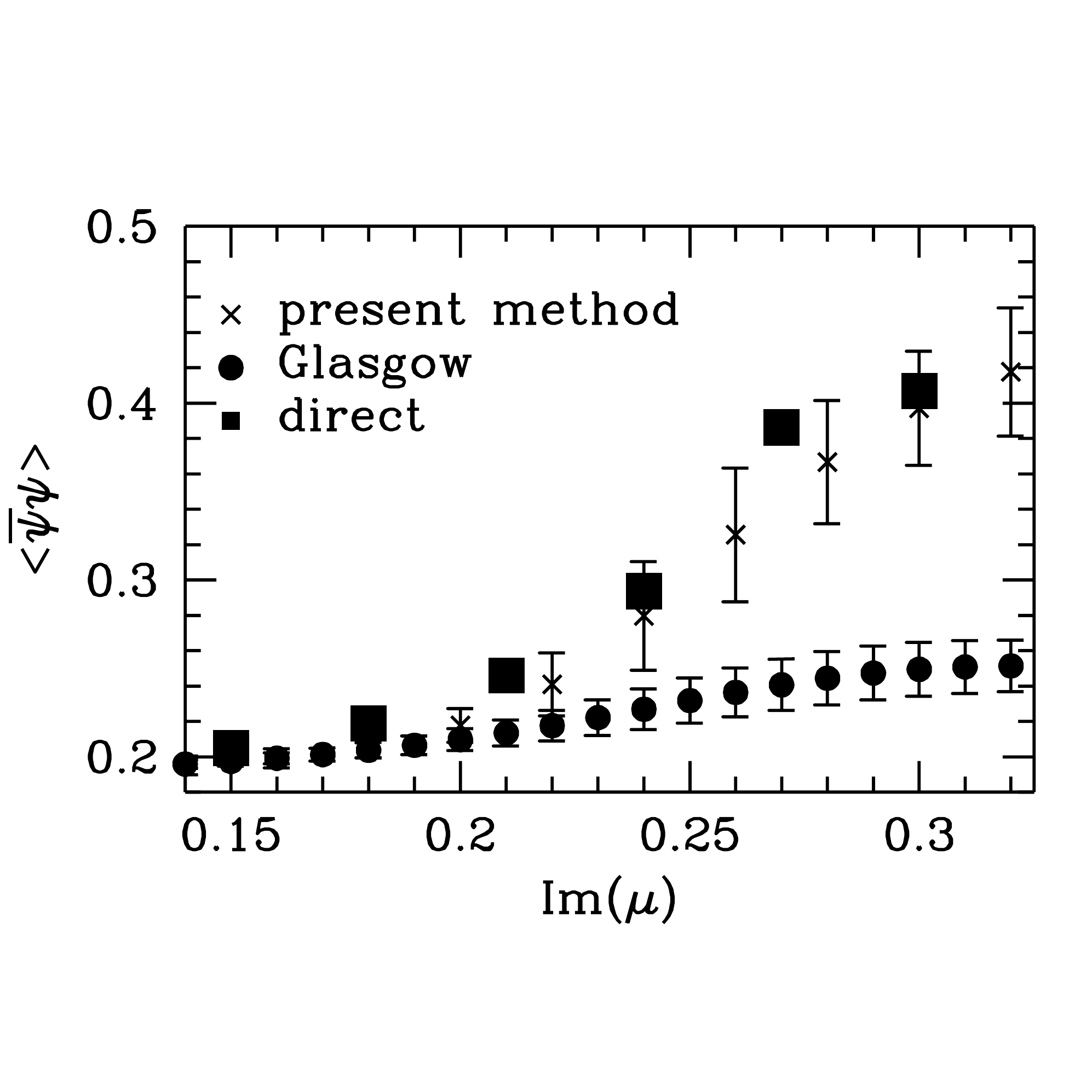}
\begin{minipage}{0.9\linewidth}
\caption{\small 
Comparison between reweighting for $\mu $ (Glasgow group) and that for $\mu, \beta$ (MPR, shown as `present method', for $6^3\times 4$ lattice)~\cite{Fodor:2001au}. The vertical and horizontal axes are the chiral condensation and imaginary chemical potential ($\mu$), respectively. We can see the improvement by the use of the MPR.
}\label{0104001phase2}
\end{minipage}
\end{figure}
\fi

The starting point is the Boltzmann factor,
\begin{align}
w(\mu, \beta) = (\det \Delta(\mu))^{N_f} e^{- \beta S_G}.
\end{align}
It is also expressed as
\begin{align}
w(\mu, \beta) = \RRW(\mu, \beta)_{(0,\beta_0)} \; w(0, \beta_0), 
\label{Oct182011eq1}
\end{align}
by using the reweighting factor $\RRW$
\begin{align}
\RRW(\mu,\beta)_{(0,\beta_0)}  = \left( \frac{\det \Delta(\mu)}{\det \Delta(0)}\right)^{N_f} e^{ - (\beta -\beta_0) S_G}.
\label{Dec082011eq1}
\end{align}
Here, the indices $(0, \beta_0)$ denote the parameters used for generating the configurations. We can generate configurations at $(\mu=0, \beta_0)$ since the weight $w(\mu=0, \beta_0)$ is real and positive there. Note that we shifted two parameters $\mu$ and $\beta$.
The partition function is 
\begin{align}
Z_{GC}(\mu, T) & = \bigintsss \calD U  \rwfac \;[\det \Delta(0)]^{N_f} e^{- \beta_0 S_G}, \nn \\
&= \int \calD U \RRW(\mu, \beta)_{(0,\beta_0)} \; w(0, \beta_0), \nn \\
&= \calN \langle \RRW(\mu, \beta)_{(0,\beta_0)} \rangle_0.
\label{eq:2017Jul18eq2}
\end{align}
The normalization factor $\calN$ does not appear in the following argument. The expectation value of $O$ is also written by using the reweighting factor,
\begin{align}
\bra O \ket & =\frac
{\ds{\int \calD U \, O \, \RRW(\mu, \beta)_{(0,\beta_0)} w(0, \beta_0)}}
{\ds{\int \calD U \, \RRW(\mu, \beta)_{(0,\beta_0)} w(0, \beta_0)}} \nn \\
&= \frac{\bra O \, \RRW(\mu, \beta)_{(0,\beta_0)} \ket_0 }{\bra  \RRW(\mu, \beta)_{(0,\beta_0)} \ket_0}.
\label{Oct052011eq1}
\end{align}
This expression is almost the same as the one in the previous section, except that the two parameters $\mu$ and $\beta$ are shifted. Refs.~\cite {Fodor:2001au, Csikor:2004ik} compared the ordinary one-parameter reweighting with MPR, for the case of the imaginary chemical potential where the sign problem does not exist and showed that larger parameter regions can be studied by using the MPR.

\subsubsection{Application to the equation of state and selection of reweighting path}

To find an optimal way to move the parameters in the MPR, some experience is needed. The MPR has been applied to some models with the first-order phase transition, and it was found that the fluctuations can be minimized along the transition line~\cite{Ejiri:2004yw}. Furthermore, Csikor et al. proposed ``overlap measure'' which is a criterion of optimal reweighting parameters~\cite{Csikor:2004ik}~\footnote{As far as we notice, a clear definition of ``overlap measure'' has not been shown in Ref.~\cite{Csikor:2004ik}.}. Ref.~\cite{Nagata:2012pc} explicitly formulated the method for finding the optimal parameters as a condition for minimizing the fluctuation of the real part of the reweighting factor. Here, we explain the essence of this method by taking the application to the equation of state (EoS) in QCD as a concrete example.

We calculate the pressure \eqref {eq:2017Mar27eq1} by using the reweighting method. We focus on the difference between the pressure at $\mu=0$ and $\mu\neq 0$, namely $\Delta p (\mu, T) = p (\mu, T)-  p (0, T)$, since the partition function has an ambiguity of normalization. In the reweighting method, we use the relation  
\begin{align}
\frac{\Delta p}{T^4} = \left(\frac{N_t}{N_s}\right)^3 \ln
\frac{\bra \RRW(\mu, \beta)_{(0,\beta_0)} \ket_0 }{\bra  \RRW(0, \beta)_{(0,\beta_0)} \ket_0}, 
\label{eq:2017Mar28eq1}
\end{align}
which gives the pressure at $(\mu, \beta)$ in terms of the configurations generated at $(0, \beta_0)$.

In principle, we can take any value of $ (\mu, \beta) $, but in practice, there are limitations due to the overlap problem and phase fluctuation. As explained in Sec.~\ref{sec:overlap_problem}, the low-probability states are truncated in actual simulations. The reweighting method cannot reproduce the ensembles in which the truncated states are dominant. Furthermore, the reweighting method fails if the phase fluctuation is severe. However, in the MPR, there is a path $\beta = \beta (\mu)$ on the $ \mu$-$\beta$ plane which improves the overlap. This kind of path selection is not possible in the reweighting in the one-dimensional parameter space. This is one of the advantages of MPR.

\iffigure
\begin{figure}
\centering
\includegraphics[width=7.5cm]{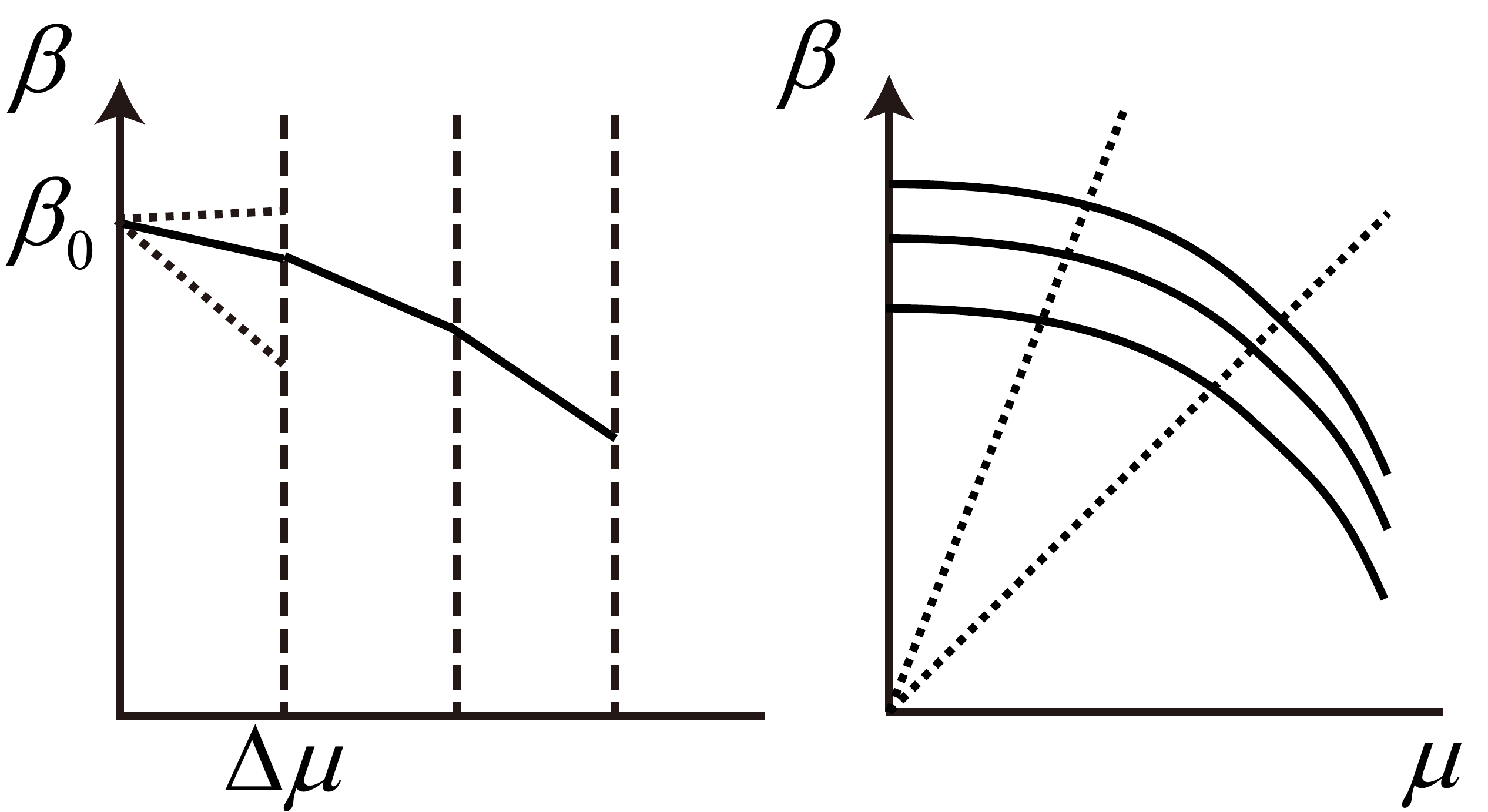}
\begin{minipage}{0.9\linewidth}
\caption{\small 
The determination of the reweighting line in the $(\mu, \beta)$ space. 
[Left] After changing $\mu$, the value of $\beta$ minimizing $X$ is determined for each $\mu$. [Right] By repeating this procedure gradually changing $\mu$, the optimal reweighting line in the $(\mu, \beta)$ space is determined. 
}\label{Nov132011fig1}
\end{minipage}
\end{figure}
\fi

There are several ways of choosing the reweighting line $\beta (\mu)$. Here we explain the basic idea by using the condition for the minimum fluctuation as an example~\cite{Nagata:2012pc}. As we have seen in Fig.~\ref{Oct142011Fig1}, the spread of the reweighting factor causes the overlap problem, hence we want to suppress it. First, let us define the fluctuation $X$ of the reweighting factor as
\begin{align}
X (\mu, \beta) = \langle (\RRW-\bra \RRW \ketSP) ^ 2 \ketSP. \label {Dec082011eq2}
\end {align}
We vary the chemical potential slightly, from $(0, \beta_0)$ to $(\delta\mu, \beta_0)$. Then the value of $\det \Delta (\delta\mu) $ in $R$ is determined. Next, we calculate $X$ varying $\beta$ and select $\beta$ that minimizes $X$. In this way, we can choose $ (\delta \mu, \beta_0 + \delta \beta)$ that suppresses the fluctuation of $R$. By repeating these processes, a path of $(\mu, \beta)$ that minimizes the value of $X$ is obtained (Fig.~\ref{Nov132011fig1}). It is possible to reweight a region away from such a line, but the deviation of overlap increases as the distance from the line increases. Note that the overlap is not complete even on this line.

To determine the value of $\beta$ that minimizes the fluctuation of $R$ for a given $\mu$, we regard $X$ as the height parametrized by $(\mu, \beta)$ and find the ``valley" on the parameter space. In the above, we explained how to search $ \beta $ that minimizes $X$ for each value of $\mu$. Instead, we can also obtain a relation between $\beta$ and $\mu$ that minimizes $X$. Under the variation of the parameters $(\mu, \beta)\to (\mu+\Delta \mu, \beta + \Delta \beta)$ with fixed $\beta_0$, the changes of $R$ and $X$ are related as $\delta X = 2 \bra \RRW \delta \RRW  \ketSP - 2\Rbar \bra \delta \RRW  \ketSP$. The fluctuation is minimum if $\delta X =0$, or equivalently 
\begin{align}
\frac{ \bra \RRW \delta \RRW  \ketSP}{\Rbar}   = \bra \delta \RRW  \ketSP.
\label{Nov302011eq3}
\end{align}
By definition, the left hand side is 
\begin{subequations}
\begin{align}
\frac{ \bra \RRW \delta \RRW  \ketSP}{\Rbar} = \bra \delta \RRW \ket 
= \frac{1}{Z}\int {\cal D} U \delta \RRW \;w(\mu, \beta).  
\end{align}%
The right-hand side is given by
\begin{align}
\bra \delta \RRW  \ketSP = \frac{1}{Z_0}\int {\cal D}U \delta \RRW \; w(0, \beta_0).
\end{align}
\end{subequations}
\eqref{Nov302011eq3} is satisfied when the two weights are equal, $w(\mu, \beta)/Z=w(0, \beta_0)/Z_0$.
This is realized when the path integral weights at the target point and the simulation point are the same. By substituting
\begin{align}
\delta \RRW(\mu, \beta) = \frac{\del \RRW}{\del (\mu/T)}  \frac{\Delta \mu}{T}
+ \frac{\del \RRW }{\del \beta } \Delta \beta
\label{Nov302011eq2}
\end{align}
to \eqref{Nov302011eq3}, we obtain
\begin{align}
\left( \left\langle \frac{\del \RRW}{\del (\mu/T)}\right\rangle - 
\left\langle \frac{\del \RRW}{\del (\mu/T)}\right\rangle_0\right)\frac{\Delta \mu}{T} = - \left( \left\langle\frac{\del \RRW }{\del \beta } \right \rangle - \left\langle\frac{\del \RRW}{\del \beta } \right \rangle_0  \right) \Delta \beta _.
 \label{Nov302011eq0}
\end{align}
\begin{subequations}
It can be simplified further by using \eqref{Oct052011eq1} as 
\begin{align}
\Delta \beta = 
 \frac{ \bra \RRW^2 a\ketSP  - \bra \RRW \ketSP\bra \RRW a\ketSP }{
 \bra \RRW^2 b\ketSP  - \bra \RRW \ketSP\bra \RRW b \ketSP }  \frac{\Delta \mu }{T}_, 
 \label{Nov302011eq1}
\end{align}%
where
\begin{align}
a & =  N_f {\rm tr} \left[ \Delta(\mu)^{-1} \frac{\partial}{\partial \mu/T} \Delta(\mu)\right], \\
b & =  S_G.
\end{align}
\end{subequations}
This expression describes a line on the $\mu$-$\beta$ plane. Along this line, the fluctuation of the reweighting factor is suppressed. \eqref{Nov302011eq1} reduces to 
\[ 
\Delta \beta = \frac{ \bra n \ket - \bra n \ketSP}{\bra S_G \ket 
- \bra S_G\ketSP}  \biggl( \frac{\Delta \mu}{T}\biggr)
\]
in the vicinity of the simulation point $(0,\beta_0)$. It was suggested that the equation of the reweighting line resembles the Clausius-Clapeyron equation in $(p,T)$ plane~\cite{Ejiri:2004yw}.

There is one important remark on the above method. The mechanism to enhance the overlap in the above method is to cancel the gauge action and the fermion action. However, this mechanism works only for the absolute value of $R$, and it does not work for the complex phase because the gauge action is real ~\cite{Ejiri:2004yw}. Therefore, the above discussion is valid only when the complex phase part of the reweighting factor is small.

\subsection{Examples of reweighting}

If the fermion determinant is real and positive, it is possible to generate gauge configurations using the importance sampling method. There are other actions that make the fermion determinant real, for instance:
\begin{itemize} 
\item QCD with imaginary chemical potential, where the quark chemical potential is pure imaginary,

\item quenched theory, where the fermion determinant is discarded,

\item phase quenched theory, where the absolute value of the fermion determinant is kept while the complex phase is discarded,

\item QCD with isospin chemical potential, where the quark chemical potentials for up and down quarks have the opposite sign ($\mu_u = - \mu_d$) in $N_f=2$ QCD,

\item two-color QCD, where the number of colors is reduced to $2$. 
\end{itemize} 
These theories are different from real QCD but still have similarities, hence they are sometimes called QCD-like theories.
Generating gauge configurations by using a QCD-like theory and reproducing QCD results by reweighting is a typical approach to avoid the sign problem. If the weight used for the importance sampling in the QCD-like theory is $w$, then the reweighting factor is
\[ R = \frac {\det \Delta (\mu) e ^ {-S_g}} {w}.\]
If the fluctuation of the complex phase of $ \det \Delta $ and of $R$  becomes severe, then the reweighting fails, whatever action is used for the configuration generation, as long as $w \in \mathbb{R}$. In other words, the range of applicability of the reweighting from the QCD-like theory is determined by the behavior of the complex phase of $\det \Delta$.

\subsubsection{Important remark on the application of MPR to the Lee-Yang zero}

\iffigure
\begin{figure}[htbp] 
\begin{center}
\includegraphics[width=8cm]{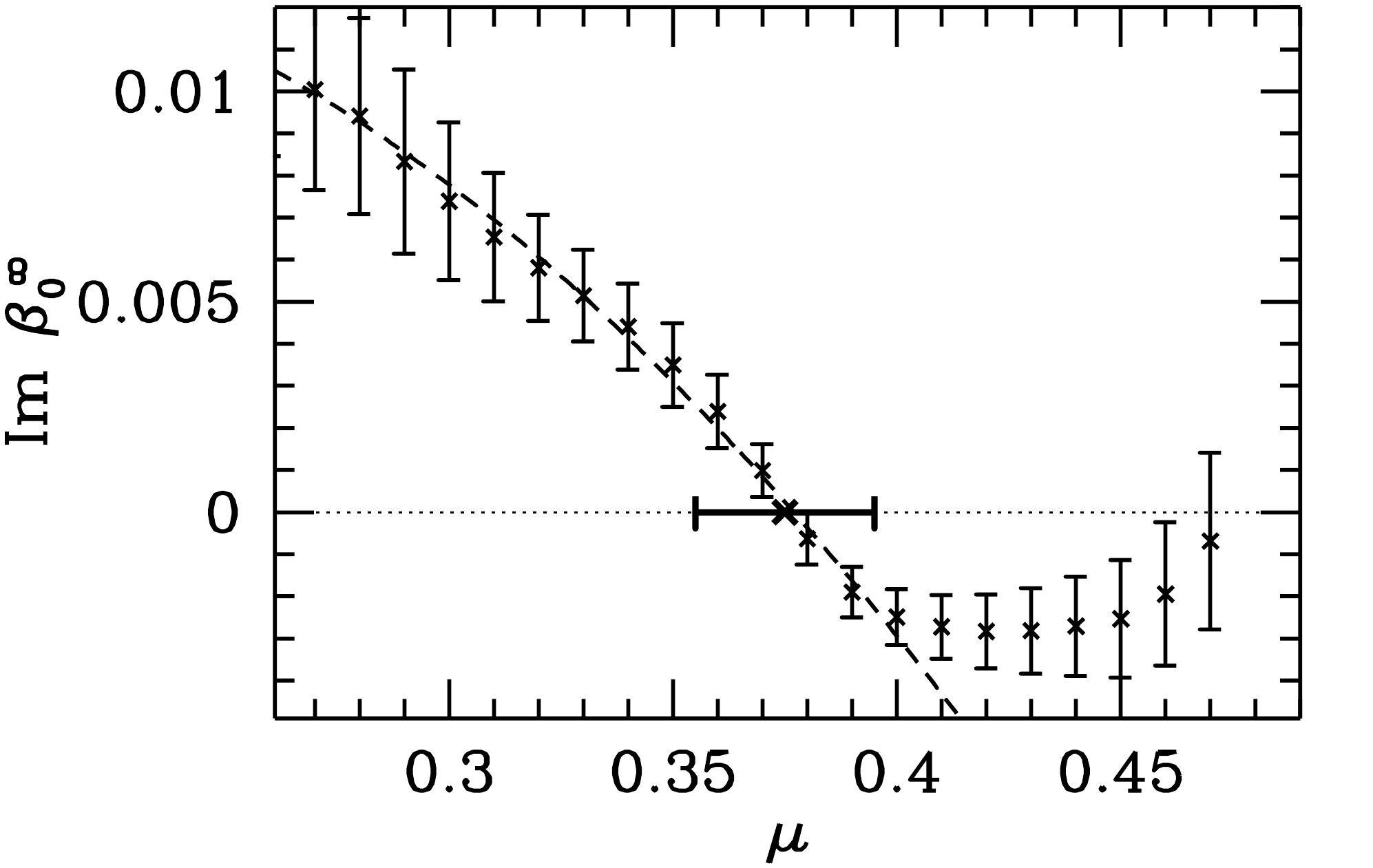}
\begin{minipage}{0.9\linewidth}
\caption{\small 
The Lee-Yang zero point in the infinite-volume limit, obtained by using MPR~\cite{Fodor:2001pe}.
The point where the partition function becomes zero is found in extended complex $ \beta $ plane. The position of the critical point on the phase diagram is obtained by converting $ \beta $ into temperature. Subsequently, the difficulty associated with the finite-size scaling was pointed out~\cite{Ejiri:2005ts}.
}
\label{0106002beta_im}
\end{minipage}
\end{center}
\end{figure} 
\fi

In Ref.~\cite{Fodor:2001pe, Fodor:2004nz}, the QCD critical point was estimated by Lee-Yang zero calculation by using the reweighting for $\mu$ and $\beta$ (Fig.~\ref{0106002beta_im})~\footnote{The zeros on the complex $\beta$ plane is called the Fisher zeros since the generalization of the zero point theorem to temperature was given by Fisher.}. However, the reweighting method sometimes leads to unphysical signals in the calculations of the Lee-Yang zeros, as pointed out in Ref.~\cite{Ejiri:2005ts}. As explained in Sec.~\ref{sec:overlap_problem}), the overlap problem becomes exponentially severe as the volume becomes large and as the parameters are shifted more. The Lee-Yang zero-point theorem states that the thermodynamic singularity representing the phase transition causes the partition function to approach zero in the thermodynamic limit, 
\[\lim_ {V \to \infty} Z \to 0. \]
(See Appendix~\ref{sec:lee-yang} for Lee-Yang zero-point theorem). The error of $Z$ increases as the lattice volume increases if the reweighting method is used so that it is difficult to decide whether the partition function is approaching zero or the error is getting too large such that $Z$ gets consistent with zero within the error bar. To identify the Lee-Yang zero point by using the reweighting method, it is necessary to investigate the volume dependence while controlling the error. Such calculation has been achieved for the Roberge-Weiss phase transition in the high-temperature region of QCD~\cite{Nagata:2014fra}, as we will explain in Sec.~\ref{sec:pt_canonical}.

Besides the methods reviewed in this section, other improved methods such as the multi-ensemble reweighting method~\cite{Ferrenberg:1989ui, Kratochvila:2005mk, deForcrand:2006ec} and the histogram method~\cite{Nakagawa:2011eu} were proposed. The density-of-state method, which is a generalization of the reweighting method, is also studied recently.
\subsection{Complexity problem: Why is the lattice calculation of finite density QCD so expensive?}
\label{sec:problem_cost}

Compared to other problems in lattice QCD research, finite-density lattice QCD requires a much larger computational cost, so people often say `finite-density QCD is heavy'. Such a large cost is needed partly because the calculation of the fermion determinant $ \det \Delta$ is costly. 
This problem appears in many methods to circumvent the sign problem, such as the reweighting method. Here, we explain why the calculation of determinant is costly and why it occurs in finite-density lattice QCD because the cost is an important factor that determines the feasible parameter range of the numerical calculations.

The calculation of the determinant is computationally expensive. A direct calculation of determinant via diagonalization or the $LU$ decomposition involves three loops regarding the matrix indices, which requires $O(V^3)$ manipulations for a matrix if rank $V$. Furthermore, for such calculations the matrix must be stored in memory, hence the necessary memory size is $O(V^2)$. If you calculate the determinant of the matrices of ranks $2, 3,$ and $4$, you can immediately understand that the calculation becomes complicated as the matrix becomes large. The fermion matrix $\Delta$ acts on the indices related to the internal degrees of freedom and lattice coordinates, and its rank is proportional to $L^4$, where $L$ is the number of lattice sites along each direction. The memory required for storing the fermion matrix is $O(L^8)$, and the computational cost is $O(L^{12})$. As the lattice size increases, both the memory size and computational cost increase rapidly.

The cost of the direct calculation of the fermion determinant strongly depends on the lattice size. In the Hybrid Monte Carlo algorithm, which is the standard method for the configuration generation in lattice QCD, this problem is avoided by using the trace of the matrix instead of the action itself. The memory required to calculate the diagonal sum is $O(V)$ and the calculation is written by a single loop so that the dependence on the lattice volume dependence is much weaker.~\footnote{
Strictly speaking, the trace of the inverse matrix of the fermion determinant is calculated, hence the inverse matrix has to be calculated.
Still, the cost is much smaller than that of the calculation of the determinant.}
Many physical quantities, such as the chiral condensate and hadron correlation functions, can also be obtained by taking the trace of the fermion matrix. In this way, the ordinary lattice-QCD calculation does not require the calculation of the fermion determinant. Therefore, the simulation at large volume is feasible.

However, then why do we have to calculate the fermion determinant when we try to avoid the sign problem by using the reweighting method? This is because the reweighting method is the extrapolation from the configuration-generation point $\mu=0$. Accurate calculation of the chemical-potential dependence is needed to maintain the correctness of the calculation by shifting $\mu$ from the configuration-generation point.
Therefore, careful treatment of the chemical-potential dependence of the reweighting factor is necessary. If we do not use the correct function, the correctness of extrapolation cannot be guaranteed. Therefore, we cannot avoid the calculation of the determinant not only in the reweighting method but also in similar methods that use the configurations generated at the sign-problem-free points. As explained in the following sections, the calculation of fermion determinants is not necessary for the Taylor expansion method and the analytic continuation from the imaginary chemical potential, but in those cases, there is a trade-off between the parameter range and the computational cost. For example, to study large chemical potential by the Taylor expansion method, it is necessary to compute the Taylor expansion of the fermion determinant to very high order, which is as costly as the exact calculation of the fermion determinant. If the Taylor expansion is truncated at a low order, the amount of calculation is small, but such approximation is valid only when the chemical potential is small.

Studies of finite-density lattice QCD are often limited to small lattice sizes because the computational cost depends on the lattice volume very strongly compared to the ordinary lattice QCD simulations. Such strong volume dependence appears because the methods for circumventing the sign problem rely on extrapolation. Although the computation of the fermion determinant is unavoidable, there is a formula that allows us to reduce the cost to some extent, as we will see in Sec.~\ref{sec:reduction}. To resolve the issue of a quick increase of the cost with volume, it is necessary to generate the configurations directly at the parameter of interest, without relying on extrapolation. In Sec.~\ref{sec:complexLangevin}, we will introduce the complex Langevin method as a recent development in this direction.

In this section, we have explained the basic ideas of the reweighting method.
We will discuss the actual applications in the next section, by comparing the reweighting method and the Taylor expansion method.

\clearpage
\section{Taylor expansion}
\label{sec:Taylor}

This section describes the Taylor expansion, which is useful when the chemical potential is small. Since the thermodynamic quantities are analytical functions of the chemical potential except for the phase transition point, the Taylor expansion is valid around $\mu=0$ as 
\begin{align}
O (T, \mu) = \sum_ {n} c_n (T) \left (\frac {\mu} {T} \right) ^ n.
\label{eq:Tayloreq1}
\end{align}
Here, the expansion coefficients are given by
\begin{align}
c_n (T) = \frac{1}{n!} \left (\frac {\partial ^ n O} {\partial \mu ^ n} \right) _ {\mu = 0}.
\end{align}
To obtain the expansion coefficients, the importance sampling (especially the HMC) is applicable, since the coefficients $c_n$ are defined at $\mu=0$. By substituting the values of $c_n$ obtained at $\mu=0$ into the expression \eqref{eq:Tayloreq1}, the expectation values at $\mu\neq 0$ can be obtained. The Taylor expansion requires the differentiability of thermodynamic quantities, and the hadron phase and the QGP phase are smoothly connected by a crossover near the temperature axis ($\mu=0$) in the QCD phase diagram. Therefore, the Taylor expansion method is indeed applicable there. In principle, $c_n$ can be calculated for any $n$, but as $n$ becomes larger, the expression becomes more complicated, and the calculation is practically hard because the statistical error increases. At present, only low-order terms have been calculated. Although the Taylor expansion method can in principle be applied within the radius of convergence, with only low-order coefficients, only the small-$\mu$ region can be studied.

The Taylor expansion has the advantage that the simulation cost does not increase quickly with the lattice size, because the calculation of determinant is not necessary. Currently, the extrapolation to the physical point has been achieved only by the Taylor expansion method and the imaginary chemical potential method explained in the next section. In this section, we introduce the idea and application of the Taylor expansion method. We discuss in which region of the QCD phase diagram the Taylor expansion method can be used reliably. Furthermore, we compare the reweighting method and Taylor expansion method and check their validities.

\subsection{Basic ideas of Taylor expansion and application to equation of state}
\label{sec:taylorvsmpr}

Let us study the $\mu$ dependence of pressure, quark number, and quark number susceptibility by using the Taylor expansion. In order to simplify the calculation in the lattice simulation, we consider the dimensionless quantity~\eqref{eq:2017Mar27eq1}, namely the pressure divided by $T^4$:
\begin {align}
\frac {p (\mu, T)} {T ^ 4} = \left (\frac{N_t}{N_s} \right)^ 3 \ln Z (\mu, T).
\label{Sep122011eq2}
\end {align}
By expanding it with respect to $\mu/T$ about $\mu=0$, we obtain
\begin{subequations}
\begin{align}
\frac{p (\mu, T)}{T ^ 4} & = \frac {p (0, T)}{T ^ 4} + \sum_ {n = 2,4, \cdots} ^ \infty c_n (T) \left (\frac {\mu}{T} \right) ^ n.
\label{eq:pres_taylor}
\end {align}
QCD has CP invariance, that is, it has a symmetry between quark and antiquark, so it can be expanded by the even powers of $\mu$.
The Taylor coefficient $ c_n $ is defined by
\begin{align}
c_n & = \frac {1}{n!} \frac {p ^ {(n)}(\mu = 0, T)}{T ^ 4}, \nn \\
& = \left. \frac{1}{n!} \left (\frac{N_t}{N_s} \right) ^ 3 T^n \frac {\del ^ n \ln Z} {\del \mu^n} \right | _ {\mu \to 0 \;.}
\label{Eq:Dec0310no1}
\end{align}
\end{subequations}
The first and second derivatives of $p$ with respect to $\mu$ are related to the quark number density $n_q$ and the quark number susceptibility $ \chi_q $, respectively:
\begin{subequations}
\begin{align}
\frac{n_q}{T ^ 3} & = \sum_ {n = 2,4, \cdots} ^ \infty n \cdot c_n (T) \left (\frac{\mu}{T} \right)^{n-1} _, \\
\frac{\chi_q}{T ^ 2} & = \sum_ {n = 2,4, \cdots} ^ \infty n (n-1) \cdot c_n (T) \left (\frac {\mu}{ T} \right)^ {n-2}.
\end{align}%
\label{Eq: 2012 Mar20eq1}%
\end{subequations}%
The coefficients $ c_1 $ and $ c_2 $ can be determined by setting $ \mu $ to zero in \eqref{eq:2017Mar29_ndens} and \eqref{eq:2017Mar29eq2}.
$ c_1 $ is zero because it is the quark number density at $ \mu = 0 $.
Because $p(\ mu)$ is an even function of $\mu$ so that the Taylor coefficient of odd order is $0$ at $\mu=0$. The rate of convergence of the series depends on temperature since $c_n$ is a function of $T$.

Before showing the simulation results of lattice QCD, let us see the behavior of $c_n$ in two common phenomenological models: the quark free gas model and the Hardon Resonance Gas (HRG) model. First, we consider the high-temperature regime. At sufficiently high temperatures, the interaction between quarks becomes small due to the asymptotic freedom, and the system can be described as a gas of free quarks.
The free energy of the quark free gas model can be described by at most a quartic function of the chemical potential~\cite{Kapusta:2006aaa}:
\begin{align}
\frac {p (\mu)} {T ^ 4} = \frac{p (0)}{T ^ 4} + c_2 \left (\frac {\mu} {T} \right) ^ 2 + c_4 \left (\frac {\mu} {T} \right) ^ 4.
\label{eq:freene_freegas}
\end{align}
In this case, $ p (\mu,T) $ can be obtained by calculating $ c_2 $ and $ c_4 $ at $\mu = 0$.\footnote{$ p(\mu,T)$ can be regarded as free energy $f(\mu,T)$ in the large volume limit.} $ c_2 $ and $ c_4 $ for free gas model are
\begin{align}
c_2 = \frac{N_f}{2}, \, c_4 = \frac{N_f}{2 \pi ^ 2}.
\end{align}

The hadron resonance gas model is a phenomenological model of QCD in the hadron phase low-density region. It reproduces the data of RHIC experiment and lattice QCD in the temperature range below $T_c$~\cite{Karsch:2003vd, Karsch:2003zq}. The free energy is expressed as
\begin{align}
\frac{p (\mu, T)} {T ^ 4} = a (T) + b (T) \biggl[ \cosh \biggl (\frac {3 \mu_q} {T} \biggr) -1 \biggr].
\end{align}
The coefficients for odd $n$ are zero, while ones for even $n$ are
\begin {align}
c_2 = \frac{9}{2} b (T),\quad
c_4 = \frac{27}{8} b (T),\quad 
c_6 = \frac{81}{80} b (T),\quad 
\cdots,\quad 
c_ { 2n} = \frac{3 ^ {2n}}{(2n)!} B (T).
\end {align}
In contrast to the free gas model, the Taylor coefficients are nonzero at $n>4$. 

\iffigure
\begin{figure}[htbp] 
\begin{center}
\includegraphics[width=8cm]{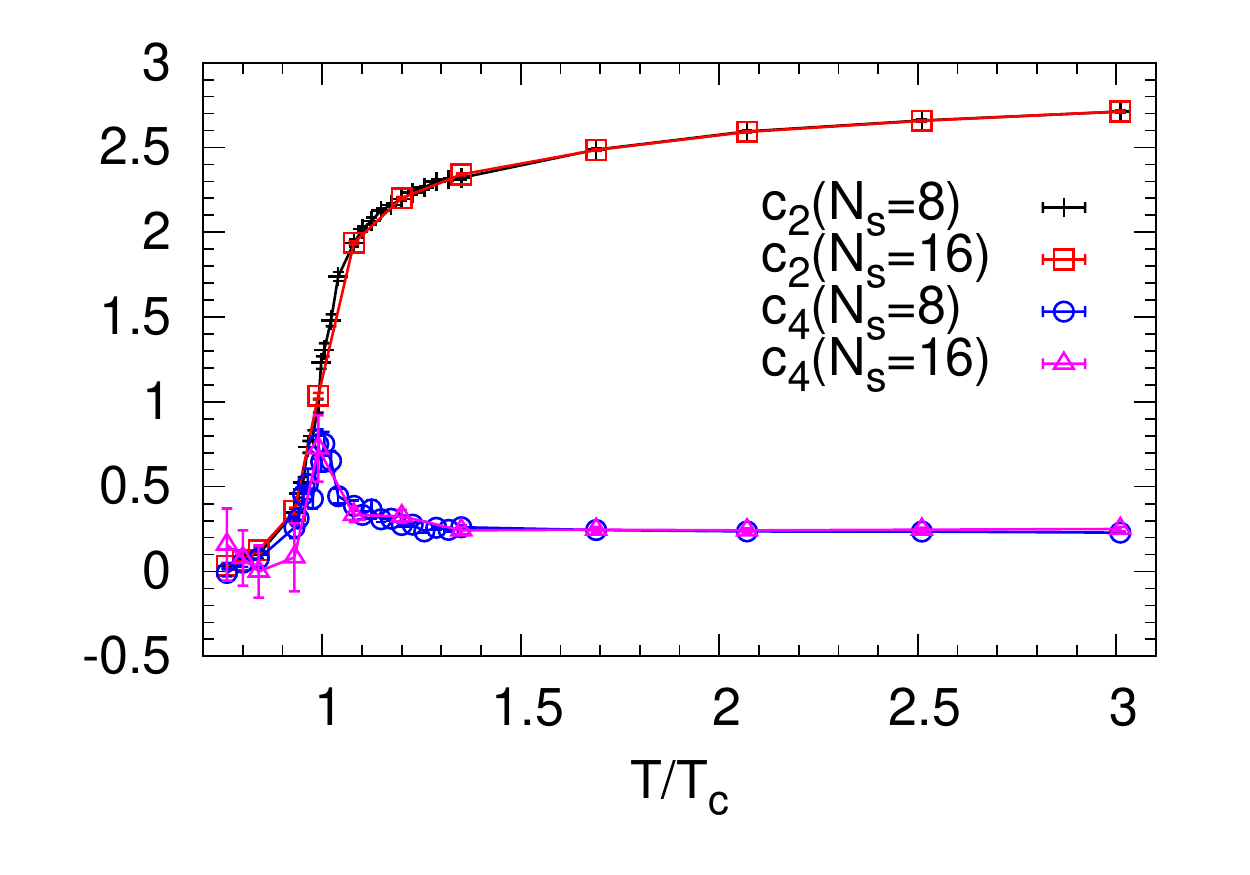}
\begin{minipage}{0.9\linewidth}
\caption{\small 
Temperature and volume dependence of the second- and fourth-order Taylor coefficients of the equation of state.
$ N_s $ is the lattice size in the spatial direction. The results for $8^3\times 4$ lattice~\cite{Nagata:2012pc} and $16^3\times 4$ lattice~\cite{Ejiri:2009hq}, 
with the same lattice parameters (except for the lattice size).}
\label{Fig:2014Mar17fig1}
\end{minipage}
\end{center}
\end{figure} 
\fi

We introduced two characteristic cases, but non-perturbative calculations of QCD are necessary to obtain $ c_n $ at a generic temperature. The Taylor coefficients are defined at $ \mu = 0 $, so they can be calculated by importance sampling (especially the HMC). As can be inferred from the \eqref{eq:2017Mar29eq2} and \eqref{eq:2017Mar29_ndens}, the coefficients at higher-order include the larger number of terms, and to obtain them both the computational cost and statistical error increase. The Taylor coefficients are given by the diagonal sums including the fermion matrix and its inverse. The diagonal sums are often calculated by using the noise method. In the noise method, the computational cost is reduced by using randomly generated vectors that satisfy the standard orthogonality condition to estimate the trace of the matrix. The number of calculations for $ tr \Delta ^ {-1} $ can be reduced in this way. On the other hand, the error due to the random numbers arises. Although such error does not matter for simple physical quantities, it does matter for the higher-order Taylor coefficients and can lead to a large error. This kind of error can be eliminated by the standard orthogonal basis for each configuration, but then the cost increases. The contraction formula can be used to accurately determine the Taylor coefficients in any order, but it requires a large cost which makes the application to large lattices difficult (see Sec.~\ref{sec:reduction}). Moreover, even if the calculation for each configuration is performed precisely, the statistical fluctuation increases with the order, hence more and more configurations are needed at higher order.

Now, we show the results of both the noise method and the exact calculation for the low-order Taylor coefficients. Fig.~\ref{Fig:2014Mar17fig1} depicts the results of $c_2$ and $c_4$,
where the data for $ N_s = 8 $ is obtained using the reduction formula, while the one for $ N_s = 16 $ is done  using the noise method. $c_2$ monotonically increases with temperature, while $ c_4 $ has a peak near $ T_c $ and then approaches a certain value at high temperature. The quark free gas model suggests that $c_2$ and $c_4$ are constant at high temperatures. The result of the lattice calculation agrees with this prediction~\footnote{
However, it is known that $ c_2 $ and $ c_4 $ deviate from the prediction by the quark free gas model. Possible reasons for the deviation are the finiteness of the temporal lattice size ($N_t$), the finite-volume effect ($N_x N_y N_z$). Another possibility is that temperature is not sufficiently high for the asymptotically free behavior to set in; the effective coupling constant $ g $ in QCD logarithmically decreases in the high-energy region as 
\[\frac{g ^ 2 (Q ^ 2)}{4 \pi} \sim \frac {6 \pi} {(33-2N_f) \ln (Q ^ 2 / \Lambda ^ 2)} \],
in the high energy region ($Q^2 \ gg 1$), and $T \sim O (T_c)$ is not sufficiently high temperature for $g\sim 0$.}. No dependence on $N_s$ (the number of lattice points in the spatial direction) can be seen for $c_2$ and $c_4$ have no dependence on $N_s$. This result is natural because $c_2$ and $c_4$ are the expansion coefficients for the equation of state normalized by the lattice volume.

Next, the results for the higher-order terms $n=6,8,10$ are shown in Fig.~\ref{fig:Taylor_coeff}. Here, we show only the result obtained by using the reduction formula. $c_6,c_8,c_ {10}$ oscillate as a function of $T$ near $T_c$, and they decay rapidly above $T_c$ and become almost zero at $T/T_c\gtrsim 1.2$. Hence, the equation of state can be approximated by a quartic function of $\mu$. This is the same as the property of the free gas model. The numerical results indicate that the high-temperature limit is not necessary for this approximation. The temperature range where the higher-order terms are vanishingly small depends on the simulation setup, usually the lower bound is $T/T_c=1.2 \sim 1.3$~\cite{Allton:2005gk, Schmidt:2006us, Ejiri:2009hq}.

\iffigure
\begin{figure}[htbp] 
\centering
\includegraphics[width=7cm]{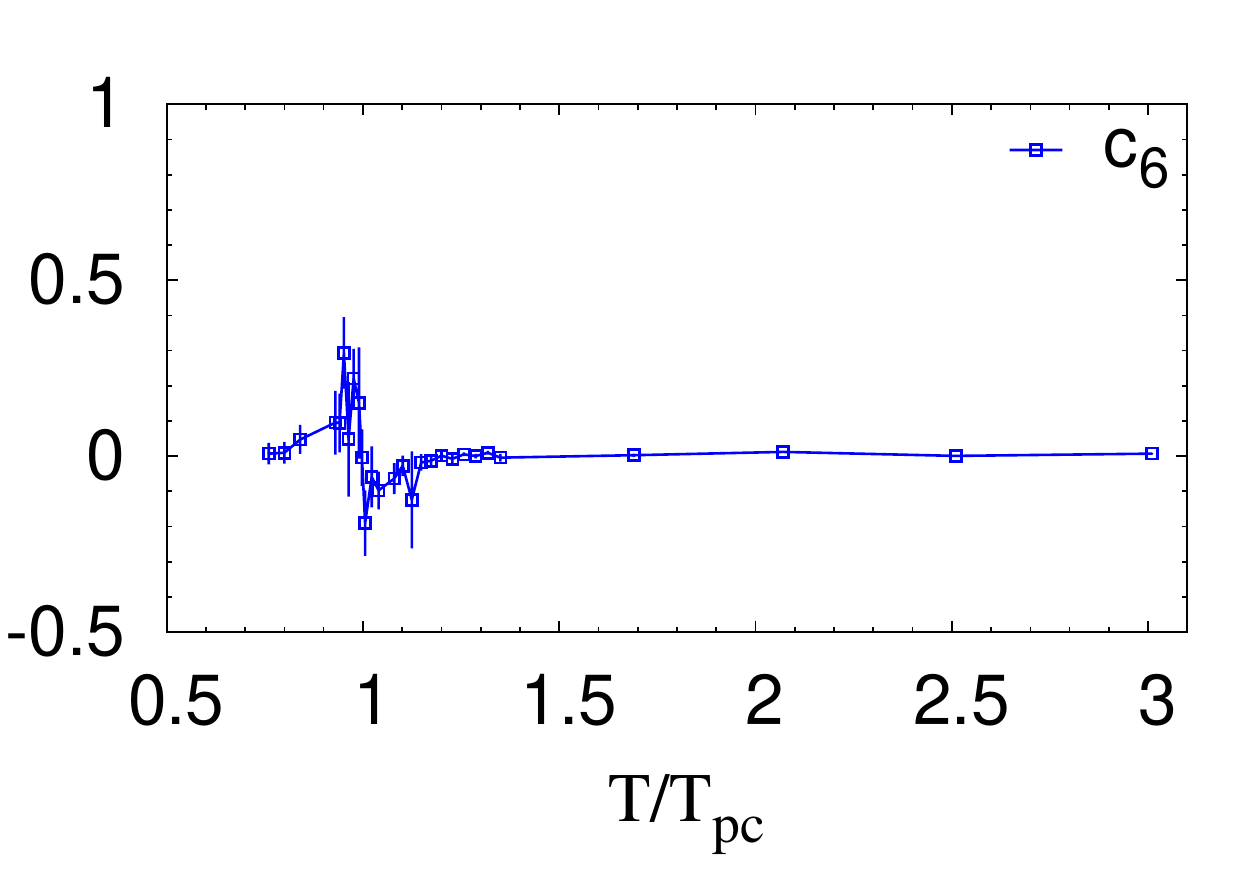}
\includegraphics[width=7cm]{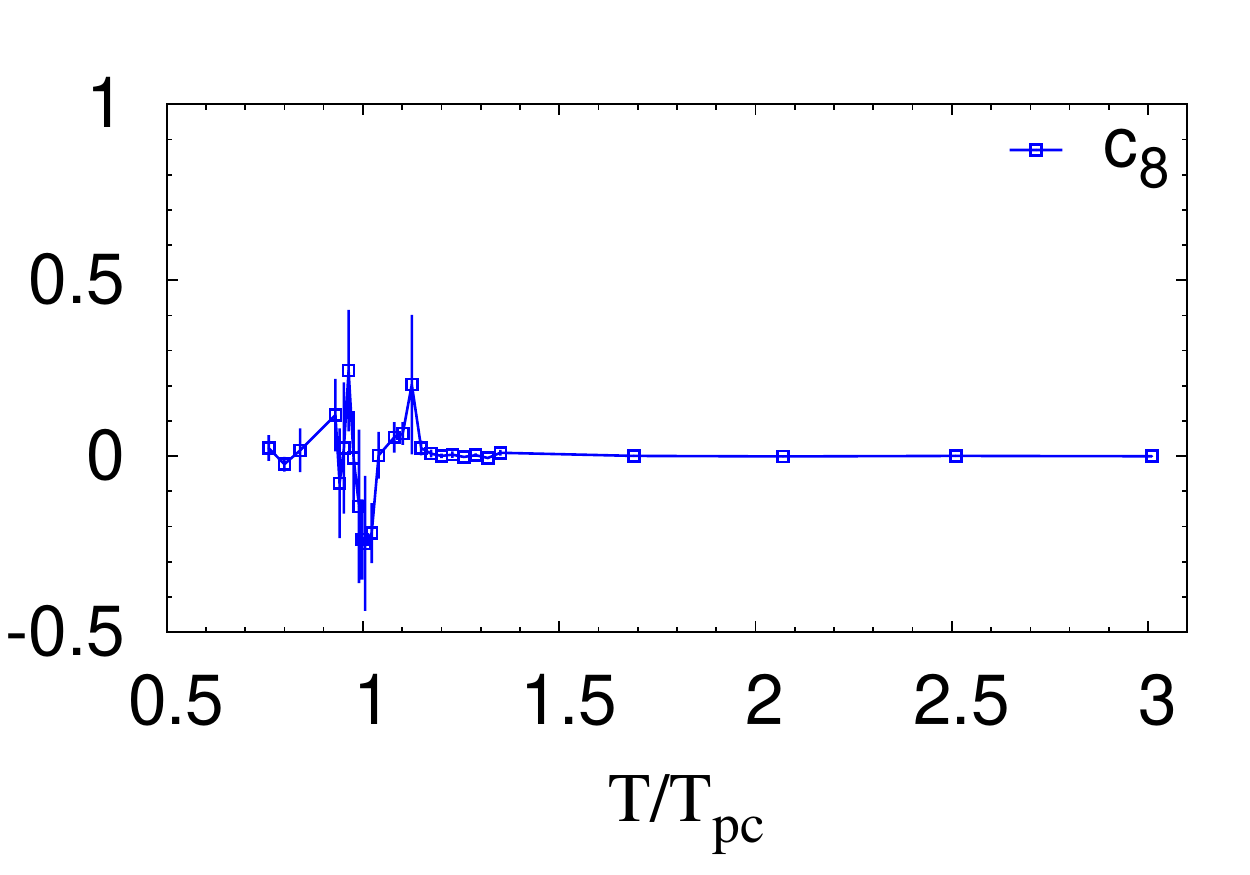}
\includegraphics[width=7cm]{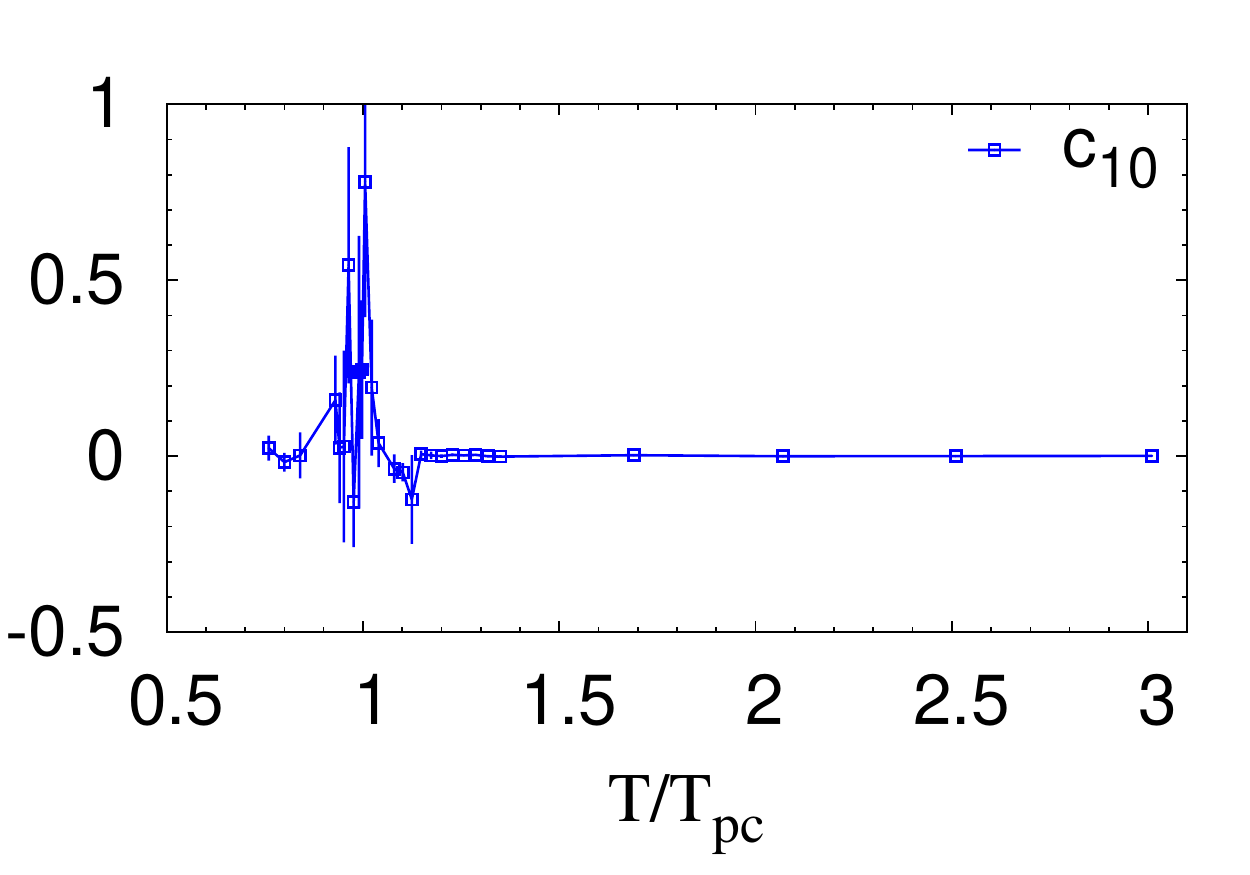}
\begin{minipage}{0.9\linewidth}
\caption{\small 
The temperature dependence of the Taylor coefficients for $8^3\times 4$ lattice~\cite{Nagata:2012pc}.
It is qualitatively consistent with the chiral effective theory~\cite{Schaefer:2009st}.}
\label{fig:Taylor_coeff}
\end{minipage}
\end{figure}
\fi

Around $T_c$ ($ 0.8 T_c \lesssim T \lesssim 1.2 T_c $), $c_n$ is wavy, it oscillates more as $n$ increases. Similar behavior can be seen in the chiral effective theory~\cite{Schaefer:2009st}. Another feature is that the higher-order terms decrease slowly. The peak height $\max_T |c_n (T)|$ is roughly $\max_T|c_n|\sim 0.5$ for $n=6,8,10$, and no significant decrease is seen. Furthermore, since the sign of $c_n$ is either positive or negative depending on $n$, it is expected that the convergence of the Taylor series (\ref{eq:pres_taylor}) is slow. Hence, the calculation of high-order terms is necessary. In addition, the error is large near $T_c$. For example, at $n=10$, the temperature dependence of $c_ {10}$ is almost invisible due to the error. To obtain higher-order terms, it is necessary to increase the statistics and the number of simulation points so that the oscillation of $ c_n $ can be correctly captured. Otherwise, it is necessary to limit $\mu/T$ to a reasonably small range, so that the truncated terms can be ignored.

In the low-temperature region, when $T$ is less than $0.7T_c\sim 0.8 T_c$, the value of $c_n$ is almost zero. It is not easy to determine the $\mu$-dependence of the equation of state because it is difficult to determine the small deviation of $c_n$ from zero precisely. The decrease of the signal-to-noise ratio is one of the difficulties in the low temperature and low-density region.

\subsection{Density dependence of EoS. Comparison between Taylor expansion and reweighting}

\iffigure
\begin{figure}
\centering
\includegraphics[width=7cm]{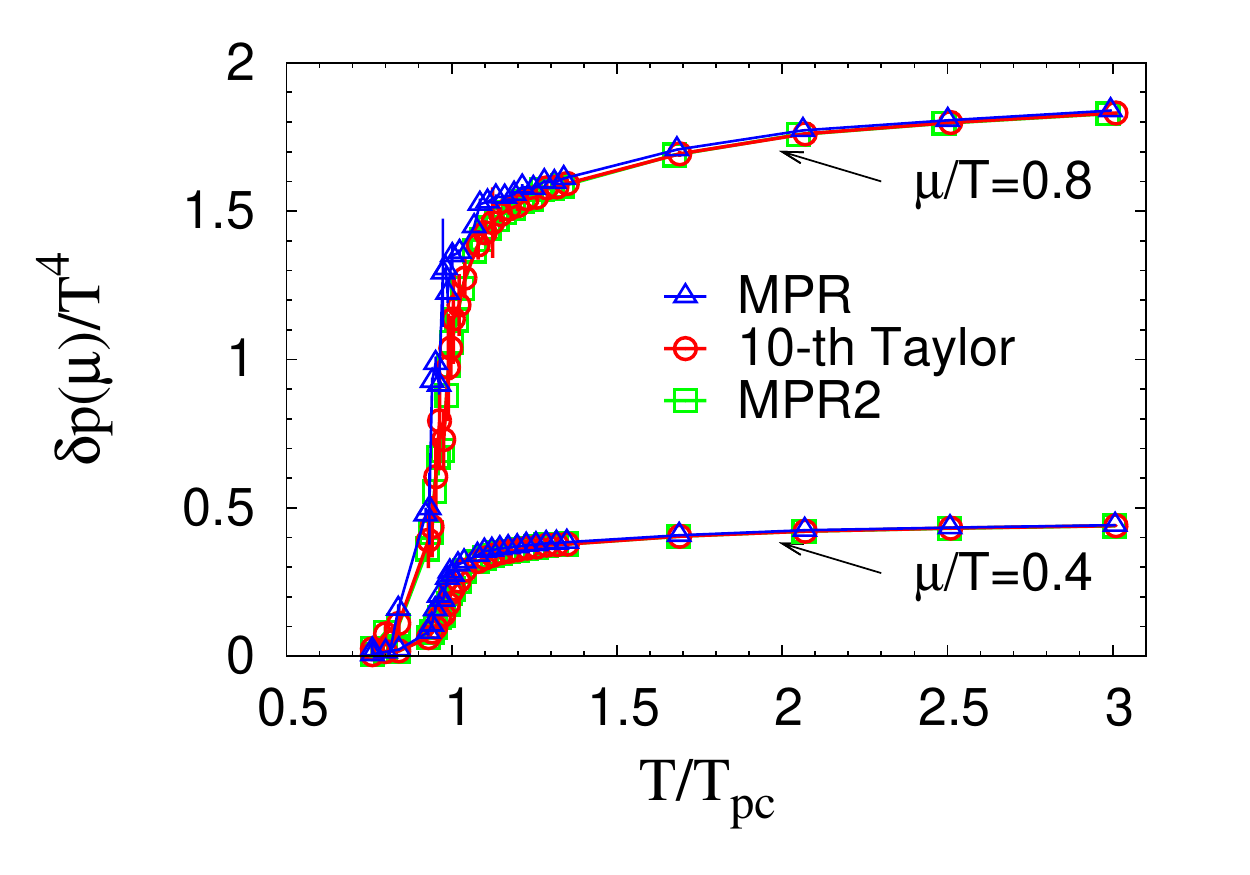}
\includegraphics[width=7cm]{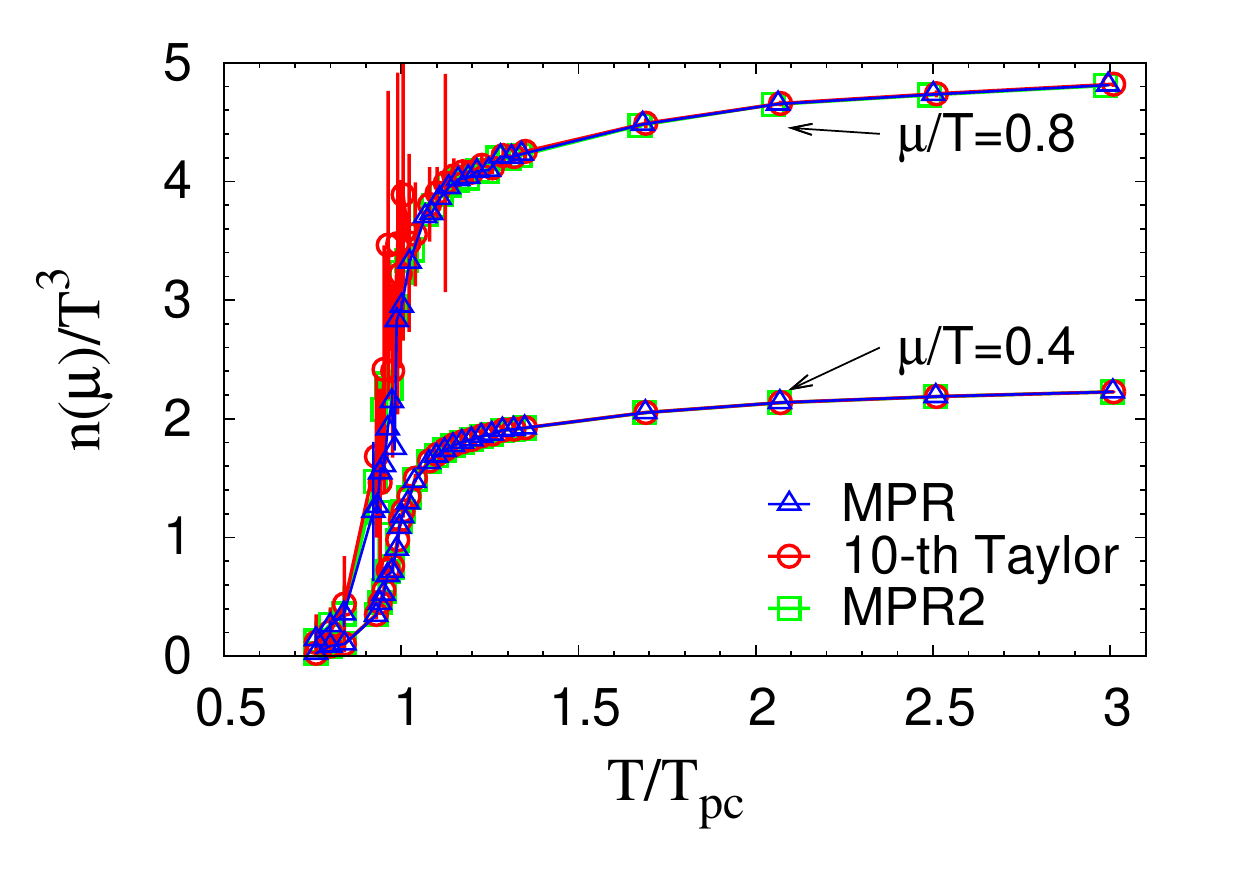}
\includegraphics[width=7cm]{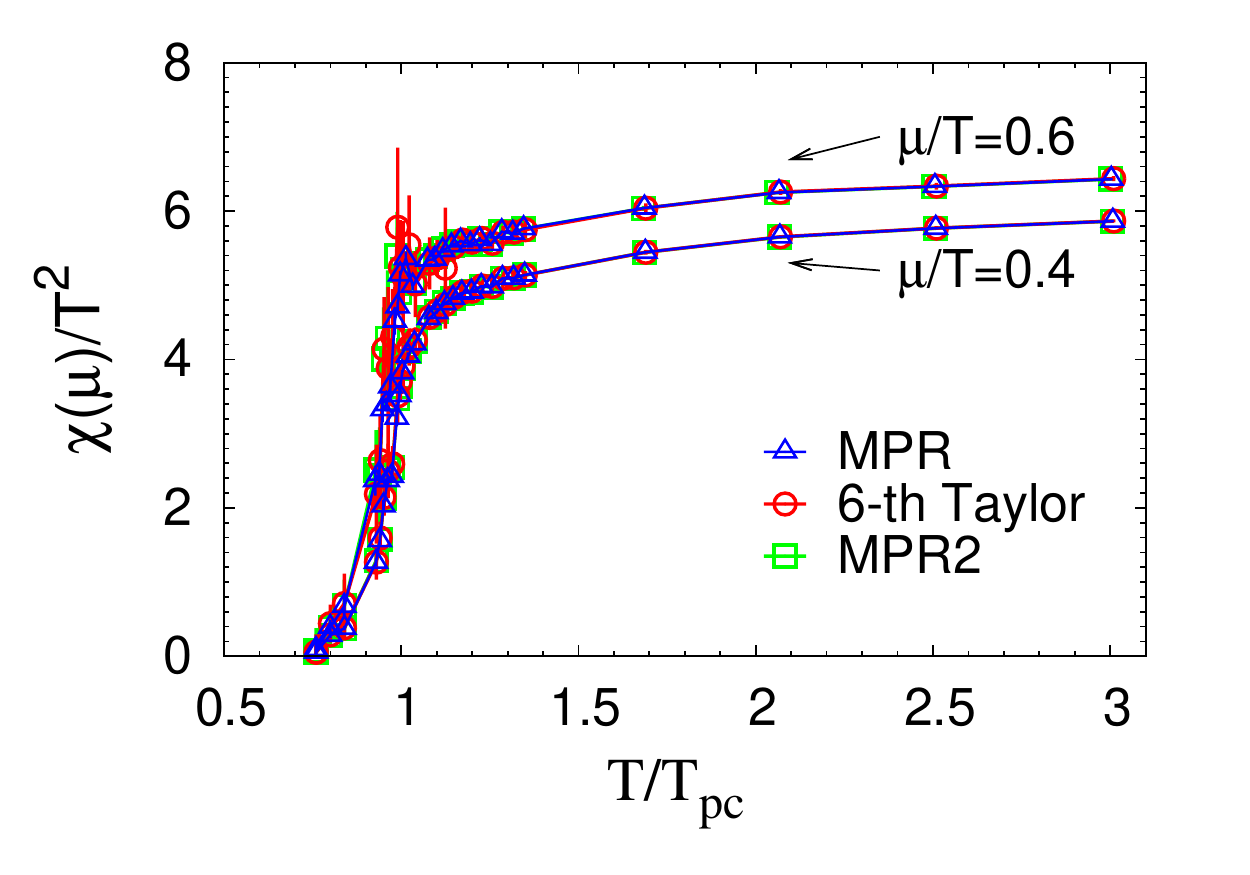}
\begin{minipage}{0.9\linewidth}
\caption{\small 
Thermodynamic quantities  for $8^3\times 4$ lattice, obtained from Taylor expansion (circle), MPR with the minimum-fluctuation condition (triangle) and MPR (square)~\cite{Nagata:2012pc}.
(In the upper left panel, $\delta p(\mu)$ denotes $p(\mu)$ in this article.)
}\label{fig:Sep262011Fig1}
\end{minipage}
\end{figure}
\fi

Now the Taylor coefficients $c_n$ are obtained, so that the pressure, quark number density, and quark number susceptibility can be calculated. In Fig.~\ref{fig:Sep262011Fig1}, these quantities calculated via the Taylor expansion method and the multi-parameter reweighting (MPR) method are compared. The highest order included in the calculation is  $n=10$ for pressure and quark number density and $n=6$ for the quark number susceptibility. Two methods are consistent up to $\mu /T \sim 0.8$ for the EoS and quark number density, and up to $\mu /T \sim 0.6$ for the quark number susceptibility. Higher-order coefficients have large error near $T_c$. The factorial factor $ (x ^ n) ^ {(m)} = n (n-1) \cdots (n-m + 1) x ^ {nm} $ makes the contribution of such higher $c_n$ becomes strong, and large error near $T_c$ can be seen in Fig.~\ref{fig:Sep262011Fig1} as well.

Based on the consistency between the results obtained by using these two methods, some of the potential biases can be excluded. While the truncation of higher-order terms of $\mu$ can introduce a bias to the Taylor expansion, the MPR has no truncation error because the fermion determinant is treated exactly. While the MPR can be biased by the overlap problem, the Taylor coefficient does not have the overlap problem because it uses the importance sampling at $\mu=0$. The agreement between two independent results with different kinds of potential biases suggests that the truncation error of the Taylor expansion and the overlap problem in MPR are negligible. The agreement can be seen up to $\mu/T=0.6\sim 0.8$. We can reasonably trust the simulation results in this parameter region.

\subsection{Discussions regarding the phase transition}
\iffigure
\begin{figure}[htbp] 
\centering
\includegraphics[width=6cm]{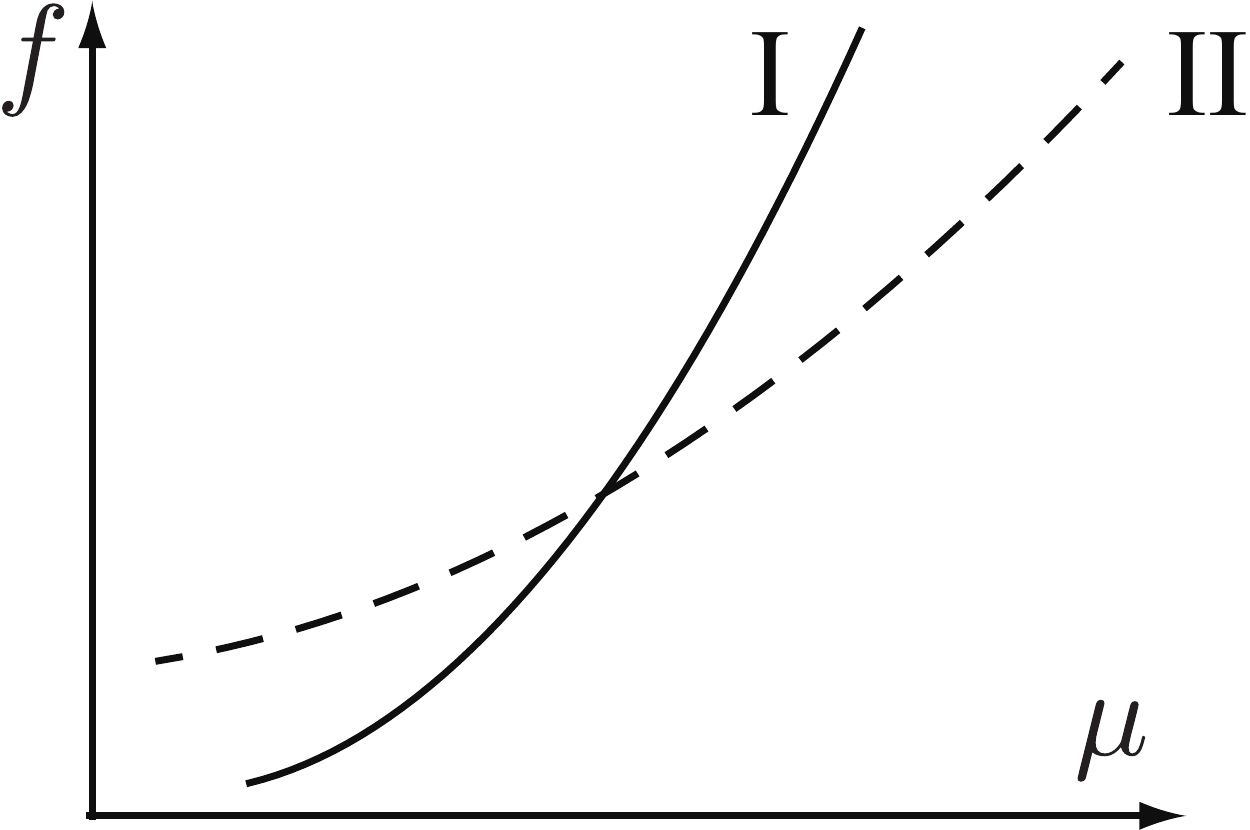}
\begin{minipage}{0.9\linewidth}
\caption{\small 
Schematic picture of the example in which the free energy has a cusp.
There are two phases (I) and (II), whose free energies are shown by solid and dashed lines. There is a possibility that the phase transition point cannot be identified even if Taylor expansion is performed in one state and the radius of convergence is determined. This situation is seen in the case of the Roberge-Weiss phase transition at an imaginary chemical potential.
}
\label{fig:PTCasp}
\end{minipage}
\end{figure}
\fi 

There are several ways to determine the radius of convergence. By using the d'Alembert method, the radius of convergence in~\eqref{eq:pres_taylor} is given by
\begin{align}
r = \lim_ {n \to \infty} \left | \frac{c_n}{c_ {n + 1}} \right |.
\label{eq:Dalambere}
\end {align}
The boundary between the QGP phase and the hadron phase is often estimated by using this criterion. There are two remarks regarding this method.

Firstly, to determine the radius of convergence, it is necessary to calculate $ c_n $ to a sufficiently high order. As we have seen in Fig.~\ref{fig:Taylor_coeff}, the Taylor series converges slowly at $ T <T_c $ where the first-order phase transition may exist. Clearly, $n=10$ is not enough. It is necessary to check whether the convergence of the expansion when calculating the radius of convergence.

Secondly, $ r $ and $ \mu_c / T $ are not always the same. Generally, we can only say $\mu_c/T\le r$ when the transition is of first order. Imagine there are two phases (I) and (II) whose free energies are given by $f_ {\rm I} (\mu)$ and $f_ {\rm II} (\mu)$ (Fig.~\ref{fig:PTCasp}). When Taylor expansion is performed assuming either state, the function only has to be defined in the region including the phase transition point, and there is no reason that the expansion fails at the phase transition point. Even if the free energies are analytic at the phase transition point, it is possible to have a cusp-type non-analyticity at the transition point. As shown in Fig.~\ref{fig:PTCasp}, even if both $ f _{\rm I}$ and $f _{\rm II} $ are well-defined at any $ \mu $, the system is in phase I when $ f _ {\rm I} <f _ {\rm II} $, in phase II for $ \mu $ such that $ f _ {\rm I}> f _ {\rm II} $, and a first-order phase transition takes place at the point where $ f _ {\rm I} = f _ {\rm II} $. In such cases, the radius of convergence of the Taylor expansion of $ f _ {\rm I}$ and $f _ {\rm II} $ are not related to the phase transition point. The Roberge-Weiss phase transition described in the next section~\cite{Roberge:1986mm} is an example of such cusp-type phase transition. The Roberge-Weiss phase transition is a phase transition in high-temperature QCD which occurs when the chemical potential is pure imaginary. The free energy in the QGP phase can be approximated by a quartic function of the chemical potential such as \eqref{eq:freene_freegas} given by the quark free gas model. In the case of \eqref{eq:freene_freegas}, of course, the radius of convergence is infinite. Still, there is a first-order phase transition, and this phase transition cannot be found by looking at the radius of convergence of the Taylor series.

With the Taylor expansion method, although it is difficult to investigate a large area of $\mu $, for small $\mu$ it is possible to perform high-precision calculations on the same level as ordinary lattice QCD simulations since the calculation method is the same as the ordinary lattice QCD simulation. An early study of the Taylor coefficient was performed for the staggered fermions by Allton et al.. It is calculated up to the $4$th order in Ref.~\cite{Allton:2002zi}, and up to the $6$th order in Ref.~\cite{Allton:2003vx, Allton:2005gk}. The calculation for Wilson fermion is performed up to the $4$th order in Ref.~\cite{Ejiri:2009hq} and up to the $10$th order in Ref.~\cite{Nagata:2012pc}. The noise method is used in all works except for Ref.~\cite{Nagata:2012pc}. In Ref.~\cite{Nagata:2012pc}, the exact calculation via the reduction formula was performed. In the calculations using the noise method, it is rather easy to increase the lattice volume while it is hard to calculate the high-order terms. On the other hand, with the reduction formula, the higher-order terms can be calculated while it is difficult to increase the lattice volume.

\clearpage
\section{Analytic continuation from imaginary chemical potential}
\label{sec:imaginary}

The sign problem arises when $ \det \Delta (\mu), (\mu \neq 0) $ takes a complex value. If the chemical potential is pure imaginary ($ \mu = i \mu_I, \mu_I \in \mathbb{R} $), the fermion determinant takes a real value. This can be easily shown as follows. 
\begin{align}
\gamma_5 \Delta^\dagger(\mu = i \mu_I) \gamma_5 & = \gamma_5 ( D_\nu \gamma_\nu + m +  i \mu_I \gamma_4)^\dagger \gamma_5, \nn \\
                                      & = \gamma_5 ( - D_\nu \gamma_\nu + m - i \mu_I \gamma_4) \gamma_5, \nn \\
                                      & = D_\nu \gamma_\nu + m + i \mu_I \gamma_4 , \nn \\
                                      & = \Delta(\mu=i \mu_I).
\label{eq:g5hermit_pi}
\end{align}
From this we can show $ (\det \Delta (\mu)) ^ * = \det \Delta (\mu) $. Hence the fermion determinant $ \det \Delta (\mu = i \mu_I) $ is real.
The expression (\ref{eq:g5hermit_pi}) is for continuous spacetime, but the same holds for the lattice as well. 

In the case of lattice QCD, the physical interpretation is also clear. The chemical potential is introduced with the link variable along the fourth direction in the fermion determinant as
\[
U_ {n 4} \to e ^ {\mu a} U_ {n 4}, 
\qquad
U_ {n 4} ^ \dagger \to e ^ {-\mu a} U_ {n 4} ^\dagger
\]
In the loop on the lattice, the difference between $ e ^ {\mu a} $ and $ e ^ {-\mu a} $ changes the relative size of the forward-loop and backward-loop winding the temporal direction.
It breaks the cancellation of their phases.
On the other hand, if $ \mu = i \mu_I $ is pure imaginary, 
\[
e ^ {-\mu a} U_ {n 4} ^ \dagger = e ^ {-i \mu_I a} U_ {n 4} ^ \dagger = (e ^ {i \mu_I a} U_ {n 4}) ^ \dagger.
\]
In this case, both forward- and backward-loops are Hermitian conjugate to each other, and the phase cancellation mechanism is restored.

Since the fermion determinant is real, the reweighting method is applicable. Then we can perform the analytic connection to real chemical potential. In this section, we explain how to investigate the QCD phase diagram based on the phase diagram at an imaginary chemical potential.

\subsection{Phase diagram with imaginary chemical potential}
\label{sec:imag_phase}
\iffigure
\begin{figure}[htbp] 
\begin{center}
\includegraphics[width=7cm]{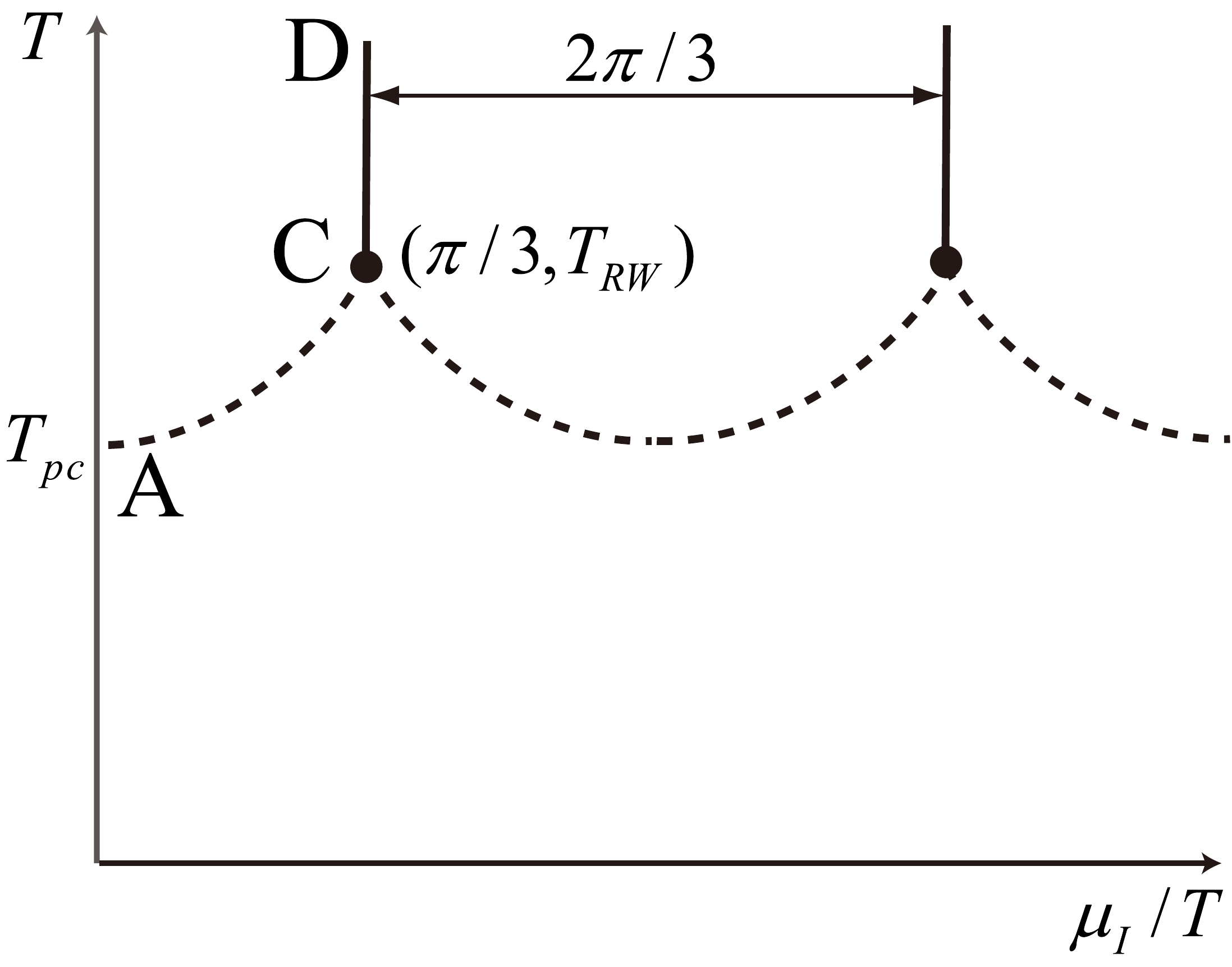}
\begin{minipage}{0.9\linewidth}
\caption{\small 
QCD phase diagram on the $\ds{\frac{\mu_I}{T}}$-$T$ plane. There exists the periodicity with respect to $\ds{ \frac{\mu_I}{T}} $, with the period $\ds{ \frac{2\pi}{3} }$. The half-line CD lies on $ \ds {\frac {\mu_I} {T} = \frac{\pi}{3}}$. AC and CD represent phase boundaries, the order of which depends on quark mass and flavor. At or near the physical point, the line segment CD indicates the first-order phase transition, at point C the second-order phase transition takes place, and the curve AC shows the crossover.
}
\label{fig:phasediagram_withmu}
\end{minipage}
\end{center}
\end{figure}
\fi 

To carry out the analytic continuation from the imaginary chemical potential, it is necessary to understand the phase diagram of the imaginary chemical potential region. This is because the analyticity of thermodynamic quantities is limited to the region without phase transition.

The qualitative aspects of the phase structure of the imaginary chemical potential region are almost completely understood. Fig.~\ref{fig:phasediagram_withmu} represents the phase diagram on the $(\mu_I / T, T)$ plane. The phase diagram depends on the quark mass $m$ and the number of flavors $N_f$. Fig.~\ref{fig:phasediagram_withmu} shows the phase diagram for a parameter set close to the real world. One of the characteristics of this phase diagram is the periodicity with respect to the imaginary chemical potential. The QCD partition function does not change when $(2\pi T i)/N_c$ is added to the chemical potential ~\cite{Roberge:1986mm},
\begin {align}
Z \biggl (\mu + \frac{2 \pi i}{N_c} T \biggr) = Z (\mu).
\end {align}
This corresponds to the periodicity of the phase diagram with respect to $ \ds{\frac{\mu_I}{T}} $ with the period $ \ds{\frac{2 \pi}{N_c}} $. 
In general, when the chemical potential is a pure imaginary number ($ \mu = i \mu_I $), the grand canonical partition function is expressed as
\[
Z ( \mu_I) = {\rm tr} e ^ {-(\hat{H} -i \mu_I \hat{N}) / T}.
\]
Because the eigenvalues of the number operator $ \hat{N} $ are integers, the periodicity with the period $2\pi$ for $ \ds {\frac{\mu_I}{T}} $ exists trivially. That the period is $ \ds {\frac{2 \pi}{N_c}} $ is related to the $\mathbb{Z}_{N_c}$ center symmetry of SU ($ N_c $) Yang-Mills theory. The center symmetry breaks explicitly when quarks are introduced, but when $ \mu_I / T $ is shifted by $ \ds{\frac{2 \pi }{N_c}} $, the breaking of $ Z_{N_c} $ and the change of chemical potential balance and cancel with each other. Therefore the invariance under the combination of $\mathbb{Z}_{N_c}$ transformation and the shift of $ \mu_I / T$ in the unit of $\ds{\frac{2 \pi }{N_c}} $ appears. This periodicity is called Roberge-Weiss periodicity (see \S.~\ref{sec:Z3RW}). Fig.~\ref{fig:phasediagram_withmu} shows the periodicity explicitly by taking the horizontal axis to be $\mu_I/T$.

Once we understand the phase structure of the first domain, $ 0 \le \mu_I / T \le 2 \pi / 3 $, the other domain can be understood using the periodicity. As shown in the figure, the first domain is divided into three areas by two phase boundaries. As in the case of $\mu\in\mathbb{R}$, there is the hadron phase at low temperature and the QGP phase at high temperature, and the curve AC represents the phase boundary. AC is known to be a crossover with a parameter set that is at or near the physical point. The property of this phase boundary depends on $m$ and $N_f$. For example, the first-order is seen in the heavy quark limit. On the other hand, the large- and small-$ \mu_I $ regions are separated by the first-order phase transition line (segment CD) if the temperature is sufficiently high. This phase transition is associated with a discontinuous change in the phase part of the Polyakov loop and is called the Roberge-Weiss (RW) phase transition. The RW phase transition is caused by a transition among the three lowest energy states when the center symmetry is spontaneously broken. Therefore, it exists only in the QGP phase where the center symmetry is spontaneously broken. The basic properties of such phase structures have been clarified by Roberge and Weiss ~\cite{Roberge:1986mm}. The lattice QCD calculation revealed the details of the phase structure.

\iffigure
\begin{figure*}[htbp]
\centering
\includegraphics[width=0.45\linewidth]{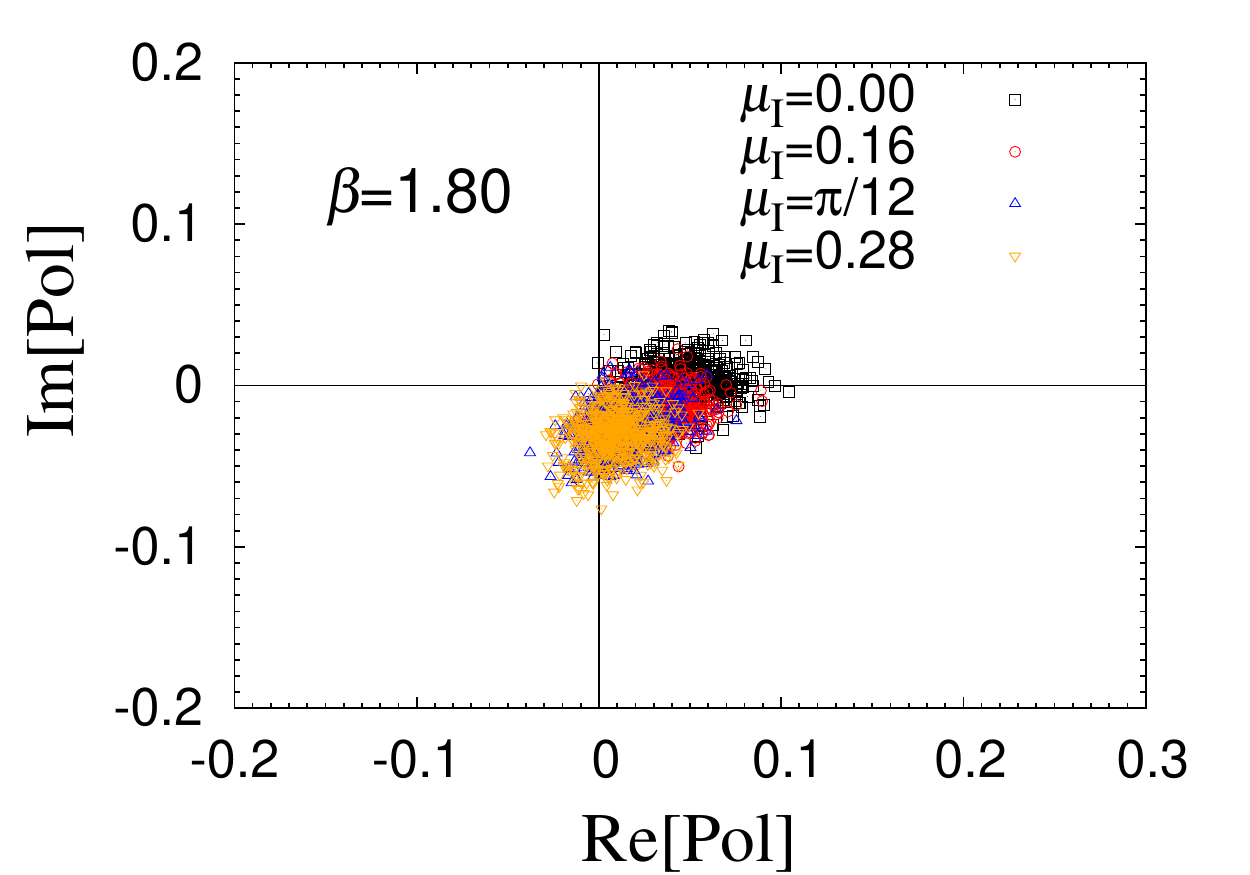}
\includegraphics[width=0.45\linewidth]{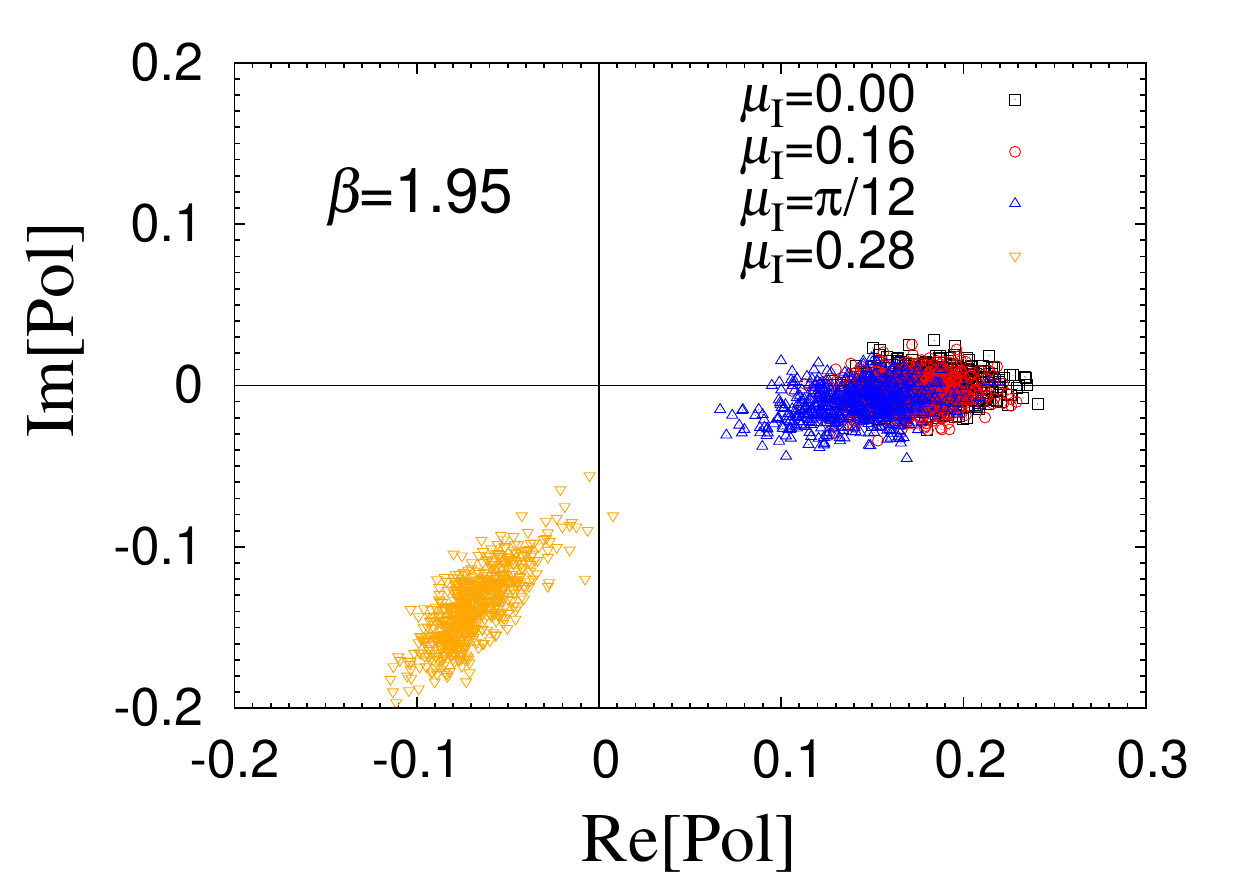}
\begin{minipage}{0.9\linewidth}
\caption{\small 
The scatter plots of the Polyakov loop~\cite{Nagata:2011yf}. [Left] $ \beta= 1.80$ (below $T_{pc}$). [Right] $\beta = 1.95$ (above $T_{RW}$). Here $\mu_I$ denotes the chemical potential in the lattice unit, i.e., $\mu_I a$. 
The lattice size is $8^3\times 4$. 
}
\label{Mar0411fig1}
\end{minipage}
\end{figure*}
\fi

To investigate this phase structure by lattice QCD simulation, it is convenient to calculate the Polyakov loop. Fig.~\ref{Mar0411fig1} shows how the scatter plot of the Polyakov loop changes as $\mu_I$ is varied.
The left figure corresponds to the hadronic phase. At $\mu_I=0$, the Polyakov loop is distributed near the origin. Although the center symmetry is explicitly broken because of the dynamical quarks, 
the effect of the breaking is not so large and the Polyakov loop takes a small value, although not zero. As $\mu_I$ is increased, the Polyakov loop moves to the bottom of the real axis. This change is smooth under the change of $\mu_I$, no discontinuity is seen. On the other hand, in the QGP phase (right figure), the approximate center symmetry is spontaneously broken, and the value of the Polyakov loop becomes large.
In this case, even if $\mu_I$ is increased, the Polyakov loop remains on the real axis up to a certain value. At $\mu_I=0.28$, the Polyakov loop jumps to the third quadrant. The $\mu_I$ where the RW phase transition occurs is given by $\ds {\frac {\mu_I}{T} = \frac{\pi}{3}}$~\cite{Roberge:1986mm}. The figure is obtained by using the $ 8 ^ 3 \times 4 $ lattice, with 
\[ \mu_I a = \frac{\pi}{3 N_t} \sim 0.26 \].
\iffigure
\begin{figure}[htbp]
\centering
\includegraphics[width=0.49\linewidth]{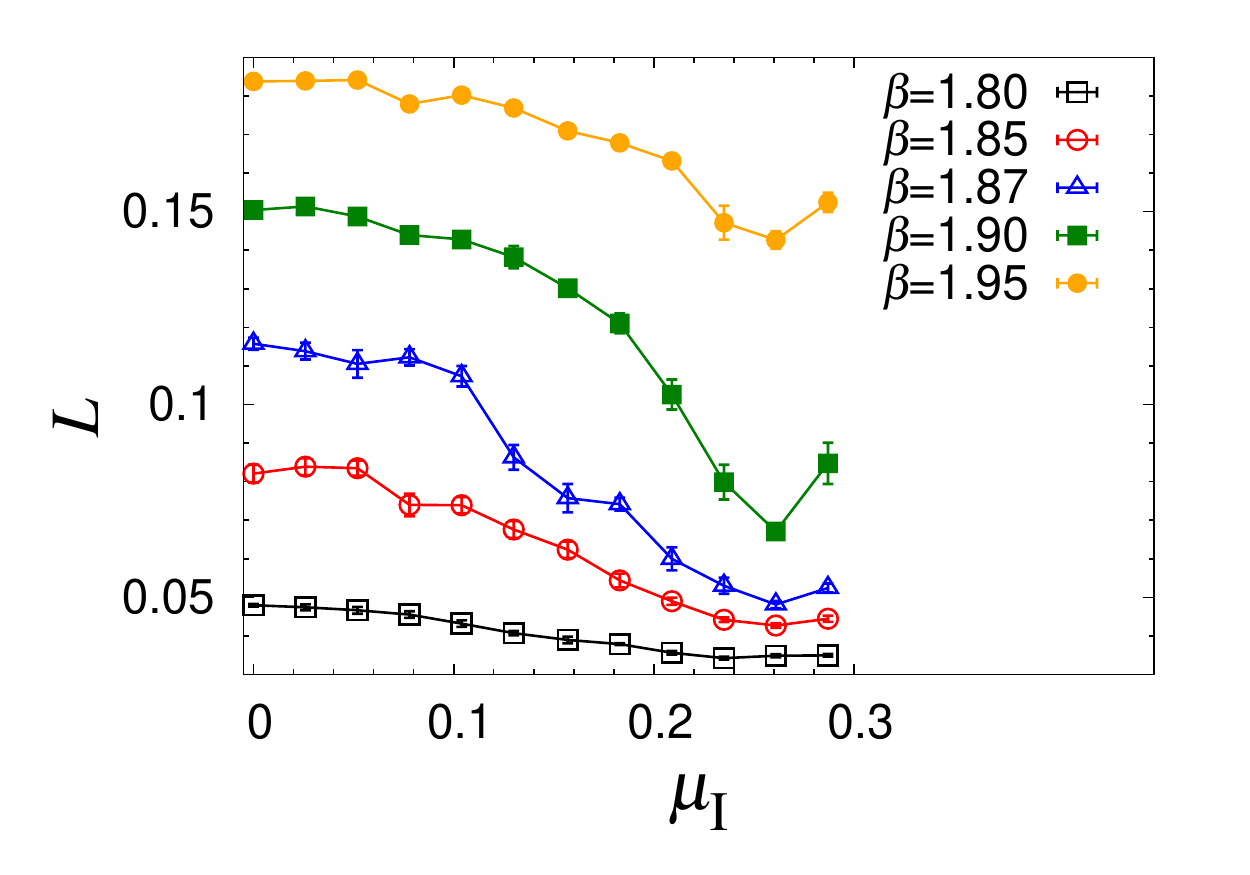}
\includegraphics[width=0.49\linewidth]{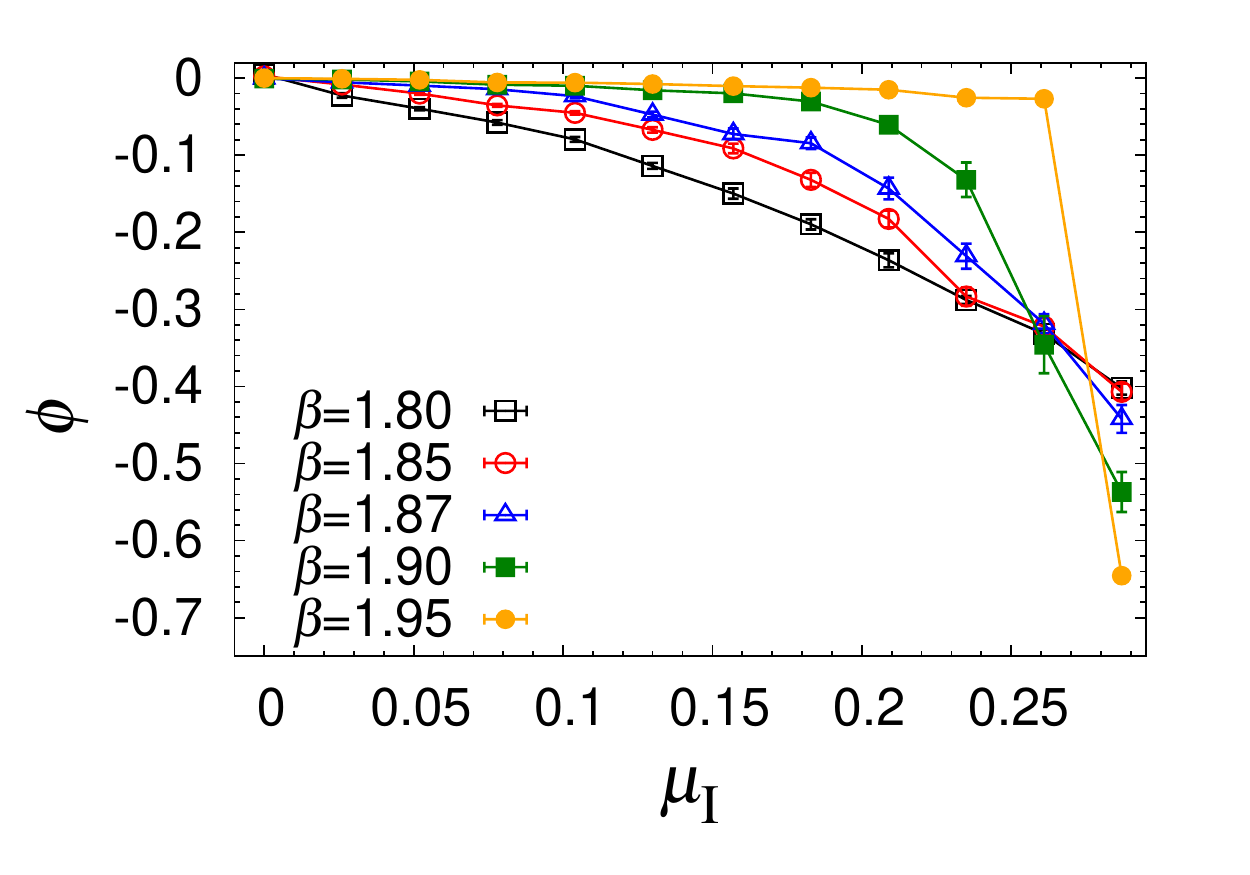}
\begin{minipage}{0.9\linewidth}
\caption{\small 
The $\mu_I$-dependence of the Polyakov loop~\cite{Nagata:2011yf}. 
[Left] The absolute value $L$. 
[Right] The phase $\phi$.
The lattice size is $8^3\times 4$. 
}\label{Jan2111fig4}
\end{minipage}
\end{figure}
\fi

Fig.~\ref{Jan2111fig4} shows the expectation value of the Polyakov loop.
As expected from the scatter plots, non-analytic behavior is observed in the phase of the Polyakov loop at high temperature ($\beta=1.95$). It indicates that in the QGP phase, the first-order phase transition occurs as $ \mu_I $ increases. This is the RW phase transition, which occurs on the line CD in Fig.~\ref{fig:phasediagram_withmu}. The existence of a critical point is suggested on $ \frac{\mu_I}{T} = \frac{\pi}{3} $ since the RW phase transition exists in the QGP phase and not in the hadron phase. This is the endpoint of the Roberge-Weiss phase transition line and is called the Roberge-Weiss endpoint (point C in Fig.~\ref{fig:phasediagram_withmu}). The phase boundary AC and RW endpoints will be explained in the following sections.

\subsection{Determination of the pseudo-critical line -- a study of the QCD phase diagram via the analytic continuation}

\subsubsection{How the analytic continuation works}
\iffigure
\begin{figure}[htbp] 
\centering
\includegraphics[width=6cm]{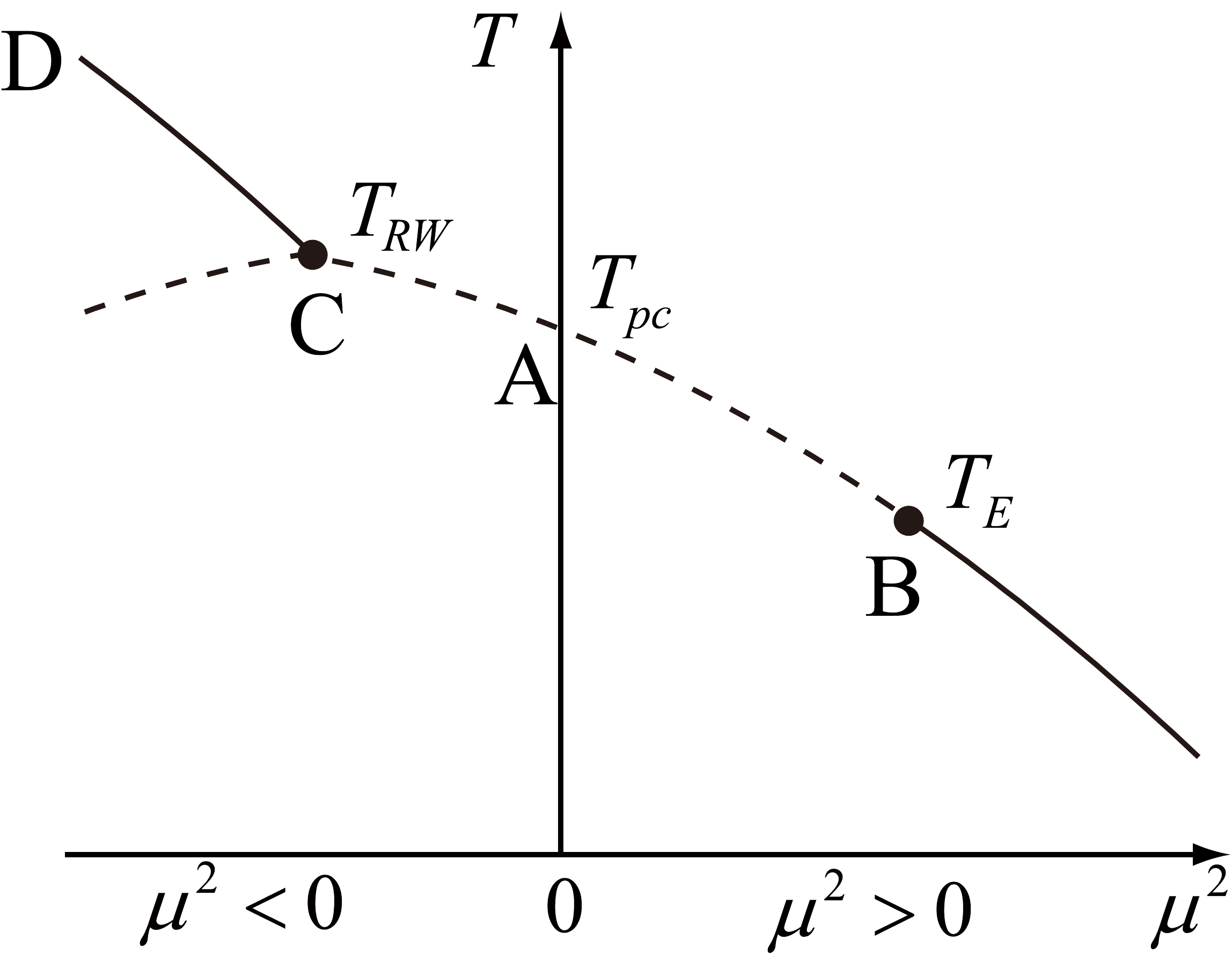}
\begin{minipage}{0.9\linewidth}
\caption{\small 
Schematic picture of the QCD phase diagram on $\mu^2$-$T$ plane.
The regimes of $\mu^2 >0, \mu^2 < 0$ correspond to the real and pure imaginary chemical potential, respectively.}
\label{fig:phasediagram_withmu2}
\end{minipage}
\end{figure}
\fi

Now, we explain how to obtain the information about the real chemical potential from the imaginary chemical potential. This idea is based on the analyticity of the thermodynamic quantities. We promote the chemical potential to a complex number $\mu\in\mathbb{R}\to\mu=\mu_R+i \mu_I\in\mathbb{C}$, and then consider the analytic continuation of the action and physical quantities, $S(\mu)\to S(\mu_R + i \mu_I)$, $\calO (\mu) \to \calO (\mu_R + i \mu_I)$. The action and physical quantities satisfy
\[
\frac {\partial S} {\partial \bar {\mu}} = 0, \frac{\partial \calO} {\partial \bar {\mu}} = 0,
\]
so that they are described by regular functions of $\mu$. The expectation value $\langle\calO\rangle$ is defined as
\[
\langle \calO \rangle = Z (\mu) ^ {-1} \int {\cal D} U \calO (\mu) e ^ {-S (\mu)}, \, \, \mu = \mu_R + i \mu_I \in \mathbb {C},
\]
hence it satisfies the Cauchy-Riemann relation except for the point with $Z=0$, which corresponds to the phase transition point. A thermodynamic quantity is expressed by a regular function of the complex chemical potential except for the phase transition point. There is no thermodynamic singularity near the temperature axis ($\mu=0$) on the QCD phase diagram for the physical setup or parameters close to it.
Therefore, the thermodynamic quantity is an analytic function near the temperature axis. Since $ \langle \calO \rangle $ is a regular function of $ \mu $, it is infinitely differentiable and the Taylor expansion can be performed,
\begin {align}
\langle \calO \rangle (\mu) = \sum_{n} c_n(T) \left( \frac{\mu}{T}\right)^n, \, \mu \in \mathbb{C}.
\end {align}
The coefficient $ c_n $ can be determined using the ordinary Monte Carlo method in the imaginary-chemical-potential region.

The phase diagrams for the real and imaginary chemical potentials are connected around $\mu=0$ via the analytic continuation. The horizontal axis in Fig.~\ref{fig:phasediagram_withmu2} is $\mu^2$. The regions $\mu^2<0$ and  $\mu^2\ge 0$ correspond to the imaginary and real chemical potential, respectively, and they are smoothly connected. In the previous section, we drew a curve going upward from point A to point C. This behavior can be understood from Fig.~\ref{fig:phasediagram_withmu2}, it is because the phase boundary continues to the real-chemical-potential region. By utilizing the analytic continuation between two regions, it is possible to approach from the imaginary chemical potential to the real chemical potential.

This method is similar to the Taylor expansion method in that both of them rely on the analyticity. However, while the Taylor expansion uses the coefficients calculated at $\mu=0$, the imaginary-chemical-potential method uses the imaginary-chemical-potential region to determine the expansion coefficients; because more information is used, the coefficients can be determined more precisely. This is one of the advantages of the imaginary-chemical-potential method. Note also that, thanks to the Roberge-Weiss periodicity, in the hadron phase, the functional form of the physical quantity can be narrowed down to some extent due to the periodicity. From a theoretical point of view, yet another advantage is that the definition of analyticity and non-analyticity becomes mathematically clear since the analyticity of the function is defined by the Cauchy-Riemann relation associated with the complex extension of the chemical potential.

\subsubsection{Determination of the pseudo-critical line via the analytic continuation}

\iffigure
\begin{figure*}[htbp]
\centering
\includegraphics[width=0.49\linewidth]{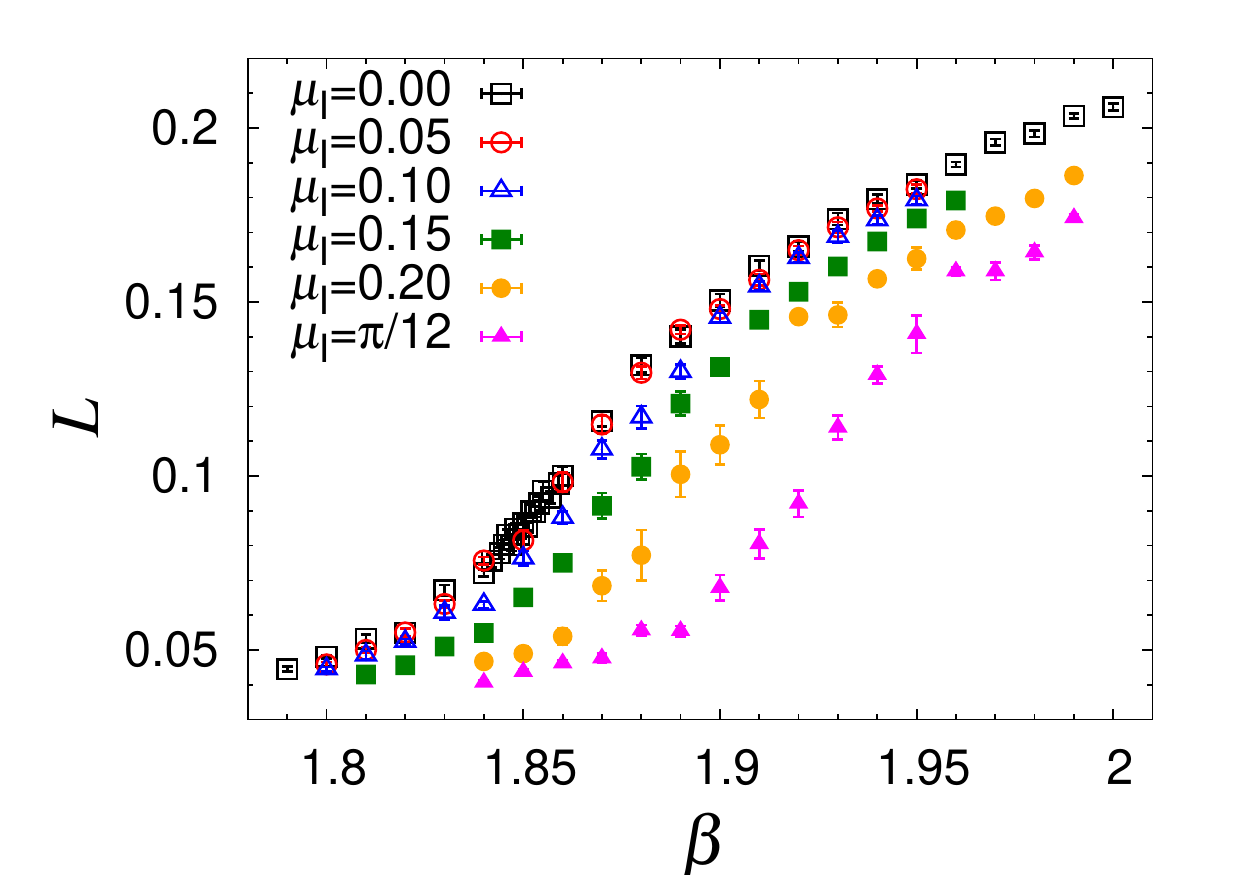}
\includegraphics[width=0.49\linewidth]{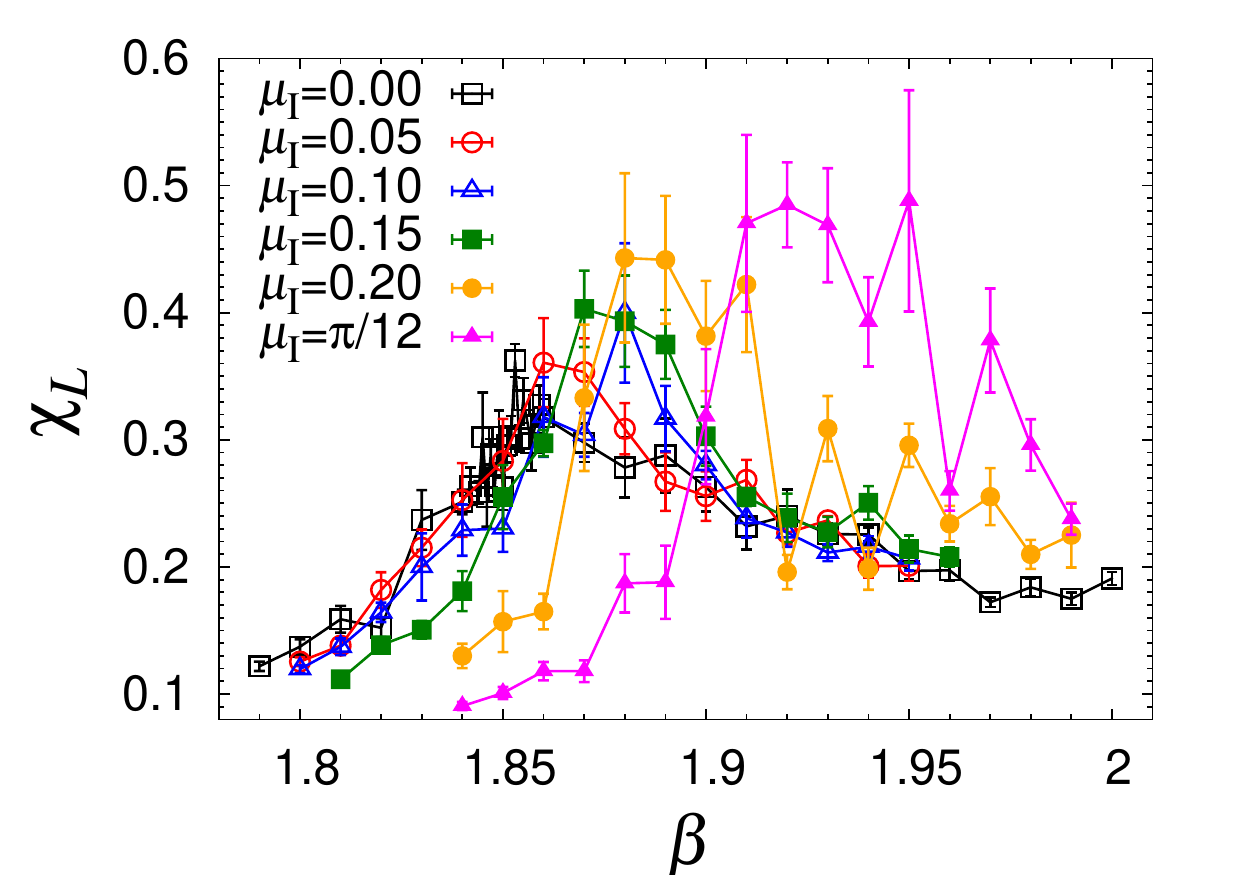}
\begin{minipage}{0.9\linewidth}
\caption{\small 
The $ \beta$ dependence of the absolute value of the Polyakov loop. The left and right panels are the expectation value and susceptibility, respectively. The values of  $\mu_I$ in the plots are the chemical potential in lattice units, i.e., $\mu_I a$. In the left panel, the slope increases in a certain range of $ \beta $, and in the same range of $\beta$, a peak appears in the right figure. By regarding the absolute value of the Polyakov loop as the order parameter of the confinement and deconfinement phases, the location of the peak can be interpreted as the phase transition temperature. The figure shows the simulation result at $N_t=4$ and $\mu_I=\pi/12$ corresponds to $\mu_I/T=\pi/3$. Except for $\mu_I=\pi/12$, the change of the Polyakov loop is smooth and can be interpreted as a crossover. $\mu_I/T=\pi/3$ is on the RW phase boundary and corresponds to the second-order transition. (Finite volume scaling is required to determine the order of the phase transition from the simulation data).
The lattice size is $8^3\times 4$. 
}\label{Jan1011fig2}
\end{minipage}
\end{figure*}
\fi

\iffigure
\begin{figure*}[htbp]
\centering
\includegraphics[width=0.49\linewidth]{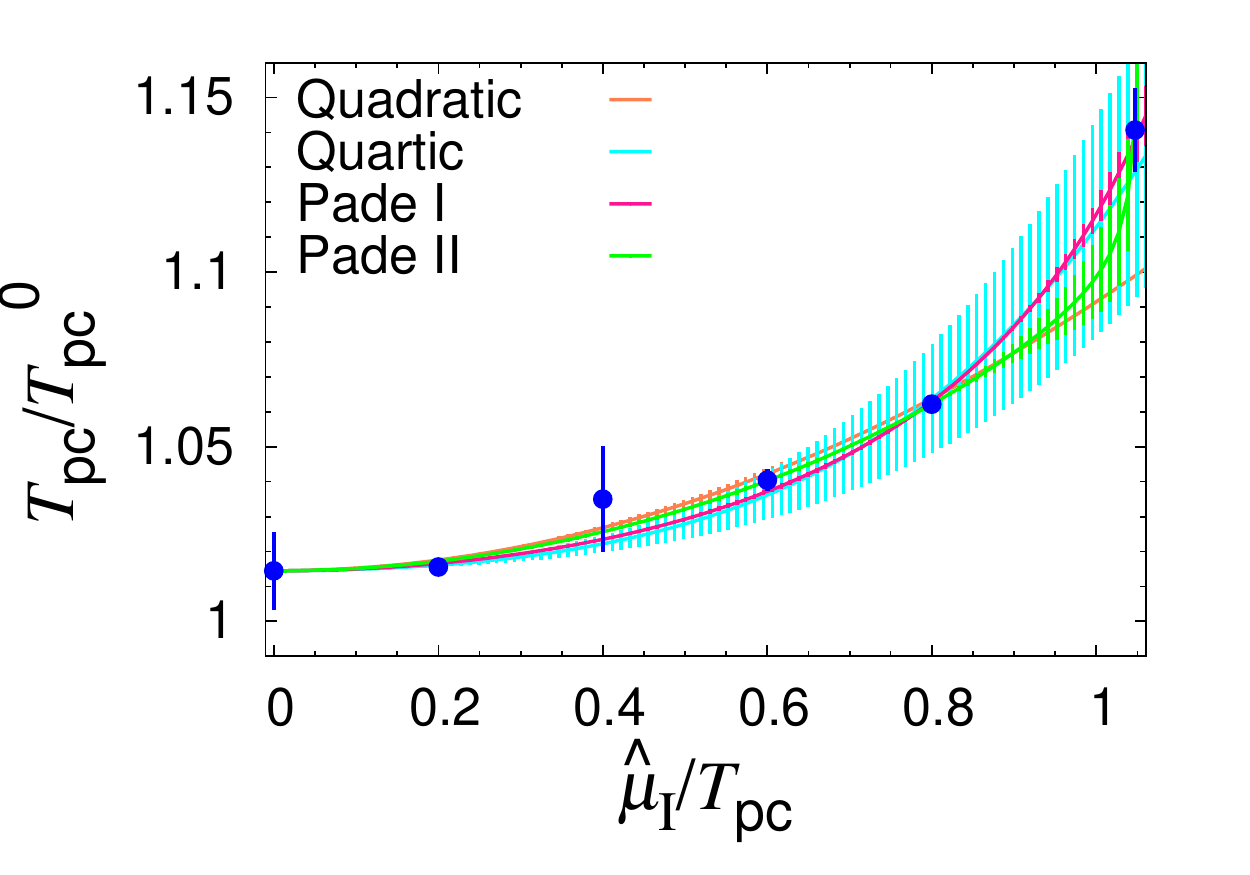}
\includegraphics[width=0.49\linewidth]{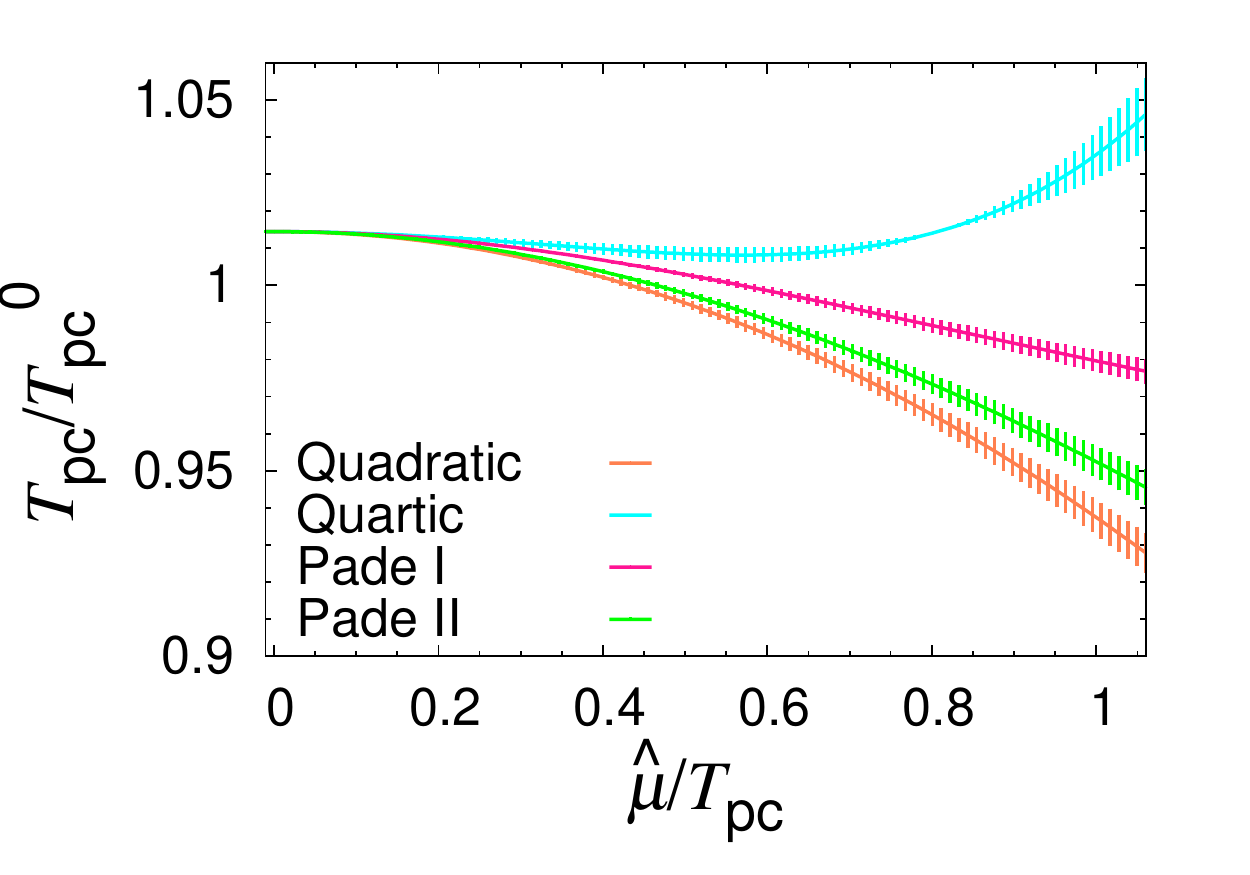}
\begin{minipage}{0.9\linewidth}
\caption{\small 
[Left] The phase boundary between the confined and deconfined phases in the imaginary chemical potential regime. The position of the peak of the susceptibility in Fig.~\ref{Jan1011fig2} is shown. The error appears when the peak is not very sharp. The solid line represents the result of the fit of the data, and the error bars are shown as well. [Right] The result after the analytic continuation to the real chemical potential regime.
The lattice size is $8^3\times 4$. 
}\label{Feb2711fig2}
\end{minipage}
\end{figure*}
\fi

One of the applications is the determination of the pseudo-critical phase transition line. The line segment AB can be obtained by finding AC on the imaginary-chemical-potential region and then analytically continue to the real-chemical-potential region.

Here, we define the phase transition line (AC) based on the peak of the susceptibility of the Polyakov loop. Although the Polyakov loop is not an exact order parameter (see \S.~\ref{sec:observables} regarding this point), it is often used as an approximate order parameter because it has a small value in the hadron phase and a large value in the QGP phase.
In Fig.~\ref{Jan1011fig2}, the temperature dependence of the expectation value and susceptibility of the Polyakov loop are shown for several values of $\mu_I/T$. Fixing $\mu_I/T$ and changing the temperature corresponds to scanning the phase diagram shown in Fig.~\ref{fig:phasediagram_withmu} along the vertical direction. Because the temperature changes monotonically with $\beta$, the horizontal axis can be regarded as the temperature but different scale (see Table~\ref{tab:avsT}). The absolute value of the Polyakov loop $L$ increases as temperature goes up, as shown in the left panel of Fig.~\ref{Jan1011fig2}.
The susceptibility $\chi_L$ shown on the right panel, and $\chi_L$ exhibits a peak, whose location coincides with the point where $L$ changes most rapidly. The position of the peak shifts to larger $\beta$, namely the higher temperature, as $\mu_I$ increases. This means that the transition temperature rises as $\mu_I$ becomes large. That the curve AC in Fig.~\ref{fig:phasediagram_withmu} goes upward from point A to point C corresponds to this result.

Next, let us determine the transition point $\beta_c$ for each value of $\mu_I$. The value of $\beta_c$ is defined by the position of the peak of the Polyakov loop susceptibility. We determine it by fitting the vicinity of the peak with an appropriate functional form $\chi_{\rm fit}(\beta,s)$. Here $s$ represents the parameter set included in the function, and the best fit value of it is determined by minimizing the difference between the function and the data, $\chi^2$.

The data points with error bars in the left panel of Fig.~\ref{Feb2711fig2} represent the location of the peak obtained by the fit. The critical temperature on the vertical axis, which can be obtained from $\beta$, in Fig.~\ref{Feb2711fig2}.
Next, we determine the phase boundary line $ T_c (\mu_I) $ using the data of $T_c$ for each $\mu_I$. The functional form is unknown; here, we assume, for example, 
\begin {align}
T_c (\mu_I) = a_0 + a_2 \mu_I ^ 2 + a_4 \mu_I ^ 4 + \cdots.
\label{eq:2017Jul07eq1}
\end {align}
The phase boundary is an even function of $\mu$ since QCD has CP invariance. The higher-order term of expansion is suppressed by $\mu_I^2<0.1$ since the domain of $\mu_I$ is $0\le\mu_I\le\pi/(3 N_t)<0.3$. Therefore, although the series converges well, it is difficult to determine the higher-order coefficients precisely. To circumvent this difficulty, a method to find higher-order terms precisely using the Pad\'e approximation assuming the rational function type $P(x)/Q(x)$ has also been proposed~\cite{Cea:2007vt}. In Fig.~\ref{Feb2711fig2}, four types of functions (quadratic, quartic, and two types of Pad\'e approximation) were used. The quadratic function has a large deviation from the data points, while the quartic and Pad\'e approximation type I show good agreement with the data.

The next step is the analytic continuation, $i \mu_I\to\mu=\mu_R+i\mu_I$. We write the phase boundary as in a power series of $\mu^2$, for example:
\begin {align}
T_c (\mu) = a_0-a_2 \mu ^ 2 + a_4 \mu ^ 4-\cdots.
\label{eq:2017Jul07eq2}
\end {align}
The same expression~(\ref{eq:2017Jul07eq2}) gives the boundary line between the hadronic and QGP phases in the real chemical potential region as well. Comparing Eqs.~(\ref{eq:2017Jul07eq1}) with \eqref{eq:2017Jul07eq2}, the sign of the coefficients in the odd-order terms of $ \mu ^ 2 $ is opposite. Due to this sign flip, $T_c$ is a decreasing function of in the small-$\mu$ region. This behavior of the hadronic / QGP phase boundary is well-known phenomenologically, and we confirmed it by the non-perturbative analysis using lattice QCD simulations in this way. The right panel in Fig.~\ref{Feb2711fig2} shows the phase boundary on the real-chemical-potential regime. Although the dependence on the fitting ansatz is invisible in the small-$\mu$ regime, in the large-$\mu$ regime we see very different results depending on the fitting ansatz; the deviation among four fitting ans\"{a}tze shown here becomes large $ \mu/T\gtrsim 0.3$. If we compare how well the data points in the imaginary-$\mu$ region can be fitted, both the quartic function and the Pad\'e approximation I work pretty well, and it is difficult to judge which one is better. Therefore, we should interpret that the analytic continuation from imaginary-$\mu$ is not reliable at $\mu/T\gtrsim 0.3$, where the real-$\mu$ region where the fit-function dependence is large.

The results shown above have been carried out by using the two-flavor improved Wilson fermions~\~cite {Nagata: 2012pc}. Let us list other relevant papers. The first lattice QCD simulations for the imaginary chemical potential were performed by D'Elia and Lombardo~\cite{D'Elia:2002gd} and by de Forcrand and Philipsen~\cite{deForcrand:2002ci} by using the four-flavor staggered fermion.
The first calculation for the Wilson fermion was performed by Wu et al.~\cite{Wu:2006su}, in which the naive Wilson fermion without clover term was utilized. The use of the Pad\'e approximation for the fit of the phase boundary was proposed in Ref.~\cite{Cea:2009ba}. After the validity of the Pad\'e approximation was verified in cases without the sign problems such as $ N_c = 2 $ and isospin chemical potential, it was applied to find the phase boundary of QCD~\cite{Cea:2010md}. Recently, a highly accurate calculation of the lowest order term, that is, the curvature $ \kappa $ of the phase boundary, has been performed. The pseudo-critical line $ T_c (\mu_B) $ is determined from the peak of chiral susceptibility,
\begin{align}
\frac{T_c (\mu_B)}{T_c (\mu = 0)} = 1-\kappa \biggl (\frac{\mu_B}{T_c (\mu_B)} \biggr) ^ 2 + \cdots,
\end{align}
where $ \kappa $ is defined as a quadratic coefficient. Here, $ \mu_B $ represents the baryon chemical potential. For $ N_f = 2 + 1 $ flavor QCD, $ \kappa = 0.018 (4) $ (HISQ / tree action discretization, lattice size $ 16^3 \times 6$, $24^3 \times 6$) ~ \cite{Cea:2014xva}, $\kappa=0.0149 \pm 0.0021$ (stout-improved staggered fermion, the continuum extrapolation from $N_t=10,12,16$)~\cite{Bellwied:2015rza} has been reported.
Bonati et al. defined $\kappa$ as 
\begin{align}
\frac{T_c (\mu_B)}{T_c (\mu = 0)} = 1-\kappa \biggl( \frac{\mu_B}{T_c (\mu_B = 0)} \biggr) ^ 2 + \cdots,
\end{align}
and reported $\kappa = 0.0135 (20)$ (stout-improved staggered fermion, the continuum extrapolation from $N_t=6,8,10,12$) \cite{Bonati:2014rfa, Bonati:2015bha}.

\subsubsection{Applications to other quantities}

\begin{figure}[htbp]
\centering
\includegraphics[width=0.5\linewidth]{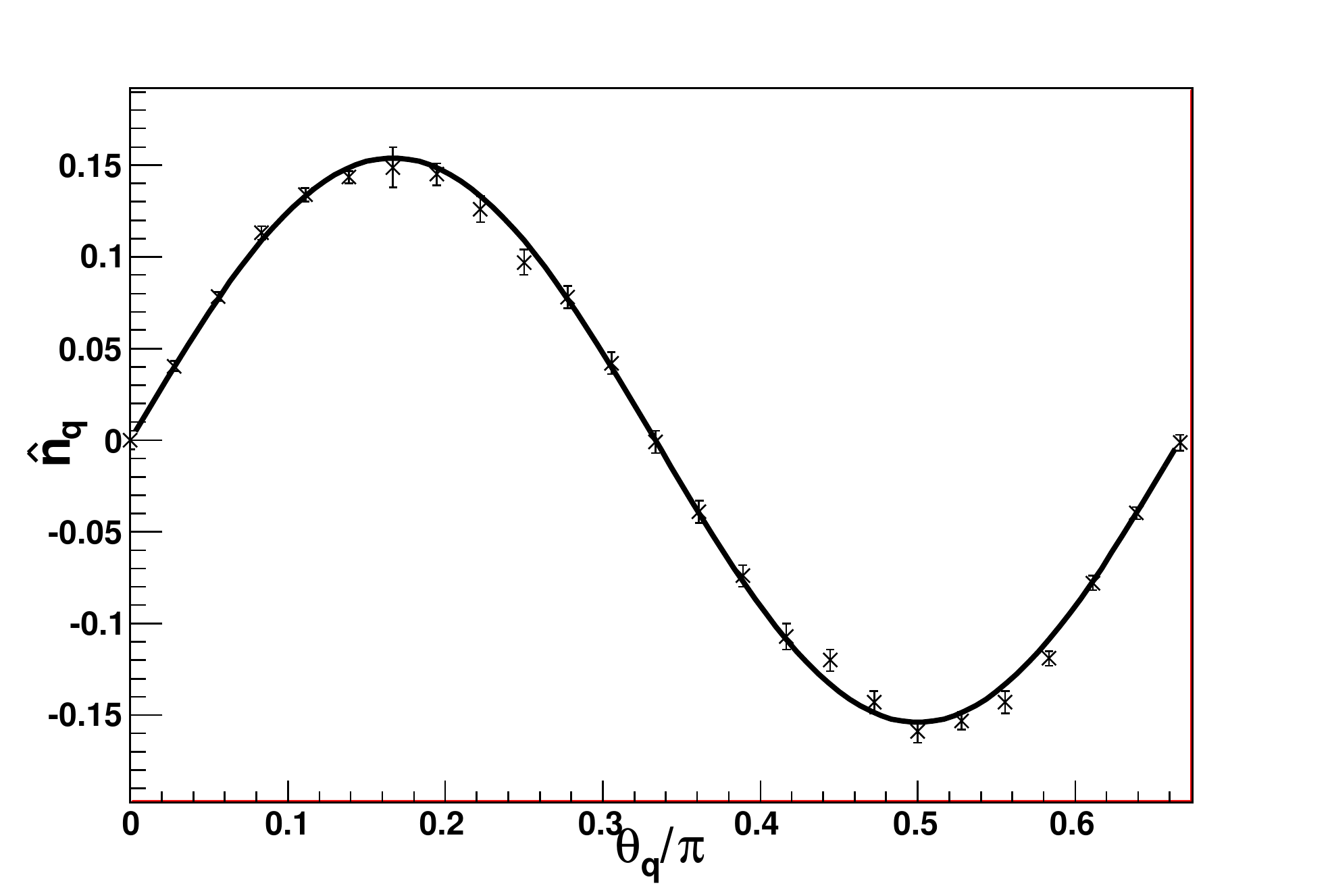}
\begin{minipage}{0.9\linewidth}
\caption{\small 
The chemical-potential dependence of the quark number density obtained by using the imaginary chemical potential at $T=0.9T_c$~\cite{D'Elia:2009tm}.
The solid curve is the same quantity obtained via the hadron resonance gas model. 
The lattice size is $16^3\times 4$. 
}\label{09041400fig1}
\end{minipage}
\end{figure}

The analytic continuation method can be applied to other quantities as well. The procedures are the same as the ones for the pseudo-critical line: the simulation at the imaginary chemical potential, the fitting of the results with some analytic function, and the analytic continuation to the real chemical potential.

This method was applied to thermodynamic quantities such as number density firstly for the $4$-flavor staggered fermions~\cite{D'Elia:2004at,D'Elia:2007ke}, then for the $2$-flavor case at $ m_ \pi \sim 280 $ MeV~\cite{D'Elia:2009tm}. In the hadron phase, the physical quantity can be expressed as a periodic function of $\mu_I/T$ due to the RW periodicity. In particular, the functional form common in the hadron resonance gas model, namely the hyperbolic function ($\cosh$), is naturally obtained as an analytic continuation of the periodic function ($\cos$). At $T\sim 0.9T_c$, lattice QCD and hadron resonance gas model agree well~(Fig.~\ref{09041400fig1}), while the deviation is seen beyond $T \sim 0.95T_c$.

The inter-quark potential and Debye screening mass have been investigated by Takahashi et al.~\cite{Takahashi:2013mja}. There is an attempt to extract the high-frequency modes of the hadron resonance gas model using the analytic continuation and explore the low-temperature region of the hadron phase~\cite{Motoki:2012en}, and the decrease of the signal-to-noise ratio at low temperature has been reported. In addition, a QCD-like theory with an exact ${\mathbb Z}_3$ symmetry was studied, for the comparison of the critical temperatures for the chiral and confinement phase transitions~\cite{Iritani:2015ara}.

For the pseudo-critical line, it is difficult to determine higher-order coefficients accurately. The same problem occurs for other quantities as well. Because only $\mu_I / T \le \pi / 3 \sim 1$ can be used for the fit, 
only $\mu/T\lesssim 1$ in the real-chemical-potential region can be studied via the analytic continuation. 

\subsection{Property of the Roberge-Weiss critical point}


\begin{figure}[htbp]
\centering
\includegraphics[width=0.5\linewidth]{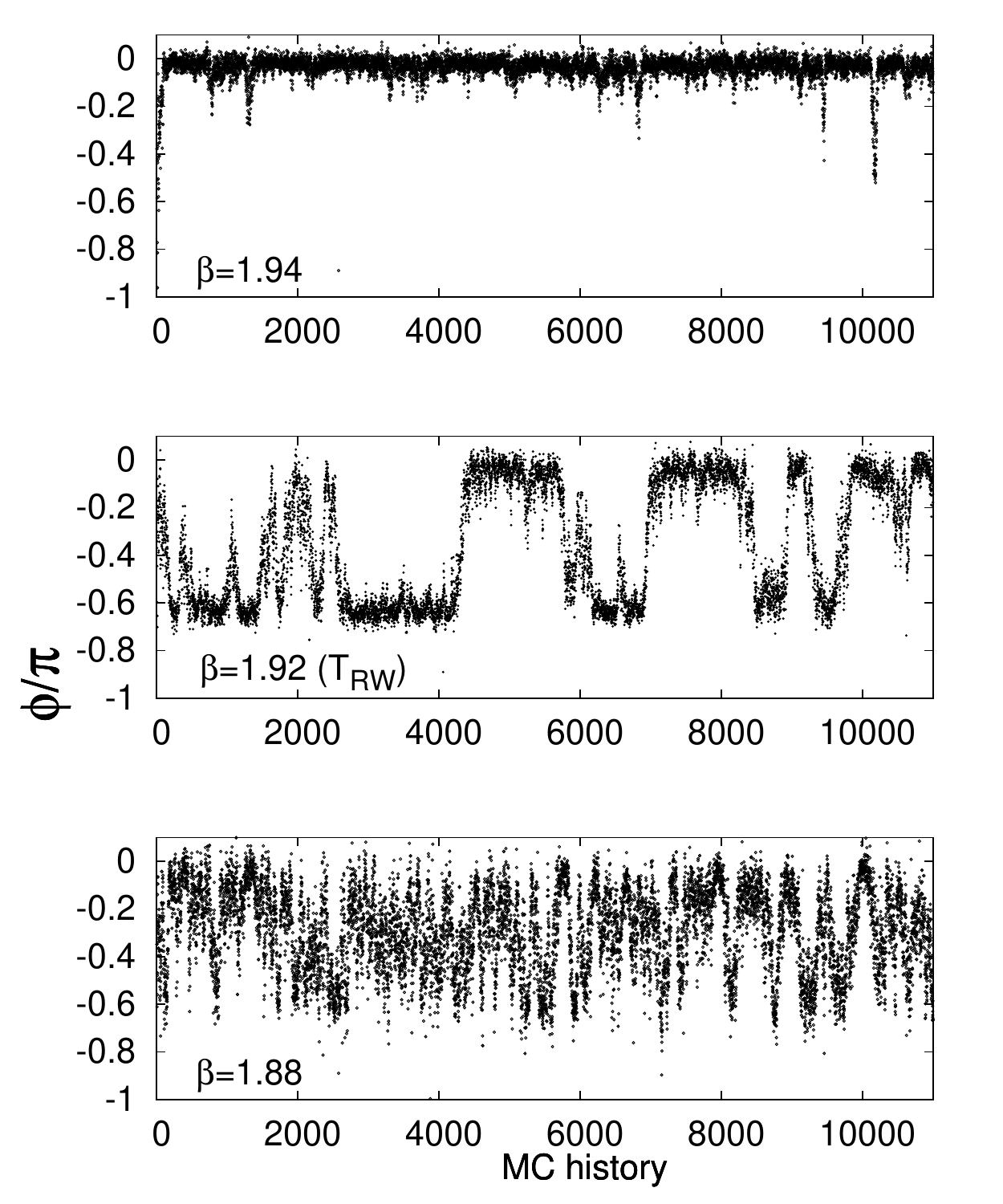}
\begin{minipage}{0.9\linewidth}
\caption{\small 
The Monte Carlo histories of the phase of the Polyakov loop $\phi$ at   $\mu_I/T = \pi/3$.
The lattice size is $8^3\times 4$. 
}\label{Sep062017fig1}
\end{minipage}
\end{figure}

\begin{figure}[htbp]
\centering
\includegraphics[width=0.5\linewidth]{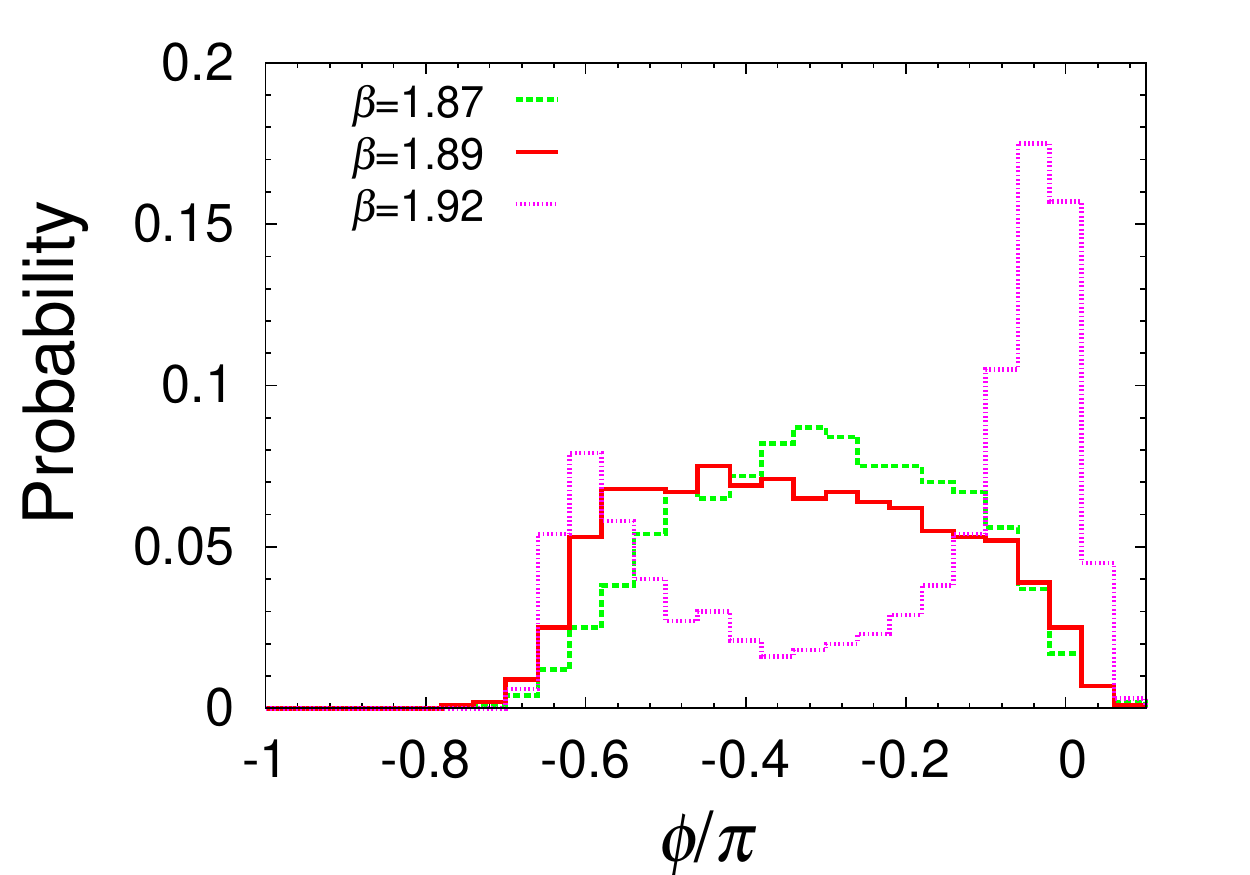}
\begin{minipage}{0.9\linewidth}
\caption{\small 
The histogram of the phase of the Polyakov loop $\phi$ at $\mu_I/T = \pi/3$.
The distribution changes with the temperature ($\beta$).
The lattice size is $8^3\times 4$. 
}\label{Jan2111fig1}
\end{minipage}
\end{figure}

We close this section with the discussion of the Roberge-Weiss (RW) endpoint, which is located at the edge of the Roberge-Weiss phase transition line. The RW endpoint is a rare example where direct analysis of the critical point on the QCD phase diagram is possible. It is also important as a place to test the lattice method to detect a critical point.
Similar to the QCD critical point, the properties of the RW endpoint depend on the quark mass and the number of flavors. One possibility is that the RW endpoint is the second-order phase transition point. Fig.~\ref{fig:phasediagram_withmu} depicts this case. Another possibility is that point C indicates the first-order phase transition. In this case, the line CD in Fig.~\ref{fig:phasediagram_withmu} extends to the curve CA.

Fig.~\ref{Sep062017fig1} shows the phase of the Polyakov loop at each step (called trajectory) of HMC simulation. Such a figure is called the Monte Carlo history, and it is useful for seeing how the physical quantity fluctuates as the virtual elapsed time in the simulation changes. When $\beta=1.94$, the phase of the Polyakov loop $\phi$ fluctuates around 0, which means that the Polyakov loop fluctuates near the real axis on the complex plane. At $\beta=1.92$, $\phi/\pi$ fluctuates around 0 or $-2/3 \sim -0.67$, which correspond to different minima, and the transitions between them occur repeatedly. The virtual elapsed time staying one minimum depends on the depth of the potential. Even if $\beta=1.94$, in principle, there should be two minima, but only one of them appears because the potential barrier is high and the transition rarely occurs once a minimum is chosen. The vacuum is selected with  a probability depending on the initial conditions of the simulation. On the other hand, at $\beta=1.88$, $\phi/\pi$ fluctuates over a wide range of $0 \sim -2 / 3$. This point is close to the RW endpoint.

Figure~\ref{Jan2111fig1} depicts the histogram generated from the Monte Carlo histories of $ \phi $. $P(\phi)$ is a unimodal distribution at low temperature $(\beta=1.87)$, a bimodal distribution at high temperature $(\beta = 1.92)$, and at the intermediate temperature $(\beta=1.89)$ it is a flat distribution with no obvious peak. The two-peak structure of the histogram at high temperature appears because the simulation samples two minima as we can see in the Monte Carlo history in the center panel of Fig.~\ref{Sep062017fig1}. Since $ \mu_I / T = \pi / N_c $ is just above the RW phase transition, the free energies at the two local minima should be equal, but in the figure, the peak on the $ \pi = 0 $ side looks high. This is probably because the number of Monte Carlo trajectories is not large enough to see sufficiently many tunnelings.  Although the simulation is trapped in one of the minima for order $ 1\,000 $ trajectories as we can see from Fig.~\ref{Sep062017fig1}, the total number of trajectories in the figure is around $ 10, 000 $, which is not large enough. If the number of trajectories is increased, the two peaks should come to the same height ~\footnote{
The histogram obtained by numerical analysis contains errors.
There are several error estimation methods for the histogram, such as the bootstrap method.}.
The bell-shaped distribution at low temperature is typical far from the phase transition. The flat structure at the intermediate temperature corresponds to the Monte Carlo history at $ \beta = 1.88 $, indicating a wide fluctuation of the phase $\phi$.

\begin{figure}[htbp]
\centering
\includegraphics[width=0.6\linewidth]{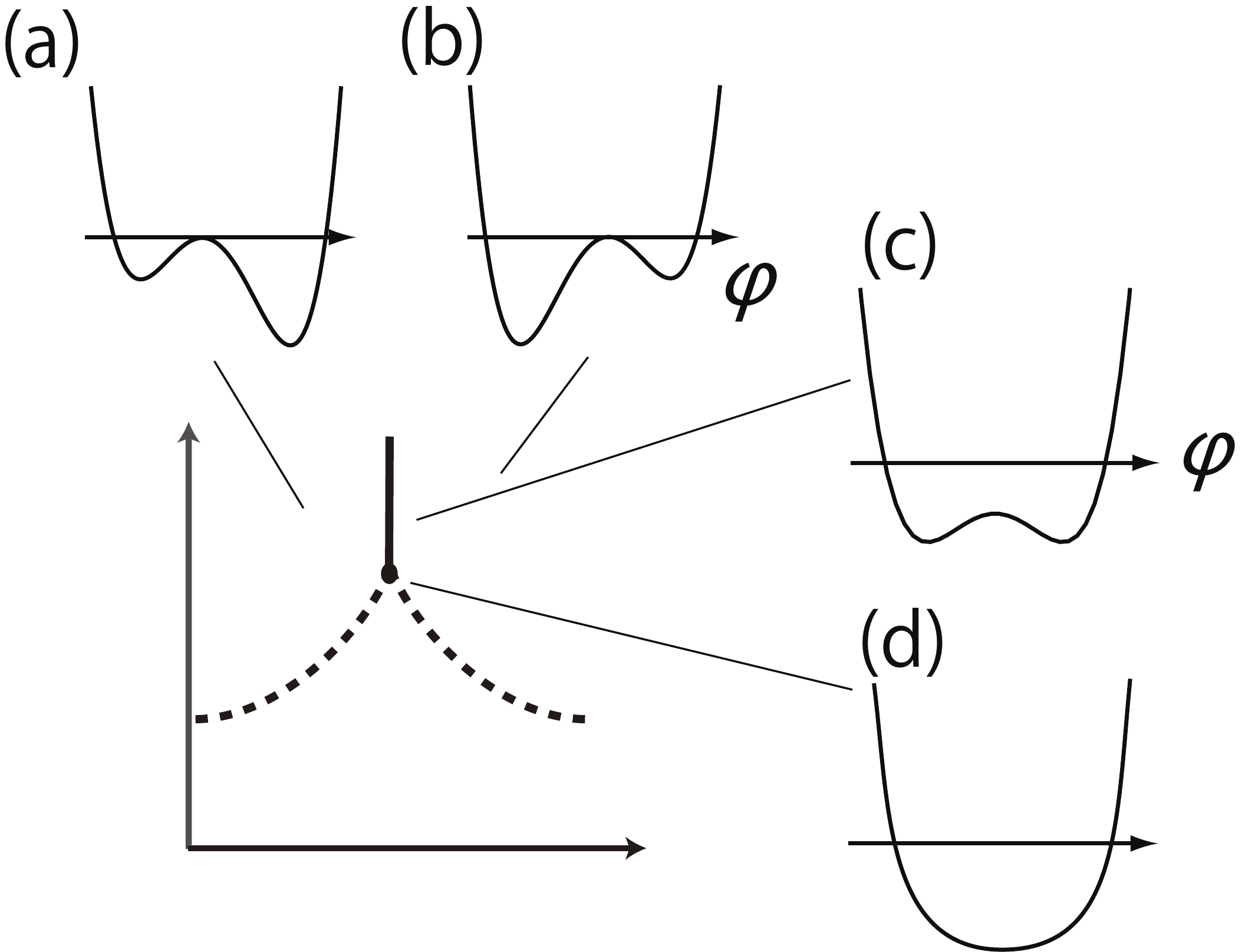}
\begin{minipage}{0.9\linewidth}
\caption{\small 
Outline of the Landau-Ginzburg potential $V(\phi)$ as a function of the phase of Polyakov loop $\phi$.
(a): high temperature and $\mu_I/T < \pi/3$, 
(b): high temperature and $\mu_I/T>\pi/3$, 
(c): $\mu_I/T = \pi/3$, and $T$ is above the RW critical point, 
(d): near the RW critical point.
}\label{Fig:2017Mar09fig1}
\end{minipage}
\end{figure}
The histogram can be interpreted as an effective potential with the phase of the Polyakov loop $\phi$ as the order parameter. The outline of the change of the Landau-Ginzburg potentials is shown in Fig.~\ref {Fig:2017Mar09fig1} in terms of $\phi$. In the QGP phase, the effective potential has three local minima. (a) and (b) of the figure show two vacua at $ \phi =0 $ and $ \phi =-2 \pi / 3 $. In the RW phase (I), the minimum value on the $ \phi = 0 $ side is the ground state, but in Phase (II), the $\phi= -2 \pi / 3 $ side becomes the ground state. Two states are degenerate at $ \mu_I / T = \pi / 3 $. From this figure, we can intuitively understand why the phase transition is of first order. Just above the RW phase transition line, the two states are degenerate, but as the temperature is lowered, a critical point (d) appears.

\begin{figure}[htbp]
\centering
\vspace{1.5cm}
\includegraphics[width=0.45\linewidth]{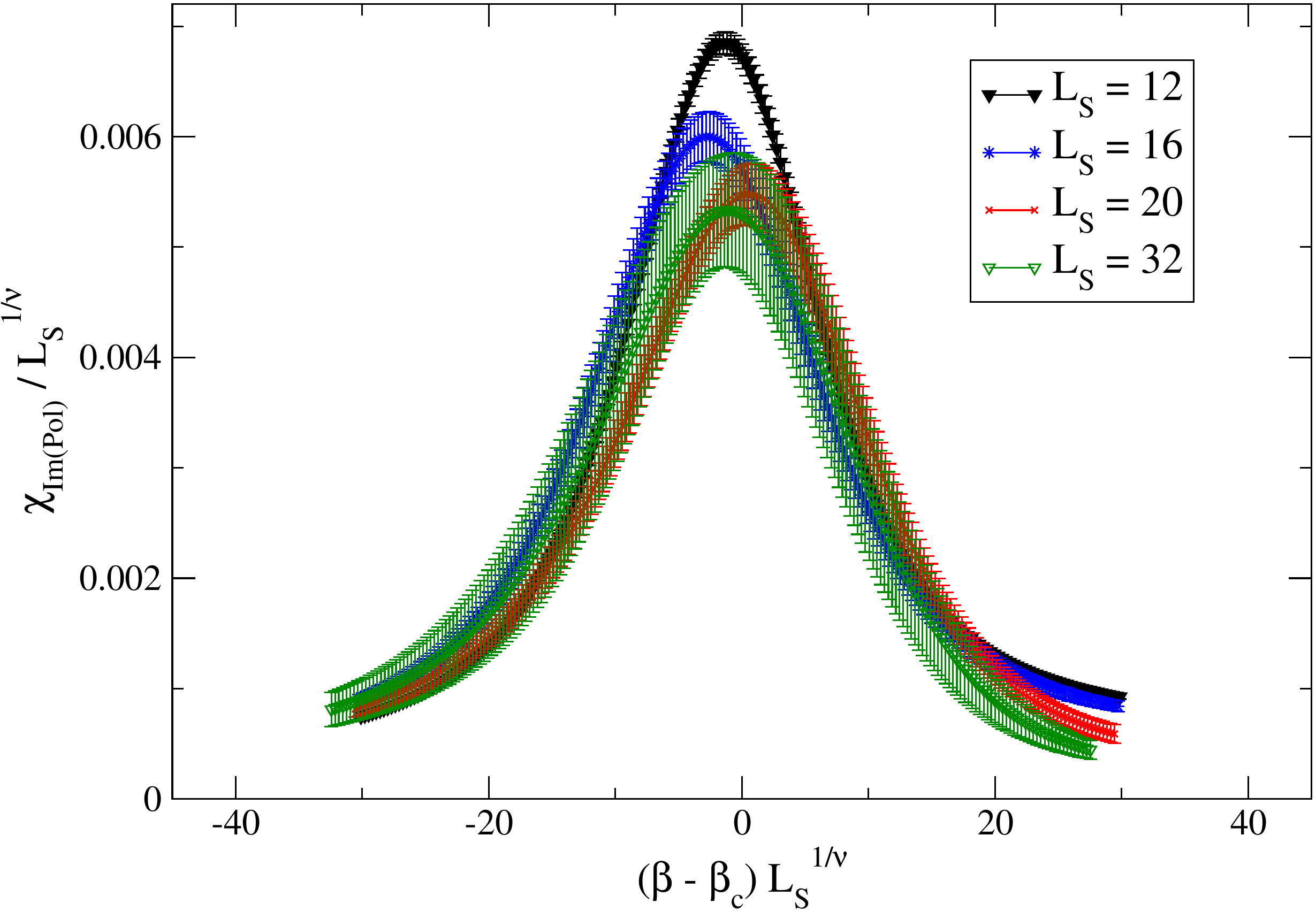}
\includegraphics[width=0.45\linewidth]{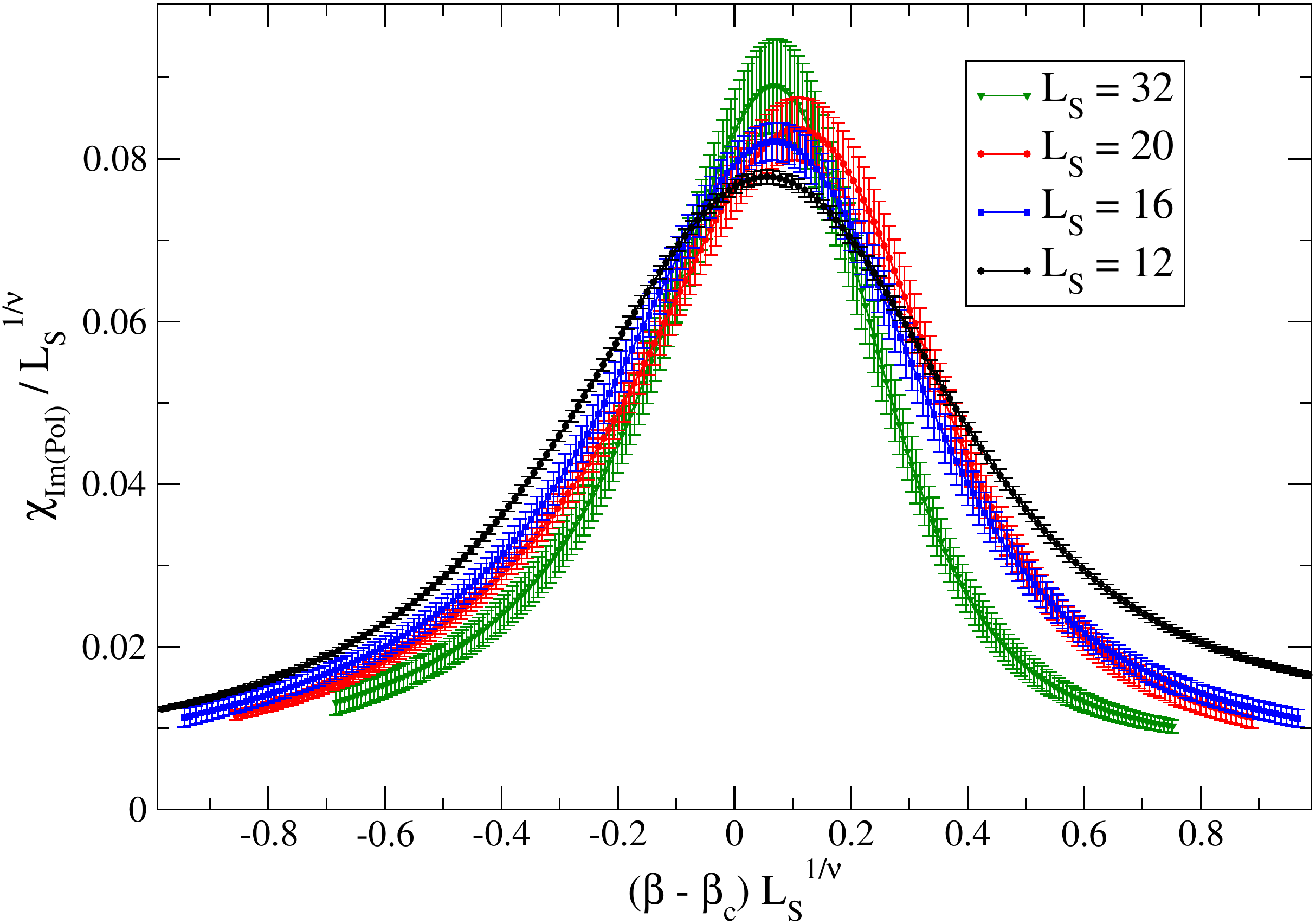}
\begin{minipage}{0.9\linewidth}
\caption{\small 
The determinations of the temperature at the RW endpoint and of the order of phase transition from the finite-size scaling for the susceptibility of the imaginary part of the Polyakov loop~\cite{D'Elia:2009qz}. The left panel is for two-flavor QCD with light quarks, scaled with the critical exponent for the first-order phase transition. The right panel is for heavy quarks (but lighter than the quenched case), scaled with the critical exponent for the second-order phase transition.
The lattice size of the temporal direction is $N_t=4$. 
}\label{09090254figs}
\end{minipage}
\end{figure}

\begin{figure}[htbp]
\centering
\includegraphics[width=0.48\linewidth]{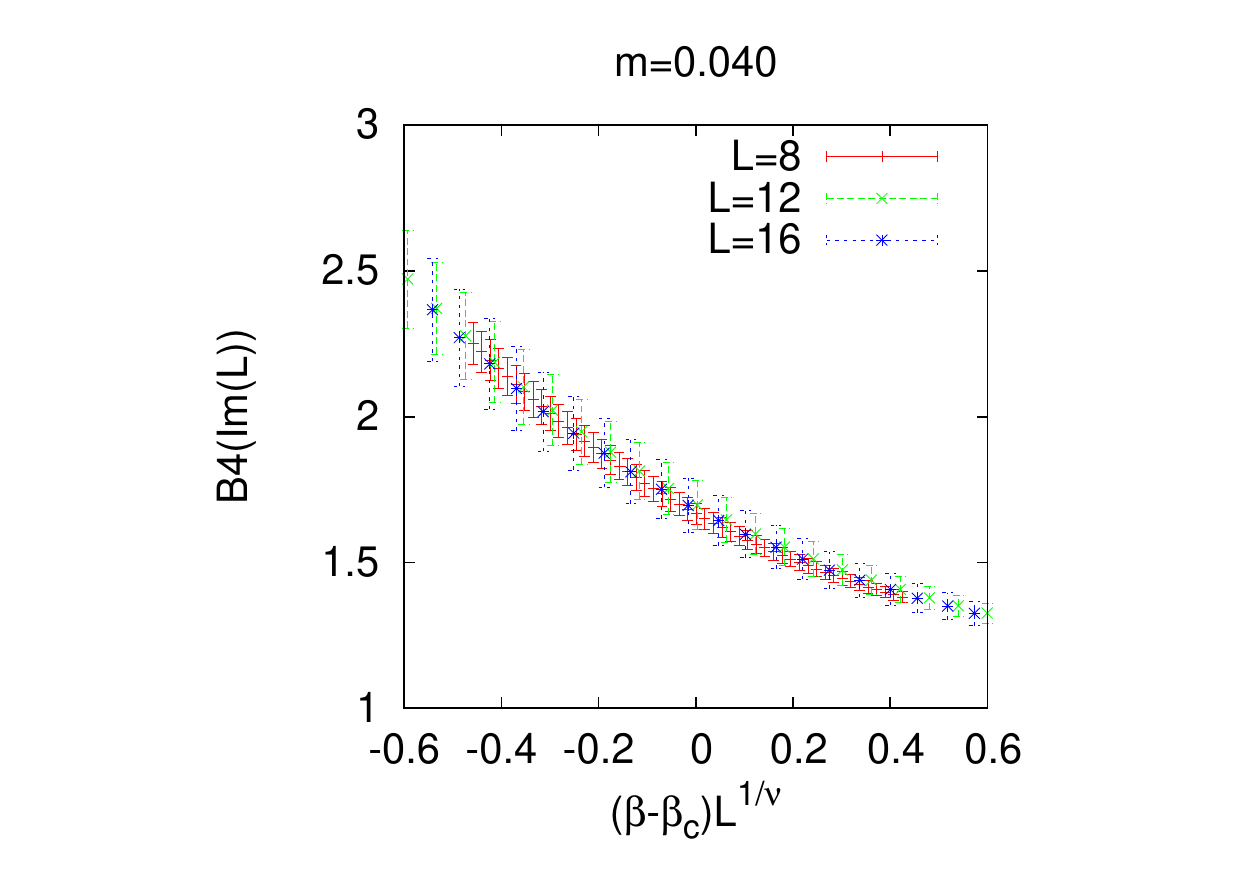}
\includegraphics[width=0.48\linewidth]{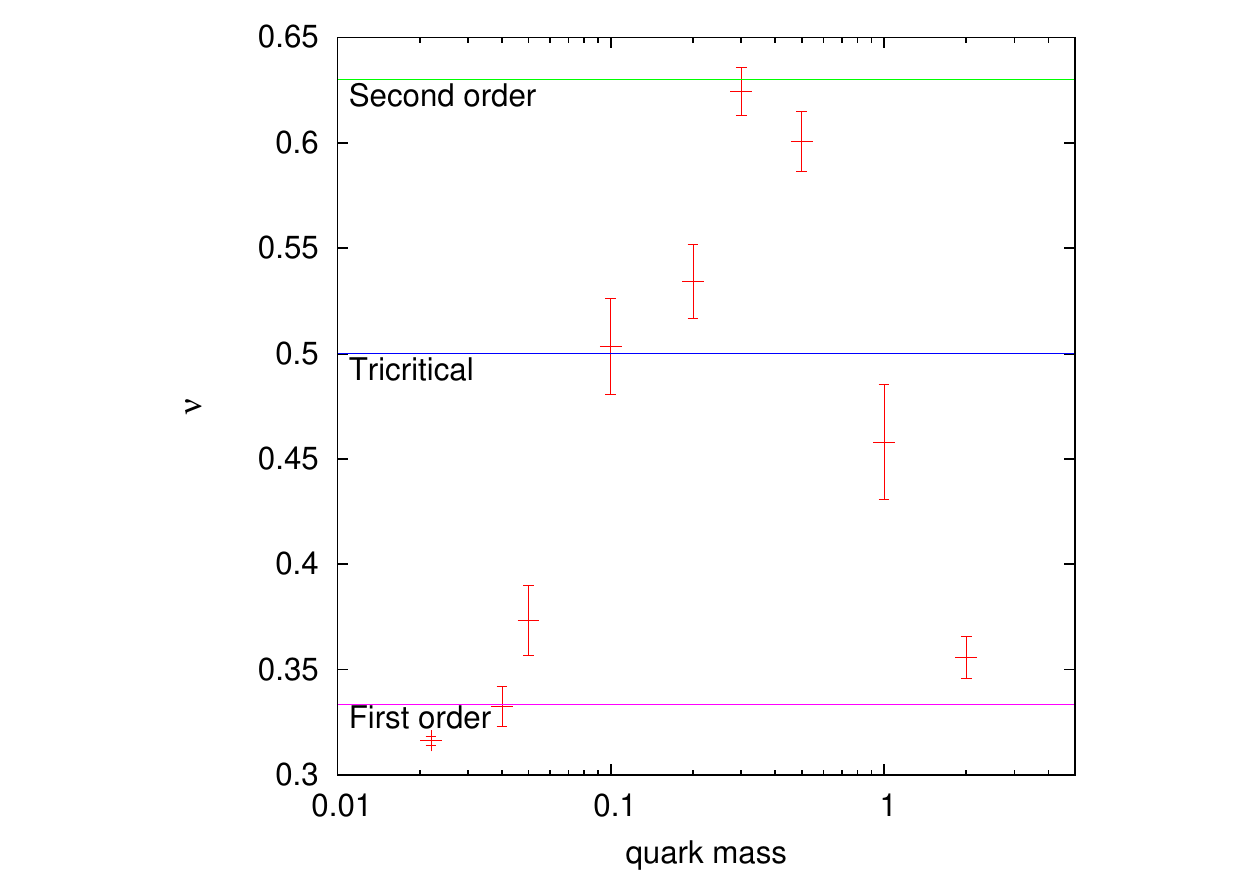}
\begin{minipage}{0.9\linewidth}
\caption{\small 
[Left] Determination of the temperature at the RW endpoint and of the order of transition from finite-size scaling for the Binder cumulant.
[Right] The quark mass dependence of the order of phase transition.
These plots were firstly shown in Ref.~\cite{deForcrand:2010he}, which considered three-flavor QCD.
$L$ denotes the spatial lattice extent $N_s$. 
The temporal lattice extent is $N_t=4$. 
}\label{10043144figs}
\end{minipage}
\end{figure}

Finite-size scaling is performed to determine the values of the critical temperature and critical exponent precisely.

Often, the finite-size scaling of the susceptibility and Binder cumulants
\[
B_4 (X) \equiv \langle (X- \langle X \rangle) ^ 4 \rangle / \langle (X- \langle X \rangle) ^ 2 \rangle ^ 2,
\]
is considered. 
At the critical point, $B_4$ does not depend on the volume. Therefore, the critical point and critical exponent can be identified by calculating $B_4$ at different volumes and by examining the temperature and density at which the results do not depend on volume. The finite-size scaling of susceptibility and $ B_4 $ has been performed for $N_f=2$~\cite{D'Elia:2009tm, Bonati:2010gi} and $N_f=3$~\cite{deForcrand:2010he}. In both cases, the RW endpoint is a first-order phase transition if the quark mass is heavy or light, while it is of second-order at the intermediate-mass (Fig.~\ref{09090254figs} and Fig.~\ref{10043144figs}).
Recently, the extrapolation to the physical quark mass has been carried out.
By using $N_f = 2 + 1$ stout Improved staggered and the tree-level Symanzik gauge action, the extrapolation to the continuum limit was performed from four data points, $ N_t = 4, 6, 8, 10 $, and the transition temperature was determined to be $ T_ {RW} = 208 (5) $ MeV, $ T_ {RW} / T_c \sim 1.34 (7) $ and the second-order transition described by the universality class with $3$-dimensional Ising model was found~\cite{Bonati:2016pwz}. Therefore, the phase diagram shown in Fig.~\ref{fig:phasediagram_withmu} seems to be realized in the imaginary chemical potential region at the physical point QCD.

The research on the large-scale simulations with the Wilson fermion is pursued by Philipsen and Pinke~\cite{Philipsen:2014rpa, Philipsen:2016hkv}, and by Wu and Meng. It is reported that the qualitative features are almost the same as the staggered fermion but the temperature at the RW endpoint is different from the staggered fermion.
It is necessary to confirm the result of the staggered fermion in the continuum limit by using other actions such as the Wilson fermion. This is an open problem because the continuum extrapolation of $ T_c $ has not yet been achieved for the Wilson fermion.

If the RW endpoint is a first-order transition, the first-order phase transition line extends to the crossover line. Then, another endpoint will appear somewhere. However, such a new critical point has not been confirmed by the lattice QCD simulation. Such a critical point may be a good testing ground for the determination of critical points avoiding the sign problem.
\clearpage
\section{Canonical method}
\label{sec:canonical}

It is difficult to use the three methods explained so far --- the reweighting, the Taylor expansion, and the analytical continuation from imaginary chemical potential --- to find the QCD critical point and to search the first-order phase transition. Although there is a claimed discovery of the QCD critical point via the reweighting method, it is the result of only one collaboration, and it is not yet an established result.

It has been reported that the canonical method is effective for the search of the first-order phase transitions. The canonical method is based on a basic technique in statistical mechanics: a grand partition function is expanded by a canonical partition function. Application of the canonical method to lattice QCD was proposed around 1990. Around 2005, de Forcrand and Kratochvila proposed a method to find the first-order phase transition by combining the canonical method with the Maxwell construction. After that, the discovery of the first-order phase transition was reported by the Ejiri and by the Kentucky groups. So far, only with the canonical method, multiple groups have reported the results with each other's regarding the first-order phase transition between hadronic and QGP phases consistently. Furthermore, the physical meaning of the canonical method is clear, and hence, with this method, we can deepen our understanding of the finite density lattice QCD. In this section, we will introduce the idea of the canonical method, the calculation method, examples of computational difficulties and ideas, and the application to the first-order phase transition.

\subsection{Idea of the canonical method}
\label{sec:canonicalidea}

The system with chemical potential can exchange particles with the heat bath. In statistical mechanics, it is called the grand canonical ensemble.
On the other hand, a system with a fixed number of particles is called the canonical ensemble. Since the grand canonical ensemble exchanges particles with the heat bath, the number of particles contained in the system changes from moment to moment and fluctuates near the expectation value. In the grand canonical ensemble at zero chemical potential, the expectation value of the number of particles is zero. However, at each instant various states appear, such as the states with no particle, one particle, one antiparticle, and so on. In other words, even if the chemical potential is zero, the system is a superposition of various states with various particle numbers. The basic idea of the canonical method is to construct a high-density state by extracting multi-particle states included as fluctuations.

First, we explain the theoretical framework. The grand canonical ensemble is represented as a superposition of the canonical ensemble.
The partition function of the grand canonical ensemble is given by
\begin{align}
Z(\mu) = {\rm tr} \, e^{- (\hat{H}-\mu \hat{N})/T}.
\label{eq:GCZ}
\end{align}
Here, $\hat{H}$ and $\hat{N}$ are the Hamiltonian and the number operator, respectively. It is assumed that $\hat{H}$ and $\hat{N}$ commute with each other. Thus, $ \dot{\hat{N}} = [\hat{H},  \hat{N}] = 0 $, and $ \hat{H} $ and $ \hat{N} $ have simultaneous eigenstates. Let $|n\rangle$ be the eigenstate of $\hat{H}$ and $\hat{N}$ satisfying\footnote{
Strictly speaking, there are multiple energy eigenstates for each $n$. 
Therefore, we should introduce another label $i$ and write the eigenvalues of $\hat{H}$ and $\hat{N}$ as
\begin{align}
\hat{H} | E_i, n_i;i \rangle &= E_i | E_i, n_i;i \rangle, \\
\hat{N} | E_i, n_i;i \rangle &= n_i | E_i, n_i;i \rangle.
\end{align}
With this notation, \eqref{eq:GCZ2} is modified to 
\begin{align}
Z(\mu) & = \sum_{n=-N}^N \sum_i\delta_{nn_i}\langle E_i, n_i;i | e^{- (\hat{H}-\mu \hat{N})/T} | E_i, n_i;i \rangle\nn \\
& = \sum_{n=-N}^N \sum_i\delta_{nn_i}\langle E_i, n_i;i | e^{- \hat{H}/T} | E_i, n_i;i \rangle e^{\mu n/T}\nn \\
& = \sum_{n=-N}^N Z_n e^{n \mu/T}.
\end{align}
} 
\begin{align}
\hat{H} | n \rangle &= E_n | n\rangle, \\
\hat{N} | n\rangle &= n| n\rangle.
\end{align}
By expanding the trace of (\ref{eq:GCZ}) with the eigenstates of the particle number operator, we obtain
\begin{align}
Z(\mu) & = \sum_{n=-N}^N \langle n | e^{- (\hat{H}-\mu \hat{N})/T} | n \rangle, \nn \\
& = \sum_{n=-N}^N \langle n | e^{- \hat{H}/T} | n \rangle e^{\mu n/T}, \nn \\
& \equiv \sum_{n=-N}^N Z_n e^{n \mu/T}
\label {eq:GCZ2}
\end{align}
Here, $N$ denotes the maximum number of particles that can be contained in the system. In continuous spacetime, there is no upper limit on the number of particles in the system, namely $N\to\infty$. On the lattice, $N$ is finite because the number of sites and the number of particles that can be on each site are finite. For QCD, $N=$ (spatial lattice volume) $\times N_c \times$ (flavor) $\times$ (spin). Negative $n$ means that the antiquark number $n$. Because of the CP invariance of QCD, Eq.~(\ref{eq:GCZ2}) is symmetric for the sign flip of $n$, and $Z_n = Z_{-n}$ is satisfied for any $n$. The grand canonical partition function is related to the Gibbs free energy $\Omega$ by $Z(\mu)=e^{-\Omega/T}$. On the other hand, $Z_n = \sum_i\delta_{nn_i}\langle E_i, n_i;i | e^{- \hat{H}/T} | E_i, n_i;i \rangle$ is expressed as $Z_n = e^{-F_n/T}$ by using the Helmholtz free energy $F_n$. Eq.~(\ref{eq:GCZ2}) can be interpreted as the expansion with respect to the quark number $n$, or with respect to the fugacity $e^{\mu/T}$, or with respect to the winding number (the number of windings in the time direction; see \S.~\ref{sec:latticeaction}).

Because $Z_n$ is a function of temperature, it needs to be determined at each temperature. How can we calculate $Z_n$? One method is to consider Eq.~(\ref{eq:GCZ2}) as a polynomial of fugacity $e^{\mu /T}$ and find its coefficients. By using the values of $Z(\mu/T)$ at multiple points $\mu/T$, such that the number of data points $(\mu/T,Z(\mu/T))$ is the same as the number of the coefficients we want, 
$Z_n$ can be determined by solving the simultaneous equations~\cite{Barbour:1988ax}~\footnote{ 
The grand partition function is regarded as a function of $\mu/T$ since the chemical potential appears as a combination of $\mu/T$ in Eq.~(\ref{eq:GCZ2}).}. $Z(\mu/T)$ can be obtained by the reweighting method except for the normalization constant.

Another method to determine $Z_n$ is to extend the chemical potential to a pure imaginary number and use the Fourier transform~\footnote{The domain of the integration $[-\pi,\pi]$ should be changed to $[-\pi/3,\pi/3]$ in QCD because of the Roberge-Weiss periodicity~\cite{Roberge:1986mm}.}.
\begin{align}
Z_n = \int_{-\pi}^{\pi} \frac{d\phi}{2\pi} Z(\phi) e^{- in \phi}, \phi = \mu_I/T.
\label{eq:canonical_Fourier}
\end {align}

Both methods are formally correct, but there is a problem with the computational accuracy of $Z(\mu/T)$ or $Z(\phi)$. In Fig.~\ref{Feb2711fig2}, we saw the ambiguity in the fitting of the data with errors. A similar problem occurs in the fit as a polynomial of the fugacity. The fit of the higher-order terms is numerically unstable and difficult. In the case of the Fourier transform, because the oscillation of the integrand becomes faster as $n$ increases, it is difficult to calculate the higher-order terms.

A more efficient way of calculation is to apply the fugacity expansion such as (\ref{eq:GCZ2}) to the fermion determinant $\det\Delta(\mu)$ rather than the partition function. Since the fugacity expansion is performed before taking the configuration average, there is no statistical error and the high-order terms can be expanded with higher accuracy. It is possible to calculate $Z_n$ with higher accuracy than the simultaneous equations and Fourier transform methods. Such an improvement in the accuracy is practically important because the lack of good accuracy at $|n|$ is a big problem in the canonical method.

The fugacity expansion of the fermion determinant is roughly classified into two types: the dimensional reduction to $\mu$ in the temporal component, and the expansion with respect to the hopping parameter in the temporal component~\footnote{
The fermion matrix is a multidimensional matrix regarding internal degrees of freedom and spacetime. Here, we perform the integration of the determinant only in the temporal direction by expressing it as a matrix with respect to time.}.
The former is called the reduction formula or propagator matrix formula. The reduction formula is used not only for application to the canonical method but also for problems in the finite density region of the hadron phase, which will be discussed in \S.~\ref{sec:silverblaze}. Hence, although the calculation is complicated, it is worth seeing the derivation. We explain the reduction formula in the next section. The reduction formula clearly expresses the chemical potential dependence of the fermion determinant, so a proper understanding of the structure of the reduction formula must be helpful in the study of the finite-density lattice QCD.  

\subsection{Reduction formula}
\label{sec:reduction}
\subsubsection{Historical remarks}

The fermion determinant $\det \Delta$ is the determinant of the multidimensional matrix whose indices contain the information of the spacetime and internal degrees of freedom. The reduction formula is a formula for analytically calculating the temporal component. It reduces the rank of the fermion matrix in the calculation of the determinant and transforms $\det \Delta$ into an analytic representation of $\mu$. This formula was firstly formulated by Gibbs using the Lanczos method~\cite{Gibbs:1986hi}. Another derivation of the formula was given by Hasenfatz and Touissaint~\cite{Hasenfratz:1991ax}. Such a simplification is possible due to the structure of the fermion matrix. Because the Dirac operator is a first-order derivative of time,  $\Delta$ as a matrix acting on the time variable is a sparse matrix. Using this structure, it is possible to calculate the temporal component of $\det\Delta$ analytically. As explained in Sec.~\ref{sec:latticeaction}, the chemical potential appears along with the link variable in the temporal direction of the fermion matrix. Therefore, the calculation of the fermion determinant in the temporal direction leads to a simplification regarding the chemical potential.

Gibbs' formula holds for the KS fermion. It cannot be applied to the Wilson fermion because the structures of the fermion matrices are different. The non-invertible blocks in the Wilson fermion matrix make it difficult to apply the formula to the KS fermion. This problem was solved by using a permutation matrix by Borici~\cite{Borici:2004bq}, and then this method was applied to the formulation of the fugacity expansion ~\cite{Nagata:2010xi, Alexandru:2010yb}. As mentioned above, the reduction formula is based on the sparseness of the fermion matrix in the temporal direction. 
If the temporal direction is kept continuous, namely if only the spatial component is discretized, the reduction formula can be derived by solving the differential equation with respect to time~\cite{Adams:2003rm}. Gattringer and Danzer gave another derivation based on the expansion with respect to the hopping term in the time direction~\cite{Danzer:2009sr, Gattringer:2009wi,Danzer:2008xs}.

\subsubsection{Reduction formula for Wilson fermion}

Below, we derive the reduction formula for the Wilson fermions following Ref.~\cite{Nagata:2010xi}. There is a good review paper~\cite{Hasenfratz:1991ax} for the formula for the KS fermions.

We reorganize the Wilson fermion matrix (\ref{Wfermion}) based on the $\mu$-dependence as
\begin{align}
\Delta &= B -  2 e^{\mu a} \kappa r_- V - 2 e^{- \mu a} \kappa r_+ V^\dagger,
\label{Eq:2014Jul16eq1}
\end{align}
where $r_\pm = (r \pm \gamma_4)/2$. We take $r=1$ following a common convention, then $r_\pm$ are the projection operators from the four-spinor representation to the two-spinor representation. In general, projection operators do not have the inverse. Hence $r_\pm$ do not have the inverse if $r=1$. The matrices $B, V, V^\dagger$ are given by
\begin{align}
B(x,x') &\equiv    \delta_{x,x'}
 - \kappa \sum_{i=1}^{3} \left\{
        (r-\gamma_i) U_i(x) \delta_{x',x+\hat{i}}
      + (r+\gamma_i) U_i^{\dagger}(x') \delta_{x',x-\hat{i}} \right\}
\nn \\
&- \delta_{x, x^\prime} C_{SW} \kappa \sum_{\mu \le \nu} \sigma_{\mu\nu} F_{\mu\nu},  \\
V(x,x') & \equiv 
 U_4(x) \delta_{x',x+\hat{4}}, \\
V^\dagger(x,x') &\equiv
  U_4^{\dagger}(x') \delta_{x',x-\hat{4}}.
\end{align}
Eq.~(\ref{Eq:2014Jul16eq1}) reorganizes the fermion matrix based on the chemical potential, and at the same time, based on the direction of propagation of the quark.

Regarding the time-component, $B$ consists of the Kronecker delta $\delta_{x'_4,x_4}$, while $V$ and $V^\dagger$ are proportional to $\delta_{x'_4,x_4+1}$ and $\delta_{x'_4,x_4-1}$, respectively\footnote{This property can easily be seen in the matrix representation. $B$ is expressed the following block-diagonal matrix,
\bea
&&\begin{array}{ccccccccccccccc}
t'\hsm =\hsm1 & & \cdots & &&&&&&&&&&&t'\hsm =\hsm N_t
\end{array}
\nn \\
B=\begin{array}{c}
t=1 \\ t=2 \\ t=3 \\
\cdot
\\
\cdot
\\
\cdot
\\
t=N_t
\end{array}
&&\left(
\begin{array}{c|c|ccc|c|c}
   B_1 & 0 & 0 & \cdots& & 0 & 0
\\ \hline
   0  &B_2 & 0 &\cdots  &  & 0& 0 
\\ \hline
  0 & 0 & B_3 & \cdots & &  & 
\\ 
 \cdots  &\cdots  &\cdots & \cdots& \cdots &\cdots  &\cdots
\\ 
   &  & &\cdots &   & 0 &0
\\ \hline
    0    & 0 & &  \cdots & 0 &B_{N_t-1}& 0 
\\ \hline
0  & 0 &  &\cdots &0& 0 &B_{N_t} \\
\end{array}
\right), \nn
\eea
while $V$ is expressed as
\bea
&&V
=
\left(
\begin{array}{c|c|ccc|c}
   0 &  U_4(1) & 0 & \cdots& &  0 
\\ \hline
   0  & 0 &  U_4 (2)&\cdots  &  &  0 
\\ \hline
  0 & 0 & 0 & \cdots & & 
\\ 
 \cdots  &\cdots  &\cdots & \cdots& \cdots  &\cdots
\\ 
   &  & &\cdots    & U_4(N_t-2) & 0 
\\ \hline
    0    & 0 & &  \cdots & 0 &  U_4(N_t\hsm-\hsm 1) 
\\ \hline
-U_4(N_t) & 0 &  &\cdots &0&  0 \\
\end{array}
\right). 
\nonumber
\eea
In this representation, the fermion matrix $\Delta$ is a band matrix with elements in the diagonal block and the ones above and below them. 
The chemical potential appears as $e^{\mu a}$ for the components above the diagonal blocks and as a $e^{-\mu a}$ for the ones below the diagonal block.} Regarding the time indices, the components
of $\Delta$ is zero except near the diagonal elements, namely the matrix is sparse. We can analytically rewrite the determinant by using sparseness. Different kinds of calculations are needed for the staggered fermion and the Wilson fermions. This is because the blocks appearing in the case of Wilson fermion contain $1\pm\gamma_4$ which are not invertible. To resolve this problem, Borici invented a method utilizing a permutation matrix~\cite{Borici:2004bq}
\begin{align}
P = (c_a r_- + c_b r_+ V z^{-1}), (z= e^{- \mu a}).
\end{align}
Here, $c_a$ and $c_b$ are nonzero constants.
$P$ is invertible while $r_\pm$ is not. It is easy to prove that $\det P = (c_a c_b z^{-1})^{N/2}, (N=4N_c N_x N_y N_z N_t) $.

Multiplying $P$ to the fermion matrix from the right, we obtain
\begin{align}
\Delta P = (c_a B r_- - 2 c_b \kappa r_+ ) + ( c_b B r_+ - 2 c_a \kappa r_-) V z^{-1}.  
\label{May0910eq1}
\end{align}
Here, we used $r_\pm r_\mp = 0, (r_\pm)^2 = r_\pm$ which holds when $r=1$. 
For simplicity, we denote\footnote{In time-plane block matrix form, the first and second terms are given by 
\begin{align}
(c_a B r_- - 2 c_b \kappa r_+ )
&= \left( \begin{array}{cccc}
 \alpha_1  & & & \\
 & \alpha_2 & & \\
 & & \ddots & \\
 & & & \alpha_{N_t}
\end{array}\right),
\label{Jun1910eq1} \\
( c_b B r_+ - 2 c_a \kappa r_-) V z^{-1}
& = \left( \begin{array}{ccccc}
 0 & \beta_1 z^{-1} & & & \\
 & 0 & \beta_2 z^{-1} & & \\
 & & 0 & \ddots & \\
 & & & \ddots & \beta_{N_t-1} z^{-1} \\
 -\beta_{N_t} z^{-1} & & & &  0
\end{array}\right),
\label{Jul05eq1}
\end{align}
respectively.}
\begin{subequations}
\begin{align}
\alpha_i &= c_a B^{ab, \mu\sigma}(\vec{x}, \vec{y}, t_i) \; r_{-}^{\sigma\nu} 
         -2  c_b  \kappa \; r_{+}^{\mu\nu} \delta^{ab} \delta(\vec{x}-\vec{y}), \\
\beta_i &= c_b B^{ac,\mu\sigma}(\vec{x}, \vec{y}, t_i)\; r_{+}^{\sigma\nu} 
U_4^{cb}(\vec{y}, t_i) -2 c_a \kappa \; r_{-}^{\mu\nu} \delta(\vec{x}-\vec{y}) U_4^{ab}(\vec{y}, t_i).
\end{align}
\end{subequations}%
$\vec{x}, \vec{y}$ denote spatial coordinates, and $\alpha_i$, $\beta_i$ describe the propagation along the spatial directions at time $t_i$.
The rank of the matrix is
\[ N_{\rm red} = N/N_t = 4 N_x N_y N_z N_c,
\]
since $\alpha_i$ and $\beta_i$ do not have a matrix structure concerning the time variable.
Hence, the rank is smaller by a factor $1/N_t$ compared to $\Delta$.

We express $\Delta P$ as a matrix acting on the time variable,
\begin{align}
  \Delta P & = \left( \begin{array}{ccccc}
  \alpha_1 & \beta_1 z^{-1} & & & \\
  & \alpha_2 & \beta_2 z^{-1} & & \\
  & & \alpha_3 & \ddots & \\
  & & & \ddots & \beta_{N_t-1} z^{-1}\\
-\beta_{N_t} z^{-1} & & & & \alpha_{N_t}
\end{array}\right),
\label{eq:2017Aug16eq1}
\end{align}
where we put $z=e^{- \mu a}$. Both block components $\alpha_i$ and $\beta_i$ are invertible. The determinant of $\Delta P$ can be rewritten by utilizing the expression (\ref{eq:2017Aug16eq1}),
\begin{align}
\Det (\Delta P) & = \left (\prod_{i = 1}^{N_t} \det (\alpha_i )\right) \det\left (1 + z^{-N_t} Q \right).
\end{align}
Here,
\begin{align}
Q = (\alpha_1^{-1} \beta_1) \cdots (\alpha_{N_t}^{-1} \beta_{N_t}),
\label {Eq:2014Jul26eq2}
\end{align}
is called the propagator matrix or the reduced matrix.

By using $\det (\Delta P) = (\det\Delta)( \det P)$, we obtain
\begin{align}
\Det \Delta & = (c_a c_b )^{-N/2} \xi^{-N_{\rm red}/2} \left(\prod_{i = 1}^{N_t} \det(\alpha_i ) \right)
\det\left (Q + \xi \right),
\label{May1010eq2}
\end{align}
where $\xi = z^{N_t}=e^{- \mu a N_t}$. By using temperature $T=1/(a N_t)$, it is expressed as $\xi=e^{-\mu/T}$. Hence $\xi$ is the fugacity.

The formula~\eqref{May1010eq2} converts the determinant of a rank-$N$ matrix $ \Delta$ of rank $N$ into the determinant of the matrix $Q + \xi$ of rank $N/N_t$. The memory required for determinant calculation and the computational time increase to $({\rm rank})^2$ and $({\rm rank})^3$, respectively. The computational cost on the right-hand side of Eq.~\eqref{May1010eq2} is considerably reduced compared to directly calculating on the left-hand side. Furthermore, in addition to reducing the amount of calculation, the chemical potential dependence of the fermion determinant is analytically given since the term $\xi$ containing the chemical potential and $Q$ and $\alpha_i$ containing the link variables are separated.

We rewrite the expression further, for the application to the canonical method. Using the eigenvalue of $Q$, which we denote by $\lambda_n, (n=1, 2, \cdots, N_{\rm red})$, $\det\Delta$ can be expressed as ~\footnote{
To calculate $\det \Delta$, we can use the $LU$ decomposition of $\Delta$, or the eigenvalue of $Q$. Using the $LU$ decomposition, the cost for one reweighting factor is smaller, but a new calculation is needed for each value of $\mu$. Such calculations are used in Ref.~\cite{Csikor:2004ik} etc. The calculation of the eigenvalues of $Q$ is more costly than the LU decomposition of $\Delta$, but once the eigenvalues are obtained, the same eigenvalues can be used for any value of $\mu$. Hence, one should choose the better one depending on the content of the calculation and the performance of the computer.}
\be
\Det \Delta = C_0 \xi^{- N_{\rm red}/2} \prod_{n=1}^{\Nred} (\xi + \lambda_n).
\label{eq:2017Mar17eq1}
\ee
Here, $C_0=(c_a c_b )^{-N/2} \left(\prod_{i = 1}^{N_t} \det(\alpha_i )\right))$.
Expanding  the product part $(\xi+\lambda_1)(\xi+\lambda_2)\cdots$  as a power of $\xi$,
\begin{align}
\Det \Delta (\mu) & = C_0 \sum_{n=-\Nred/2}^{\Nred/2} c_n (e^{\mu/T})^n.
\label {Jun1410eq1}
\end{align}
Eq.~(\ref{Jun1410eq1}) has the same form as Eq.~\eqref{eq:GCZ2} in the grand partition function. We can see that $c_n$ corresponds to the $n$-quark part of the fermion determinant. $\Nred/2$ is given by 
\[
\Nred/2 = 2 N_c N_x N_y N_z = (\mbox{spinor}) \times (\mbox{color}) \times (\mbox{volume}).
\]

\subsubsection{Physical meaning of the reduction formula}

\iffigure
\begin{figure}[htbp] 
\centering
\includegraphics[width=7cm]{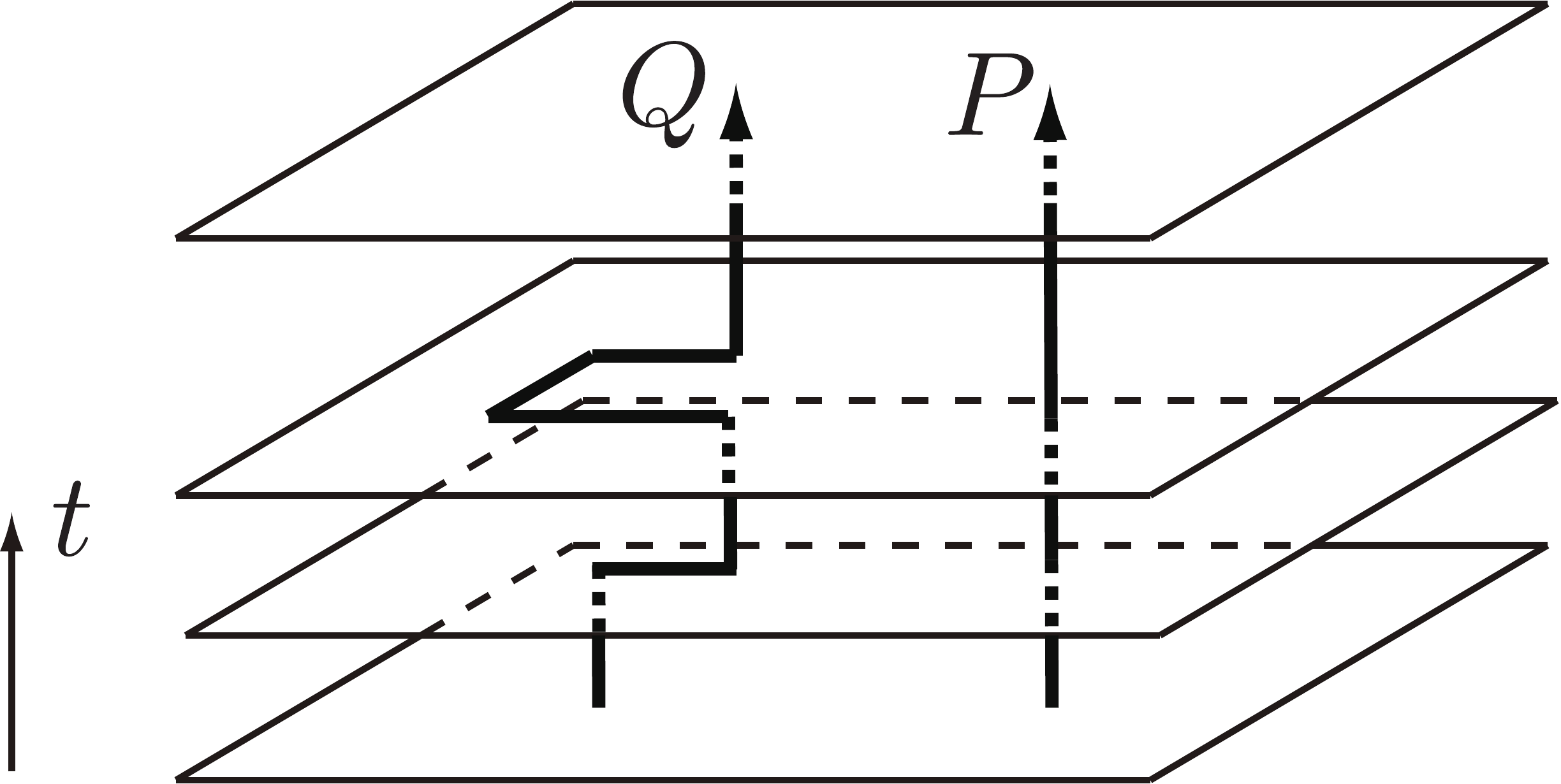}
\begin{minipage}{0.9\linewidth}
\caption{\small 
Schematic picture of the reduction matrix $Q$. The matrix $Q$ represents the propagation of the quark line from initial to final times. It is a process similar to the Polyakov lone $P$, which is drawn by the straight line.}
\label{Fig:2013Sep28eq1}
\end{minipage}
\end{figure} 
\fi

Now let us interpret $Q$ diagrammatically. $\alpha_1$ describes the motion of quark at a fixed time slice $x_4=t=1$. $\alpha^{-1}$ can be expanded formally as a power of $B$, which corresponds to the expansion with respect to the hopping along the spatial direction. $\beta_1$ looks similar to $\alpha_1$, except that it includes the link variable $U_4$. $\beta_1$ represents the move from the $t=1$-plane to $t=2$-plane since $U_4(\vec{x}, t=1)$ expresses the propagation from $(\vec{x}, t=1)$ to $(\vec{x}, t=2)$. Therefore, $\alpha_1^{-1} \beta_1$ describes a diagram that moves to $t=2$ after moving on the $t=1$ plane. $Q$ is obtained by repeating such move $t=1$ to $t=N_t$, so $Q$ describes the diagram in which the quark propagates in the time direction.

$(\alpha_i)^{-1} \beta_i$ represents the time evolution of quarks, and using the Hamiltonian, which is the time evolution operator, $(\alpha_i)^{-1} \beta_i \propto e^{ -H (t_{i+1}-t_i) }$. This is called the transfer matrix in the lattice model. $Q$ is the product of them and can be written as $Q\propto e^{- H (t_{N_t}-t_1)} = e^{-H a N_t} = e^{-H/T}$.
As a support for this interpretation, we can show that $Q$ gives Polyakov line in the heavy quark limit $\kappa\to 0$. Indeed, in the limit $\kappa\to 0$, $Q$ is given by
\begin{align}
Q = \left( \begin{matrix}
(2 \kappa)^{N_t} & 0 \\
0 & (2 \kappa)^{-N_t}
\end{matrix}
\right) U_4(t_1) U_4(t_2) \cdots U_4(t_{N_t}).
\label{Eq:2014Jul29eq1}
\end{align}
The first matrix on the right-hand side is the term that acts on the spinor component. The rest of the right hand side, $U_4(t_1) U_4(t_2) \cdots U_4(t_{N_t})$, expresses the Polyakov line (\ref{eq:Ploop}). The free energy of quark in the heavy quark limit ($m\to \infty$), $F_{m\to \infty}$ is related with the expectation value of the Polyakov loop as $\langle \tr P \rangle \propto e^{-F_{m\to \infty}/T}$. $P=\prod U_4(t_i)$ can be written as $P \propto e^{- H t}$ for Hamiltonian $H(m\to \infty)$ in the heavy quark limit. The Polyakov loop is defined by the heavy quark limit, which does not include a dynamical quark that propagates in the spatial direction, while $Q$ can be defined for any quark mass so that it also includes the propagation in the spatial direction. In this sense, $Q$ is a generalization of the Polyakov line, and it turns out that it corresponds to $\langle\tr Q \rangle\propto e^{-F/T}$, where $F$ denotes the free energy of quark and without taking the heavy quark limit. In Sec.~\ref{sec:silverblaze}, we will discuss the low-temperature region of the hadron phase, where the physical interpretation of $Q$ plays an important role in taking the zero-temperature limit and associating the chemical potential with the $\pi$ meson mass. Moreover, $Q$ is the product of transfer matrices and is also called the propagator matrix due to the similarity to the propagator. It represents the propagation of quarks, but it should be noted that the diagram described is different from the quark propagator $\Delta^{-1}$.

\subsection{Calculation of canonical partition function}

We apply the reduction formula to the calculation of the canonical partition function~\cite{Barbour:1990tb,Hasenfratz:1991ax}. By substituting Eq.~(\ref{Jun1410eq1}) to $\det \Delta$, the grand partition function $Z(\mu)$ is given by
\begin{align}
Z(\mu) &= \int {\cal D} U  C_0 \sum_{n=-\Nred/2}^{\Nred/2} c_n e^{ n \mu/T}  e^{-S_g}  , \nn \\
      &= \sum_{n=-\Nred/2}^{\Nred/2} \biggl( \int {\cal D}U  C_0 c_n e^{-S_g} \biggr) e^{n \mu/T}.
\label{Eq:2014Jul27eq2}
\end{align}
Here, we assume $N_f=1$ for simplicity. In the case of $N_f$ flavors, we take the $N_f$-th  power of the determinant and then expand (\ref{eq:2017Mar17eq1}) in terms of the fugacity. At that time, there is no difference in the calculation except that $\Nred$ in (\ref{Eq:2014Jul27eq2}) is multiplied by $N_f$. Equations (\ref{eq:GCZ2}) and (\ref{Eq:2014Jul27eq2}) for $Z(\mu)$ must be equal to each other at any $\mu$. Therefore, the canonical partition function is given by 
\begin{align}
Z_n = \int {\cal D} U C_0 c_n e^{-S_g}.
\end{align}

Here, the coefficient $c_n$ is a complex number, so that the sign problem appears again and the importance sampling with $c_n e^{-S_g}$ cannot be performed. To avoid this problem, the reweighting method is applied:
\begin{align}
Z_n & = \int {\cal D} U C_0 c_n e^{-S_g}, \nn \\
    & = \bigint sss {\cal D} U \frac{C_0 c_n}{\det \Delta(0)} \det \Delta(0) e^{-S_g}, \nn\\
    & = Z_0 \left \langle \frac {C_0 c_n} {\det \Delta(0)} \right \rangle_0.
\label{eq:2017Jul13eq1}
\end{align}
If we generate gauge configurations at $\mu=0$ and determine the expectation value of $C_0 c_n /\det\Delta(0)$, the canonical partition function for the $n$-quark system can be obtained. The normalization of  $Z_0$ cannot be determined, thus, there is an ambiguity of the magnitude of $Z_n$. However, since the overall factor $Z_0$ does not depend on $n$, in the discussion of the $n$-dependence of $Z_n$, this ambiguity does not cause a problem. On the other hand, since $Z_0$ depends on temperature, we have to remove the $Z_0$ dependence from the discussion of the temperature dependence of $Z_n$. The reweighting method is available as long as the configurations can be generated. Thus, other theories may be used for the configuration generation. However, the overlap problem and phase fluctuation problems always appear in the reweighting method. This point will be considered later together with the data from lattice QCD.

\subsubsection{Triality}

The fugacity expansion of the fermion determinant includes contributions such as $c_1$ and $c_2$ whose quark number is not a multiple of 3. These contributions are not zero for each gauge configuration, but it can be shown that $Z_1$ and $Z_2$ are zero after taking the configuration average~\cite{Roberge:1986mm}. That is, the grand partition function can be described by the sum of the contributions with integer baryon numbers. This property is called the triality. It is widely believed that the triality suggests global color neutrality in quark matter.

As explained in Sec.~\ref{sec:imag_phase}, the grand canonical partition function is invariant under the shift of the imaginary part of the chemical potential
\[ \frac{\mu_I}{T}  \to \frac{\mu_I}{T} + \frac{2\pi}{N_c}.
\]
It implies the RW periodicity,
\[
Z( \mu  ) = Z\biggl( \mu + \frac{2\pi T i}{N_c} \biggr).
\]
Expanding this equation in terms of the fugacity, we obtain
\[
\sum_n Z_n e^{ n \mu/T} =  \sum_n Z_n e^{ n \mu/T} e^{ i 2n \pi/3}.
\]
The contributions in the left- and right-hand sides cancel each other if $n\equiv 0\mod 3$, and the remaining terms become
\[
\sum_{n\in \{ n|n\not\equiv 0\rm mod 3\}} Z_n e^{ n \mu/T} (1 - e^{ i 2n \pi/3}) = 0.
\]
In order for this equation to hold for any $\mu$,
\begin{align}
Z_n = 0,\ n \not\equiv 0\mod 3 
\label{Eq:2014Jan08eq1}
\end{align}
has to hold, namely the canonical partition function $Z_n$ vanishes if $n$ is not a multiple number of $3$. Therefore, the grand canonical partition function consists only of the sectors with $n\equiv 0\mod 3$.

The contributions that make the baryon number rational, such as $c_1$ and $c_2$, are not zero for each gauge configuration. It can naturally be understood by imagining that hadrons can exist on the boundary and only some of the hadrons can be inside the system. The above formula shows that such contributions disappear after taking the average. However, as we will see later, in actual data, $Z_1$, etc., may deviate significantly from zero. The reason for such deviations should be the lack of sufficiently large statistics. However, as far as we notice, nobody has demonstrated that such contributions vanish in the limit of infinitely large statistics.

By using Eq.~(\ref{Eq:2014Jan08eq1}), we can rewrite (\ref{eq:GCZ2}) as a series of the baryon number $n_{\rm B} = n/3$, instead of a series of the quark number $n$:\begin{align}
Z(\mu) &= \sum_{n \equiv 0 ({\rm mod} 3)} Z_n \xi^n
        = \sum_{n_{\rm B}} Z_{n_{\rm B}} \xi_{\rm B}^n, \,\, \xi_{\rm B}=e^{3 \mu/T }=\xi^3.
\label {Eq:2014Dec31eq1}
\end{align}

Although the discussion above can be applied not only to the partition functions but also to other physical quantities ${\cal O}$, the situation changes depending on how ${\cal O}$ transforms under $Z_3$. If ${\cal O}$ is $Z_3$-invariant, we can straightforwardly extend the above discussion, and then only sectors where $n$ is a multiple of $3$ contribute to the expectation value of physical quantity. On the other hand, if ${\cal O}$ is not $Z_3$-invariant, the terms in the fermion determinant whose quark number $n$ are not multiple of $3$ can contribute~\cite{Kratochvila:2006jx}.

\subsubsection{Canonical partition function of QCD}

Now, let us determine the canonical partition function $Z_n$. Here, we introduce a non-perturbative calculation via lattice QCD simulation and an analytic calculation in the high-temperature limit.

First, we analytically derive the canonical partition function in the high-temperature limit. The grand canonical partition function is given by $Z=e^{-F/T}$, where $F$ denotes Gibbs energy. Using $f(\mu)$ in Eq.~(\ref{eq:freene_freegas}), the Gibbs energy is expressed as $F(\mu) = V f(\mu)$, where $V$ is the spatial volume. By substituting $Z(\mu) = e^{-V f(\mu)/T} $ into (\ref{eq:canonical_Fourier}) and performing Fourier integration, we obtain 
\begin{align}
Z_n = \frac{3}{2\pi} \int_{-\pi/3}^{\pi/3} d \theta\; e^{T^3 V g(\theta) + in\theta} , \ \ (n \equiv 0 \ \mod \ 3), (\theta = \mu_I/T).
\label {Eq:2014 Mar16eq3}
\end{align}
Here, we defined $g(\theta) = f(\mu)/T^4$. This $g$ is expanded as $g(\theta) = c_0-c_2 \theta^2 + c_4 \theta^4$. It has one local maximum at $\theta=0$ and two local minima at $\theta = \pm \sqrt{c_2/(2 c_4)}$. As shown in Fig.~\ref{Fig:2014Mar17fig1}, $c_2>c_4$ holds in the high temperature region. In the SB limit, $\sqrt{c_2/(2 c_4)} = \sqrt{2}\pi$~\cite{Kapusta:2006aaa} holds; the lattice data gives $\sqrt{c_2/(2 c_4)} \sim \sqrt{5}$~\cite{Allton:2005gk,Miao:2008sz,Nagata:2012pc,Ejiri:2009hq}; either way, we can assume that the two local minima are outside the domain of integration $[-\pi/3, \pi/3]$. Therefore, $g(\theta)$ is an upwardly convex function in the interval $\theta \in [-\pi/3, \pi/3]$, which has a maximum at $\theta=0$. The peak becomes sharper as the volume increases and $\theta=0$ dominates the integration since the exponent is proportional to the volume. Therefore, the saddle point approximation can be used and $Z_n$ is obtained as
\begin{align}
Z_n = C e^{-n^2/(4T^3V c_2)},
\label {Eq:2014Apr21eq1}
\end{align}
where $\ds{C = \frac{3}{2\pi} \sqrt{\frac{\pi}{T^3 V c_2}} e^{T^3 V c_0}}$. The details of the calculation are given in Appendix~\ref{Sec:incompletegamma}. If the temperature is sufficiently high, the canonical partition function can be obtained without performing complicated calculations such as the fermion determinant, just by using \eqref{Eq:2014Apr21eq1}.

\begin{figure}[htbp] 
\centering
\includegraphics[width=7.5cm]{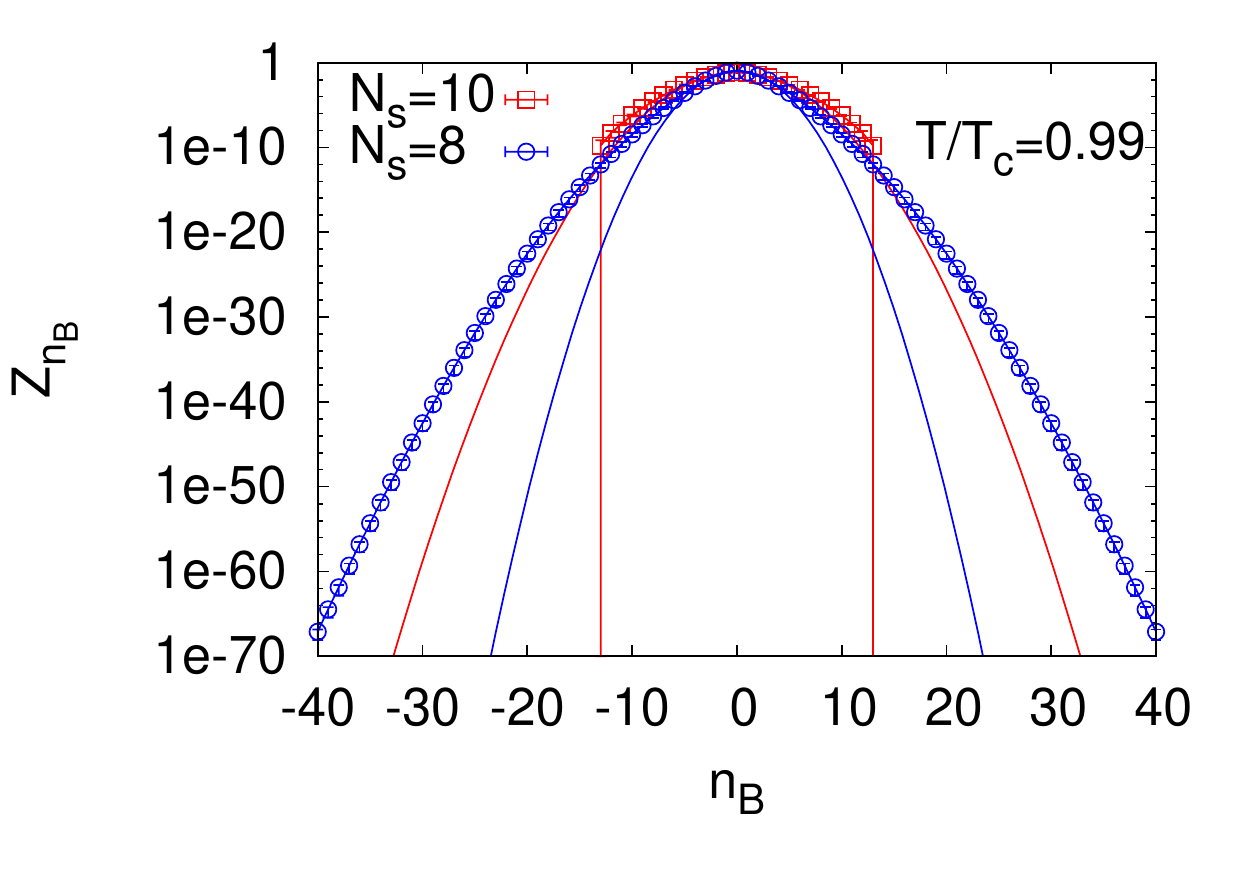}
\includegraphics[width=7.5cm]{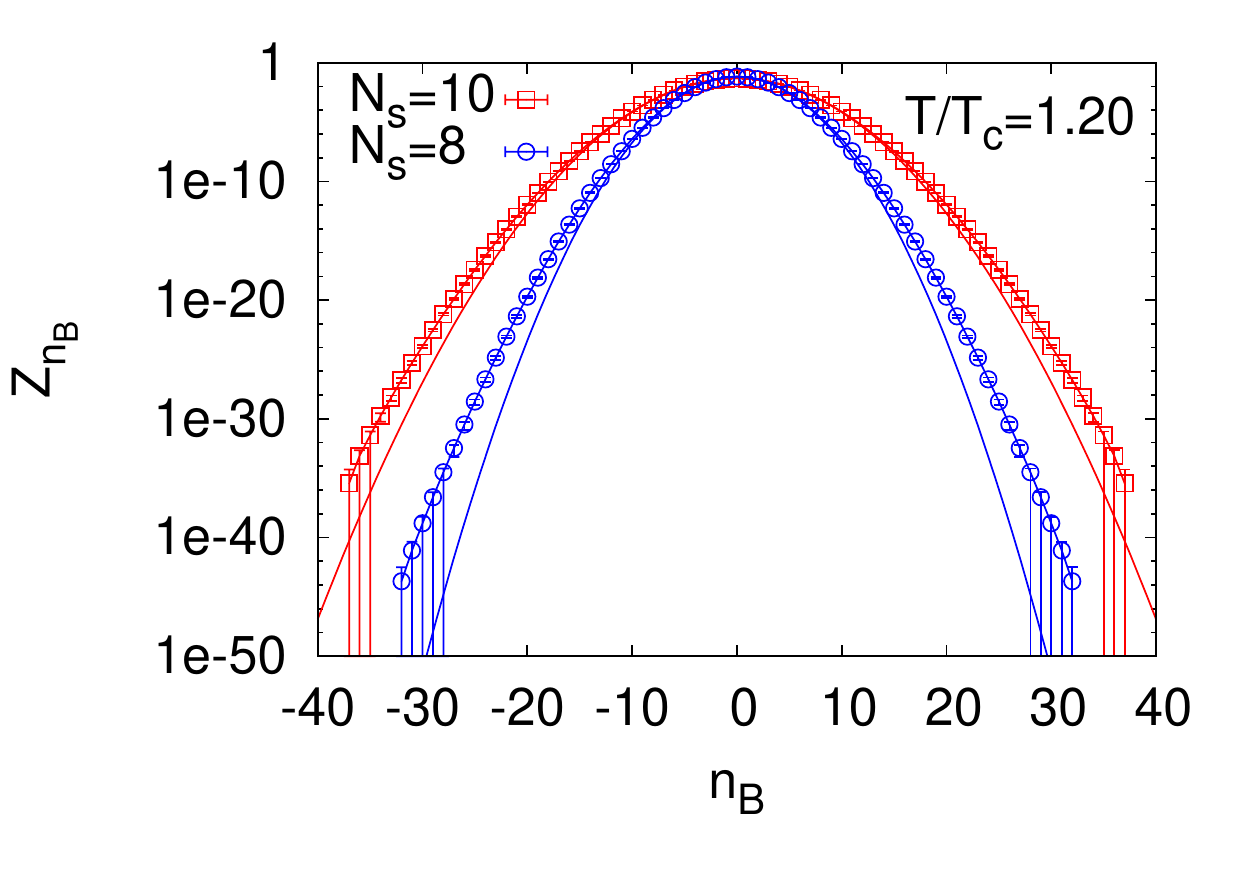}
\begin{minipage}{0.9\linewidth}
\caption{\small 
The canonical partition function as a function of the baryon number $n_B$, obtained by using the canonical method. [Left] $T/T_c=0.99$, [Right] $T/T_c=1.20$. The lattice size is $N_s=8,10$ and $N_t=4$. The solid curve denotes the result of the saddle point approximation, and the value of $c_2$ obtained by the lattice QCD simulation was used. We plot only the data in the regime where the real part of $Z_n$ is positive. The missing data points indicate the regime where the sign problem is severe and the real part of $Z_n$ became negative.
}
\label{Fig:2014Mar17fig2}
\end{minipage}
\end{figure} 

The numerical calculation of $Z_n$ is tricky.
The basic strategy is as follows:
\begin{enumerate}
\item Generate configurations at $\mu=0$.
\item Calculate the eigenvalues of $Q$ in Eq.~(\ref{Eq:2014Jul26eq2}).
\item Find $c_n$ by expanding Eq.~(\ref{eq:2017Mar17eq1}).
\item Obtain Eq.~(\ref{eq:2017Jul13eq1}).
\end{enumerate} 
The expansion in step 3 is a simple operation, but in the actual calculation, we often encounter the cancellation of significant digit and rounding errors, since this operation involves the sum and difference of numbers with different signs and orders. Another common problem is the overflow/underflow. That is, the number exceeds the upper limit of the double-precision variable if there are too many terms. The former problem is  solved by using recursive calculation~\cite{Nagata:2010xi}. 
The method which utilizes the Vandermonde determinant seems to be less precise when the rank exceeds $100$. As for the latter problem, it is essential to use variables beyond double precision. Common libraries are GNU libraries and FMLIB constructed by Smith~\cite{Web:FMLIB}. Ref.~\cite{Nagata:2010xi} used an original library with double precision to keep the calculation accuracy. Although such devices are needed, it is a powerful approach because it can be applied to the canonical method and the reweighting method once $\lambda_n$ and $c_n$ are obtained.

Fig.~\ref{Fig:2014Mar17fig2} depicts the canonical partition function as a function of the baryon number. The left and right panels show the results of $T\sim T_c$ and $T> T_c$, respectively. The data points with error bars denote the numerical results of lattice QCD, and the solid line presents the Gaussian distribution~(\ref{Eq:2014Apr21eq1}) obtained by the saddle point approximation. To fix the value of $c_2$ in the Gaussian distribution,  the lattice QCD results shown in Fig.~\ref{Fig:2014Mar17fig1} was used.

As mentioned in \S.~\ref{sec:canonicalidea}, $Z_n$ is symmetric under the changing of the sign of $n$. The data correctly reproduces this property. $Z_n$ decreases exponentially as $|n|$ increases. Because $Z_n = e^{-F_n/T}$, it means that the Helmholtz free energy $F_n$ of the $n$ particle system increases as the number of particles increases. The $n$ dependence of $Z_n$ or $F_n$ depends on the dynamics of the system, and changes depending on the quark mass, lattice coupling constant, temperature, etc. If we look at the $n$-dependence carefully, we notice that the high-temperature and low-temperature behaviors are different. At high temperature ($T/T_c=1.20$), $Z_{n_{\rm B}}$ agrees well with the Gaussian distribution in the region where $n_{\rm B}$ is small. The agreement becomes better at a larger lattice size. As $n_{\rm B}$ increases, it deviates from the Gaussian approximation because of the contribution of the oscillation part $e^{in \theta}$ in the Fourier integral. On the other hand, at low temperature ($T/T_c=0.99$), the lattice results do not agree with the Gaussian distribution. This is because the conditions for deriving the Gaussian distribution are not satisfied near $T_c$. This difference in the $n$-dependence of $Z_n$ at high and low temperatures has an important physical meaning. Let us consider two typical situations, $Z_n \propto e^{-a n^2}$ and $Z_n \propto e^{-a | n|}$. If $Z_n$ is Gaussian, then the fugacity series $\sum e^{- an^2} e^{-n \mu/T}$ converges for any $\mu$. The radius of convergence is infinite, and there is no discontinuity concerning $\mu$. On the other hand, if t$Z_n \propto e^{-a | n|}$, the radius of convergence of the fugacity series $\sum e^{- a|n|} e^{-n \mu/T}$ is $\mu=Ta$.~\footnote{In the actual calculation, the lattice size is finite so that the grand canonical partition function becomes a finite polynomial and converges for any value of $\mu$.} An example of the latter case is the system of free particles.  In this case, $a$ is the particle mass $m$, and the number of particles in the system rapidly increases if the chemical potential exceeds the mass of the particles. That $Z_n$ increases as $N_s$ increases means it is easier to increase the number of particles in the system if the volume of the system is larger.

Next, let us consider the density accessible by using the canonical method. The baryon density $\rho_B$ is defined as $\rho_B = n_B / (N_sa)^3$ by using baryon number $n_B$ and lattice volume $(N_s a)^3$. The lattice spacing $a$ is a function of the lattice coupling constant $\beta$. The value of $a$ must be determined by inputting a physical quantity, but here we give a rough estimate of the baryon density assuming $a$ to be $0.2$ fm. Regarding the case of $N_s=10$, the extent along the spatial direction is $N_sa \sim 2$ fm, the volume is $(N_s a)^3 \sim 8$ fm$^3$, and the density is $\rho_B \sim \frac{1}{8} n_B $fm$^{-3}$. In $n_B=2$, it becomes $\rho_B \sim 0.25$ fm$^{-3}$, while in $n_B=8$, it becomes $\rho_B \sim 1.0$ fm$^{-3}$, In both cases, the density is higher than the standard nuclear density, $0.17$ fm$^{-3}$. We can also estimate the nucleon exclusion volume. The volume of the nucleon is $4 \pi (0.87)^3/3 \sim 2.7$ fm$^3$ since the nucleon charge radius is $0.87$ fm. If there are three nuclei, the exclusive volume occupied by the nuclei becomes approximately the same as the lattice volume. Thus, the wave functions of the nucleons begin to overlap. In the present calculation, the quark mass is larger than the actual physical value and the ultraviolet cut is small so that the interaction between quarks is over-estimated. Therefore, the nucleon radius is likely to be smaller than the physical value. Assuming the nucleon charge radius for this parameter set is about $0.5$ fm, the nucleon volume is $0.5$ fm$^3$. Therefore, if there are about 20 nucleons, then the nucleon wave functions will overlap each other. It is a rough estimate and hence we should not take it too seriously, but still, we expect that the system including $10\sim 20$ baryons corresponds to a high-density region. If the density is sufficiently high, we may be able to see the phase transition from the hadron phase to the QGP phase. The method of determining the phase transition point will be explained in the next section.

We close this section with two remarks on the canonical method. The first remark is that the precision goes down as $n$ increases. The absolute value of coefficient $c_n$ in the fugacity expansion becomes smaller as $n$ increases ~\cite{Nagata:2010xi}. Furthermore, the statistical error of $Z_n$ becomes large since $c_n$ is a complex number. Thus, the relative error becomes very large. In Fig.~\ref{Fig:2014Mar17fig2},  the data points in the large $n$ region are missing because the real part of $Z_n$ becomes negative due to severe phase fluctuation. Note that the canonical partition function should be real and positive, and negative $Z_n$ is an unphysical result associated with the sign problem. This problem is especially severe in the hadronic phase, and occurs even for relatively small $n$. To study the first-order phase transition line, it is necessary to increase the density at the temperature corresponding to the hadron phase. In that case, the problem is how well this phase fluctuation is under control. The basic strategy to this problem is to increase the number of statistics. A method utilizing multiple ensembles ~\cite{Kratochvila:2005mk} and a two-step accept/reject method~\cite{Li:2008fm} are proposed as well.

The second remark is on the volume dependence of the computational cost. The canonical method requires the calculation of the eigenvalues of the reduced matrix. The rank of $Q$ depends on the spatial volume so that the computational cost increases rapidly with the increase of the lattice volume. As explained in Sec.~\ref{sec:problem_cost}, the cost for calculating determinant grows rapidly with rank, so that the application of the canonical method to a large lattice is currently difficult. In Ref.~\cite{Ejiri:2008xt}, a method using the saddle point method is proposed. The calculation is carried out when the lattice size is relatively large, namely $N_s=16$.

\subsection{The study of phase transition within the canonical method}
\label{sec:pt_canonical}

In this section, we introduce several criteria to find the phase transition point: the Lee-Yang zero theorem, which has a beautiful mathematical structure; the Maxwell construction, which is practically useful; and the method based on the higher-order fluctuation, which is related with the Beam Energy Scan experiment. 

\subsubsection{Lee-Yang zero theorem}

The fugacity expansion of the grand partition function~\eqref{eq:GCZ2} can be regarded as a polynomial of $e^{\mu/T}$. At a finite lattice size, the fugacity polynomial~(\ref{eq:GCZ2}) is just a finite sum and converges uniformly for any value of the chemical potential, and hence, thermodynamic singularity never shows up. The Lee-Yang zero-point theorem~\cite{Yang:1952be, Lee:1952ig} relates the phase transition and the behavior of the root of the fugacity polynomial in the thermodynamic limit. If we consider Eq.~\eqref{eq:GCZ2} as a polynomial of $e^{\mu/T}$, the number of the roots is the same as the order of the polynomial. Because the coefficients $Z_n$ are real positive, the roots are not real positive; they are distributed on the complex plane excluding the positive real axis. Such points on the complex plane that satisfy $Z(\mu) = 0$ are called the Lee-Yang zero points. Lee and Yang discovered that the distribution of the zero points approaches asymptotically close to the positive real axis in the thermodynamic limit and causes the singularity in thermodynamic quantities. (Lee-Yang zero-point theorem will be explained in detail in Sec.~\ref{sec:lee-yang}).

Since the Lee-Yang zero-point theorem is formulated by using the fugacity expansion of the grand canonical partition function, it can easily be combined with the canonical method. The Lee-Yang zero-point calculation of QCD via the canonical method was performed by the Glasgow group including Barbour in the 1990's~\cite{Barbour:1997ej,Barbour:1997bh,Barbour:1991vs,Barbour:1999mc}. They reported that a phase transition occurs at $\mu = m_\pi/2$ at zero temperature. This is an unphysical result, but it turns out that it is not a problem only in the canonical method, rather it is a difficult problem in the analyses at the low-temperature and finite-density regions which occurs in various approaches. This point will be explained in detail in Sec.~\ref{sec:silverblaze}.

Fodor and Katz have investigated the Lee-Yang zero point (strictly Fisher zero point~\cite{Fisher:0}) by extending the temperature to the complex plane and applying the MPR, and found the QCD critical point~\cite{Fodor:2001pe}. The calculation by Fodor and Katz had a strong impact because it was the first example to find the QCD critical point in lattice QCD. However, Ejiri pointed out that, in their method,  the error increases in the thermodynamic limit. No other group could reproduce the result so far, and hence, the validity of the Lee-Yang zero-point calculation via the MPR remains controversial.

\subsubsection{The Lee-Yang zeros at high temperature}

Here, we introduce the calculations of the Lee-Yang zero points at the high-temperature region of QCD, which are reliable. To verify the reliability of numerical calculation, an approximate solution at high temperature will be also derived~\cite{Nagata:2014fra}.

By using the approximate solution \eqref{Eq:2014Apr21eq1} of $Z_n$ for the high-temperature QCD, $Z(\mu)$ is expressed by
\begin{align}
Z(\mu) &= C \sum_{{n_{\rm B}}=-\infty}^\infty e^{- 9 {n_{\rm B}}^2/(4T^3 V c_2) + 3 {n_{\rm B}} \mu/T}.
\label{Eq:2014Apr21eq2}
\end{align}
By using the theta function,
\[ 
\vartheta(z, \tau) = \sum_{n=-\infty}^{\infty} e^{\pi i n^2 \tau + 2 \pi i n z},
\]
\eqref{Eq:2014Apr21eq2} can be written as 
\begin{align}
Z(\mu) \propto \vartheta(z, \tau).
\end{align}
Here, $z$ and $\tau$ denote
\begin{align}
2\pi i z  = 3 \frac{\mu}{T},
\quad
\pi i \tau = - \frac{9}{4T^3 V c_2},
\end{align}
respectively.

The zero points of the theta-function $\vartheta(z,\tau)$ are given by
\[
 z = \frac{2k+1}{2} + \frac{2\ell+1}{2} \tau. 
\]
Here, $k, \ell \in\mathbb{Z}$ are associate with the pseudo-double periodicity of the theta-function. The zero points of \eqref{Eq:2014Apr21eq2} are expressed as 
\begin{align}
\frac{\mu}{T} = \frac{(2k+1) \pi i}{3} - \frac{3(2 \ell +1)}{4 T^3 V c_2}.
\label{Eq:2014Apr28eq1}
\end{align}

On the complex $\xi$ plane, the zero points given by (\ref{Eq:2014Apr28eq1}) are distributed on three radial line segments $\arg \xi = \Im (\mu/T) = \pi/3, \pi,$ and $5\pi /3$. The Lee-Yang zero points near the RW phase transition point, $(\Re(\mu/T), \Im(\mu/T)) = (0, (2k+1)\pi/3)$, are given by
\begin{align}
\frac{\mu}{T} = \frac{(2k+1) \pi i}{3} \pm \frac{3}{4 T^3 V c_2}.
\label {Eq:2014 Mar22eq1}
\end{align}
They asymptotically approach the RW phase transition point; the distance from the transition point is $\sim 1/V$. It suggests that the RW phase transition is the first-order phase transition.

Next, let us consider how to find the Lee-Yang zeros numerically. \eqref{Eq:2014Dec31eq1} contains a negative power term, and obviously $\xi_{\rm B}=0$ does not satisfy $Z(\mu)=0$. Therefore, we can factor out the largest negative power such that
\[
Z(\mu) = Z_{-N} \xi^{-N} (\xi^{2N} + c_1 \xi^{2N-1} + \cdots + 1)
\]
then the zero-points of $Z(\mu)=0$ are those of $\xi^{2N} + c_1 \xi^{2N-1} + \cdots + 1=0$.  
Here, $\xi_{\rm B}$ stands for $\xi$ and $c_i$ denotes a renormalized coefficient in which the coefficient of the first term in the canonical partition function is normalized as $1$. Thanks to the symmetry, the highest-order coefficient and the lowest-order coefficient are the same. We denote the roots as $\alpha_i$, and then $Z(\mu)$ can be expressed as
\[
Z(\mu) = Z_{-N} \xi^{-N} \prod_i (\xi -\alpha_i ).
\]
If $\alpha$ is a root, $1/\alpha$ is also a root ~\footnote{
It comes from the CP invariance, $Z(\mu)=Z(-\mu)$, which is related to the exchange of quarks and antiquarks. If $\xi=e^{-\mu/T}=\alpha$ is a root, then $1/\xi =e^{\mu/T}= 1/\alpha$ is also the root .}.
Therefore, if all the roots inside or outside the unit circle on the complex $\xi$ plane are found, the rest can be found from the symmetry. Note that the fermion determinant is real positive on the unit circle on the complex $\xi$ plane, so there is no Lee-Yang zero there. The simplest way to find the roots of a polynomial is a sequential computation that iterates the root search and factorization. This method can be accurate to some extent if the order of the polynomial is small, but as the order becomes higher the numerical error comes from factorization accumulates and the accuracy decreases. The validity of the root calculation can be verified by recalculating the coefficients in the polynomial from the roots and comparing them with the original coefficients.

The cut-Baumkuchen method (cBK method)~\cite{Nakamura:2013ska} is a highly accurate method that combines the residue theorem and the divide-and-conquer method. Taking the derivative of the logarithm of the ground canonical partition function $Z$, we obtain
\[
\frac{Z'}{Z} = \sum_i \frac{1}{\xi-\alpha_i}.
\]
By considering the line integral along the closed curve $C$ that surrounds $\alpha_i$ on the complex plane, the residue theorem tells us 
\[
\frac{1}{2\pi i}\int_C \frac{1}{\xi-\alpha_i} =1.
\]
The cBK method starts with a path that includes all poles and repeats the domain decomposition and the residue calculation. The line integral tells us how many poles are surrounded by the curve, and hence, we can narrow down the location of the poles with arbitrary accuracy.

We should also care about the statistical error of $Z_n$. Until now, we assumed that the value of $Z_n$ is fixed to a single value in either the factorization method or the cBK method. There is an error in $Z_n$ obtained via simulation, hence there should be an error in the location of the zero point as well. To detect the phase transition via the Lee-Yang theorem, we have to see whether the zero point approaches the positive real axis on the complex plane in the thermodynamic limit. However, the increase of volume causes an increase of the error for $Z_n$, which may make it difficult to determine the behavior of the zero point~\cite{Ejiri:2004yw}.

In Ref.~\cite{Nagata:2014fra}, an error estimation method using the bootstrap method is proposed. $N$ configurations are extracted randomly from $N$ original gauge configurations, allowing duplication, and calculate $Z_n$ for them. By using these $Z_n$, the zero points are determined.  Allowing duplication again, we randomly extract $N$ configurations, calculate $Z_n$, and calculate the zero points. By repeating this many times and finally taking the average value of the zero points, the error of the zero points can be estimated. This cBK+Bootstrap method requires a large number of calculations because it repeats the calculations of $Z_n$ and Lee-Yang zeros many times. However, by obtaining the fugacity coefficient $c_n$ of the fermion determinant in advance, the calculation can be performed efficiently.

\iffigure
\begin{figure}[htbp] 
\includegraphics[width=8cm]{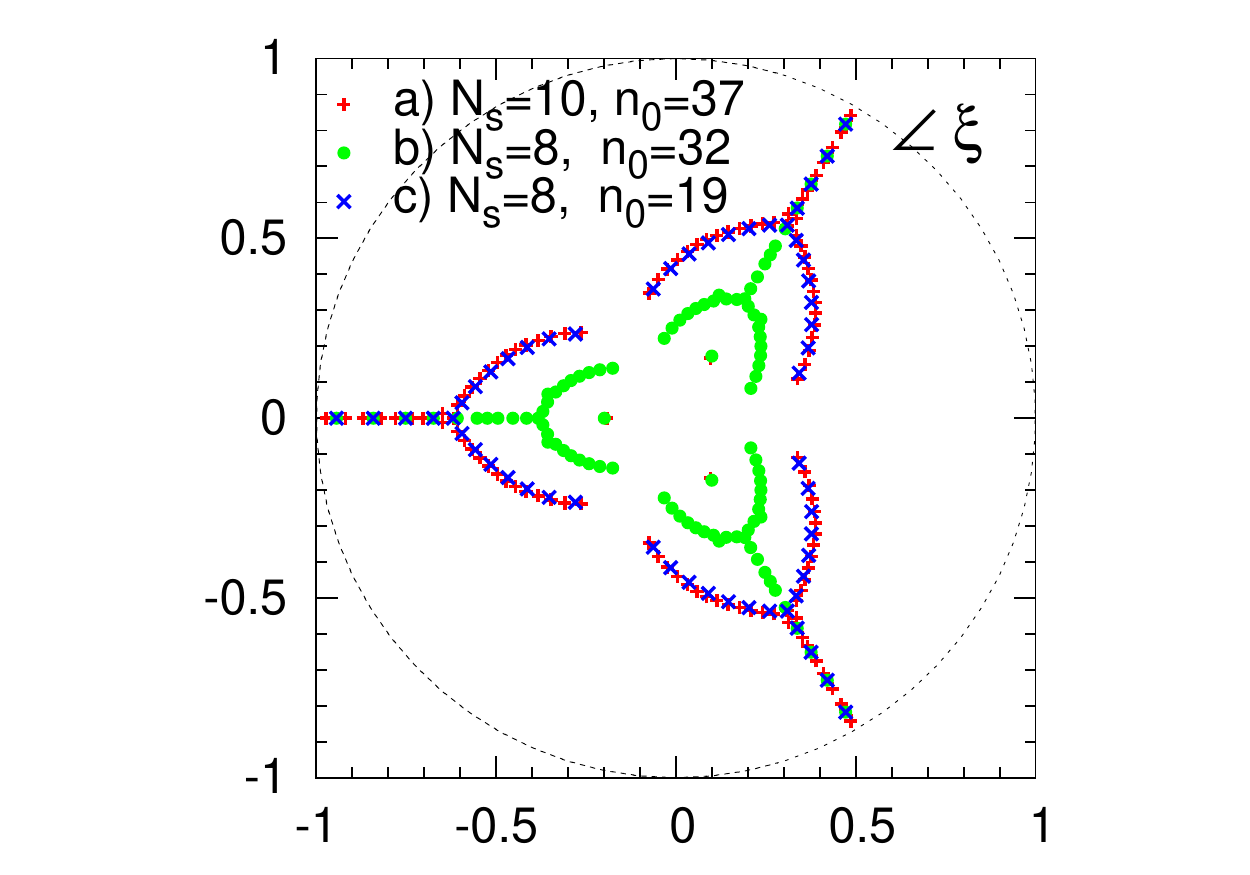}
\includegraphics[width=8cm]{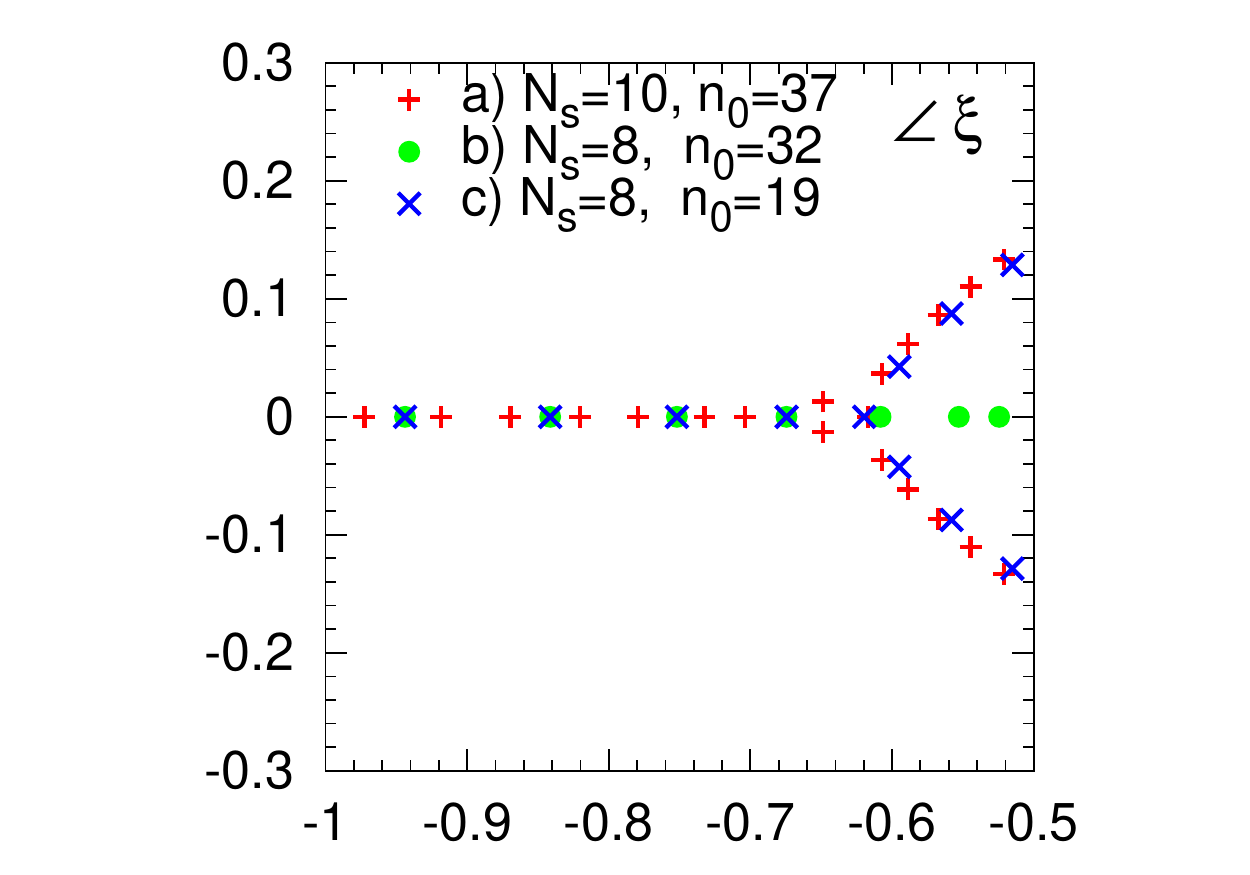}
\begin{minipage}{0.9\linewidth}
\caption{\small 
The distribution of Lee-Yang zeros on the complex fugacity plane at high temperature ($\beta=1.95, T/T_c=1.20$), for one Bootstrap sample. [Left] Lee-Yang zeros inside the unit circle. [Right] The region around the negative real axis. Red, green and blue points are the results of (a) $(n_0, N_s) =(37, 10)$, (b) $(32, 8)$, (c) $(19, 8)$, respectively. The Lee-Yang zeros exist outside the unit circle as well, according to the symmetry $\xi\leftrightarrow 1/\xi$.
The temporal lattice extent is $N_t=4$. 
}
\label{Fig:2014Mar17fig3}
\end{minipage}
\end{figure} 
\fi

The Lee-Yang zero points obtained by the cBK+Bootstrap method are shown in Fig.~\ref{Fig:2014Mar17fig3}. The phase fluctuation of the coefficient of $Z_n$ becomes more severe as $n$ increases, and eventually, the real part can become negative. Such negative $Z_n$'s are artifact due to the sign problem since $Z_n$ must be real and positive. Such terms may lead to artificial singularities, so they must be removed. $n_0$ in the figure is the maximum number of baryons for which $\{Z_{n_B}>0~|~\forall |n_B| <n_0\}$, and terms exceeding this value are discarded. To see the convergence of the fugacity polynomial and the lattice volume dependence, the Lee-Yang zeros are calculated for three cases: (a) $(n_0, N_s) =(37, 10)$, (b) $(32, 8)$, (c) $(19, 8)$. For $N_s=10$, the maximum possible value is $n_0=37$ (see Fig.~\ref{Fig:2014Mar17fig2}). For $N_s=8$, in the case of (b), the maximum possible value is $n_0=32$. For (c), we took $n_0=19$ (though larger $n_0$ could be used) because with this choice the maximum number of baryons per volume is the same as  the case of $N_s=10$: $n_0=19\simeq 37\times(8/10)^3$.

In Fig.~\ref{Fig:2014Mar17fig3}, the zeros are aligned on three straight lines $\arg \xi = \pi/3, \pi,$ and $5\pi/3$  in the region near the unit circle, as expected from Eq.~\eqref{Eq:2014Apr28eq1}. Each straight line splits into two curves as it approaches the origin. Similar behavior has been reported in the first calculation of the distribution of the Lee-Yang zeros by Barbour et al.~\cite{Barbour:1991vs}~\footnote{Barbour et al. plot the zero points on the $e^{\mu}$-plane. By taking the power of $N_t$ of their result, the distribution on the $e^{\mu /T}$ plane can be obtained.}. Even if $n_0$ increases from $19$ to $32$ with $N_s=8$, the zeros near the unit circle $(0.6 <|\xi| <1)$ do not move. It suggests that the fugacity polynomial has converged in this region. In the region near the origin, the zeros move with $n_0$, therefore we cannot tell whether the fugacity series has converged. This is because the stronger dependence on the large $n$ states and the stronger effect of truncation appear near the origin.

\iffigure
\begin{figure}[htbp]
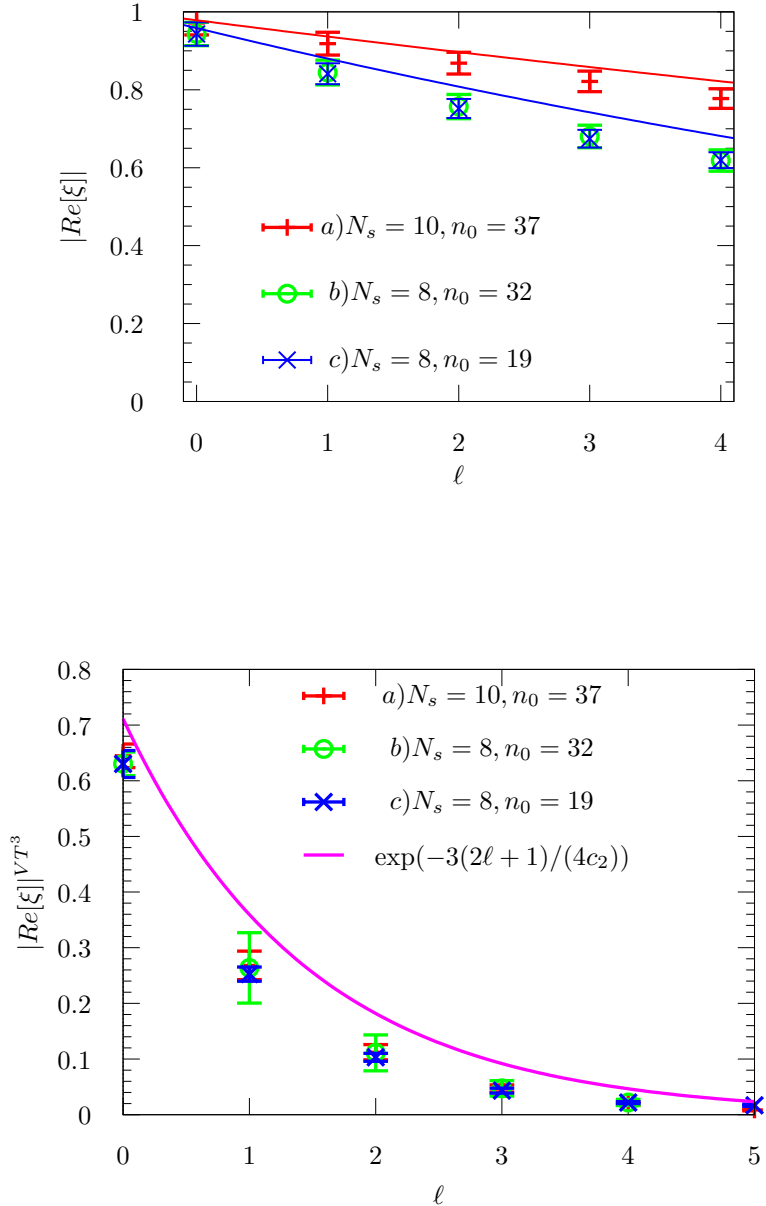
 
\centering
\input{inputf}
\input{inputf_vol} 
\vspace{-1.5cm}
\caption{\small 
[Top] $|\Re \, \xi|$ for Lee-Yang zeros on the negative real axis near the unit circle, where $\xi = \exp(\mu/T)$. Red, green, and blue symbols denote Lee-Yang zeros for (a) $N_s = 10$ and $n_0 = 37$, (b) $N_s = 8$ and $n_0 = 32$, and (c) $N_s = 8$ and $n_0 = 19$, respectively. The curves represent predictions from the Jacobi theta function $\exp (-3(2l +1)/(4T^3 V c_2))$ for $N_s = 10$ (red) and $N_s = 8$ (green and blue). [Bottom] Volume-independent combination $|\Re \xi|^{V T^3}$ .  The errors are estimated by the bootstrap method, with 1000 Bootstrap samples. The curve represents $\exp(-3(2l+ 1)/(4c_2))$. Note that the analytical result is drawn as a continuous curve, although it exists only at  $\ell \in \mathbb{Z}$.
The temporal lattice extent is $N_t=4$. 
 }
\label{Fig:2014Sep03fig2}
\end{figure} 
\fi

Fig.~\ref{Fig:2014Mar17fig3} shows the Lee-Yang zeros obtained from one Bootstrap sample. We carry out the same calculation for multiple bootstrap samples and estimate the error by taking the average value of the zeros from different samples. At high temperatures, the zero points distributed on the negative real axis hardly change depending on the bootstrap samples, while the zero points in the two bifurcated regions do vary depending on the bootstrap sample~\cite{Nagata:2014fra}. The numerical and analytical results for the stable zero points are shown in Fig.~\ref{Fig:2014Sep03fig2}. In the figure, $|\re\, \xi|$ and $|\re\, \xi|^{VT^3}$ are plotted for the zero near the unit circle. The lattice QCD and analytical approach are almost consistent with each other. From this finite-size scaling of the Lee-Yang zeros, it can be seen that the Roberge-Weiss phase transition is of first order.

\iffigure
\begin{figure*}[htb] 
\centering
\includegraphics[width=13cm]{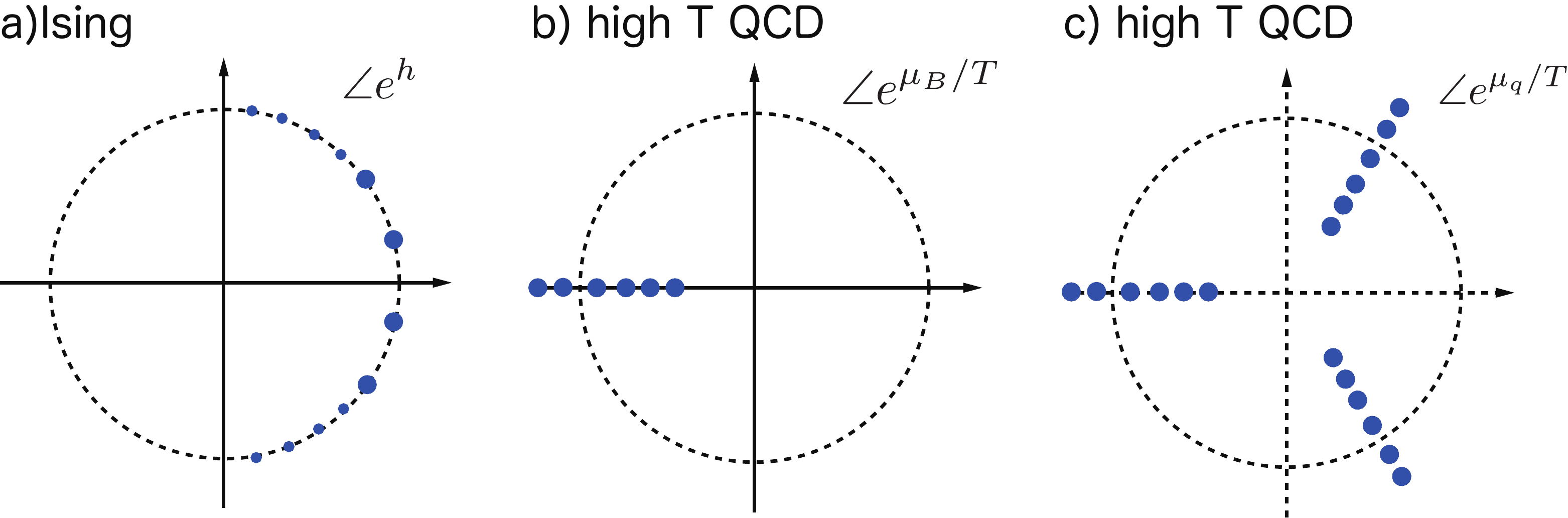}
\begin{minipage}{0.9\linewidth}
\caption{\small 
Schematic figures for the distribution of Lee-Yang zeros in several cases. 
(a) Ising models on the complex plane for $e^{h}$, where $h$ is the external magnetic field in Ising models, (b) QCD on the complex plane for baryon fugacity, and (c) QCD on the complex quark fugacity plane. Dotted circles denote the unit circle. Case (b) can be generalized to free fermion theories.
}
\label{Fig:2014Sep21fig1}
\end{minipage}
\end{figure*} 
\fi

Fig.~\ref{Fig:2014Sep21fig1} contains interesting results regarding the Lee-Yang zeros. Lee and Yang showed that the Lee-Yang zeros are distributed on the unit circle for the Ising model (see the left panel in Fig.\ref{Fig:2014Sep21fig1}). Asano obtained similar results for spin $3/2$ particles~\cite{Asano:1968aaa}. On the other hand, when the canonical partition function follows the Gaussian distribution, as shown in \eqref{Eq:2014Apr28eq1}, the zeros are aligned on the line segment orthogonal to the unit circle. The Gaussian distribution of the canonical partition function follows from the free gas approximation (see \eqref{Eq:2014Apr21eq1}), hence this result can be regarded as an approximate solution of the Lee-Yang zero points for the free Fermi gas (see the central and left panels in Fig.~\ref{Fig:2014Sep21fig1}). The distribution of the Lee-Yang zeros in QCD is different from the Ising model. When the number of flavors is even, it can be clearly shown that there is no zero on the unit circle, because the fermion determinant is real for the pure imaginary chemical potential and hence the QCD action is real everywhere on the unit circle.

Ref.~\cite{Nakamura:2013ska} studied the Lee-Yang zero-points in the hadron phase. However, the error of the locations of the zero points has not yet been estimated. In the hadron phase, the phase fluctuation is severe, and it is difficult to calculate $Z_n$ in the large $n$ region. Highly accurate calculations of the Lee-Yang zeros are an important future task.

\subsubsection{Usage of Maxwell-construction}

\iffigure
\begin{figure}[htbp] 
\centering
\includegraphics[width=13cm]{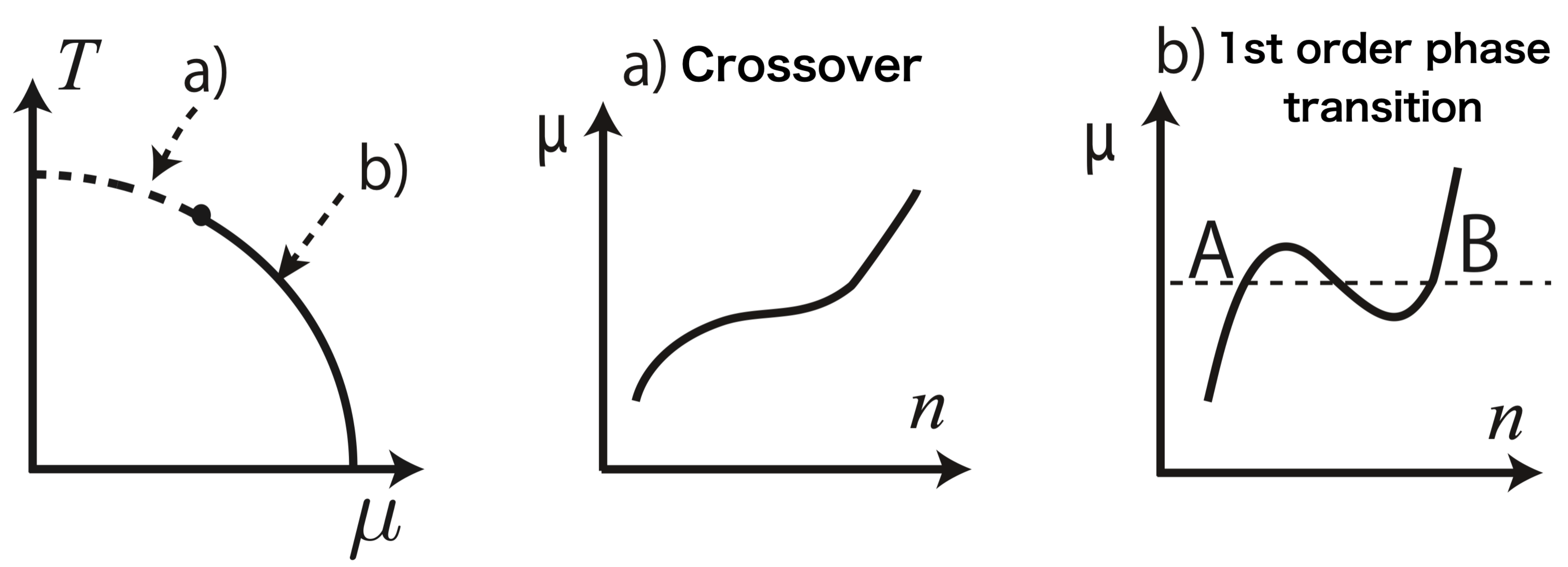}
\begin{minipage}{0.9\linewidth}
\caption{\small 
The $\mu$-$n$ curve and the Maxwell construction. The left panel is the schematic picture of the QCD phase diagram. (a) is the $\mu$-$n$ curve for the theory with crossover regime or in the QGP phase. (b) is the $\mu$-$n$ curve for the theory with first-order phase transition.  In the case of (b), if two areas enclosed by the dotted and solid lines are equal, then the free energies of A and B are also equal. Therefore, this gives the first-order phase transition point. This way of finding the phase transition is called Maxwell construction.  
}
\label{Fig:2017Apr06_MaxwellConstruction}
\end{minipage}
\end{figure} 
\fi

The most successful approach to the hadron-QGP phase boundary is the Maxwell construction~\cite{Kratochvila:2004wz,Kratochvila:2005mk,deForcrand:2006ec}. The Maxwell construction is well known as a method to determine the liquid-gas phase transition point from the van der Waals equation of state. This method can be combined with the canonical method.

The chemical potential is the free energy needed for adding one particle to the system so that it can be expressed by using the canonical partition function as 
\begin{align}
\mu & = F_{n+1}-F_n \nn ,\\
    & =-T (\ln Z_{n+1}-\ln Z_n).
\end{align}
Within each phase, due to the thermodynamic stability of the system, $\partial \mu/\partial n >0$ must be satisfied. Thus, if $n$ increases, then $\mu$ increases (See the middle panel in Fig.~\ref{Fig:2017Apr06_MaxwellConstruction}). On the other hand, if there is a first-order phase transition, the behavior is as shown on the right panel in the figure. In the right panel, $\mu$ depicts an S-curve behavior as a function of $n$. The first increase of the number of particles describes the increase as a function of the chemical potential in the gas phase. The second increase represents the increase in the liquid phase. The intermediate region with $\partial \mu/\partial n <0$ is thermodynamically unstable. Physically, it represents the first-order phase transition via the coexistence of the gas and liquid phases. This first-order phase transition takes place when the free energies of the gas and liquid phases become equal. The chemical potential at this point is determined so that the areas of the two regions enclosed by the dotted line and the curve are equal. This is called Maxwell's equal-area law or Maxwell construction, which gives the phase transition point $\mu_c$.

\iffigure
\begin{figure}[htbp] 
\centering
\includegraphics[width=8cm,angle=270]{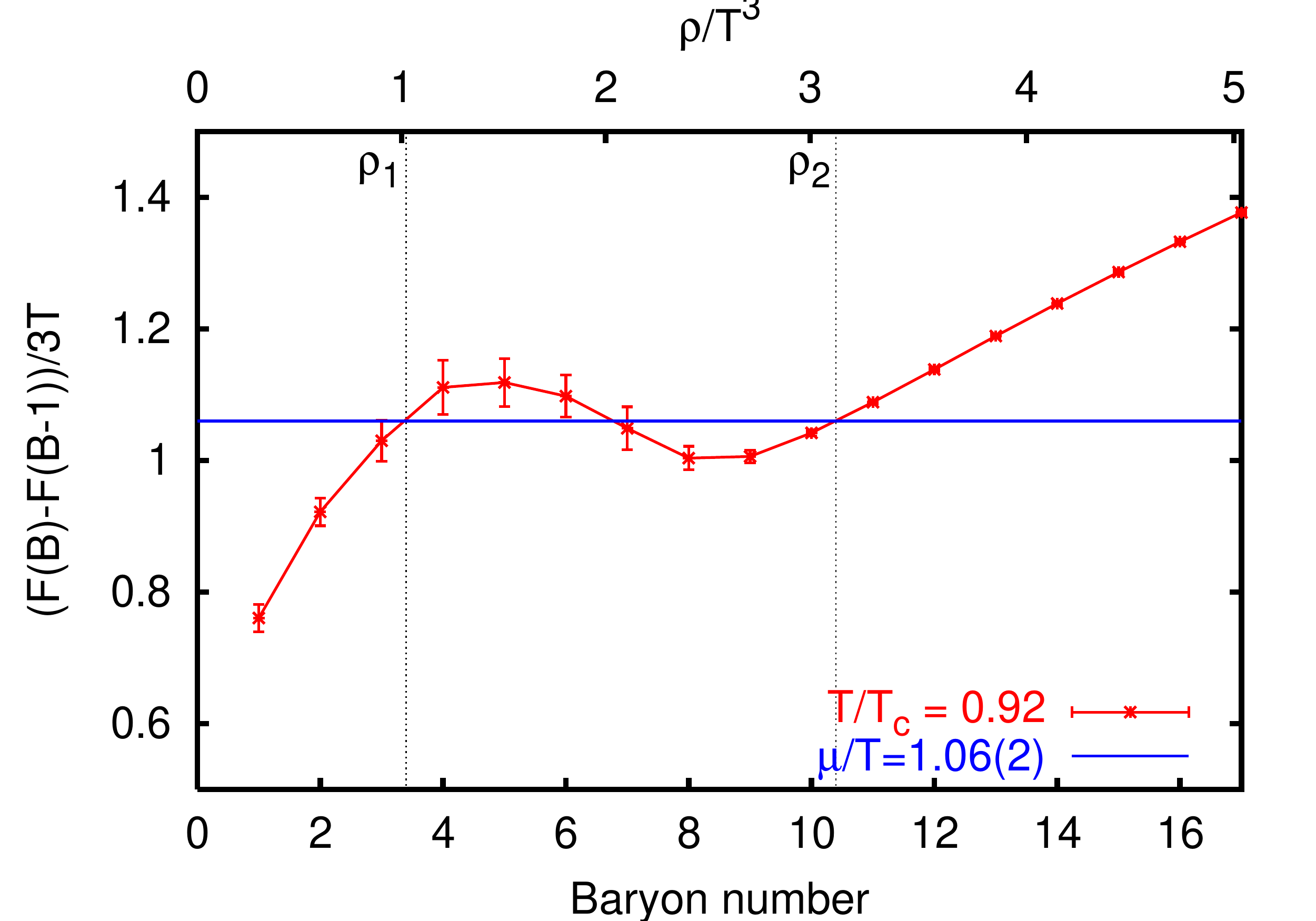}
\begin{minipage}{0.9\linewidth}
\vspace{1.0cm}
\caption{\small 
First successful determination of the first-order phase transition at finite density, via the combination of the canonical method and the Maxwell construction~\cite{Kratochvila:2005mk}.
The lattice size is $6^3\times 4$. 
 }
\label{0509143_MaxwellConstruction}
\end{minipage}
\end{figure} 
\fi

de Forcrand and Kratochvila determined the $\mu-n$ curve by using the canonical method and then used the Maxwell construction to find the first-order phase transition line between hadron and QGP phases~\cite{Kratochvila:2004wz,Kratochvila:2005mk,deForcrand:2006ec} (Fig.~\ref{Fig:2017Apr06_MaxwellConstruction}). Their calculation was for $N_f=4$ staggered fermions. Ref.~\cite{Ejiri:2008xt} applied the same method to $N_f=2$ staggered fermions. Since the canonical partition function $Z_n$ is calculated by using the reweighting method, the overlap problem appears. The lack of good enough accuracy becomes more problematic in the hadron phase, so it is necessary to find a way to determine $Z_n$ accurately in the region where $n$ is large in the hadron phase to identify the QCD critical point. The Kentucky group has proposed a method of accepting/rejecting gauge configurations in two steps for this problem. They applied the canonical method to $N_f=2, 3, 2+1, 4$ Wilson fermions~\cite{Li:2008fm,Li:2007bj,Li:2011ee,Alexandru:2005ix,Meng:2008hj}. In the case of $N_f=3, 2+1, 4$, the QCD critical point and the first-order phase transition line are obtained. On the other hand, in the case of two flavors, the first-order phase transition line is not found up to about $0.9T_c$.

So far, only via the combination of the canonical method and Maxwell construction method, multiple independent groups reported the discovery of the QCD critical point. The main reason why this method is more suitable for finding phase transitions than other methods may be that this method can detect phase transitions by using only some specific sectors of the canonical partition function. For example, the multi-parameter reweighting method uses the grand canonical partition function, which is equivalent to the sum of all particle numbers. On the other hand, in Maxwell construction, it is not necessary to take the sum of the fugacity series, so that the error can be suppressed.
\subsubsection{Fluctuation of the baryon number}

An important quantity in the relativistic heavy-ion collision experiment is the high-order moment of the conserved quantity such as the baryon number. High-order momenta get attention as a signal of the critical point in BES experiments, because it shows characteristic behavior near the QCD critical point~\cite {Stephanov:1998dy,Hatta:2003wn,Stephanov:2008qz,Stephanov:2011pb}. The particle number produced in relativistic heavy-ion collision experiments agrees well with a simple statistical model parameterized by the temperature and chemical potential~\cite{Cleymans:2005xv}. It is believed that the temperature and chemical potential at the chemical freeze-out point, where the particle production processes in the fireball generated by heavy-ion collisions end, can be obtained from the number of particles measured after heavy-ion collisions.  

From the past data analysis, the temperature and chemical potential in chemical freezing are known to be increasing and decreasing functions of the collision energy, respectively. By changing the collision energy, the temperature and density of the fireball can be controlled to some extent. The BES experiment utilizes this property for the study of the QCD phase diagram.

The higher moments are defined by $\langle (\delta n)^m \rangle$, where $n$ is the particle number operator and $\delta n = n-\langle n \rangle$. In particular, the variance $\sigma^2$, skewness $S$, and kurtosis $\kappa$ are defined by
\begin{subequations}
\begin{align}
\sigma^2 &= \langle (\delta n)^2 \rangle, \\
S &= \frac{\langle (\delta n)^3\rangle }{\sigma^3},\label{eq:skewness}\\
\kappa &= \frac {\langle (\delta n)^4 \rangle }{\sigma^4}-3, \label{eq:kurtosis}
\end{align}
\end{subequations}
respectively. These moments are related to the shape of the distribution, and the second moment represents the spread of the distribution, as is well known. The third moment is $0$ if the distribution is symmetric around the expectation value. Thus, it represents the left-right asymmetry. The fourth moment represents the sharpness of the distribution, and in the above definition (subtracting $3$) the Gaussian distribution gives $\kappa=0$. These higher-order fluctuations change significantly near the critical end point~\cite{Stephanov:2008qz, Asakawa:2009aj, Stephanov:2011pb}.

\iffigure
\begin{figure}[htbp]
\centering
\includegraphics[width=0.48\linewidth]{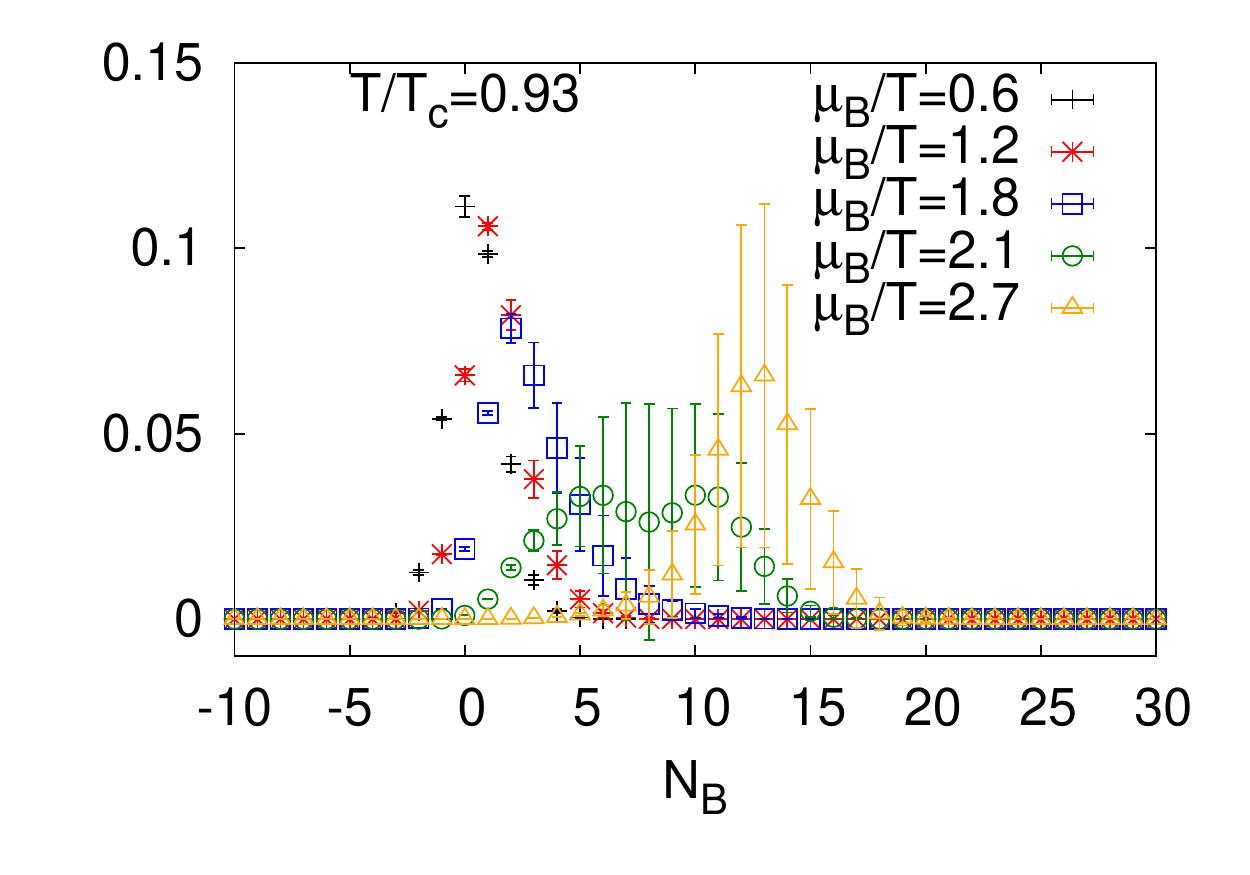}
\includegraphics[width=0.48\linewidth]{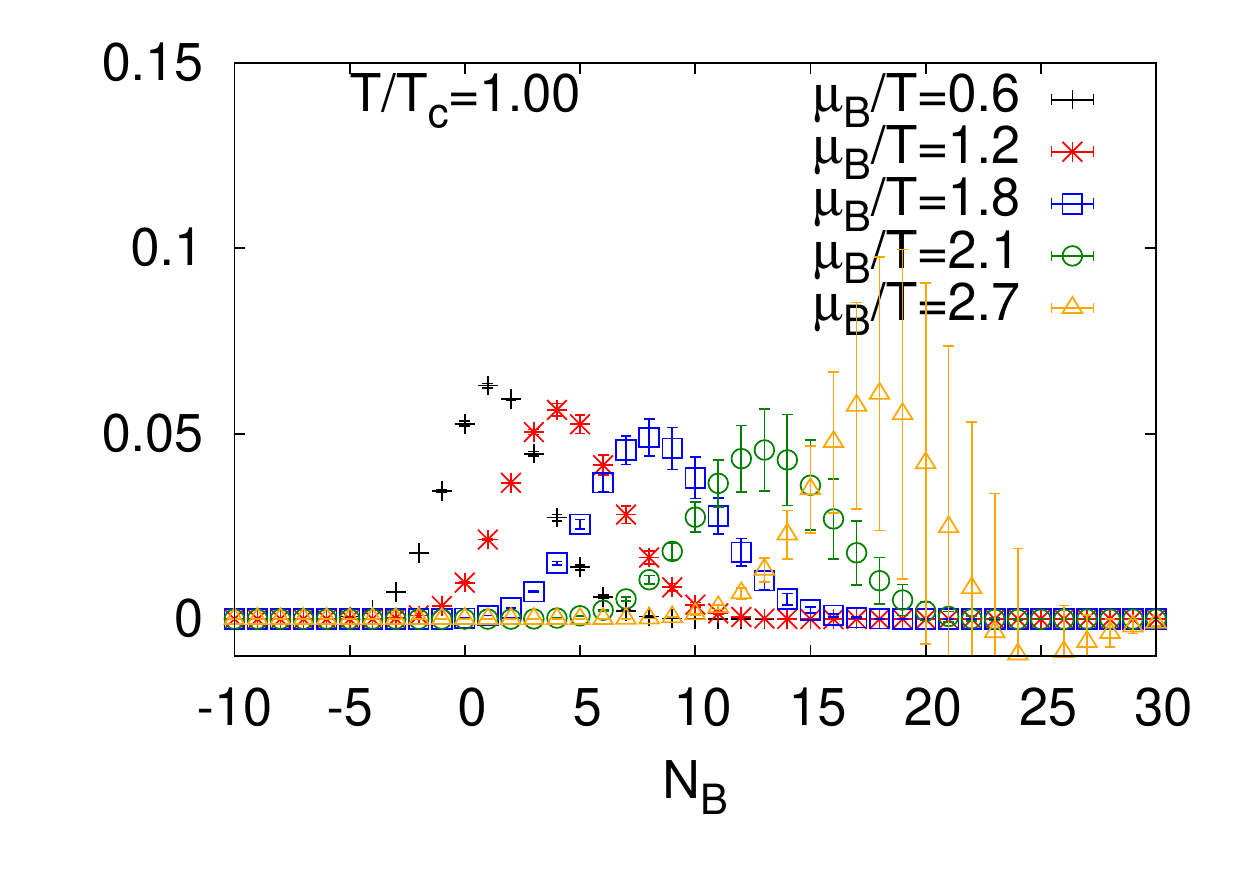}
\includegraphics[width=0.48\linewidth]{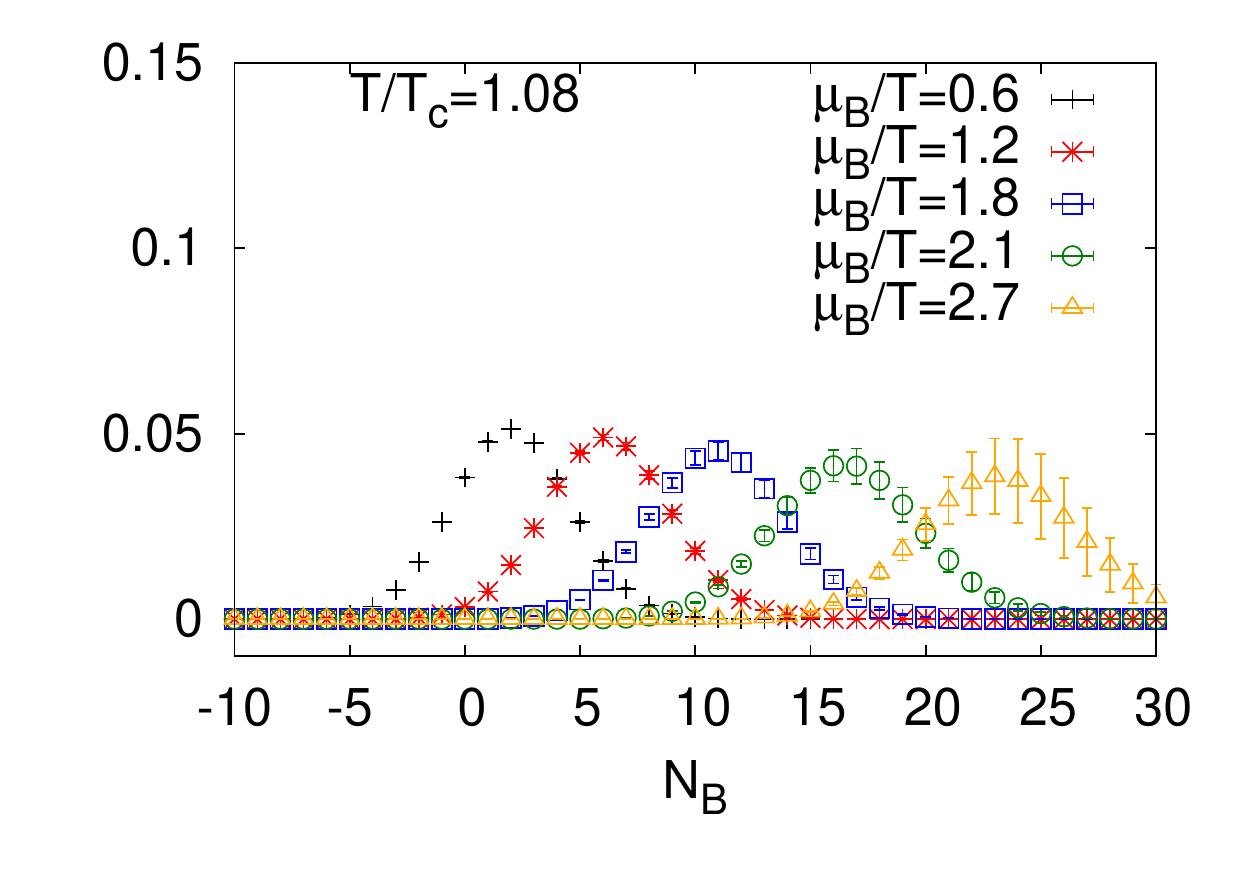}
\begin{minipage}{0.9\linewidth}
\caption{\small 
$\frac{Z_n e^ { n \mu /T} }{\sum Z_n e^ { n \mu /T} }$ as a function of the baryon number, for several values of the chemical potential, at $T/T_c = 0.93, 1.0$ and $1.08$.
}\label{Fig:2012Oct17fig2}
\end{minipage}
\end{figure}
\fi

\iffigure
\begin{figure}[htbp]
\centering
\includegraphics[width=0.48\linewidth]{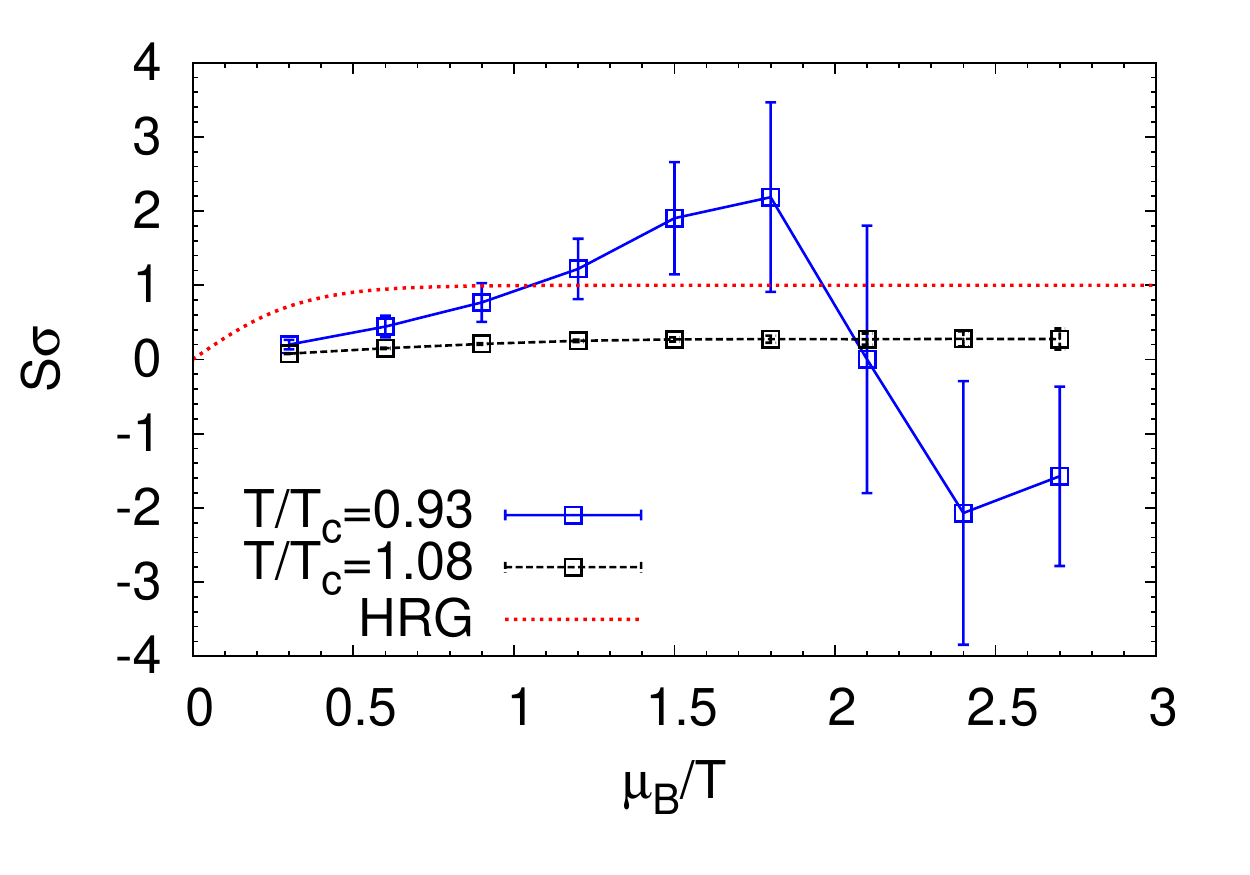}
\includegraphics[width=0.48\linewidth]{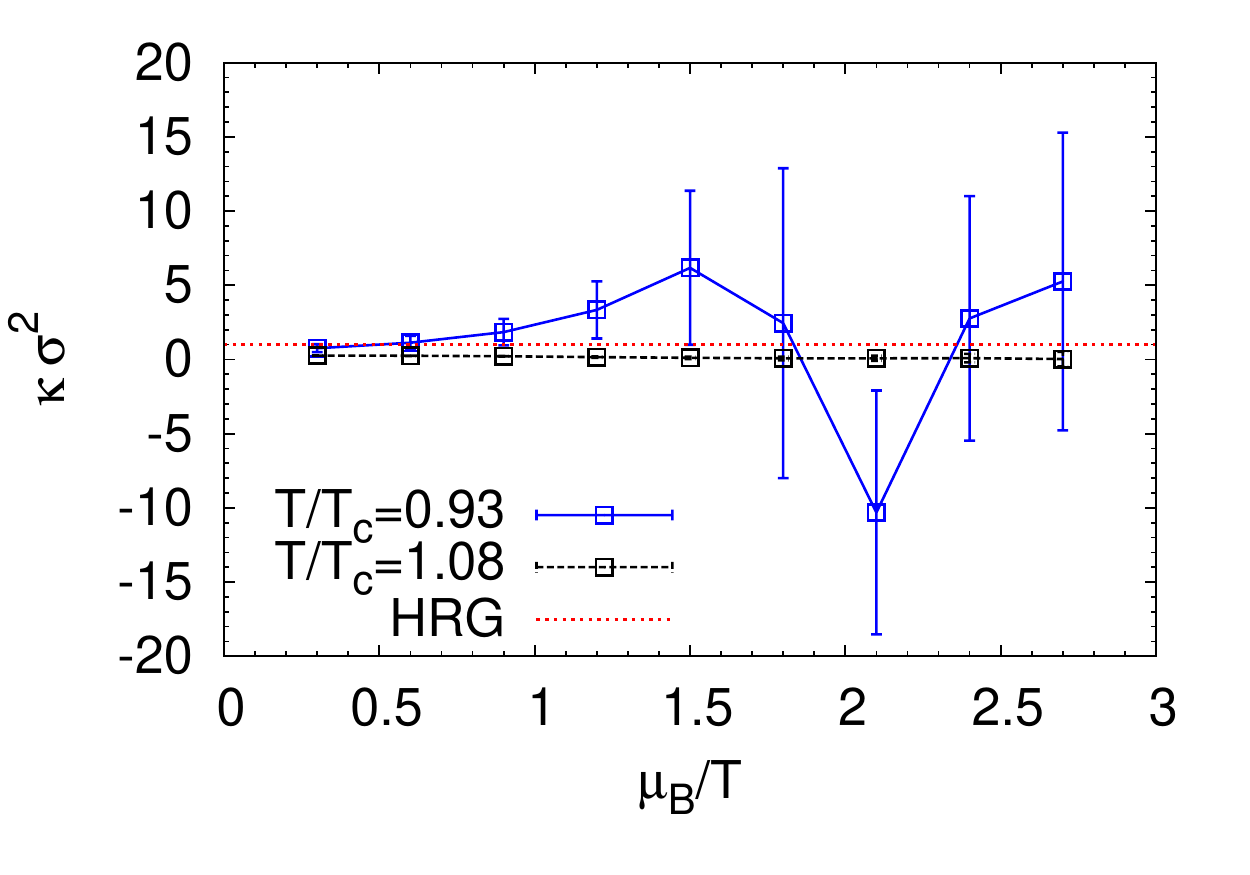}
\begin{minipage}{0.9\linewidth}
\caption{\small 
Higher moments vs. chemical potential.
}\label{Fig:2017Sep13fig2}
\end{minipage}
\end{figure}
\fi

We consider the parameter dependence of higher-order moments by using the canonical partition function. In Fig.~\ref{Fig:2012Oct17fig2}, $Z_n e^{ n \mu/T}/\sum Z_n e^{ n \mu/T}$ is shown for three different temperatures. At high temperatures ($T/T_c=1.08$), the shape of the distribution does not change with $\mu$. As mentioned above, $Z_n$ can be approximated by the Gaussian distribution at high temperature: 
$Z_n \propto e^{-a n^2}$. Then
\[
Z_n e^ {n \mu /T} \propto e^{-a (n-\frac{1}{2a}\mu /T )^2}.
\]
Therefore, the distribution remains Gaussian even if $\mu/T$ is changed, and only the position of the peak of moves. The higher-order moments do not change. On the other hand, at the temperature corresponding to the hadron phase, the shape of the distribution changes with $\mu$. At $\mu_B=0$, the distribution is left-right-symmetric about $n=0$. As $\mu_B$ is increased, the tail of the distribution extends toward larger $n$ (see the right panel of the figure). A flat structure with no peak appears at $\mu_B/T=2.1$, and the tail of the distribution extends to the left at $\mu_B/T=2.7$. Those tails are captured by the skewness, and the flat structure appears as the negative kurtosis. These higher moments are shown in Fig.~\ref{Fig:2017Sep13fig2}. Thus, at low temperatures, the canonical partition function indicates the deviation from the Gaussian distribution, which affects higher-order moments.

Here, it is worth mentioning that the distributions at different values of $\mu$ are obtained from the same set of $Z_n$'s, in each plot of Fig.~\ref{Fig:2012Oct17fig2}. For example, $Z_{N_B}$ are the same for $\mu_B /T = 0.6$--$2.7$ at $T/T_c=0.93$. The only difference is the fugacity factor $\exp (n_B \mu_B/T)$. As described at the beginning of this section, gauge configurations are generated at $\mu=0$, and the high-density state can be reconstructed from the fluctuations.

Because the canonical method is based on general features of statistical mechanics, it applies to the experimental data in principle. If the number of particles contained in the system is measured and the probability that the $n$-particle state is observed is calculated, then it gives the normalized $Z_n$. Of course, such a calculation is impractical for an ideal statistical dynamical system containing many particles of the order of the Avogadro's number. However, the number of particles generated in the relativistic heavy-ion collision experiment is of the order of 1000, which is not too big. The particles that entered the measuring instrument are counted and the particle number distribution is obtained~\cite{Aggarwal:2010wy}. In Ref.~\cite{Nakamura:2013ska}, the canonical method is applied to the RHIC experimental data. Identifying the particle number distribution obtained in the heavy-ion collision experiment with $\{ Z_n e^{n\mu/T} \}$ and by assuming the CP invariance, $\mu/T$ and $Z_n$ can be determined simultaneously. The value of $\mu/T$ obtained in this manner agrees with the value at the chemical freezing~\cite{Nakamura:2013ska}. Moreover, by removing the nucleon mass from $Z_{N_B}$ in the hadron phase, the binding energy of the nucleus $\Delta E(N_B)$ defined by 
\[ \Delta E(N_B) =-T \ln Z_{N_B}-N_B m_N(\mu_B)\qquad (T\to 0)\]
can be calculated. Such an application is difficult at this moment due to a large computational cost. If the problem of computational resources can be resolved, a derivation of the Bethe-Weizs\"acker formula based on lattice QCD will be doable in the future.

\iffigure
\begin{figure}[htbp] 
\centering
\includegraphics[width=8cm]{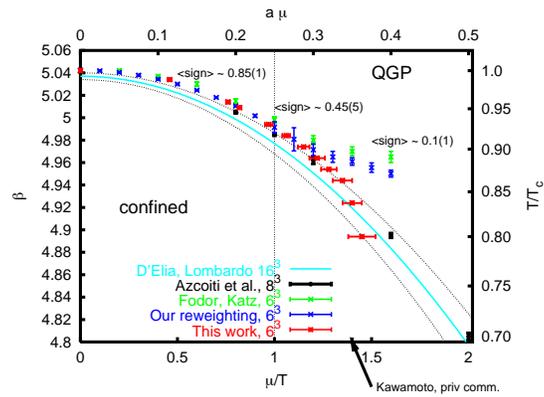}
\begin{minipage}{0.9\linewidth}
\vspace{1cm}
\caption{\small 
The phase diagram in the $T$-$\mu$-plane obtained by several methods. This plot has firstly shown in Ref.~\cite{Kratochvila:2005mk}.
In the small $\mu/T$ regime, different methods lead to the same result.
In the large $\mu/T$ regime, there is no consensus. 
}\label{0509143phasediagplusstr}
\end{minipage}
\end{figure} 

The methods introduced so far --- the reweighting method, Taylor expansion, the analytic continuation from the imaginary chemical potential, and the canonical method --- give consistent results in the low-density region near $T_c$; see Fig.~\ref{0509143phasediagplusstr}, which was presented originally in Ref.~\cite{Kratochvila:2005mk,deForcrand:2010ys}. This agreement suggests that we can trust the results in this parameter region. Although there are remaining tasks such as the continuum limit, large-volume limit, and the extrapolation to the physical quark mass, we can expect steady progress.

\clearpage
\section{Difficulty in the high-density region of the hadron phase}
\label{sec:silverblaze}

In the previous sections, we have seen that the methods to circumvent the sign problem based on the importance sampling are useful in high-temperature, low-density regions. On the contrary, those methods do not work in the low-temperature hadron phase. Behind the difficulty in the low-temperature region is the early-onset problem: when the quark chemical potential reaches $m_\pi/2$ (where $m_\pi$ is the meson mass) at zero temperature, intense phase fluctuations set in. This problem was discovered by Gibbs in the derivation of the propagator matrix and then reported in a lattice QCD simulation by the Glasgow group. Due to the early-onset problem, it is difficult to go beyond $\mu=m_\pi/2$ by using the gauge configurations generated at $\mu<m_\pi/2$. Therefore, it is impossible to study the hadron phase in the finite density region corresponding to the nucleus or the neutron star by the methods explained so far or their minor improvements. This presents the limitation of the methods explained in Sec.~\ref{sec:oldmethod} and suggests that we have to invent a new method to study the high-density regions. In this section, the problems in the study of the hadron phase in the finite density region will be explained in detail. 

\subsection{The early onset problem}

We consider the vacuum at finite volume $V$ ($T=0, \mu=0$). If we add one fermion with mass $m$, then the free energy of the system goes up by $m$. The chemical potential needed for this is $\mu=m$ because it is the free energy required to add one particle to the system. At $T=0$, if $\mu$ is less than $m$, no particles can be added to the system, so that the number of particles does not change. Namely, 
\[
\langle n \rangle = 0 \; \mbox{for} \; \mu <m
\quad
{\rm at}
\quad
T=0.
\]
The minimum free energy required to raise the baryon number $\langle n_B \rangle $ in the hadron phase is equal to the ground state nucleon mass $m_N$. Thus, for baryon chemical potential, $\mu_B$,
\[
\langle n_B \rangle = 0 \; \mbox{for} \; \mu_B <m_N
\quad
{\rm at}
\quad
T=0,
\]
or for the quark chemical potential, $\mu$,
\[
\langle n_q \rangle = 0 \; \mbox{for} \; \mu <\frac{1}{3} m_N
\quad
{\rm at}
\quad
T=0.
\]
Therefore, if $\mu$  is turned on at zero temperature, the quark number starts to increase at $\mu=m_N/3$. This is a phenomenologically natural expectation.

In the mid-1980s, at the beginning of the study of finite density lattice QCD, the quenched simulations have been performed. Then it was reported that the quark density starts to increase at $\mu=m_\pi/2$, which is different from the expectation~\cite{Barbour:1986jf}. This is different from empirical understanding, hence there must be some mistake in the calculations. The Glasgow group called it the early onset problem, meaning that the threshold of $\mu$ for the rise of the quark number is smaller than the phenomenological expectation. They tried to solve this problem from the 1980s to the 1990s. One of the weaknesses of lattice QCD calculation is that, when the result is wrong, it is difficult to pinpoint the reason. Although there are similar difficulties in any method, in the case of lattice QCD, various tricks are used to reduce the computational cost, and one has to check them one by one. Furthermore, there are various possible biases in a finite density lattice QCD, including the discretization error, the finite lattice size effect, deviations from realistic values of parameters such as heavy quark mass, and some common errors such as small statistics and under-estimate of the numerical errors. In addition, the method to avoid the sign problem might have failed. The Glasgow group made various attempts to avoid the early-onset problem, but the problem was not resolved and the cause was not identified. Their results are summarized in Ref.~\cite{Barbour:1997ej}.

It seems that the early-onset problem was initially recognized as a mere calculation error. Several studies from the 1980s to recent years have revealed that the occurrence of this phenomenon is related to the fermion action of the lattice QCD ~\cite {Gibbs:1986hi, Stephanov:1996ki,Cohen:2003kd, Nagata:2012tc, Nagata:2012ad, Nagata:2012mn}. Gibbs pointed out that in 1986, that is almost same time that the Glasgow group encountered the early onset problem in their quenched simulation, the increase in the number of quarks occurs at $\mu=m_\pi/2$~\cite{Gibbs:1986hi} by investing the relationship between the spectrum of the propagator matrix and the hadron mass.
In the late 1990s, Stephanov applied the replica method to the chiral random matrix theory and pointed out that the chiral symmetry is restored at $\mu>0$ in the quench approximation (note that $m_\pi=0$ in the chiral limit)~\cite{Stephanov:1996ki}. In 2003, Cohen coined the name ``Silver Blaze'' taking a metaphor from an episode of Sherlock Holmes, by referring to the behavior that did not appear even though the chemical potential is introduced to Lagrangian. He studied the relationship between the Silver Blaze phenomenon in the finite isospin QCD (i.e., QCD with an isospin chemical potential) and the spectrum of the QCD Hamiltonian~\cite{Cohen:2003kd}. He conjectured that the early onset and the cancellation of complex phases are needed to ensure the Silver Braze concerning the baryon chemical potential. The works of Stephanov and Cohen are theoretical; the early-onset problem does not come from lattice artifacts such as discretization errors and finite-volume effects, or a flaw in the numerical simulation. For the case of the baryon chemical potential, the author and collaborators derived the zero-temperature limit of the fermion determinant on the lattice, and showed that the fermion determinant does not depend on the chemical potential at $\mu<m_\pi/2$~\cite{Nagata:2012tc,Nagata:2012ad,Nagata:2012mn}.
All above suggest that the phase fluctuation of the fermion determinant is related to the early onset problem.

Recently, new path integral methods such as the complex Langevin method and Lefschetz thimble method have been developed. Even these methods encounter troubles at $\mu=m_\pi/2$; the complex Langevin method has the problem of wrong convergence, and the Lefschetz thimble method has a problem associated with the summation of multiple thimbles ~\cite{Tanizaki:2015rda}. These problems come from the same fermion determinant.

In the following sections, we explain the properties of the fermion determinant in the low-temperature finite-density region, which causes the early onset problem. Then, we describe the limitation of the methods introduced in this chapter.

\subsection{Zero-temperature limit of the fermion determinant}

\subsubsection{Property of the reduction formula for the fermion determinant}
\iffigure
\begin{figure}[htbp] 
\centering
\includegraphics[width=8cm]{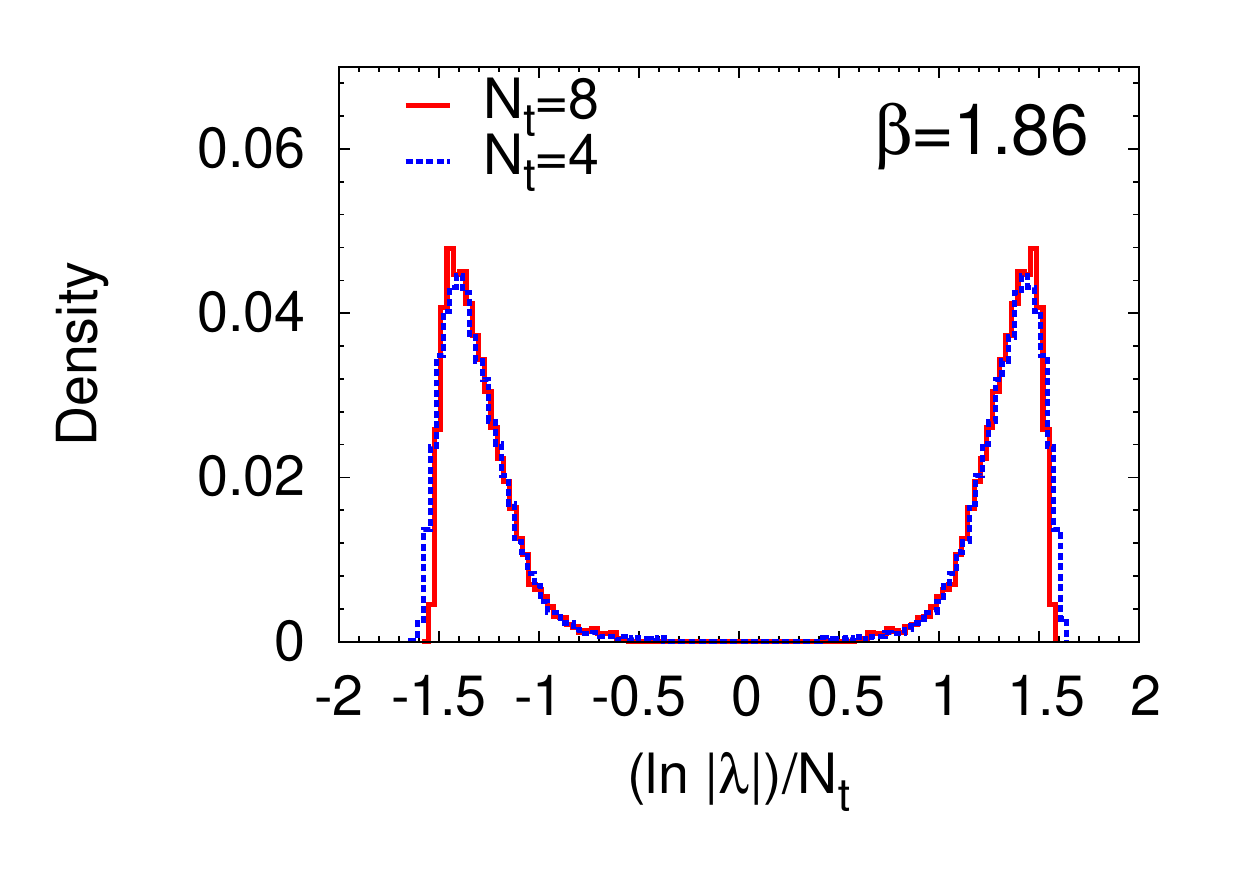}
\begin{minipage}{0.9\linewidth}
\caption{\small 
Histogram of the eigenvalue distribution as a function of $\frac{\ln |\lambda|}{N_t}$, at $m_\pi/m_\rho=0.8$~\cite{Nagata:2012tc}. The angular direction of $\lambda$ is integrated out. The blue and red lines are $N_t = 4$ and $8$, respectively.
Except for $N_t$, the same lattice parameters are used, including the spatial lattice extent $N_s=8$.  
}\label{Fig:2012Jan01fig2}
\end{minipage}
\end{figure} 
\fi

To investigate the early onset problem, we derive the fermion determinant in the zero-temperature limit by using the reduction formula.

We denote the eigenvalue of $Q$ as $\lambda=|\lambda|e^{ i\theta}$, and the eigenvalue density on the complex plane as $\rho(|\lambda|, \theta)$. The density $\rho(|\lambda|)$ along the radial direction of $\lambda$ is expressed by
\[
\rho(|\lambda|) = \int d\theta \rho(|\lambda|, \theta).
\]
This is shown in Fig.~\ref{Fig:2012Jan01fig2}. Fig.~\ref{Fig:2013Sep29fig1} shows the magnitude of eigenvalue,  $|\lambda_n|$, in the increasing order. By using these plots, we will explain three important properties of the reduction matrix $Q$:
\begin{description}

\item[(i) Symmetry]  If $\lambda$ is the eigenvalue of $Q$, then $1/\lambda^*$ is also the eigenvalue of $Q$.

\item [(ii) Gap] The eigenvalue distribution of $Q$ has a gap near $|\lambda | \sim 1$ ($\rho(|\lambda | \sim 1) = 0 )$.

\item[(iii) $N_t$ scaling] The density of $|\lambda|^{1/N_t}$ does not depend on $N_t$. That is, the size of $\lambda$ follows certain scaling law.
\end{description}

Firstly, (i) is an exact property associated with the $\gamma_5$-Hermiticity of the fermion matrix, which can easily be proven ~\footnote{ 
If $\det \Delta(\mu) = 0$, then $\det \Delta^\dagger(-\mu)=0$ follows from Eq.~\eqref{eq:g5ahrelation2}. According to the reduction formula, if $\det (Q + e^{-\mu/T}) = 0$ holds, then $\det (Q^\dagger + e^{\mu^*/T})=0$ holds as well. Therefore, if $Q$ has an eigenvalue $\lambda$, $1/\lambda^*$ is also an eigenvalue of $Q$. Different proofs can be found in Ref.~\cite{Fodor:2007ga,Alexandru:2010yb}.}. The locations of the two symmetric peaks in Fig.~\ref{Fig:2012Jan01fig2} are related via this symmetry. On the complex $\lambda$ plane, this symmetry relates inside and outside of the unit circle. If we denote the eigenvalues inside the unit circle are ${\cal S}=\{\lambda_n | |\lambda_n|<1, n = 1, 2, \cdots, \Nred/2\}$ and set the external eigenvalue to ${\cal L}=\{\lambda_n | |\lambda_n|>1, n = 1, 2, \cdots, \Nred/2\}$, then the eigenvalues of ${\cal S}$ and ${\cal L}$ form a pair.

The property (ii) can be understood by noticing that $\rho(\lambda) = 0$ in the region of $-0.5 \lesssim \frac{\ln |\lambda_n|}{N_t} \lesssim 0.5$ in Fig.~\ref{Fig:2012Jan01fig2}. By characterizing the size of the gap by the eigenvalue $\max_{\lambda_n\in \calS} |\lambda_n|$ closest to the unit circle, the gap size is related to the $\pi$ meson mass. We will explain it later.

To see property (iii), we note that the distributions of $\ln |\lambda|/N_t$ at $N_t=4$ and  $N_t=8$ shown in Fig.~\ref{Fig:2012Jan01fig2} coincide well. Hence, $\rho( |\lambda|^{1/N_t})$ is independent of $N_t$. The eigenvalue spectra in Fig.~\ref{Fig:2013Sep29fig1} show this property more directly. With this scaling, $\lambda_n\in {\cal S}$ can be expressed as $\lambda_n = \exp(-\epsilon_n a N_t + i\theta_n(N_t))$, where $\epsilon_n$ is a quantity that does not depend on $N_t$. The pair of eigenvalues, $\lambda_n \in {\cal L}$, can be written  as $\lambda_n = \exp( \epsilon_n a N_t + i\theta_n(N_t))$. Furthermore, by using $T=(aN_t)^{-1}$, we obtain $\lambda_n = \exp(\pm \epsilon_n /T + i\theta_n(N_t))$. The scaling with respect to $N_t$ can be interpreted as the scaling with respect to temperature~\footnote{
Ideally, we should check this scaling at even larger values of $N_t$. However, it is difficult due to the required numerical accuracy. Because the reduction matrix $Q$ is the $N_t$-th power of the transfer matrix, 
the condition number is given by 
\[ \frac{| \lambda_{\rm max}|}{| \lambda_{\rm min}|} = \left(\frac{a}{b}\right)^{N_t}.
\]
by using the maximum and minimum eigenvalues of the transfer matrix, $a$, and $b$. As $N_t$ increases, the condition number increases and it becomes harder to determine the eigenvalues accurately. Furthermore,  the convergence of iterative methods such as the CG method is slow because $Q$ is a dense matrix. Therefore, a non-perturbative verification of the scaling law up to larger values of $N_t$ is a numerically challenging task. So far, an improved method utilizing some approximations for the estimate of the eigenvalues of $Q$ has been proposed in ~Ref.~\cite{Fodor:2007ga}, and an application of the residue theorem has been proposed in Ref.~\cite{Futamura:2014kga}.}.

\begin{figure}[htbp] 
\centering
\includegraphics[width=7.5cm]{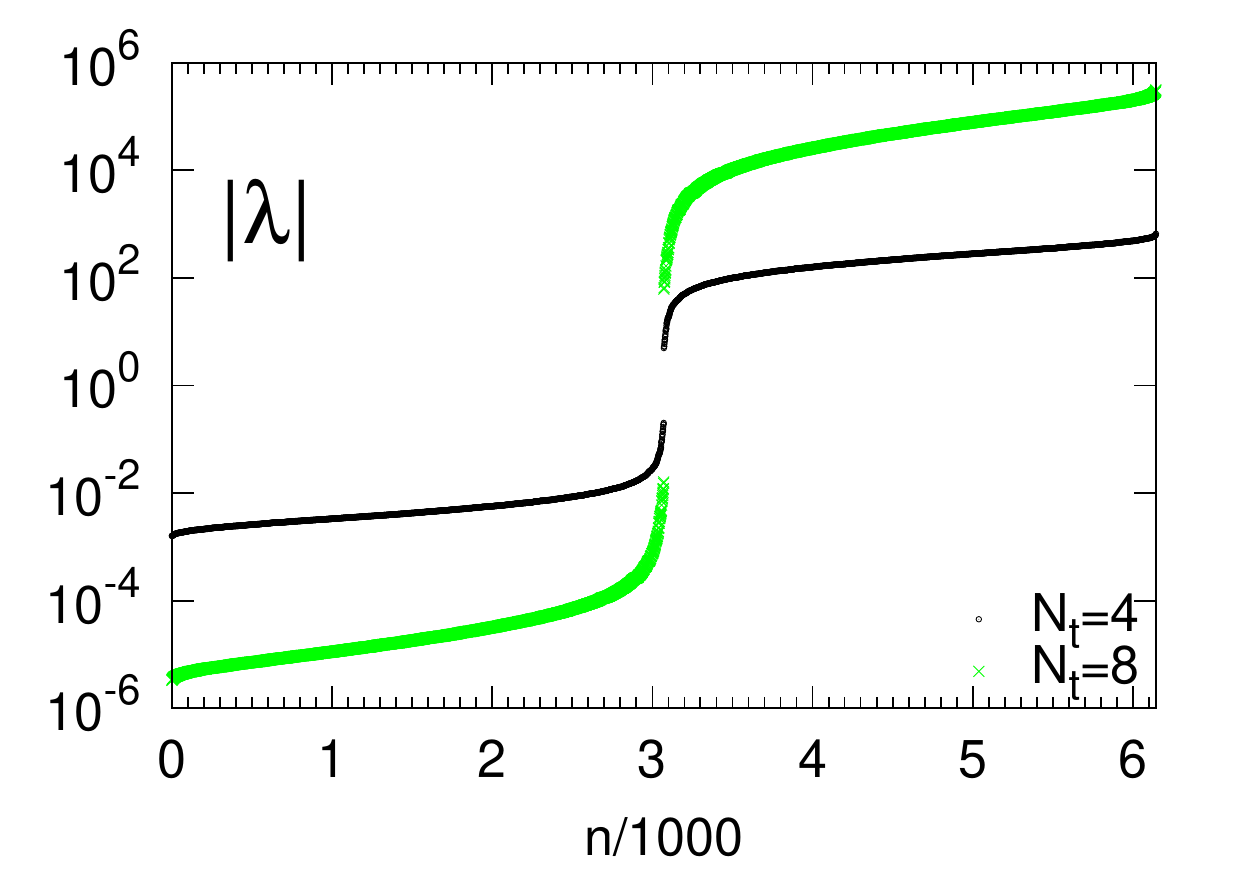}
\includegraphics[width=7.5cm]{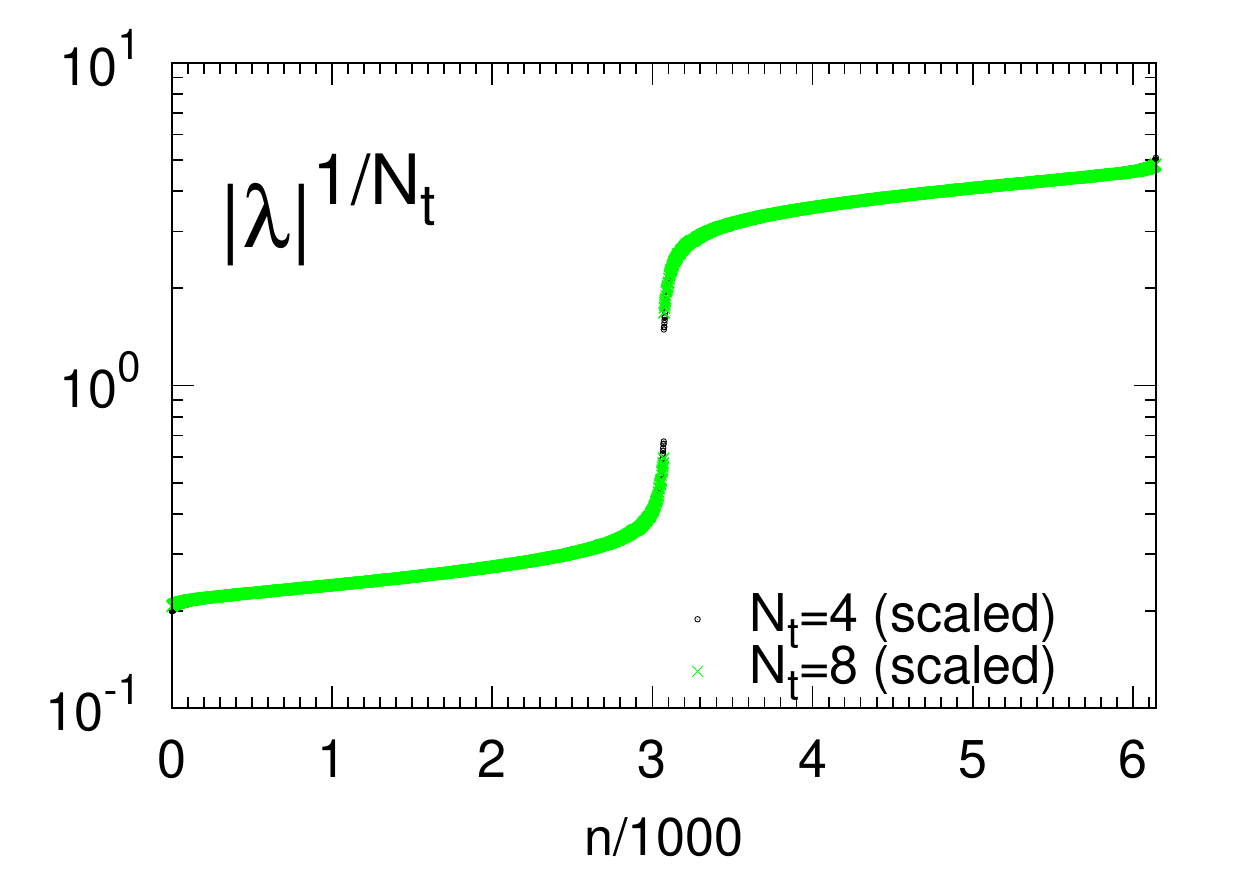}
\begin{minipage}{0.9\linewidth}
\caption{\small 
The spectrum of $Q$. The magnitudes of $\Nred = 8^3 \times 12 = 6144$ eigenvalues are plotted in increasing order.
The left and right panels are for $|\lambda_n|$ and $|\lambda_n|^{1/N_t}$, respectively.  }
\label{Fig:2013Sep29fig1}
\end{minipage}
\end{figure} 
\fi

These three properties can be understood naturally from the properties of the reduction matrix $Q$ explained in Sec.~\ref{sec:reduction}. $Q$ describes the Euclidean time evolution of quarks, $Q=e^{-H/T}$. By comparing  $\lambda_n = e^{ \pm \epsilon_n /T + i \theta_n(N_t)} $ and $Q=e^{-H/T}$, we find that $\pm \epsilon_n + i \theta_n(N_t) T$ correspond to the eigenvalues of the Hamiltonian. Here, $Q$ is a function of gauge configurations, and $H$ and $\lambda$ also depend on the gauge configuration. Thus, $\lambda_n$ is an eigenvalue for a Hamiltonian defined by a gauge configuration. The real parts of eigenvalues belonging to ${\cal S}$ and ${\cal L}$ are $\pm\epsilon_n$. They have the same magnitude, and their signs are the opposite. By interpreting $\lambda_n$ as the eigenvalue of the Hamiltonian, this symmetry can be interpreted as the symmetry between the positive and negative energy states, namely the symmetry between quarks and antiquarks. The existence of the gap means that there is a minimum value in the energy eigenvalue $\epsilon_n$. Moreover, the relation $Q = e^{- H/T}$ is nothing but the $N_t$-scaling law.

\subsubsection{$\mu$ independence of fermion determinant at zero temperature}
\label{sec:2013Oct06sec1}
 
Let us derive the fermion determinant in the zero-temperature limit, $N_t\to\infty$, by using the property of the eigenvalues of the reduction matrix~\footnote{Here, we assume that the time direction is discretized. See Ref.~\cite{Adams:2003rm,Adams:2004yy} for the case of the continuum time, and Ref.~\cite{Cohen:2003kd} for the continuum theory.}.

The starting point is the fermion determinant in terms of the eigenvalues of the reduction matrix, 
\begin{align*}
\det \Delta(\mu) & = C_0 \xi^{-\Nred/2} \det ( Q + \xi) , \\
                 & = C_0 \xi^{-\Nred/2} \prod_{i=1}^{\Nred} ( \lambda_n + \xi), \\
                 & = C_0 \xi^{-\Nred/2} \prod_{\lambda_n \in \calS} ( \lambda_n + \xi) \prod_{\lambda_n \in \calL} ( \lambda_n + \xi).
\end{align*}
The eigenvalues belonging to ${\cal L}$ can be expressed by using the eigenvalues belonging to ${\cal S}$, due to the symmetry explained above, as $\lambda_n \to 1/ \lambda_n^*$. Hence
\begin{align*}
\det \Delta(\mu) & = C_0 \xi^{-\Nred/2} \prod_{\lambda_n \in \calS} ( \lambda_n + \xi) ( 1/\lambda_n^* + \xi).
\end{align*}
Note that the product is taken for the eigenvalues inside the unit circle $\lambda_n \in \calS$.
It can be rewritten as 
\begin{align}
\det \Delta(\mu) = C_0 \prod_{n=1}^{\Nred/2} (\lambda_n^*)^{-1}(1+\lambda_n\xi^{-1}) (1+\lambda_n^*\xi).
\label{Eq:2012Apr22eq2}
\end{align}
By using the $N_t$ scaling law $\lambda_n = \exp(- \epsilon_n /T + i\theta_n)$ and $\xi=e^{-\mu/T}$,
we obtain
\begin{align}
\det \Delta(\mu) = &C_0 \prod_{n=1}^{\Nred/2} (\lambda_n^*)^{-1}(1+e^{-(\epsilon_n - \mu)/T+i\theta_n}) 
    (1+e^{-(\epsilon_n + \mu)/T-i\theta_n}).
\label{Eq:2012Apr22eq3}
\end{align}

Next, we fix the lattice spacing $a$ and the spatial lattice size $N_s^3$, and take the zero-temperature limit $N_t\to \infty$~\footnote{The following discussion may depend on the order of limits. The verification of this is currently difficult because we do not know how the eigenvalues of the reduction matrix change as $a\to 0$ and $V\to 0$, which are necessary for the precise understanding of the continuum limit and the large-volume limit.}. In the following, we assume $\mu>0$. Because Eq.~\eqref{Eq:2012Apr22eq3} contains only the eigenvalues in ${\cal S}$, $\epsilon_n (=-T \ln| \lambda_n|)$ is positive for all $n$, and in the limit of $N_t \to \infty$ we obtain $e^{-(\epsilon_n + \mu)/T} \to 0$:
\[
\lim_{N_t \to \infty} \det \Delta(\mu) \to C_0 \prod_{n=1}^{\Nred/2} (\lambda_n^*)^{-1}(1+e^{-(\epsilon_n - \mu)/T+i\theta_n}).
\]
The $T\to 0$ limit of $1+e^{-(\epsilon_n-\mu)/T}$ depends on the sign of the exponent. For a given value of $\mu$, if $\epsilon_n$ is greater than $\mu$ then $(1+e^{-(\epsilon_n-\mu)/T+i\theta_n})$ converges to $1$, otherwise we can ignore $1$ and approximate it with $e^{-(\epsilon_n-\mu)/T+i\theta_n}$. Therefore, the $\mu$ dependence of the fermion determinant at $T\to 0$ arises from the terms with $\epsilon_n<\mu$. If $\mu$ is less than the minimum value of $\epsilon_n$, namely $\epsilon_{\rm min} =- T \max_{\lambda_n\in \calS} |\lambda_n|$, $1+e^{-(\epsilon_n-\mu)/T}$ converges to $1$ for all $n$. Thus, we find that
\begin{align}
\lim_{N_t \to \infty} \det \Delta(\mu) \to C_0 \prod_{n=1}^{\Nred/2} (\lambda_n^*)^{-1}
\qquad
(\mu <\epsilon_{\rm min}).
\label{Eq:2012Apr22eq4}
\end{align}
Note that Eq.~\eqref{Eq:2012Apr22eq4} does not contain $\mu$.
In this way, we see that the fermion determinant has no chemical potential dependence at $T=0$ and $\mu<\epsilon_{\rm min}$.

This  $\mu$-independence appears also in the Fermi distribution. To see it explicitly, let us consider the quark number operator,
\begin{align}
\hat{n}=\frac{T}{V_s} (\det\Delta(\mu))^{-1} \frac{\del\det\Delta(\mu)}{\del\mu}
\qquad
 (V_s=N_s^3).
\end{align}
By substituting $\lambda_n=e^{-\epsilon_n/T+i\theta_n}$ and $\xi=e^{-\mu/T}$, we obtain
\begin{align}
\hat{n} = \frac{1}{V_s} \sum_{n=1}^{\Nred/2}
\left(
\frac{1}{1+e^{(\epsilon_n-\mu)/T-i\theta_n}}
-\frac{1}{1+e^{(\epsilon_n+\mu)/T + i\theta_n}}
\right).
\label {Eq:2012 Apr24eq1}
\end{align}
This has the same structure as the Fermi distribution. The coefficient of the second term is negative, indicating that this term corresponds to the antiquark state.
In the zero-temperature limit, the first term becomes zero if $\mu<\epsilon_n$ and does not contribute to the number of particles, while it becomes non-zero and contributes to the number of particles if $\mu>\epsilon_n$. This can be interpreted that the energy level $\epsilon_n$ is smaller than the chemical potential that is occupied by a quark. $\epsilon_{\rm min}$ is the minimum energy, and if $\mu$ is smaller, no level can be occupied.

Note that the discussion so far was for operators. $Q$ is a function of gauge configuration so that $\lambda_n$ also depends on the gauge configuration. Among the three properties of $Q$ described above, the symmetry of eigenvalue comes from $\gamma_5$-Hermiticity, and the $N_t$ scaling is because $Q$ represents the propagation in the time direction. Both properties hold independently of the gauge configuration. On the other hand, the value of $\epsilon_{\rm min}$ depends on the gauge configuration, and hence the threshold at which the $\mu$ dependence disappears depends on the gauge configuration.

What exactly is the value of $\epsilon_{\rm min}$? Phenomenologically, $\epsilon_{\rm min}=m_N/3$ would be expected, but the actual value is $\mu=m_\pi/2$. As we explained at the beginning of this chapter, it is first pointed out by Gibbs, and he predicted that
\begin{align}
a m_\pi =-\frac{1}{N_t} \max_{\lambda_n \in S} \ln |\lambda_n|^2,
\label{eq:2017Jul19eq1}
\end{align}
when he derived the formula of the propagator matrix~\cite{Gibbs:1986hi}. This relation has been applied to a numerical calculation by Fodor, Szab\'o, and T\'oth~\cite{Fodor:2007ga}. The author and collaborators directly took the zero-temperature limit of the fermion determinant.

\subsubsection{Relation between the threshold value and hadron mass}
\label{sec:2013Oct06sec2}

$\epsilon_{\rm min}$ and $m_\pi$ are related because the propagation of quarks along the time direction is related to both the pion propagator and the quark number operator.

$Q$ represents a diagram in which quarks propagate from time $0$ to $N_t$. We can represent $Q(t)$ as a quark field at time $t$ as $Q \propto q(N_t) \Gamma q^\dagger (0)$, where $\Gamma$ is a function of link variables. $Q^\dagger$ describes backward propagation, which is $Q^\dagger \propto q(0) \Gamma q^\dagger (N_t)$. Therefore, we see that $QQ^\dagger$ corresponds to the propagator of the meson. If we take a large lattice extent $N_t$ in the temporal direction, then the lowest energy state, that is, the contribution from the $\pi$ meson, becomes dominant:
\begin{align}
\bra {\rm tr} Q Q^\dagger \ket \stackrel {N_t \to \infty}{=} C e^{- m_\pi a N_t}.
\label {Eq:2013Sep21eq1}
\end{align}
In Eq.~\eqref{Eq:2013Sep21eq1}, only exponentially decaying modes are considered.
Thus, only $\lambda_n \in \calS$ need be considered among the eigenvalues of $Q$:
\begin{align}
\bra {\rm tr} Q Q ^\dagger \ket = \left \langle \sum_{ \lambda_n \in {\cal S}} | \lambda_n|^2 \right \rangle.
\end{align}
Furthermore, by substituting the $N_t$-scaling $\lambda_n = e^{- \epsilon_n a N_t+ i\theta_n}$,
\begin{align}
\bra {\rm tr} Q Q^\dagger \ket = \left \langle \sum_{ \lambda_n \in {\cal S}} e^{- 2 \epsilon_n a N_t} \right \rangle.
\end{align}
If we take $N_t\to \infty$, the eigenvalue corresponding to $\epsilon_{\rm min}$ becomes dominant:
\begin{align}
\lim_{N_t \to \infty} \bra {\rm tr} Q Q^\dagger \ket = \left \langle e^{- 2 \epsilon_{\rm min} a N_t} \right \rangle.
\end{align}
Therefore, in the $N_t \to \infty$ limit we obtain the following relationship,
\begin{align}
\bra e^{- 2 \epsilon_{\rm min} a N_t} \ket = C e^{- m_\pi a N_t}.
\end{align}
Although $\epsilon_{\rm min}$ can take different values depending on the gauge configurations, the contribution from the minimum value remains in the zero-temperature limit. Therefore, we obtain
\begin{align}
\epsilon_{\rm min}=\frac{1}{2}m_\pi.
\end{align}
Combined with Eq.~(\eqref{Eq:2012Apr22eq4}, the quark number operator is zero for any gauge configuration if $\mu$ is less than $m_\pi/2$.

Naively, we would expect a similar relation between $\epsilon_{\rm min}$ and the nucleon mass via a similar calculation, but this is not the case. The baryon consists of three quarks, so it corresponds to $Q^3$, 
\begin{align}
\bra {\rm tr} Q^3 \ket = C e^{-m a N_t}. 
\end{align}
In the limit of $N_t\to \infty$, the ground state nucleon remains on the right-hand side, so we can replace  $m$ with $m_N$. 
By rewriting the left hand side in terms of the eigenvalues, we obtain 
\begin{align}
\bra {\rm tr} Q^3 \ket = \left \langle \sum_{ \lambda_n \in {\cal S}} e^{- 3 \epsilon_n a N_t + 3 i \theta_n N_t} \right \rangle.
\label{Eq:2013Oct01eq1}
\end{align}
In the case of meson, the phases are canceled between $Q$ and $Q^\dagger$; but in the case of baryon, they are not canceled. Because of the phase fluctuations, the sum of eigenvalues is not approximated by keeping $\epsilon_{\rm min}$. Therefore, the relation of $\epsilon_{\rm min} = M_N/3$ does not hold.

\subsection{ Properties of QCD at $\mu>m_\pi/2$ and the limitation of the simulation methods based on importance sampling}

From the discussion so far, if $T=0$ and $\mu<m_\pi/2$, the fermion determinant has no chemical potential dependence and the quark number density is $0$ at the operator level. If $\mu$ exceeds $m_\pi/2$, the quark number can have a non-zero value at the operator level. However, physically, the increase in quark number should not occur until $\mu$ exceeds $M_N/3$.

The key to making these two features consistent with each other is the fact that the quark number operator is complex at nonzero $\mu$. Even if each gauge configuration has a non-zero value, the expectation value can be zero due to the cancellation between complex phases. In other words, below and above $\mu=m_\pi/2$ different mechanisms set the expectation value of quark number $\langle \hat{n} \rangle$ to zero:
\begin{itemize}
\item $\mu<m_\pi/2$: $\hat{n}=0$ for each gauge configuration
\item $m_\pi/2 <\mu <m_N/3$: Although $\hat{n} \neq 0$ for each gauge configuration, the expectation value $\langle\hat{n} \rangle$ is zero
\end{itemize}
Of course, the latter is merely a conjecture and we do not know how this can happen. Ipsen and Splittorff have invented a simple model in which the cancellation does occur~\cite{Ipsen:2012ug}. It is not known how the phase is distributed at $\mu = m_\pi/2$ in QCD and whether the expectation value of the quark number becomes zero.

The fact that severe phase fluctuations show up at $\mu = m_\pi/2$ is almost equivalent to the failure of the reweighting method at this point.
We denote the real weight used for the configuration generation as $w$, then the reweighting factor is expressed by
\[
R = \frac{ (\det\Delta (\mu))^{N_f} }{w}.
\]
The phase in the numerator fluctuates significantly at $\mu=m_\pi/2$ in the hadron phase, while 
the denominator $w$ for any weight is always positive in the configuration generation.
Therefore, the reweighting factor becomes
\[
\langle R \rangle \simeq 0
\]
at $\mu=m_\pi/2$, and the reweighting method fails. 
Regarding Taylor expansion and analytic continuation from imaginary chemical potential, it may work beyond $\mu=m_\pi/2$ since there is no non-analyticity at $\mu=m_\pi/2$ and the thermodynamic analyticity is guaranteed. However, at low temperatures, the signal-to-noise ratio decreases, which makes the numerical calculation very difficult. Because such difficulties were hidden behind the early onset problem, it could not be solved within the framework of the reweighting method.

\iffigure
\begin{figure}[htbp] 
\centering
\includegraphics[width=7.5cm]{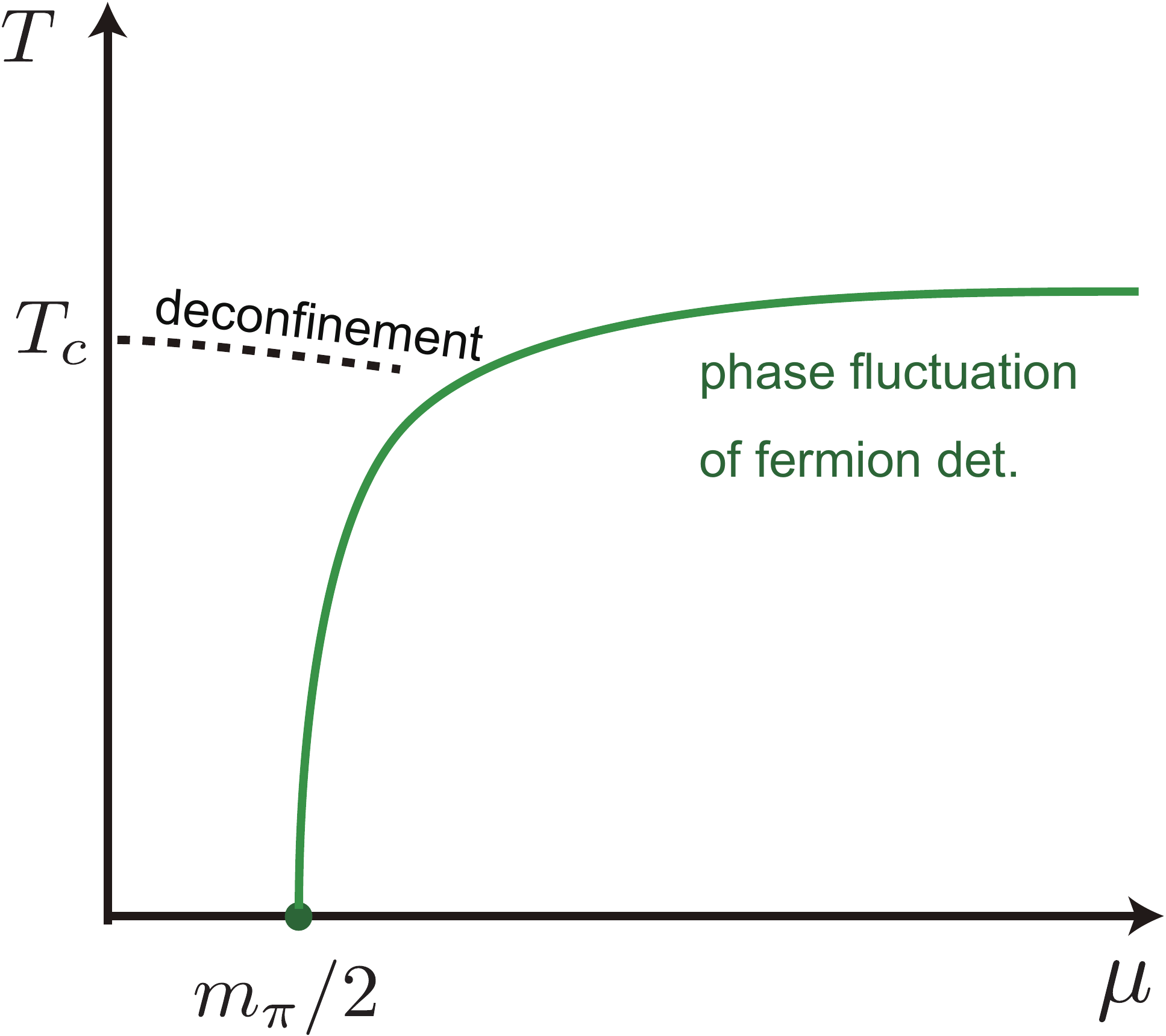}
\begin{minipage}{0.9\linewidth}
\caption{\small 
The region where the early-onset problem occurs in the QCD phase diagram. This region coincides with the pion-condensed phase when the horizontal axis is taken to be the isospin chemical potential. Phase fluctuation is severe and $\langle R \rangle_0$ becomes zero on the curve. Therefore, it is difficult to examine the right side of the curve using the configurations generated on the left side of the curve.
}
\label{fig:2017Sep14fig1}
\end{minipage}
\end{figure} 
\fi

So far we focused on the $T=0$ case, but the early-onset problem likely exists even at finite-temperature, finite-density region of the hadron phase ($0< T< T_c$). Almost the same argument can be applied, except that the dependence on the chemical potential changes from a step function to a smooth function. The property of the fermion determinant changes when the pions condense (with the isospin chemical potential). The methods explained in this section cannot be used beyond the curve in Fig.~\ref{fig:2017Sep14fig1}.

Our discussions above have been based on the reduction matrix. We also mention the analyses based on the fermion matrix itself. At $\mu=0$, the fermion matrix $\Delta = D + m $ has the eigenvalue $m \pm i \lambda, (\lambda \in \mathbb{R})$, due to the anti-Hermiticity of $D$. 
They are distributed on the straight line with real part $m$ and imaginary part $\pm\lambda$ on the complex plane. When the chemical potential is turned on, the eigenvalues of $\Delta$ become generic complex numbers, which are distributed on the complex plane two-dimensionally, since the chemical potential violates the anti-Hermiticity of the Dirac operator. As $\mu$ increases, the distribution spreads and eventually some eigenvalues become zero at $\mu=m_\pi/2$~\cite{Barbour:1986jf,Davies:1990qk}. In the hadron phase,  the same problem occurs even if $T \neq 0$ since the zero eigenvalue density $\rho(0)$ of the Dirac operator is nonzero, $\rho(0)\neq 0$, due to the Banks-Casher relation.\footnote{
Strictly speaking, one has to consider a generalized version of the Banks-Casher relation that is applied to complexified configurations generated by the CLM. The first qualitative discussion was given in Ref.~\cite{Splittorff:2014zca}. A precise quantitative formulation of the generalized Banks-Casher relation was given in ref.~\cite{Nagata:2016alq}, where it was also confirmed in the chiral Random Matrix Theory. 
} In the QGP phase, a gap opens in the Dirac spectrum due to the restoration of chiral symmetry, so that the zero eigenvalues in $\Delta$ do not appear.
Note that the calculation of the inverse matrix $\Delta^{-1}$ is needed in the gauge-configuration generation process of lattice QCD. Usually, the conjugate gradient method (CG method) is used for this. The number of iterations in the CG method depends on the ratio of the maximum eigenvalue to the minimum eigenvalue of the matrix. If the zero eigenvalues appear, then the CG method ceases to work. Therefore, the computational cost for the phase quench simulation becomes larger as one goes closer to the solid line in Fig.~\ref{fig:2017Sep14fig1}.

The regions where the early onset problem occurs include regions corresponding to standard nuclear densities and neutron stars.
Furthermore, Hidaka and Yamamoto derived the QCD inequality under the assumption that the disconnected diagram can be ignored, and
pointed out that the QCD critical point may exist inside this region~\cite{Hidaka:2011jj}.

Let us return to the discussion of the sign problem. As we already mentioned, if the quark chemical potential becomes greater than half of the $\pi$ meson mass in the hadron phase, then severe phase fluctuations occur and the methods to avoid the sign problem using the gauge configurations generated at $\mu=0$ fails (almost) certainly. There are other configuration generation methods such as the quench approximation, the finite isospin chemical potential, the phase quench, the imaginary chemical potential, but most likely all these methods break down when the phase fluctuations of the fermion determinant and the quark number operator show up. Of course, the reweighting and Taylor expansion are the correct methods in principle, so strictly speaking, we should say that the amount of calculation required to go beyond $\mu=m_\pi/2$ is enormous and the calculation fails in practice. We mentioned in Sec.~\ref{sec:solve_signproblem} that the sign problem is a problem of computational cost. This problem does occur in QCD.

At the beginning of this paper, we have made some scientific questions about the QCD phase diagram. The conclusion in this chapter is that solving these problems is extremely difficult as long as we work in the framework of the importance sampling. Therefore, a new approach to the sign problem is needed to solve the open problems regarding the baryonic matter.

\bibliographystyle{utphys}
\bibliography{ref_list}

\newcommand{\Sbar}{\bar{S}}
\newcommand{\im}{{\rm Im}}
\newcommand{\bardel}{\bar{\partial}}
\newcommand{\barK}{\bar{K}}
\newcommand{\barS}{\bar{S}}
\newcommand{\barV}{\bar{V}}

\newcommand{\sign}{{\rm sign}}

\newpage
\chapter[Solution to sign problem 2: Complex Langevin]
{Solution to sign problem 2: Configuration generation method applicable to complex actions}
\label{sec:newmethod}

In the previous section, we discussed several methods to circumvent the sign problem based on the importance sampling. They are effective at small chemical potential, but they do not work at a large chemical potential. 

Recently, the simulation techniques applicable to complex actions are studied actively. Such methods include the complex Langevin method, Lefschetz thimble method, tensor network method and the dual variable method. 
The level of difficulty of the sign problem depends on the theory;  sometimes such new methods are useful, sometimes not.   
The study of the hadronic phase of QCD is still challenging. 

In this chapter we study the only example of such new method which has been applied to QCD so far: the complex Langevin method. 
We explain its properties, applications, and open problems. 

\section{Complex Langevin method}
\label{sec:complexLangevin}

The complex Langevin method (CLM) provides us with a possible solution to the sign problem. The original Langevin method for the real actions was proposed by Parisi and Wu~\cite{Parisi:1980ys} as a numerical method to perform path integral by utilizing the Langevin equation. (Below, we sometimes call it `real Langevin method' to distinguish it from the CLM.) This is also called the stochastic quantization method because it uses the Langevin equation which is a stochastic process. While the Markov Chain Monte Carlo is based on the Metropolis steps (accept or reject proposed configurations), in the stochastic quantization configurations are generated following the motion in the phase space described by a stochastic differential equation. The Langevin method may be regarded as a simplified analog of Hybrid Monte Carlo, which does not use the Metropolis test.

The CLM was proposed in the early 1980s by Parisi and Klauder~\cite{Parisi:1984cs,Klauder:1983sp}, as a generalization of the real Langevin method applicable to theories with complex actions. The key observation is that the Langevin method does not use the Metropolis step with `probability' $e^{-S}$, and hence, there is no apparent obstacle toward the generalization to the complex action.  However, as realized soon after the proposal, sometimes the CLM gives a wrong answer (the problem of `wrong convergence'). For this reason, until recently the CLM was not regarded as a useful technique. Recently, there is a revived interest in the CLM because Aarts, Seiler, and Stamatescu identified the mechanism of the wrong convergence in 2009~\cite{Aarts:2009uq}. The CLM was applied to the high-temperature, high-density region of QCD in 2013~\cite{Sexty:2013ica}. The application to the low-temperature, high-density region is still challenging, and being studied intensively.

In this chapter, we discuss the CLM method in detail. We start with the original Langevin method for real actions in Sec.~\ref{sec:reallangevin}. 
Then we formulate the CLM, explain the justification condition needed to avoid the wrong convergence, and review more recent developments such as the gauge cooling. For readers interested in stochastic quantization, we recommend review papers such as Ref.~\cite{Damgaard:1987rr}. 
\subsection{Langevin method for real actions}
\label{sec:reallangevin}

Firstly we explain the Langevin method for real actions~\cite{Parisi:1980ys}. We consider the path integral of a system with real variables $x_i \in \mathbb{R}, (i=1, 2, \cdots N)$ and the real action $S=S(x) \in \mathbb{R}$, 
\begin{align}
Z=\int \prod_i dx_i \eqspc e^{- S(x)}. 
\label{eq:2016Jan10eq1}
\end{align}
We introduce a new variable $t$ and assume that the dynamical variables $x_i$ 
follows the Langevin equation 
\begin{align}
\bibun{x_i}{t} = - \bibun{S}{x_i} + \eta(t), 
\label{eq:2016Jan10eq2}
\end{align}
as a function of $t$.
While the parameter $t$ plays the role of time in the Langevin equation, it is not physical time; it is called the `fictitious time' or `Langevin time'. $\eta$ is a stochastic variable, which follows the Gaussian distribution at each fictitious time $t$. The average over $\eta$ (`noise average') of observables at each Langevin time $t$ is defined by 
\begin{align}
\nsave{O(t)} = \frac{\int \calD \eta \eqspc O(t) e^{-\frac{1}{4} \int_0^t d\tau \eqspc\eta(\tau)^2}}{
\int \calD \eta \eqspc e^{-\frac{1}{4} \int_0^t d\tau \eqspc\eta(\tau)^2}}.  
\label{eq:2017Jul20eq1}
\end{align}
We can see that the following relations hold: 
\begin{subequations}
\begin{align}
\nsave{\eta(t)} &= 0, 
\label{Eq:2015Aug07eq1a}\\
\nsave{\eta(t_1) \eta(t_2)} &= 2 \delta(t_1-t_2).
\label{Eq:2015Aug07eq1b}
\end{align}
\end{subequations}

Equation~\eqref{eq:2016Jan10eq2} is the Langevin equation in the phase space. If the stochastic variable $\eta$ is not there, the solution of this equation converges to the classical solution with the minimum action. Due to the existence of the stochastic variable, quantum fluctuation is introduced.

In numerical simulations, we need to discretize the Langevin time $t$. When the step size for $t$ is $\epsilon$, the Langevin equation and the distribution of the stochastic variable are\footnote{Occasionally, a different normalization $\eta' = \sqrt{\epsilon}\eta$ is used. 
In that case the distribution is $e^{-\eta^2/4}$. Note that the variance depends on the normalization. }   
\begin{align}
x(t+\epsilon) &= x(t) - \bibun{S}{x} \epsilon + \eta(t) \eqspc \epsilon, \\ 
\int \calD \eta \eqspc e^{-\frac{1}{4} \int_0^t d\tau \eta(\tau)^2} 
& = \int \prod_i d\eta(t_i) \eqspc e^{-\frac{1}{4} \epsilon \sum_i  \eta(t_i)^2}.  
\label{eq:2016Jan09eq1}
\end{align}

So far, we have seen how the configurations are generated via the Langevin equation. At this stage, it is not clear whether the configurations obtained in this way give the correct distribution that reproduces the path integral. Below, we confirm that the correct distribution is obtained. Namely, we show that the expectation values obtained by the Langevin method, 
\[
\langle O \rangle_{\rm LM} = \langle O \rangle_\eta \]
and the expectation value defined by path integral, 
\[
\langle O \rangle_{\rm phys} = Z^{-1} \int {\calD} U O e^{ - S}, 
\]
are the same. Because the proof is rather long, we show only the essence here, leaving several steps to Appendix~C.

We introduce the probability distribution of the dynamical variables and the Fokker-Planck equation that describes the time evolution of the probability distribution. Let us use $x^{(\eta)}$ to denote the dynamical variable obtained by solving the Langevin equation and $x$ to denote the coordinate in the phase space. The expectation value of the observable at the Langevin time $t$, denoted by $O(x^{(\eta(t))})$,
can be written as 
\begin{align}
O(x^{(\eta(t))}) = \int dx \, O(x) \,\delta (x - x^{(\eta(t))}). 
\label{eq:2017Jul20eq2}
\end{align}
Note that the variable $x^{(\eta)}$ at time $t$ explicitly depends on the noise $\eta(t')$ at earlier time $t'$ ($t'<t$ for the continuum Langevin equation, and $t'\le t-\eps$ for the discretized version). Now we substitute eq.~\eqref{eq:2017Jul20eq2} to eq.~\eqref{eq:2017Jul20eq1}. 
Noticing that $x$ in the right-hand side of eq.~\eqref{eq:2017Jul20eq2} is an integration variable and that $O(x)$ does not depend on $\eta$, we can take $O(x)$ out of the average over $\eta$. 
Then we obtain~\footnote{
Let $\omega(\eta)$ be the probability distribution of the noise
and $Z$ be the normalization factor. 
Eq.~(\ref{Eq:2015Sep05eq1}) is obtained as follows:
\begin{align}
\langle O(x^\eta)\rangle &= Z^{-1} \int d\eta \, O(x^\eta) \,\omega(\eta), \nn \\
&= Z^{-1} \int d\eta \left( \int dx  \, O(x) \delta(x-x^\eta)\right) \omega(\eta), \nn \\
&= Z^{-1} \int dx  \, O(x)  \left( \int d\eta \, \delta(x-x^\eta) \omega(\eta) \right), \nn \\
&= \int dx  \, O(x)  P(x). \nn 
\end{align}
}  
\begin{align}
\nsave{O\left(x^{(\eta)}(t)\right)} = \int dx \, O(x) \, P(x;t), 
\label{Eq:2015Sep05eq1}
\end{align}
where $P(x; t)$ is defined by 
\begin{align}
P(x;t) = \nsave{ \delta\left(x-x^{(\eta)}(t)\right)}. 
\end{align}
This $P(x;t)$ is the probability that a point moving in the phase space following the Langevin equation is at a point $x$ at a Langevin time $t$.

Because $x^{(\eta)}$ changes with the Langevin time $t$, the probability distribution $P(x;t)$ changes as well, following the Fokker-Planck equation
\begin{align}
\bibun{P}{t} = \bibun{}{x}\biggl( \bibun{}{x} + \bibun{S}{x}\biggr) P.  
\label{Eq:2015Ag13eq1}
\end{align}
The derivation of the Fokker-Planck equation is given in Appendix~\ref{sec:FP_disc}. If the stationary point $\partial S/ \partial x = 0$ exists, the probability distribution $P(x;t)$ converges to the equilibrium solution around the stationary point at a sufficiently late Langevin time $t$. We will show that the equilibrium solution is 
\begin{align}
\lim_{t\to\infty} P(x;t) \to \frac{1}{Z} e^{-S}. 
\label{eq:2015Sep05eq2}
\end{align}
For that purpose, let us define the Fokker-Planck Hamiltonian $\HFP$ and its wave function $\psi(x;t)$ by 
\begin{subequations}
\begin{align}
\psi(x;t) & = P(x;t) e^{\frac12 S(x)}, \\
\HFP & = - \biggl( \delx - \frac{1}{2}S' \biggr) \biggl(\delx + \frac12 S' \biggr), 
\end{align}
\end{subequations}
We can easily confirm that 
\begin{align}
- \partial_t \psi  & = \HFP \psi. 
\end{align}
Moreover, we can easily check that $\psi(x;t) = e^{-\frac12 S}$ is the eigenstate of $\HFP$ with the eigenvalue zero, and that $H_{\rm FP}$ is Hermitian (see Appendix~\ref{sec:stationarysolution}). Furthermore, $\HFP$ is positive-semidefinite, i.e.~the eigenvalues are non-negative.\footnote{
(Proof) By using $A=\delx + \frac12 S'$, the Fokker-Planck Hamiltonian can be written as $\HFP=A^\dagger A$. Therefore, for any vector $v$, 
\begin{align}
v^\dagger \HFP v &= v^\dagger A^\dagger A v \nn \\
                &= \vert\vert A v\vert \vert^2 \ge 0.
\label{eq:2019Oct09eq1}
\end{align}
Let $v$ be an eigenvector of $\HFP$, whose eigenvalue is $\lambda$. 
Then, 
\begin{align*}
v^\dagger \HFP v &= \lambda v^\dagger v \nn \\
                &= \lambda \vert \vert v \vert \vert^2 .
\end{align*}
The left-hand side is non-negative due to \eqref{eq:2019Oct09eq1}. By noticing that $\vert \vert  v \vert \vert^2 \ge 0$, we conclude 
\[ \lambda \ge 0 .\]
This argument holds for all eigenvectors, and hence, all eigenvalues of $\HFP $ are non-negative.}
$\psi$ can be written as a linear combination of the eigenstates of $\HFP$, and only the zero-mode survives as $t\to \infty$. 
Therefore, 
\[ 
\lim_{ t\to \infty} \psi(x;t) \to e^{-\frac{1}{2} S}, 
\]
and the relation \eqref{eq:2015Sep05eq2} holds for the probability distribution $P(x;t)$.

By sending $t$ to infinity in \eqref{Eq:2015Sep05eq1}, and combining it with \eqref{eq:2015Sep05eq2}, we obtain
\begin{align}
\lim_{t \to \infty} 
\nsave{O\left(x^{(\eta)}(t)\right)} &= \int dx \, O(x) \, \lim_{t \to \infty } P(x;t), \nn \\
& \to \frac{1}{Z}  \int dx \, O(x) \, e^{-S}. 
\end{align}
The right-hand side is the expectation value of the observable defined in the path integral formalism. In this way, at sufficiently late Langevin time, the configuration generated by using the Langevin method reproduces the probability distribution $e^{-S}$ and the expectation value $\nsave{O\left(x^{(\eta)}(t)\right)}$ agrees with the one defined in the path integral formulation.

The noise average \eqref{eq:2017Jul20eq1} is obtained by generating many configurations at the same Langevin time, i.e., multiple measurements are done at the same Langevin time. The same expectation value can be obtained also by measuring at multiple Langevin times, but only one measurement at each time. In the latter, the average over the Langevin time is taken. These two methods should give the same answer if the ergodicity is assumed, and hence, the expectation value in the Langevin method can also be defined by 
\begin{align}
\langle O \rangle = \frac{1}{T-t_0} \int_{t_0}^T dt O(t). 
\label{eq:2017Apr17eq1}
\end{align}
$t_0$ is taken to be sufficiently large so that the system is in equilibrium after this Langevin time. As $T\to\infty$, the average over the Langevin time converges to the expectation value defined in the path integral method.

\subsection{Complex Langevin Method}\label{subsec:CLM}

Next, we consider the case that $x \in \mathbb{R}, S(x)\in \mathbb{C}$ in \eqref{eq:2016Jan10eq2}. (In case that the dynamical variables are complex, we separate them into real and imaginary parts and interpret them as real variables.) An apparent problem is that, because the action is complex, the drift term $(\partial S/\partial x)$ is also complex, and hence, the real variables $x$ develop the imaginary part during the Langevin-time evolution. This problem can be circumvented by extending the real variables to complex. This is the complex Langevin method proposed by Parisi and Klauder~\cite{Parisi:1984cs,Klauder:1983sp}

\subsubsection{Complexified  theory}

Let us promote the real variables $x_i \in \mathbb{R}$ to complex variables $z_i  =x_i+\sqrt{-1}y_i\in\mathbb{C}$, and analytically continue the action as $S(x) \mapsto S(z)$. Note that the Cauchy-Riemann equation $\partial S(z)/\partial \bar{z}=0$ is satisfied. Associated with this complexification, the Langevin equation is modified as  
\begin{align}
\frac{d z_i}{dt} = - \frac{\partial S(z)}{\partial z_i} + \eta(t),
\label{eq:complexlangevin}
\end{align}
where the stochastic variable $\eta$ becomes complex as well:
\begin{align}
\frac{\partial z }{\partial t} = - \frac{\partial S}{\partial z} + c_R \eta_R + i c_I \eta_I.
\label{Eq:2015Aug04eq1}
\end{align}
Here $\eta_{R}$ and $\eta_{I}$ are real-valued stochastic variables, and the real-valued parameters  $c_R$ and $c_I$ are required to satisfy $c_R^2-c_I^2=1$. 

The real and imaginary parts of \eqref{Eq:2015Aug04eq1} are
\begin{subequations}
\begin{align}
\frac{\partial x}{\partial t} &= - \Re \left[ \frac{\partial S}{\partial z} \right] + c_R \eta_R = K_x + c_R \eta_R, \\
\frac{\partial y}{\partial t} &= - \Im \left[ \frac{\partial S}{\partial z} \right] + c_I \eta_I = K_y + c_I \eta_I.
\end{align}
\label{Eq:2015Aug04eq6}
\end{subequations}
The stochasric variables follow the Gaussian distribution 
\begin{align}
P(\eta) =\normns^{-1} e^{-\frac{1}{4} \int d\tau (\eta_R^2 + \eta_I^2)},
\label{Eq:2015Aug04eq2}
\end{align}
where $\normns$ is a normalization factor
\begin{align}
\normns=\inteta \, \Gdistns. 
\label{Eq:2015Aug04eq3}
\end{align}
The expectation value of an observable $f$ is defined by 
\begin{align}
\nsave{f} = N^{-1}_\eta  \inteta\, f \Gdistns.  
\label{eq:obs_clm}
\end{align} 

The phase space becomes larger associated with the complexification. 
The dynamical variables move in this bigger, complex phase space following the complex Langevin equation. The probability that the dynamical variable is at a point $(x, y)$ at a Langevin time $t$, which is denoted by $P(x,y;t)$, is 
\begin{align}
P(x,y;t)= \nsave{ \delta(x-x^{(\eta)}(t)) \, \delta(y-y^{(\eta)}(t))}. 
\end{align}
By using $P(x,y;t)$ we can write the expectation value of $f$ as
\begin{align}
\langle f \rangle_\eta = \int dx \, dy \, P(x,y;t) \, f. 
\end{align}
Note that $P$ is real and non-negative by definition, and it is not a regular function.\footnote{Real-valued functions cannot satisfy the Cauchy-Riemann equation, unless they are constant.} This is the reason we used the notation $P(x,y;t)$ rather than $P(z;t)$.

\subsubsection{The problem of wrong convergence}

Now we extend the Langevin method to complex-valued actions. The next step is to show the validity of this complex Langevin method. Different from the real-valued action, $P(x,y;t)$ is not guaranteed to converge to $e^{-S}$. For the real-valued action, the convergence is guaranteed because of the Hermiticity of the Fokker-Planck Hamiltonian $\HFP$ and the existence of the zero mode. For the complex-valued action, $\HFP$ is not necessarily Hermitian, and hence, the convergence to the zero mode is not guaranteed. Sometimes the right answer is obtained, sometimes not.

In actual simulations, this problem is seen in two forms: the first case is the {\it runaway}, namely the distribution $P(x,y;t)$ does not converge. The second case is the {\it wrong convergence}. The runaway can occur in the real Langevin method as well if the theory does not satisfy the stability condition $V(x)  \to \infty$ for $ |x| \to \infty$. When the action is complexified, the stability condition can easily be lost, and the runaway problem is common. Often, the runaway can be circumvented by taking the step size  $\epsilon$ sufficiently small. When doing this, $\epsilon$ can be taken to be small all the time, or only when the drift term is large~\cite{Aarts:2009dg}. The latter is called the adaptive step size method. 

It is easy to detect the runaway because the observables diverge. On the other hand, it is difficult to detect the wrong convergence. Recently, Aarts, Seiler, and Stamatescu studied the condition for the correct convergence, i.e.~the condition for the observable obtained by the complex Langevin method and the path integral to agree. They found that the correct result is obtained when the two conditions are satisfied~\cite{Aarts:2009uq}. In the next subsection, we will see these conditions. 

\subsection{The conditions for the justification of the complex Langevin method}

\subsubsection{Fokker-Planck-like equation}

Firstly we derive the equation describing the time evolution of physical quantity $f$. Here we use the continuum Langevin equation, rather than the discrete Langevin equation. (For the latter, see Appendix~C). We consider the time derivative of $\langle f \rangle_\eta$, 
\begin{align}
\frac{d}{dt} \langle f \rangle_\eta = \left\langle \frac{\partial x}{\partial t} \frac{\partial f}{\partial x}
+\frac{\partial y}{\partial t} \frac{\partial f}{\partial y}\right\rangle_\eta.
\label{Eq:2015Aug04eq5}
\end{align}
By substituting \eqref{Eq:2015Aug04eq6} to \eqref{Eq:2015Aug04eq5}, we obtain 
\begin{align}
\frac{d}{dt} \langle f \rangle_\eta = \left\langle (K_x + c_R \eta_R) \frac{\partial f}{\partial x}
+ ( K_y + c_I \eta_I )  \frac{\partial f}{\partial y}\right\rangle_\eta. 
\label{Eq:2015Aug04eq7}
\end{align}
From \eqref{eq:obs_clm}, we see that 
\begin{align}
\langle g \eta \rangle_\eta = \frac{2}{\eps} \biggl\langle 
 \frac{\delta g(x) }{ \delta \eta(t)} \biggr\rangle_\eta
\end{align}
holds for $\eta = \eta_R, \eta_I$~\cite{Damgaard:1987rr}. For the $x$-component, 
\begin{align}
\langle \etaR g \rangle_\eta = \frac{2}{\eps} \biggl \langle 
 \frac{\delta g }{ \delta \etaR(t)} \biggr\rangle_\eta.
\end{align}
Here, a relation 
\begin{align}
 \frac{\delta g }{ \delta \etaR(t)} & = \frac{\delta x }{ \delta \etaR(t)}  \frac{\delta g }{ \delta x} , \nn \\
 & = \frac{\eps \,  c_R }{2}  \frac{\delta g }{ \delta x}
\end{align}
holds, we can see that 
\begin{align}
\left\langle (K_x + c_R \eta_R) \frac{\partial f}{\partial x}\right\rangle_\eta
& = \left\langle \biggl(K_x + c_R^2 \bibun{}{x} \biggr) \frac{\partial}{\partial x} f \biggr)\right\rangle_\eta.
\label{eq:2016Jan12eq1}
\end{align}
A similar result is obtained for the $y$-component, and hence, the time evolution of $f(x,y)$ is described by 
\begin{subequations}
\begin{align}
\llangle \bibun{f}{t} \rrangle_\eta &= \nsave{L f}, 
\nonumber\\
L  &= [ (K_x + c_R^2 \partial_x) \partial_x + (K_y + c_I^2 \partial_y) \partial_y].  
\label{Eq:2015Aug12eq1}
\end{align}
\end{subequations}
The solution is formally given by 
\begin{align}
f(x,y;t) = e^{t L} f(x,y;0). 
\end{align}

Up to now, we did not assume any particular property of $f$. Next let us consider the case that $f$ is a holomorphic function of $z$, i.e., 
\begin{align}
\frac{\partial f}{\partial \bar{z}}=0. 
\label{eq:2016Jan11eq1}
\end{align}
By rewriting the derivative with respect to $x$ and $y$ by using $z$ and $\barz$,~\footnote{
Because of $z= x+ iy$ and $\bar{z}= x - iy$, we obtain $\frac{\partial}{\partial x} = 
\frac{\partial z}{\partial x} \frac{\partial}{\partial z} +
\frac{\partial \bar{z}}{\partial x} \frac{\partial}{\partial \bar{z}}  = 
 \frac{\partial}{\partial z} + \frac{\partial}{\partial \bar{z}}$
and
$\frac{\partial}{\partial y} = 
\frac{\partial z}{\partial y} \frac{\partial}{\partial z} +
\frac{\partial \bar{z}}{\partial y} \frac{\partial}{\partial \bar{z}} 
 = 
i \frac{\partial}{\partial z} -i  \frac{\partial}{\partial \bar{z}}$.
}
and by using \eqref{eq:2016Jan11eq1}, we can rewrite the first term in $L$ as
\begin{align}
(c_R^2 \partial_x + K_x) \partial_x f 
& = \left(c_R^2  \frac{\partial z}{\partial x} \partial_z + c_R^2 \frac{\partial \bar{z}}{\partial x} \partial_{\bar{z}} + K_x\right) 
\left(\frac{\partial z}{\partial x} \partial_z + \frac{\partial \bar{z}}{\partial x} \partial_{\bar{z}} \right)f,  \nn \\
& = \left(c_R^2 \partial_z + K_x\right) 
\left( \partial_z  \right)f, \nn \\
& = (c_R^2 \partial_z^2 + K_x \partial_z)f.
\end{align}
The second component can also be calculated in the same manner: 
\begin{align}
(c_I^2 \partial_y + K_y) \partial_y f & = \left(c_I^2 i \partial_z + K_y \right) \left( i \partial_z  \right)f \nn \\
& = (- c_I^2 \partial_z^2 + i K_y \partial_z) f.
\end{align}
By summing them, we obtain 
\begin{align}
[(c_R^2 \partial_x + K_x) \partial_x + (c_I^2 \partial_y + K_y) \partial_y ] f 
 &= (c_R^2 \partial_z^2 + K_x \partial_z)f + (- c_I^2 \partial_z^2 + i K_y \partial_z) f \nn \\
 &= \biggl[ \biggl( (c_R^2- c_I^2) \partial_z + (K_x + i K_y) \biggr) \partial_z \biggr] f \nn \\
 &= \biggl[ (\partial_z + K ) \partial_z \biggr] f .
\end{align}
Here we used $K_x + i K_y = K (= - \delz S)$ and $c_R^2 - c_I^2 = 1$. Therefore, the following relation holds for holomorphic quantities: 
\begin{align}
L f &= \tilde{L} f, \\
\tilde{L} &= [  \partial_z^2 + K_z \partial_z].
\label{eq:2016Jan11eq2}
\end{align}

Next, we derive the equation describing the time evolution of the probability distribution $P$, from the time evolution of $f$. In order to obtain the equation for generic physical quantities, we start with \eqref{Eq:2015Aug12eq1} which does not assume the holomorphy:
\begin{align}
\nsave{L f} = \int \, dx \, dy \, \biggl[ \biggl( 
(K_x + c_R^2 \delx) \delx + (K_y + c_I^2 \dely) \dely \biggr) f \biggr]  \, P(x,y;t). 
\end{align}
We replace the derivatives acting on $f$ with the ones acting on $P$ by integrating by parts, just as we did for the real Langevin dynamics. Note that, when integrating by parts, the surface term in a large-$|y|$ region does not necessarily vanish. If the probability distribution vanishes sufficiently fast as $|y|$ increases, we obtain 
\begin{align}
\nsave{L f} = \int \, dx \, dy \, f \biggl( 
\delx (- K_x + c_R^2 \delx)  + \dely ( -K_y + c_I^2 \dely) \biggr)  \, P(x,y;t).
\end{align}
Note that the factor $-1$ appears associated with each integration by parts. 

Therefore, the time evolution of the probability distribution $P$ is described by 
\begin{align}
\bibun{P}{t}&=L^T P, \\
L^T      &=[ \partial_x(c_R^2 \partial_x - K_x) + \partial_y(c_I^2 \partial_y - K_y)]. 
\label{Eq:2015Aug12eq2}
\end{align}
This is called the Fokker-Planck-like equation.

\subsubsection{The justification condition for complex Langevin method}

Unlike the Fokker-Planck equation, the Fokker-Planck-like equation is not Hermitian. Hence the equilibrium solution of \eqref{Eq:2015Aug12eq2} is not necessarily $e^{-S}$. However, when a few conditions are satisfied, it can be shown that the expectation value obtained via the complex Langevin method and that obtained via path integral are the same~\cite{Aarts:2009uq}. Let us derive those conditions for justification. 
We introduce a new parameter $\tau \, (0\le \tau \le t)$, and define $F(t,\tau)$ by
\begin{align}
F(t,\tau) \equiv \int P(x,y;t-\tau) \, \calO (x, y;\tau) dx \, dy. 
\end{align}
Here, $\calO(x, y; \tau)$ represents an observable evolving with time.
$F(t,0)$ is 
\begin{align}
F(t,0)= \int P(x,y;t) \,  \calO (x, y; 0) \, dx \, dy, 
\end{align}
namely it is the expectation value obtained via the complex Langevin method. 
$F(t,t)$ is 
\begin{align}
F(t,t)= \int P(x,y;0) \calO(x, y;t) \, dx \, dy. 
\label{Eq:2015Aug13eq2}
\end{align} 
The proof by Aarts and collaborators~\cite{Aarts:2009uq} consists of the following two steps:
\begin{enumerate}
\item[(a)] $F(t, t)$ can be written as 
\begin{align}
F(t,t) &= \int \rho(x;t) \calO(x;0) dx, 
\end{align} 
by using $\rho(x;t)$ which follows the Fokker-Planck equation. 

\item [(b)] $F(t,\tau)$ does not depend on $\tau$, and hence, $F(t, t) = F(t, 0)$ holds. 
\end{enumerate}
By combining these two steps, the equivalence between the complex Langevin method and path integral is guaranteed.

First, we prove (a). We assume that the probability distribution is localized near $y=0$ at $t=0$:
\begin{align}
P(x,y;0) = \rho(x;0) \delta(y).
\label{Eq:2015Aug13eq3}
\end{align}
This condition is satisfied just by choosing the initial condition appropriately.

By substituting \eqref{Eq:2015Aug13eq3} to \eqref{Eq:2015Aug13eq2}, 
we obtain 
\begin{align}
F(t,t) &= \int \rho(x;0) \delta(y) \calO(x, y;t) \, dx \, dy.
\end{align}
If $\calO$ is holomorphic and written as a function of $x+iy$ as $\calO(x+iy)$, its time evolution is described by \eqref{eq:2016Jan11eq2}.
By using the formal solution $\calO(z;t) = e^{ t \tilde{L}} \calO(z;0)$ we obtain 
\begin{align}
       F(t,t)  &= \int \rho(x;0) \delta(y) e^{t \tilde{L}}\calO(x+iy;0) \, dx \, dy.
\end{align}
Because $\calO$ is an analytic function of $z=x+iy$, we write it as a power series of $z$ as $\calO(x+iy;0) = \sum_n a_n z^n$. Then
\begin{align}
   \int \rho(x;0) \delta(y) \tilde{L} \calO(x+iy;0) \, dx \, dy & = 
   \int \rho(x;0) \delta(y) \left[ ( \delz + K_z) \delz \sum_n a_n z^n  \right] dx \, dy , \nn \\
&=   \int \rho(x;0) \delta(y)  \sum_n a_n  n \left( (n-1) z^{n-2} + K_z  z^{n-1} \right)  dx \, dy . 
\end{align}
Because the integrand contains $\delta(y)$, we can easily perform the integral with respect to $y$, and obtain 
\begin{align}
   \int \rho(x;0) \delta(y) \tilde{L} \calO(x+iy;0) \, dx \, dy &=   \int \rho(x;0)   \sum_n a_n  n( (n-1) x^{n-2} + K_x  x^{n-1})  dx.
\label{eq:2017Jul21eq1}
\end{align}
Here, $K_x = K_z|_{y\to 0}$. We assumed that the action is holomorphic as well, so that this limit can be taken safely. By rewriting the right hand side of \eqref{eq:2017Jul21eq1} by using $L_0 = [\delx + K_x] \delx$, we obtain 
\begin{align}
   \int \rho(x;0) \delta(y) \tilde{L} \calO(x+iy;0) \, dx \, dy &=   \int \rho(x;0)  L_0 \calO(x) dx. 
\end{align}
Similar relations hold for higher powers of $\tilde{L}$, and we obtain
\begin{align}
 F(t,t) &= \int \rho(x;0) e^{t L_0} \calO(x;0) dx.
\end{align}
From the stability of the theory along the $x$-direction at $y=0$, the surface term is zero we integrate $x$ by parts, and we obtain 
\begin{align}
F(t,t) &=  \int e^{t L_0^T} \rho(x;0) \calO(x;0) dx,
\label{Eq:2015Aug13eq4}
\end{align}
Here $L_0^T$ represents the transpose of $L_0$ defined by 
\begin{align}
L_0^T &= \delx [ \delx + (\delx S(x))]. 
\end{align}
$e^{t L_0^T} \rho(x;0)$ is the distribution at time $t$. 
By writing it as 
\begin{align}
\rho(x;t) &= e^{t L_0^T} \rho(x;0),
\label{Eq:2015Aug13eq5}
\end{align}
$F(t,t)$ can be expressed as 
\begin{align}
F(t,t) &= \int \rho(x;t) \calO(x;0) dx.
\end{align}
Here $\rho(x;t)$ satisfies
\begin{align}
\bibun{\rho(x;t)}{t} &= L_0^T \rho(x;t). 
\label{Eq:2015Aug13eq5_sub1}
\end{align}
This is the Fokker-Planck equation \eqref{Eq:2015Ag13eq1}. Therefore, $\rho(x;t) \to e^{-S}$ holds as $t\to \infty$.\footnote{
In fact, there is a subtlety here because the real part of the eigenvalues of the Fokker-Planck Hamiltonian are not guaranteed to be positive. However, it has been argued that that is actually the case and $e^{-S}$ is obtained as far as the CLM simulation converges to a stable equilibrium state, which is also demonstrated in a simple model~\cite{Nishimura:2015pba}.
} Therefore, $F(t,t)$ converges to the expectation value defined via the path integral:
\begin{align}
F(t,t)\ \to \int dx \, O \, e^{-S} 
\qquad
(t\to\infty).  
\end{align}

Next, we show (b). The derivative of $F$ with respect to $\tau$ is 
\begin{align}
\bibun{}{\tau} F(t,\tau) &= \int \bibun{}{\tau} P(x,y,t-\tau) \calO(x+iy, \tau) dx dy 
+ \int P(x,y,t-\tau) \bibun{}{\tau}\calO(x+iy,\tau)dxdy, \nn \\
 &= - \int  L^T P(x,y,t-\tau) \calO(x+iy, \tau) dx dy + \int P(x,y,t-\tau) L \calO(x+iy;\tau)dxdy.  
\label{Eq:2015Aug12eq3}
\end{align}
If we can integrate by parts, two terms cancel with each other, and 
\begin{align}
\bibun{}{\tau} F(t,\tau)=0
\end{align}
can hold. 
Then, $F(t,\tau)$ does not depend on $\tau$, and $F(t,0)=F(t,t)$ holds.  

If (a) and (b) hold, the expectation values calculated via the complex Langevin method agree with the ones defined by using the path integral formulation. From the derivation above, we can see the conditions for the complex Langevin method to be justified: firstly, the surface term has to vanish for the integration by parts for $y$, for that the probability distribution $P(x,y;t)$ has to vanish not just as $|x|\to \infty$ but also as $|y|\to \infty$. Secondly, the observables under consideration and the action have to be holomorphic, i.e., they have to satisfy the Cauchy-Riemann relation concerning the complexified variables.   

These justification conditions enable us to judge if the results obtained via the complex Langevin method are correct. However, it is not so easy to see whether these conditions are satisfied in actual simulations. 
Especially, when there are many variables, the dimension of the phase space is large, then it is not easy to visualize $P$ and confirm that it vanishes at infinity. To resolve such difficulties, Ref.~\cite{Nagata:2016vkn} reformulated the conditions introduced by Aarts and collaborators, and gave justification conditions in a practically more useful manner, by using the probability distribution of the drift terms.

Importantly, the conditions for justification give us a clear guideline toward the improvement of the complex Langevin method: we should remove the factors which can potentially spoil steps (a) and (b). As we have seen, two factors can break the justification condition: the spread of $P(x,y;t)$ along the $y$-direction, called the excursion problem~\cite{Aarts:2011ax}, and the breaking of the holomorphy, which is called the singular drift problem. The excursion problem arises because the complexified direction can be unstable. This problem appears in many theories. The singular drift problem often appears in the theory with the fermion determinant $\det \Delta$, because the analyticity of the effective action $S \propto \log \det \Delta$ can be broken when the determinant becomes zero. Those problems can appear when the complex Langevin method is applied to lattice QCD. To resolve the excursion problem, the gauge cooling method was introduced, with which the simulation at the high-density QGP region was achieved. In the next section, we proceed with the application to QCD. We will explain the problems we encounter there, and as a resolution, we explain the gauge cooling method.     
\subsection{Application to lattice QCD}

\subsubsection{Real-valued action}
Let us consider the real-valued action first. As in the previous section, we denote the Langevin time, the discretization step of the Langevin time, and the QCD action as $t$, $\epsilon$, and $S$, respectively. We treat the link variables as functions of the Langevin time and use the Langevin equation describing the Langevin-time evolution 
\begin{align}
U_{x\mu}(t+\eps) = e^{iX} U_{x\mu}(t). 
\label{eq:2017Apr18eq2}
\end{align}
(See Sec.~\ref{sec:QCDLangevinequation}.)
Here $X$ is given by
\begin{align}
X &= \sum_a \lambda_a \biggl[ \biggl( - D_{ax\mu} S \biggr) \eps  + \sqrt{2 \eps} \, \eta_{ax\mu}\biggr], 
\label{eq:2016feb25eq2}
\end{align}
where $\eta$ is the stochastic variable and $\lambda_a$ is the Gell-Mann matrix. The derivative $D_{ax\mu}$ is defined by 
\begin{align}
D_{ax\mu} f( \link{x}{\mu}) = \lim_{\delta\to 0} \frac{ f( e^{ i \delta \lambda_a} \link{x}{\mu}) - f( \link{x}{\mu})}{\delta}.  
\label{eq:2016feb25eq1}
\end{align}
For the fermion determinant $S_f = - \frac{N_f}{4} \ln \det \Delta$, this derivative leads to
\begin{align}
X_f  
     & = \frac{N_f}{4} \sum_{a=1}^{N_c^2-1} \tr \left[ \Delta^{-1} D_a \Delta \right] \lambda_a \epsilon. 
\label{eq:2017apr18eq1}
\end{align}
While this definition is sufficient for theoretical considerations, for actual simulations, the bilinear noise method~\cite{Batrouni:1985jn} is used because the calculation of \eqref{eq:2017apr18eq1} is time-consuming. In the bilinear noise method, a new Gaussian noise $\eta$ is introduced, and the trace is replaced with
\begin{align}
\tr [\Delta^{-1} D_{ax\mu} \Delta]  & =  \eta^\dagger \Delta^{-1} D_{ax \mu} \Delta \eta, \nn \\
    & =  \psi^\dagger  D_{ax \mu} \Delta \eta. 
\end{align}
Here $\psi$ is defined by $\psi^\dagger = \eta^\dagger \Delta^{-1} $, and it is calculated by using the CG method. While the number of uses of CG solver for the calculation of \eqref{eq:2017apr18eq1} is the same as the rank of the fermion matrix the bilinear noise method requires only one use of the CG solver. Therefore, the computational cost is drastically reduced. 

\subsubsection{Complex-valued action}

Next, we consider the complex-valued action. When the dynamical variable was real, say $x\in\mathbb{R}$, it was complexified as $x\to z=x+iy$, the action and observables are analytically continued to the complex plane, and the complex Langevin dynamics was formulated. 
The dynamical variables in lattice QCD are the unitary link variables. Each link variable can be described by using 8($=N_c^2-1$) real variables. For example, an element near the identity can be written as $U=e^{i \omega_a \lambda_a}$ with 8 parameters $\omega_a$ ($a=1, 2, \cdots, 8)$. By complexifying these 8 parameters ($\omega \in \mathbb{R} \to \mathbb{C}$), the link variable can be complexified. Then the link variable turns from SU(3) to ${\rm SL(3,}\mathbb{C})$. To emphasize that the variables are promoted to complex numbers, we use different symbols for the original and complexified variabels~\cite{Nagata:2015uga}:
\begin{align}
U_{x\mu} \in {\rm SU(3)} \to \calU_{x\mu} \in {\rm SL(3,}\mathbb{C}). 
\end{align}
The complexified link variable $\calU_{x\mu}$ is not unitary, and $(\calU_{x\mu})^\dagger = (\calU_{x\mu})^{-1}$ does not hold. Furthermore, because the imaginary part of $\omega_a$ is in $(-\infty, \infty)$, the link variable is noncompact along this direction.

Next, we perform the analytic continuation to the action $S$ and the observable $O$. For that purpose, we write $U^\dagger$ in $S$ and $O$ as $U^{-1}$, and replace $U$ and $U^{-1}$ with $\calU$ and $\calU^{-1}$, respectively. For example, the plaquette is analytically continued to a complex function as 
\begin{align}
U_{\mu\nu}(n) = U_{n,\mu} \, U_{n+\hat{\mu},\nu} \, U_{n+\hat{\nu},\mu}^\dagger \, U_{n,\nu}^\dagger 
\quad \to \quad \calU_{\mu\nu}(n)  = \calU_{n,\mu} \,  \calU_{n+\hat{\mu},\nu} \, \calU_{n+\hat{\nu},\mu}^{-1} \, \calU_{n,\nu}^{-1}. 
\end{align}
We do the same complexification on the fermion determinant and the observables. 

The complex Langevin equation is obtained by replacing $U$ in \eqref{eq:2017Apr18eq2} with $\calU$:
\begin{align}
{\cal U}_{x\mu}(t+\eps) = e^{iX} {\cal U}_{x\mu}(t).
\label{eq:2017Apr18eq3}
\end{align}
\eqref{eq:2017Apr18eq3} may look the same as \eqref{eq:2017Apr18eq2}, 
but the link variables in the former contain more degrees of freedom due to the breaking of the unitarity condition associated with the complexification. \eqref{eq:2017Apr18eq3} contains the Langevin equation for the additional degrees of freedom, which correspond to $y$ in the case of scalar field. 
\subsection{Gauge Cooling method}

The complexified link variables are noncompact. In the complex Langevin simulation of lattice QCD, a flow along this noncompact direction can appear. This is a counterpart of the flow along the $y$-direction in the scalar case. When the flow along this noncompact direction is too large, 
one of the justification conditions --- $P(x, y; t) \to 0 $ as $|y|\rightarrow \infty$ --- can be violated. The gauge cooling method~\cite{Seiler:2012wz} was invented to overcome this problem.  

The original lattice QCD action is invariant under 
\begin{align}
U_{n\mu} \to g_n U_{n \mu} g_{n+\hat{\mu}}^\dagger, \quad g_n \in \SUN{3}. 
\end{align}
On the other hand, the complexified QCD action is invariant under a larger gauge transformation, 
\begin{align}
\calU_{n\mu} \to g_n \calU_{n \mu} g_{n+\hat{\mu}}^{-1}, \quad g_n \in\SLNC{3}. 
\label{eq:2017Aug13eq1}
\end{align}

One of the conditions for justification of complex Langevin is that the probability distribution does not spread along the imaginary directions associated with the complexification. Such spread in lattice QCD is quantified as the deviation from SU(3). The unitarity norm~\cite{Seiler:2012wz}, which characterizes the deviation of the link variables from SU(3), is defined by 
\begin{align}
\calN_{\rm u} = \frac{1}{4V} \sum_{n, \mu} \tr \left[ \, \calU_{n\mu} \, \calU_{n,\mu}^\dagger + (\calU_{n,\mu} \, \calU_{n,\mu}^\dagger)^{-1} - 2 \times \1 \, \right]. 
\end{align}
The Hermitian conjugate is taken for the complexified link variable ${\cal U}$. When $\calU \in \SUN{3}$, 
\[ \calU^\dagger = \calU^{-1}, \]
and hence $\calN_{\rm u}=0$.
On the other hand, when $\calU \notin \SUN{3}$, $\calU^\dagger \neq \calU$ and hence $\calN_{\rm u} > 0$. The unitarity norm gives the distance from the unitary slice. The main idea of gauge cooling is to control the unitarity norm by using the complexified gauge transformation. 

The unitarity norm is invariant under SU(3), but not under $\SLNC{3}$. Therefore, by using the complexified gauge transformation with $g_n \in \SLNC{3}$ the unitarity norm can be varied. By choosing $g_n$ appropriately, the norm can always be decreased. Let us consider a gauge transformation at a point $n$, generated by $g_n$. This gauge transformation acts on all links connected to this point. By picking up the terms in the definition of the unitarity norm which is affected by this gauge transformation, we obtain 
\begin{align}
\calN_{\rm u}(n) & = \sum_{\mu} \biggl[ \, \calU_{n,\mu} \, \calU_{n,\mu}^\dagger + (\calU_{n,\mu} \, \calU_{n,\mu}^\dagger)^{-1} 
      + \calU_{n-\hat{\mu},\mu} \, \calU_{n-\hat{\mu},\mu}^\dagger + (\calU_{n-\hat{\mu},\mu} \, \calU_{n-\hat{\mu},\mu}^\dagger)^{-1} \biggr].
\end{align}
The link variables are transformed as 
\begin{align}
\calU_{n,\, \mu}  &\to \calU_{n, \,  \mu}' = g_n \, \calU_{n,\, \mu} \, , \\
\calU_{n-\hat{\mu},\, \mu} &\to \calU_{n-\hat{\mu},\, \mu}' = \calU_{n-\hat{\mu},\, \mu} \, g_n^{-1} \, . 
\end{align}
In order to suppress the unitarity norm, we use the gauge transformation along the noncompact directions. Therefore, by using an Hermitian matrix $G$($G^\dagger = G$) and a real parameter $\alpha$, we define $g_n$ as 
\begin{align}
g_n = e^{ \alpha G}. 
\end{align}
For an infinitesimal transformation $g_n \sim 1 + \alpha G$, the change of the unitarity norm is 
\begin{align}
\delta \calN_u(n) &= 
\sum_{\mu} \biggl[ \, \calU_{n,\mu} \, \delta \calU_{n,\mu}^\dagger + (\delta \calU_{n,\mu}) \, \calU_{n,\mu}^\dagger
+ (\calU_{n,\mu} \,  \delta\calU_{n,\mu}^\dagger)^{-1} + ( ( \delta \calU_{n,\mu}) \, \calU_{n,\mu}^\dagger)^{-1} \nn \\
& + \calU_{n-\hat{\mu},\mu} \, \delta \calU_{n-\hat{\mu},\mu}^\dagger + ( \delta \calU_{n-\hat{\mu},\mu} )\, \calU_{n-\hat{\mu},\mu}^\dagger 
+ (\calU_{n-\hat{\mu},\mu} \, \delta \calU_{n-\hat{\mu},\mu}^\dagger)^{-1} + ( (\delta \calU_{n-\hat{\mu},\mu}) \, \calU_{n-\hat{\mu},\mu}^\dagger)^{-1} 
\biggr]. 
\end{align}
If we choose $G_n$ as 
\begin{align}
G_n = - \left\{ \sum_\nu \left[ \calU_{n \nu} \calU_{n \nu}^\dagger - ( \calU_{n\nu}^\dagger)^{-1} \calU_{n\nu}^{-1} 
- \calU_{n-\hat{\nu}\, \nu}^\dagger \calU_{n-\hat{\nu}\, \nu} + \calU_{n-\hat{\nu}\, \nu}^{-1} (\calU_{n-\hat{\nu}\,\nu}^\dagger)^{-1} \right] \right\},  
\end{align}
$\delta \calN_u\le 0$ holds, and the unitarity norm decreases. 

Now we have constructed a gauge transformation under which the unitarity norm decreases. The reduction of the unitarity norm by such gauge transformation is called gauge cooling. The idea of the gauge cooling method is to perform gauge coolings after the update of the link variables via the Langevin dynamics. Gauge coolings can be done multiple times. In the actual simulation, the number of the gauge coolings and the parameter $\alpha$ are chosen such that the unitarity norm is minimized.

Because the gauge cooling was introduced based on a heuristic argument, there were criticisms regarding its validity. By now, the validity has been proven for gauge-invariant quantities~\cite{Nagata:2015uga,Nagata:2016vkn}. The gauge cooling method can be valid for other theories with many symmetry generators, including the random matrix theory~\cite{Nagata:2015ijn,Nagata:2016alq}.

\iffigure
\begin{figure}[htbp] 
\centering
\includegraphics[width=9cm]{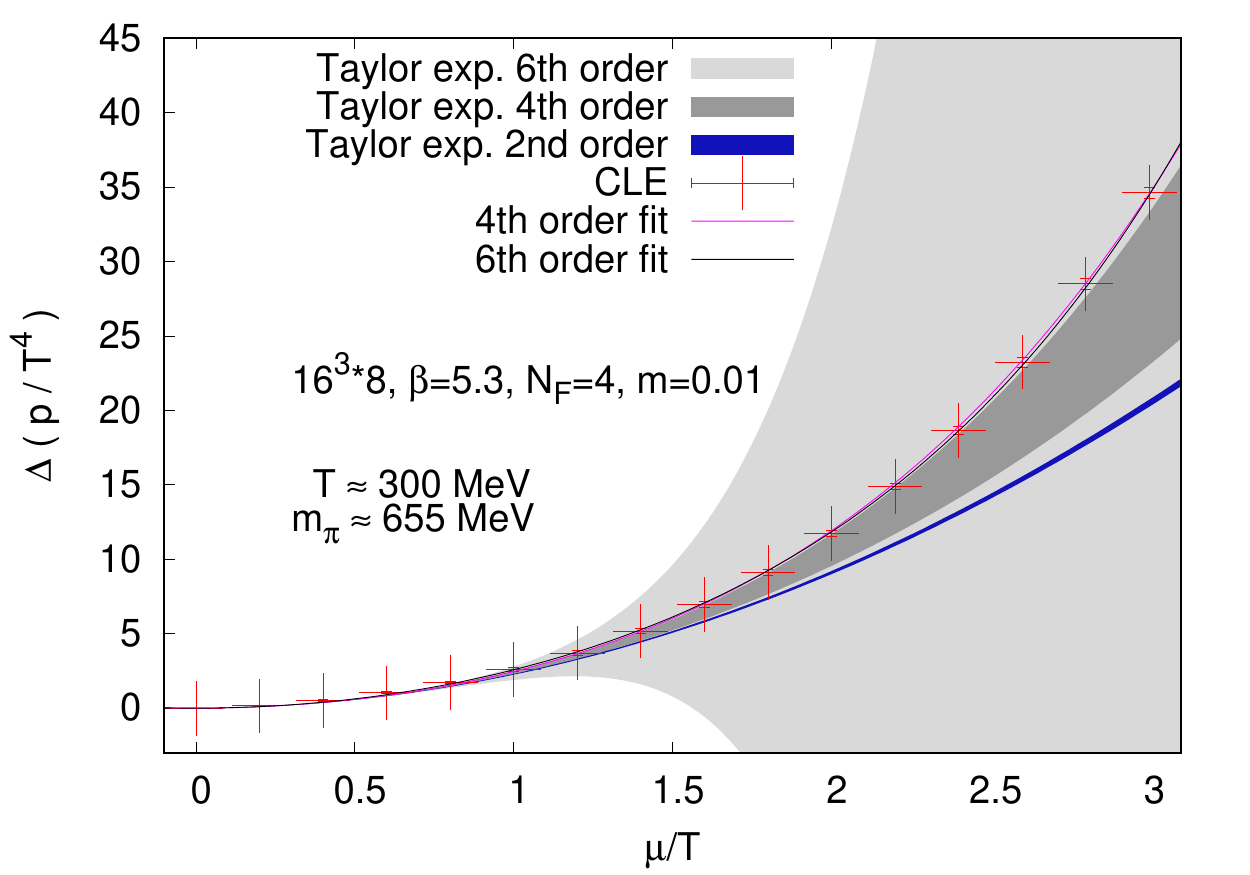}
\begin{minipage}{0.9\linewidth}
\caption{\small 
Pressure in QCD calculated via the complex Langevin method (CLE) and the Taylor expansion (Taylor exp.)~\cite{Sexty:2019vqx}. Two methods are consistent at $\mu/T<1$. As $\mu/T$ becomes large, the error bar spreads in the Taylor expansion method, unlike the complex Langevin. 
}
\label{Sexty2019vqxfigpres}
\end{minipage}
\end{figure}
\fi

The gauge cooling method was introduced in Ref.~\cite{Seiler:2012wz}. It was shown that, even when the usual complex Langevin simulation ceases to work, the gauge cooling can stabilize the simulation. Later, the simulations for heavy dense QCD and high-temperature high-density QCD were achieved~\cite{Sexty:2013ica,Fodor:2015doa,Sexty:2019vqx}, up to $\mu/T=  5 \sim 10$. In Ref.~\cite{Fodor:2015doa}, the result of the complex Langevin simulation was compared to that of the phase quench simulation combined via reweighting, and the agreement was observed. Recently the equations of state in the high-temperature high-density region were calculated as well~\cite{Sexty:2019vqx}, up to $\mu/T=3\sim 4$, which goes far beyond the limitations of the other methods introduced in the previous section (Fig.~\ref{Sexty2019vqxfigpres}). Improved gauge actions are used, and the simulation is going closer to the continuum limit.  It seems that the justification conditions are satisfied to some extent. 

Below, we summarize the advantages of the complex Langevin method, 
compared to the methods based on the  importance sampling. 

\begin{description}

\item[Larger chemical potential]
Some methods rely on the extrapolation regarding the chemical potential, such as the reweighting, canonical method, and Taylor expansion which use the configurations at $\mu=0$, and the analytic continuation from the imaginary chemical potential. These methods become less effective as the chemical potential increases. The CLM does not rely on extrapolation, rather the simulation is done at each value of $\mu$. Therefore, there is no issue associated with the extrapolation regarding $\mu$.

\item[Volume dependence of the computational cost] 
The methods based on the importance sampling and the extrapolation regarding the chemical potential require the calculation of the fermion determinant, which leads to a quick increase of the simulation cost at a large volume. In the CLM, the simulations are performed directly at the parameters of interest, and hence, the calculation of the fermion determinant is not needed, and the volume dependence of the computational cost can be drastically reduced. For this reason, while the methods based on the importance sampling are not practically applicable at large lattice volume except for small $\mu$, the CLM enables us to study a large lattice volume, and the continuum limit may eventually be achieved even in the high-density region. 

From the point of view of the computational techniques, the algorithms used in the CLM are  the same as the ones used for usual lattice QCD simulations. Therefore, many techniques developed in the past can be utilized.

\item[Phase fluctuation] 
As one can easily see in the reweighting method, the methods based on the importance sampling do not work when the phase fluctuation is violent. On the other hand, because the CLM does not use \eqref{Oct052011eq1}, it is not spoiled by a violent phase fluctuation, and hence, it can be used even when the average phase becomes almost zero. 

\item[Judging the validity] 
Another advantage of the CLM is that the conditions for justification are given. There are sources of the systematic errors associated with other methods, for example, the overlap problem for the reweighting method and the truncation error for the Taylor expansion. It is difficult to estimate such systematic errors quantitatively, and hence, it is difficult to judge the validity of the simulations. On the other hand, the justification conditions for the CLM are characterized quantitatively to some extent. Therefore, the validity of the simulations can be checked based on the results of the simulations themselves.

\end{description}

The advantages listed above are essential. Indeed, the CLM enabled the simulation at large chemical potential, which is far beyond the limitations of the other methods ($\mu/T \sim 1$).  

While the CLM is a very useful method, a few problematic points were found. Firstly, the application to the high-density region of the hadronic phase is difficult, because of the breakdown of the analyticity, which is one of the justification conditions. We will discuss this issue in detail in the next section.  

Secondly, instability of the simulation was found when the lattice coupling constant $\beta$ is small~\cite{Fodor:2015doa}. The origin of this problem has not been completely identified. Recently, it was suggested that the problem is related to the calculation of the inverse of the fermion matrix~\cite{Bloch:2017jzi}. It has also been reported that the unitarity norm increases as $\beta$ decreases~\cite{Kogut:2019qmi}.

More researches on the CLM are needed, for example, regarding autocorrelation. Empirically, it appears that the autocorrelation in the CLM simulations is much longer than that in the HMC simulations. 
 
\subsection{Application to low-temperature region and the singular drift problem}

\subsubsection{Singular drift problem in hadron phase}

The gauge cooling enabled us to study high-temperature high-density QCD. It is very nice if this method can be applied to the low-temperature region as well. However, it turned out that the problem of non-analyticity arises in the hadron phase. The origin of this problem is the zero of the fermion determinant at $\mu = m_\pi/2$, which we have seen in the previous chapter. It appears that this problem is very deeply rooted. Here, we explain how the complex Langevin method fails in the low-temperature region of QCD, then we discuss recent attempts to avoid this obstacle. 

This issue was first reported by Mollgaard and Splittorff.
They applied the CLM to the chiral random matrix theory (ChRMT)
and showed that the phase-quenched solution is obtained~\cite{Mollgaard:2013qra}. Because the phase-quenched solution is different from the right solution, their result suggests the wrong convergence. Greensite~\cite{Greensite:2014cxa} showed the same for an effective model which utilizes the Polyakov loop. Later, Mollgaard and Splittorff invented a method that uses the polar coordinate, with which they obtained the correct convergence~\cite{Mollgaard:2014mga}. They compared the distribution of the eigenvalues of the fermion matrix in two cases (wrong and correct convergences) and found that some eigenvalues are close to zero when the wrong convergence takes place.  

At first, it was believed that the reason for the failure is that the logarithm used in the fermion effective action $S_f \propto \log \det \Delta$ breaks the analyticity. However, Nishimura and Shimasaki~\cite{Nishimura:2015pba} showed that the Langevin equation can be written down without introducing the effective action, by choosing the drift term in the integral \[ Z  = \int dx \, w(x) \] as 
\begin{align}
v(x) = w(x)^{-1} \frac{\partial w(x)}{\partial x}, 
\end{align}
and showed that the trouble comes from the configurations with $w=0$. 

\iffigure
\begin{figure}[htbp] 
\centering
\includegraphics[width=11cm]{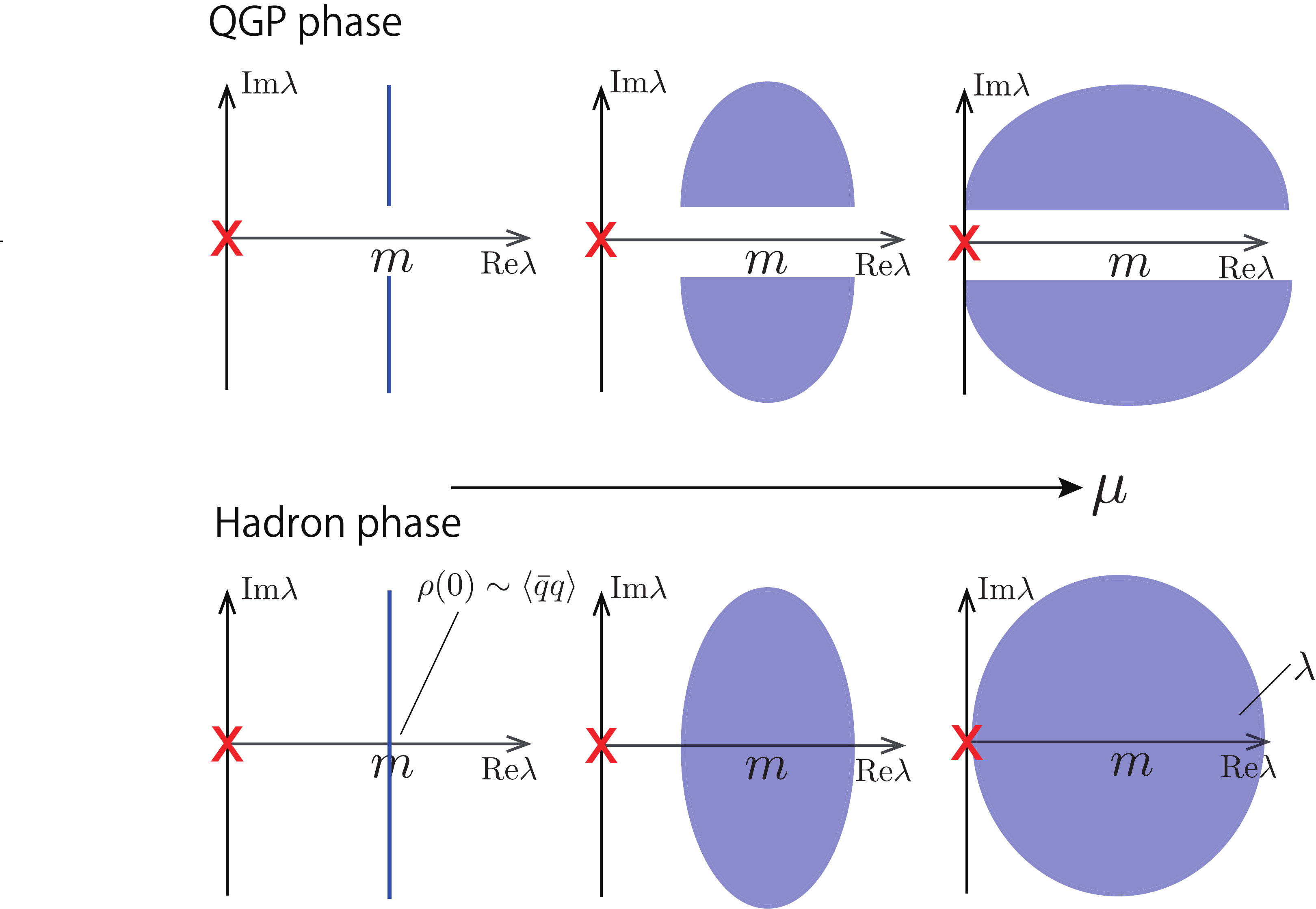}
\begin{minipage}{0.9\linewidth}
\vspace{1.0cm}
\caption{\small 
The change of the eigenvalues of the fermion matrix, as the chemical potential is varied. Different behaviors are seen in the hadron phase and the QGP phase. When the eigenvalue reaches the origin shown by a red $\times$, the fermion determinant becomes zero, and the analyticity of the action is lost. 
}
\label{fig:2017Sep17fig1}
\end{minipage}
\end{figure}
\fi

The CLM does not work for ChRMT because the fermion matrix $\Delta=D+m$ can have a zero eigenvalue which breaks the analyticity of the action, and then one of the conditions for justification is broken. Because the Dirac operators in ChRMT and QCD have the same properties regarding the anti-Hermiticity, the Banks-Casher relation, and so on, QCD has the same problem. To understand this, let us look at the fermion matrix $\Delta = D + m$ and the Dirac operator $D$more closely.

The eigenvalues of the Dirac operator form pairs of pure-imaginary eigenvalues. Therefore, when $\mu = 0$, the eigenvalues of the fermion matrix $D+m$ are $ \pm i \lambda + m, (\lambda \in \mathbb{R})$. Namely, they are distributed on a line in a complex plane (real part $=m$). The density of the eigenvalues of $D$, which we denote by $\rho(\lambda)$, and the chiral condensate $\Sigma$ satisfy 
\begin{align}
\Sigma = \frac{\pi}{V} \rho(0)
\end{align}
due to the Banks-Casher relation~\cite{Banks:1979yr}. Therefore, in the hadron phase $\rho(0) \propto \Sigma \neq 0$ due to the chiral symmetry breaking, and in the QGP phase $\rho(0) \propto \Sigma = 0$. 
The gap exists only in the QGP phase.

When the chemical potential is introduced to the Dirac operator, the eigenvalues become complex. The eigenvalues of $\Delta$ are not on the line (real part $=m$) anymore, it spreads two-dimensionally. The width of the distribution becomes wider as the chemical potential becomes larger. In the QGP phase, because the eigenvalue distribution is gapped, even with large chemical potential the fermion matrix does not have a zero eigenvalue. On the other hand, in the hadron phase, the fermion matrix can have a zero eigenvalue due to the absence of the gap. While the Langevin equation itself has a pole in either phase, in the QGP phase no configuration makes the action singular. On the other hand, in the hadron phase, configurations with singularities in the action can appear, and they break the justification condition for the CLM to be valid.  

That the fermion matrix is causing the problem at $\mu=m_\pi/2$ suggests the connection to the early-onset problem. That the wrong result is close to the phase-quenched one is similar to the early-onset problem, too.   

\subsubsection{Toward a solution to the singular drift problem}
A big challenge toward the complete understanding of the QCD phase diagram is the finite-density region of the hadron phase. We will review recent studies regarding this problem.

For the singular drift problem in ChRMT, Mollgaard and Splittorff showed that the use of the polar coordinate for the complex variable $z$ ($z=r e^{i \theta}$), rather than the Cartesian coordinate, can resolve the issue~\cite{Mollgaard:2014mga}
~\footnote{
Here, `the complex variable' is the one of the ChRMT from the beginning, it is not the complexification of a real variable. }.
Physically, at the low-density region of the hadron phase, there should be no dependence on the chemical potential. When the polar coordinate $z=r e^{ i\theta}$ is used, and the chemical potential takes the form $z e^{\mu}$, the imaginary part of $\theta$ which appears after the complexification cancels the contribution from $e^{\mu}$, so that the Silver-Blaze-like chemical-potential-independence is obtained. This is an important demonstration of a solution to the singular drift problem. In this polar-coordinate method, chemical potential and the imaginary part of $\theta$ cancel with each other. This can happen when $\theta$ is a U$(1)$ phase, which is not the case for QCD.

A generalization of the gauge-cooling method proposed in Refs.~\cite{Nagata:2015ijn,Nagata:2016alq} can be applied to more generic theories. Although the gauge-cooling method was originally applied to gauge theory, it can be applied to more generic theories, the only requirement is the number of symmetry generators is close to the number of dynamical variables. Furthermore, the norm used for the cooling does not have to be the unitarity norm. In Refs.~\cite{Nagata:2015ijn,Nagata:2016alq}, a method was proposed based on
the $U(N)\times U(N)$ symmetry of ChRMT and the anti-Hermiticity norm
\begin{align}
\calN_{\rm a.h.} & = c\, \tr ( D + D^\dagger ) (D + D^\dagger)^\dagger,  
\end{align}
and the correct solution was obtained for ChRMT. 

Two methods explained above --- the use of the polar coordinate, and the generalized gauge cooling --- avoid the pole of the fermion drift by changing the distribution of the Dirac eigenvalues, such that the conditions for justification of the CLM are not broken. However, isn't it problematic to change the distribution of Dirac eigenvalues? Regarding this, the author received two kinds of questions frequently: (1) why the gauge transformation can change the distribution, which is gauge-invariant, and (2) whether the change of the eigenvalue distribution can be justified. The answer to (1) is that, although the Dirac eigenvalues are invariant under the gauge transformation, the configurations before and after the gauge transformation are different, and hence, the eigenvalues after the next Langevin step can change. Question (2) is related to the notion of the distribution of the Dirac eigenvalues in the CLM, and we have to discuss the entire distribution and the zero-mode density separately. Firstly, the entire distribution is not a holomorphic observable, and hence, it is not physical in the CLM simulation. (The distribution has to be real and non-negative, which cannot satisfy the Cauchy-Riemann relation.) Such non-holomorphic quantity cannot be obtained by the CLM, which can calculate only the holomorphic quantities. In other words, the distribution of the Dirac eigenvalues obtained via the CLM simulation is different from the physical eigenvalue distribution. The gauge-cooling method or the use of the polar coordinate system modifies such unphysical eigenvalue distribution, they are not changing the physical eigenvalue distribution. Therefore, that the (unphysical) eigenvalue distribution changes in the CLM is not a bug. There is an important lesson here: the Banks-Casher relation connects the density of zero eigenvalues to a holomorphic observable, chiral condensate. Hence, if the CLM functions correctly, neither the chiral condensate nor the zero-mode density can be changed by the gauge cooling. This means that, as $m$ becomes small in the hadron phase, the configuration with zero fermion determinant appears inevitably.

Let us summarize the argument above. The distribution of the Dirac eigenvalues is not physical in the CLM simulation, and hence, that it changes under the gauge cooling is not a problem. This is an advantage that allows us to modify the distribution via gauge cooling. The connection between the zero-mode density and the chiral condensate is an advantage that makes the calculation of chiral condensate in the CLM simulation possible, but at the same time, this connection indicates the limitation of the CLM: configurations with zero fermion determinant cannot be eliminated by the gauge cooling. 

There are ongoing studies of the singular drift problem in the hadron phase of lattice QCD. The application of the method based on the anti-Hermiticity norm was discussed in Ref.~\cite{Nagata:2016mmh}. In QCD, the anti-Hermiticity norm and the unitarity norm are not too different. While the method based on the anti-Hermiticity norm works well in ChRMT, in the hadron phase of QCD neither the anti-Hermiticity norm nor the unitarity norm is under good control. Presumably, the reason is the difference in the bosonic part of the action. In ChRMT, the bosonic part is Gaussian, which stabilizes the system and eliminates the excursion problem. Hence, we can focus on the singular drift problem, by using the anti-Hermiticity norm. In QCD, however, there is an instability toward the noncompact direction, which leads to the excursion problem. The use of the anti-Hermiticity norm makes it difficult to control the excursion problem. In the hadron phase of QCD, both the singular drift problem and the excursion problem are difficult issues.   

On another note, a justification condition that utilizes the probability distribution of the drift term has been derived~\cite{Nagata:2016vkn}. Previously, some connections between the probability distribution of the observables and the justification condition of the CLM were known. In Ref.~\cite{Nagata:2016vkn}, the justification condition is defined quantitatively, as a behavior of the tail of the distribution. Combined with the fact that the drift term appears in any CLM simulation of any theory, this condition is practically useful and convenient. 

In Ref.~\cite{Ito:2016efb,Ito:2016hlj}, a combination of the CLM and a deformation of the theory is proposed as a solution to the singular drift problem. When the simulation with $\det \Delta$ is impossible due to the existence of the zero-mode in $\Delta$, the fermion matrix is deformed $\Delta \to \Delta + \alpha \Gamma$, where $\Gamma$ is an appropriate matrix and $\alpha$ is the deformation parameter. Then the simulation is performed at $\alpha \neq 0$, where the singular drift problem is absent, and then the extrapolation $\alpha \to 0$ is taken. For the extrapolation, only the range of $\alpha$ where the CLM is functioning correctly should be used. Such range is determined by using the probability distribution of the drift term~\cite{Nagata:2016vkn}. In this way, the right answer was reproduced for an SO(10)-symmetric matrix model. Recently, this method has been applied to the hadron phase of QCD as well~\cite{Nagata:2018mkb}.

Hayata, Tanizaki, and Hidaka compared the CLM and the Lefschetz thimble method by using the one-site Hubbard model as an example. 
In the low-temperature finite-density region, they observed that multiple saddles exist around a singular point and that the CLM ceases to be valid when those saddles become important. Furthermore, they showed that this problem can be avoided to some extent by combining the CLM and the reweighting method~\cite{Hayata:2015lzj}. Their result presumably relies on the flow structure of the theory. Because the flow structures of QCD and one-site Hubbard model are likely to be different,\footnote{
The periodic structure of the saddles in the one-site Hubbard model arises probably because the exponent of the transfer matrix is the dynamical variable. In QCD, the transfer matrix itself is a dynamical variable, and hence, it is not clear whether the same periodic structure exists. } it is not clear how many portions of their findings can be related to QCD. Other recent attempts include the combination of the CLM and Lefschetz thimble method~\cite{Tsutsui:2015tua,Nishimura:2017vav} or the CLM and reweighting~\cite{Bloch:2017sfg,Bloch:2017ods}. 

While various methods were proposed, all of them have pros and cons, and there is no established approach to the finite-density region of the hadron phase of QCD. The search for better methods is ongoing.

\bibliographystyle{utphys}
\bibliography{ref_list}

\chapter{Concluding Remarks}
\section*{}

In this article, we have seen the history of finite-density lattice QCD and sign problem, from its birth to recent developments. We reviewed the ways to circumvent the sign problem based on the importance sampling, such as the reweighting method, Taylor expansion method, imaginary chemical potential method, and the canonical method. We also reviewed the complex Langevin method, which allows us to generate configurations even when the action is complex. For these approaches, we explained the basic ideas, applications, cautionary points, and limitations. We could not review all the proposals because there are vast numbers of them; we believe this review is useful when the readers try to understand those other methods, including the phase-quench simulation method, density of state method, Lefschetz thimble method, and tensor network method.  

We briefly comment on the current status and open problems. 
\begin{description}

\item[High temperature ($T>T_c$)] 
The sign problem is relatively mild. Because the equations of state can be well approximated by low-order polynomials of the chemical potential, the simulation is relatively easy, up to $\mu/T\sim 1$. The continuum limit has been obtained via the Taylor expansion method and the imaginary chemical potential method. As for the reweighting method and the canonical method, the calculation is difficult when the lattice volume is large. The complex Langevin method has been successful up to $\mu/T\sim 4, 5$, and the simulations are getting more accurate.

\item[Near $T_c$]
The phase fluctuation is large, and higher-order terms of the chemical potential are required to approximate the equations of state. Because the convergence of the Taylor series is slow, the Taylor expansion method and imaginary potential method are useful only when the truncation at low order is justified, and hence, it is not clear whether they can be used for the search of the QCD critical point. Although the discovery of the QCD critical point via the MPR and Lee-Yang zero method has been reported, there remain some subtleties. The complex Langevin method has not been successful either, because of the difficulty associated with the singularity and the instability at small $\beta$. Probably, the most successful approach so far is the canonical method, with which multiple groups reported the discovery of the QCD critical point. However, all simulations were performed on small lattice sizes, and it is important to go to a larger lattice size.

\item[Low temperature]
There is no $\mu$-dependence up to $\mu<m_\pi/2$, and violent phase fluctuation sets in at $\mu=m_\pi/2$. It is hard to study this parameter region by using the methods based on the importance sampling. To apply the complex Langevin method, one has to solve the difficulty associated with the singularity. Several approaches, for example, the one based on the deformation of the theory, are being pursued.

\end{description}

This summary is based on the author's current understanding. Perhaps the parameter regions currently out of reach can be studied in the future, or perhaps the results currently believed to be correct turn out to be wrong. In any case, intensive research by many researchers will be needed for a complete understanding of the QCD phase diagram.

Solving the sign problem means reducing the computational cost for the path integral as much as possible by inventing more efficient numerical techniques. Because it is a problem regarding the numerical integration, probably there is no generic solution, and one has to choose the appropriate method depending on the property of the integrand.

Is there a minimum computational cost for the path integral? If there is, how much? This question is related to the `$P$ versus $NP$ problem', which is an unsolved problem in computer science regarding the computational complexity~\cite{Benenti:2012book}. When a problem about the system of size $x$ is given, the class of problems which can be solved within the time of order of a polynomial of $x$ is $P$. The class of problems which cannot be solved within polynomial time, but once the solution is provided its validity can be checked within polynomial time, is $NP$. Obviously, the class $NP$ includes the class $P$, because if a problem can be solved within the polynomial-time then the correctness of the solution can be checked within a polynomial time by actually solving the problem. However, it is not known whether $P$ includes $NP$. Simply put, the $P$ versus $NP$ problem is whether the scale of the problem $N$ increases the time to solve the problem increases as a polynomial of $N$ or it can increase exponentially. Although the $P$ versus $NP$ problem has not been solved yet, many people believe that $P$ and $NP$ are different and there is a problem that cannot be solved within a polynomial time. We do not know the minimum computational cost for the path integral, but it may increase exponentially when the sign problem is severe. If this is the case, then a generic solution to the sign problem that can work for any theory within a polynomial time would be impossible.

Quantum entanglement may be a good characterization of the computational cost. The computational cost required for the tensor network approach, which is a promising way to solve the sign problem, is known to be dependent on the amount of quantum entanglement in the system. Quantum entanglement quantifies the effective degrees of freedom in a quantum vacuum. That this quantity is a measure of the computational cost required for the numerical solution appears to be natural and suggestive. The determination of quantum entanglement is a hard calculation in itself, and the research on quantum entanglement in lattice gauge theory is still at its infancy~\cite{Buividovich:2008kq,Nakagawa:2009jk,Nakagawa:2011su,Itou:2015cyu,Aoki:2016lma}. Such a new viewpoint may lead to the solution of the sign problem.         

\bibliographystyle{utphys}
\bibliography{ref_list}

\appendix

\chapter{Appendix for Chapter~2 (Lattice QCD)}

In this section, we briefly explain how to construct the lattice QCD action. There are good textbooks regarding this subject, suitable for the interested readers. 

\section{QCD in continuous spacetime}

First, let us recall the Lagrangian of the gauge theory continuous Minkowski spacetime. The QED Lagrangian with gauge field $A_\mu, (\mu=0, 1,2,3)$ and fermion field $\psi$ is 
\begin{align}
{\cal L} = - \frac14 F_{\mu\nu} F^{\mu\nu} + \bar{\psi} (i \Slash{D} - m) \psi.
\end{align}
Here $D_\mu$ is the covariant derivative $D_\mu = \partial_\mu + ig A_\mu$, and $g$ is the gauge coupling constant.

We denote the gluon, which is the gauge field in QCD, by $A_\mu^a, (a = 1, 2, \cdots 8)$. The component of $A_\mu^a$ is described by the gauge Lagrangian, just like photon. The Lagrangian is 
\begin{align}
{\cal L} = -\frac14 \sum_{a=1}^{8} F_{\mu \nu}^a F^{\mu\nu a} + \bar{\psi} (i \Slash{D} - m) \psi.
\end{align}
The covariant derivative and the field strength are given by 
\begin{align}
D_\mu & = (\del_\mu + i g A_\mu^a t^a),  \\
F_{\mu\nu}^a &= \del_\mu A_\nu^a - \del_\nu A_\mu^a + g f^{abc} A_\mu^b A_\nu^c.
\end{align}
$t^a$ is the generator of SU(3). By using $t^a$ to write the field strength as
\begin{align}
F_{\mu\nu} &= \sum_a F_{\mu\nu}^a t_a, 
\end{align}
we obtain 
\begin{align}
F_{\mu \nu} &= \frac{1}{ig}[ D_\mu, D_\nu] = \del_\mu A_\nu - \del_\nu A_\mu+ ig [ A_\mu, A_\nu].
\end{align}

By using a formula for the generators $\tr t^a t^b = \frac{1}{2} \delta_{ab}$, 
\begin{align}
\tr \left[ F_{\mu \nu} F^{\mu\nu} \right] & = \sum_{a,b} F_{\mu\nu}^a F^{\mu\nu b} \, \tr \, \left[ t^a t^b \right], \nn \\
& = \frac12 \sum_{a=1}^{8} F_{\mu\nu}^a F^{\mu\nu a}. 
\end{align}
Hence, the QCD Lagrangian is written also as 
\begin{align}
{\cal L} = - \frac{1}{2} \tr F_{\mu \nu} F^{\mu\nu} +  \bar{\psi} (i \Slash{D} - m) \psi.
\end{align}

Next, we go to the Euclidean signature. The time variable $t$ is analytically continued as $t\to -i x_4, (x_4\in \mathbb{R})$. The derivative becomes $\partial_t \to i \partial_4$. The Euclidean action is 
\begin{align}
S_E  = \int d^4 x_E \left( \frac{1}{2} \tr F_{\mu \nu} F_{\mu\nu} + \bar{\psi} (\Slash{D} + m) \psi\right), 
\end{align}
and the path-integral weight is 
\begin{align}
e^{ i S} \to e^{ -S_E}. 
\end{align}
\section{Lattice QCD}
We show that the continuum limit of the plaquette action \eqref{eq:gauge_action}gives the right gauge action. The plaquette $U_{\mu\nu}(n)$, which starts from the lattice point $n$ and go around $\mu, \nu$ direction, can be expanded as 
\begin{align}
U_{\mu\nu}(n) &= e^{i g a^2 F_{\mu\nu}},  \nn \\
 & = \1+ i g a^2 F_{\mu\nu} + \frac12 (i g a^2)^2 F_{\mu \nu}^2 + \cdots
\end{align}
when $a\ll 1$, $U_{\mu\nu}(n)$ and $F_{\mu\nu}$ are $3\times 3$ matrices. By removing the identity matrix, we obtain 
\begin{align}
\1 - U_{\mu\nu}(n) = - i g a^2 F_{\mu\nu} + \frac12 g^2 a^4 F_{\mu \nu}^2 - \cdots. 
\label{eq:2017Mar26eq1}
\end{align}
The link variable is unitary ($U^{-1} = U^\dagger$) and $F_{\mu\nu}$ is Hermitian ($F^\dagger = F$). The eigenvalues of a Hermitian matrix are real, and hence, the trace of a Hermitian matrix is also real. By taking the trace of  
\eqref{eq:2017Mar26eq1} and keeping the real part, 
\begin{align}
{\rm Re} {\rm tr} \left[\1 - U_{\mu\nu}(n) \right] =  \frac12 g^2 a^4  {\rm tr} F_{\mu \nu}^2 + O (a^6) 
\label{eq:2017Mar26eq2}
\end{align}
Note that the odd powers disappear. By summin over $\mu, \nu$ and $n$, we obtain 
\begin{align}
\sum_{n} \sum_{\mu,\nu } 2 {\rm Re}\, {\rm tr} [ \1 - U_{\mu \nu}(n)] 
= g^2 \sum_{n} \sum_{ \mu, \nu}   a^4 {\rm tr}  F_{\mu \nu}^2 (1 + O(a^2)).
\end{align}
We further use 
\begin{align}
a^4 \sum_{n} = \int d^4x
\end{align}
and obtain 
\begin{align}
\sum_{n} \sum_{ \bra \mu, \nu \ket} 2 {\rm Re} \, {\rm tr} [ \1 - U_{\mu \nu}(n)] = 
 g^2 \int d^4x \, {\rm tr}  F_{\mu \nu}^2 (1 + O(a^2)).
\end{align}
In the right hand side, Einstein's summation convention with respect to $\mu$ and $\nu$ is assumed. Hence the gauge action on lattice is given by 
\begin{align}
S & = \int d^4x \left( \frac12 {\rm tr} \, F_{\mu \nu}^2  \right)= 
 \frac{1}{g^2} \sum_{n} \sum_{\mu, \nu} {\rm Re} \,{\rm tr} [ \1 - U_{\mu \nu}(n)]. 
\end{align}

The first term on the right-hand side is a constant proportional to the volume. This term does not affect the expectation values, so often this term is neglected and the gauge action is written as  
\begin{align}
S & = - \frac{1}{g^2} \sum_{n} \sum_{\mu \neq \nu} {\rm Re} \,{\rm tr} [ U_{\mu \nu}(n)], \\
  & = - \frac{1}{2g^2} \sum_{n} \sum_{\mu \neq \nu} \,{\rm tr} [ U_{\mu \nu}(n) + U_{\mu \nu}(n)^\dagger].
\end{align}
Also, the lattice coupling constant is introduced as $\beta = 2 N_c/ g^2$. 
\section{Anti-Hermiticity of Staggered fermion}
\label{sec:anti_hermite_KS}
Let us write the second term in the definition of the staggered fermion \eqref{Stfermion} as $D_{xy}$. This $D$ is the Dirac operator on a lattice. The Dirac operator in the continuum theory satisfies the anti-Hermiticity $D^\dagger = - D$. We show that the staggered-type Dirac operator on lattice satisfies the lattice version of the anti-Hermiticity $D_{yx}^\dagger = - D_{xy}$. 

By taking the Hermitian conjugate of $D_{xy}$, we obtain 
\begin{align}
D_{yx}^\dagger = \sum_{\nu} \frac{\eta_\nu(y)}{2a} \left[ 
e^{\mu \delta_{4\nu}} U_\nu(y)^\dagger \delta_{y+\hat{\nu}, x }
- e^{- \mu \delta_{4\nu}} U_\nu(y-\hat{\nu}) \delta_{y-\hat{\nu},x} 
\right]. 
\end{align}
Because $\delta_{y+\hat{\nu}, x }= \delta_{y, x -\hat{\nu}}$, the first term can be written as $U_\nu^\dagger (y) = U_\nu^\dagger(x-\hat{\nu})$.
In a similar manner, we can rewrite the second term as $U_\nu(y-\hat{\nu})= U_\nu(x)$. Therefore, 
\begin{align}
D_{yx}^\dagger = \sum_{\nu} \frac{\eta_\nu(y)}{2a} \left[ 
- e^{- \mu \delta_{4\nu}} U_\nu(x) \delta_{y,x+\hat{\nu}} + 
e^{\mu \delta_{4\nu}} U_\nu(x-\hat{\nu})^\dagger \delta_{y, x-\hat{\nu} }
\right]
\end{align}
By using $\eta_\nu(x+\hat{\nu}) = \eta_\nu(x)$, $\eta_\nu(x-\hat{\nu}) = \eta_\nu(x)$, we obtain 
\begin{align}
D_{yx}^\dagger &= \sum_{\nu} \frac{\eta_\nu(x)}{2a} \left[ 
- e^{- \mu \delta_{4\nu}} U_\nu(x) \delta_{y,x+\hat{\nu}} + 
e^{\mu \delta_{4\nu}} U_\nu(x-\hat{\nu})^\dagger \delta_{y, x-\hat{\nu} }
\right]. 
\end{align}
Therefore, if $\mu = 0$, we can see the anti-Hermiticity 
\begin{align}
D_{yx}^\dagger&= -D_{xy}. 
\end{align}
We can also see that the anti-Hermiticity is lost if $\mu \neq 0$. By generalizing the expressions above slightly, we can easily see the anti-Hermiticity for pure-imaginary $\mu$. 
\section{Haar measure}
\label{sec:haarmeasure}

We give a brief summary of the integral on a group manifold.~\footnote{See Sec.~3.1.2 of Ref.~\cite{Gattringer:book}.}. An element $U$ of a compact Lie group $G$ can be parameterized by real parameters $\omega^{(k)}$. The number of parameters is the same as the number of the generators. The Haar measure associated with $U=U(\omega)$ is defined by 
\begin{align}
dU = c\sqrt{\det [ g(\omega)]} \prod_k d\omega^{(k)}. 
\end{align}
A constant $c$ is determined such that $\int dU=1$. $g$ is the metric tensor given by 
\begin{align}
g(\omega) = \tr \biggl[ \frac{\partial U(\omega)}{\partial \omega^{(n)}} U(\omega)^{-1}
\biggl( \frac{\partial U(\omega)}{\partial \omega^{(m)} }U(\omega)^{-1} \biggr)^{-1} \biggr].  
\end{align}
In the case of the unitary group ($U^\dagger = U^{-1}$), 
\begin{align}
g(\omega) = \tr \biggl[ \frac{\partial U(\omega)}{\partial \omega^{(n)} }
 \frac{\partial U(\omega)^\dagger}{\partial \omega^{(m)}}  \biggr].  
\end{align}

For example, we write down the Haar measure of U(1) group explicitly. 
By writing a group element $U=e^{i \omega}$, we obtain $\partial_\omega U = i U$, hence we immediately get $g(\omega) = 1$.
Therefore, 
\begin{align}
dU = c d \omega.
\end{align}
From the normalization convention $\int dU \, 1 = 1$,
we obtain $c = 1/ 2\pi$. Therefore, the Haar measure of U(1) group is 
\begin{align}
dU = \frac{d\omega}{2\pi}.
\end{align}
\section{Properties of the Dirac operator with complexified gauge field}

We study what happens to the Dirac operator when the gauge field is complexified. Firstly, let us see the property of the Dirac operator before the gauge field is complexified. The Dirac operator is defined by 
\begin{align}
\Delta = \gamma_\mu D_\mu + \mu \gamma_4.  
\label{eq:2016feb22eq1}
\end{align}
Here, $D_\mu$ is the covariant derivative, which satisfies $D_\mu^\dagger = - D_\mu$. In the Euclidean spacetime $\gamma_\mu^\dagger=\gamma_\mu$ holds, hence 
\begin{align}
\Delta^\dagger(\mu) &= - \gamma_\mu D_\mu + \mu \gamma_4, \nn \\ 
                    &= - (\gamma_\mu D_\mu - \mu \gamma_4), \nn \\ 
                    &= - \Delta(-\mu).  
\label{eq:2016feb22eq2}
\end{align}
Hence 
\begin{align}
\gamma_5 \Delta^\dagger(\mu) \gamma_5  &=  \gamma_\mu D_\mu - \mu \gamma_4, \nn \\ 
                    &= \Delta(-\mu).  
\label{eq:2016feb22eq3}
\end{align}

On the other hand, from \eqref{eq:2016feb22eq1} we can see that 
\begin{align} 
\{D, \gamma_5\}=0
\end{align}
holds regardless of the value of $\mu$. This anti-commutation relation holds because $\Delta$ has one power of $\gamma_\mu$.

When the gauge field is complexified, the anti-Hermiticity of the covariant derivative is lost: $D_\nu^\dagger \neq - D_\nu$.
Therefore, 
\begin{align}
\Delta^\dagger(\mu) &= - (\gamma_\mu D_\mu^\dagger - \mu \gamma_4), \nn \\ 
 & \neq - \Delta(-\mu), \\
\gamma_5 \Delta^\dagger(\mu) \gamma_5 &=  (\gamma_\mu D_\mu^\dagger - \mu \gamma_4), \nn \\ 
 & \neq  \Delta(-\mu). 
\end{align}
On the other hand,
\begin{align}
\{ \Delta(\mu), \gamma_5 \} = 0
\end{align}
still holds, because it is based only on the property of the $\gamma$ matrix.
In summary, the Dirac operator has three types of important properties shown in Table~\ref{tab:cond_delta}: 
\begin{table}[tbh]
\begin{center}
\caption{\small Relations satisfied by the fermion matrix, and when it holds}
\label{tab:cond_delta}
\begin{tabular}{l|lll}
\hline
Relation        & Before complexification ($\mu=0$) & Before complexification ($\mu \neq 0) $ & After complexification\\
\hline
$\Delta^\dagger = - \Delta$       & holds   & does not hold & does not hold \\
$\Delta^\dagger(\mu) = - \Delta(-\mu)$       & holds    & holds & does not hold  \\
$\gamma_5 \Delta^\dagger \gamma_5 = \Delta$ & holds   & does not hold & does not hold \\
$\gamma_5 \Delta^\dagger(\mu) \gamma_5 = \Delta(-\mu)$ & holds    & holds & does not hold \\
$\{\Delta, \gamma_5\}=0$               & holds    & holds  & holds\\
\hline
\end{tabular}
\end{center}
\end{table}

Next, we study the symmetries of the eigenvalues controlled by the anti-Hermiticity, $\gamma_5$-Hermiticity, and anticommutation relation.

\underline{Anti-Hermiticity}

Suppose $D | u\ket = \lambda |u \ket$ holds. Then
\begin{align}
\bra u | D | u \ket = \lambda.  
\end{align}
By taking the Hermitian conjugate, we obtain
\begin{align}
( \bra u | D | u \ket) ^\dagger = \bra u | D^\dagger  | u \ket, 
\end{align}
whose the left and right hand sides are
\begin{align}
(l.h.s) = ( \lambda)^\dagger = \lambda^*, \\
(r.h.s) = \bra u | (-D) |u \ket = - \lambda
\end{align}
Hence $\lambda^* = - \lambda$ holds, namely $\lambda$ is pure imaginary.

\underline{$\gamma_5$-Hermiticity}

Because of 
\begin{align}
& \det (\Delta - \lambda I) = 0, \\
& \to  \det ( \Delta^\dagger - \lambda^* I ) = 0 , \\
& \to \det (\Delta - \lambda^* I ) = 0,
\end{align}
eigenvalues appear in pairs $(\lambda, \lambda^*)$. The distribution is symmetric under the reflection with respect to the real axis.

\underline{Anti-commutation relation}

Because of $\gamma_5 \Delta \gamma_5 = - \Delta$, 
\begin{align}
&\Delta | v_n \ket = \lambda_n | v_n \ket, \\
&\to \gamma_5 \Delta \gamma_5 \, \gamma_5 |v_n \ket = \lambda_n \gamma_5 | v_n \ket, \\
&\to  \Delta | u_n \ket = - \lambda_n |u_n \ket. 
\end{align}
Therefore, eigenvalues form pairs $(\lambda_n, - \lambda_n)$, and the eigenvectors are related by $|u_n\ket = \gamma_5 |v_n\ket$.

\underline{Symmetry of the Wilson fermion}

The Wilson fermion is defined by
\begin{align}
\Delta_W = \Delta_0 + a D_\mu D_\mu. 
\end{align}
Due to the second term on the right-hand side, the Wilson fermion is not anti-Hermitian. Therefore, the eigenvalue of the Wilson Dirac operator is complex.

On the other hand, the $\gamma_5$-Hermiticity is satisfied: 
\begin{align}
\gamma_5 \Delta_W(\mu)^\dagger \gamma_5 
& = \gamma_5 \Delta(\mu)^\dagger \gamma_5 + a \gamma_5 D_\mu D_\mu  \gamma_5, \nn \\
& =  \Delta(-\mu) + a D_\mu D_\mu, \nn \\
& = \Delta_W(- \mu).  
\end{align}

\chapter{Appendix for Chapter~3}

\section{$Z_3$-transformation and Roberge-Weiss periodicity}
\label{sec:Z3RW}
We consider the $Z_3$-transformation which acts on all the $U_{n4}$'s at a time slice $t=t_i, (i=1, \cdots N_t)$:
\[ 
U_{n \mu} \to U_{n \mu}' = \omega U_{n \mu},  (\omega \in Z_3, \mu=4). 
\]
The gauge action \eqref{eq:2017Jun07eq2} is invariant under this $Z_3$-transformation because $\omega$ and $\omega^{-1}$ from $U_{n\mu}$ and $U_{n\mu}^\dagger$ cancel with each other. 

The fermion action breaks this symmetry explicitly. By using the reduction formula, we can see how the fermion determinant transforms under $Z_3$. In \eqref{Eq:2014Jul26eq2}, only the $\beta_i$-part is affected, as $\beta_i \to \omega \beta_i$. 
Hence $Q$ and the determinant transform as 
\[ 
Q \to \omega  Q
\]
and
\[ 
\det \Delta(\mu) \to \det \Delta(\mu)  = C_0 \xi^{ -\Nred/2}  \det ( \omega Q + \xi).  
\]

In order to restore the $Z_3$-invariance, we vary 
$\xi = e^{- \mu/T}$ as $\xi\to\omega \xi$, then 
\[
\det \Delta (\mu ) = C_0 \xi^{ -\Nred/2} \det \omega (Q+ \xi). 
\]
Note that the rank of the determinant is proportional to $N_c=3$, 
and we used $\omega^{N_c}=1$ to show the invariance of the determinant. 
Because $\omega \xi = e^{ 2\pi k /3 i + \mu/T}, k\in \mathbb{Z}$, this is proof of the Roberge-Weiss periodicity on a lattice. 
\clearpage
\section{Lee-Yang zero theorem}
\label{sec:lee-yang}
Lee-Yang zero theorem was derived by Lee and Yang around 1950. 
Back then, people were discussing how the non-analyticity of a phase transition can appear in the thermodynamic limit. Lee-Yang zero theorem answers this problem.\footnote{
The original references~\cite{Yang:1952be,Lee:1952ig} are written very clearly, and it is worth reading them not just to learn the theorem but also as a good example of the way scientific papers should be written. Ref.~\cite{Blythe:2003aaa} is also a good introduction. It discusses the application to non-equilibrium systems as well.
}

We consider the grand canonical partition function~\eqref{eq:GCZ2}:
\begin{align}
Z(\mu) = \lim_{N\to\infty} \sum_{n=-N}^N Z_n \xi^n. 
\label{Eq:2014Mar16eq0}
\end{align}
Here $Z_n$ is real and positive for any $n$, and due to the symmetry between particle and anti-particle $Z_n = Z_{-n}$ is satisfied. $N$ is the largest possible number of particles in the system, which is finite on a finite lattice but becomes infinite in the thermodynamic limit. The Helmholtz free energy $f (n)$ of the $n$-particle system is given by $Z_n = \exp(-V f (n) /T)$. Note that $Z_n$ and $Z(\mu)$ are functions of temperature $T$. 

If $N$ is finite, \eqref{Eq:2014Mar16eq0} is analytic. Lee and Yang considered how the grand canonical partition function \eqref{Eq:2014Mar16eq0} can develop the non-analyticity. They discovered that the behavior of the zeros of the partition function in the complex plane is the key. 

Because \eqref{Eq:2014Mar16eq0} is a polynomial of $\xi$ of order $2N$, (when $N$ is finite) there are $2N$ roots. By writing them $\xi_i$, we obtain
\begin{align}
Z(\mu) = \lim_{N\to \infty} Z_{-N} \xi^{-N}  \prod_{i=1}^{2N}
\left(1-\frac{\xi}{\xi_i}\right).
\label{Eq:2014Mar16eq1}
\end{align}
These roots $\{\xi_i\}$ are called Lee-Yang zeros. Since \eqref{Eq:2014Mar16eq0} is nonzero for any real and positive $\xi$, the roots $\xi_i$ cannot be real and positive. Furthermore, from the symmetry $Z_n = Z_{-n}$, if $\xi_i$ is a Lee-Yang zero then $1/\xi_i$ is also Lee-Yang zero. Therefore, Lee-Yang zeros are distributed in the complex plane, except for the positive part of the real axis, and the zeros inside and outside the unit circle form pairs.   

Physically, the chemical potential $\mu$ has to be real, and the fugacity $\xi$ has to be real and positive. If the partition function $Z(\mu)$ has a non-analyticity at real and positive $\xi$, the phase transition takes place. No singularity is allowed at finite $N$, but in the thermodynamic limit $N\to\infty$ the Lee-Yang zeros can approach the positive part of the real axis and develop the non-analyticity there. This is the Lee-Yang theorem.

\begin{table}[tbh]
\begin{center}
\caption{Thermodynamics vs electrostatic potential problem}
\label{tab:EManalogy}
\begin{tabular}{c|c}
\hline 
$\phi$ & electrostatic potential\\
$ \nabla \phi$ & electric field\\
$\xi_i$ & location of electric charges \\
\hline
\end{tabular}
\end{center}
\end{table}
The relationship between Lee-Yang zeros and the non-analyticity can be understood intuitively based on the electrostatic analogy~\cite{Lee:1952ig}.
When the fugacity $\xi$ is promoted to complex and the analytic continuation of the free energy $f$ is considered, 
$\phi \equiv {\rm Re}\, f$ is 
\begin{align}
\phi(\xi)
 = - \frac{T}{V} \sum_{i=1}^{2N} \ln|\xi-\xi_i| - \frac{T}{V} \ln Z_{N} + \frac{NT}{V} \ln |\xi|. 
\end{align}
By taking the derivative of $\phi$ with respect to $\xi$, we obtain 
\begin{align}
\nabla_\xi^2
\phi(\xi) = - 2\pi \frac{T}{V} \sum_{i=1}^{2N}
\delta^{(2)}
(\xi-\xi_i), 
\label{Eq:2015Jan04eq1}
\end{align}
where $\nabla_\xi \equiv ( \partial/(\partial\,\Re\,\xi), \partial/(\partial\,\Im\,\xi))$.
Here we used
\begin{align}
\nabla^2
 \ln |z| = \left( \frac{\partial^2 }{\partial x^2} + 
\frac{\partial^2}{\partial y^2} \right) \ln | x+ iy| = 2 \pi \delta(x) \delta(y). 
\end{align}
The equation \eqref{Eq:2015Jan04eq1} is the Poisson equation in the two-dimensional electrostatic potential problem.
$\phi$, $\nabla \phi$, $\xi_i$ and \eqref{Eq:2015Jan04eq1} correspond to the electrostatic potential, electric field, the location of the electric charge, and the Gauss's law, respectively (Table~\ref{tab:EManalogy}).

Imagine the situation that the point charges are distributed discretely on a line in a two-dimensional space. Between the charges, the potential and the field are analytic functions. Suppose the number of point charges increases and eventually, the distribution becomes continuous. Across such one-dimensional charge distribution, the potential is continuous, while its derivative --- the electric field --- is discontinuous.  

The electrostatic analogy for the Lee-Yang zeros means the mathematical correspondence between the singularities in thermodynamics and the electrostatic problem. When the distribution of the Lee-Yang zeros becomes dense, $\nabla_\xi \phi$, which is the counterpart of the electric field, changes discontinuously across the line of Lee-Yang zeros. Such condensation of Lee-Yang zeros in the complex fugacity plane can cause a phase transition.

Lee and Yang showed that the Lee-Yang-zeros of the spin-$\frac{1}{2}$ Ising model on the complex $e^h$-plane are distributed only on the unit circle~\cite{Lee:1952ig}.~\footnote{Lee-Yang zero theorem can mean two things: the general relation between zeros and phase transition, and a specific example of the Ising model. Sometimes the latter is called Lee-Yang zero circle theorem, to distinguish it from the former. }. The distributions of the Lee-Yang zero are known also for the spin-$\frac{3}{2}$ model~\cite{Asano:1968aaa}, Blume-Capel model, Potts model~\cite{Biskup:2000prl1}, and high-temperature QCD~\cite{Nagata:2014fra}. Fisher proposed that temperature can be complexified, and the zero, in that case, is called Fisher zero. 
 
Multiple Lee-Yang zero points can appear in the complex plane. As the thermodynamic limit is taken, one or more of them can approach the real axis and cause a phase transition. The order of phase transition depends on the way the zeros condense. 
For a first-order transition, the scaling on the real axis is described by $1/V$. For a second-order transition, various scalings can appear depending on the critical exponents~\cite{Itzykson:1983gb,Blythe:2003aaa,Stephanov:2006dn}. 

\section{Fourier Integral}
\label{Sec:incompletegamma}

We derive \eqref{Eq:2014Apr21eq1}. Firstly, in the integral, we can replace $e^{ i n \theta}$ with $\cos n \theta$. By expanding $\cos n \theta$ with respect to $\theta$, we obtain 
\begin{align}
Z_n\propto \int_{-\pi/3}^{\pi/3} d\theta\,e^{-a\theta^2}\cos n\theta
=\sum_{k=0}^\infty  \frac{(-1)^k n^{2k}}{(2k)!}I_k~.
\label{incgammaeq1}
\end{align}
Here we used $a=V T^3 c_2$ and
\begin{align}
I_k=\int_{-\pi/3}^{\pi/3} d\theta\,e^{-a\theta^2}\theta^{2k}. 
\end{align}
By using the gamma function $\Gamma(z)$and the incomplete gamma function $\Gamma(z,p)=\int_p^\infty e^{-t} t^{z-1}dt$, $I_k$ can be expressed as 
\begin{align}
I_k = \frac{1}{a^{k+1/2}}\Bigl(\Gamma(k+1/2)-\Gamma(k+1/2,a \pi^2/9)\Bigr).
\end{align}
Because $\Gamma(z,p)$ approaches zero as $p\propto V \to \infty$, 
$I_k$ can be written using only the gamma function in the thermodynamic limit. By using an identity 
\begin{align}
\frac{\Gamma(k+1/2)}{(2k)!}=\frac{\sqrt{\pi}}{4^k k!}~,
\end{align}
and then taking the sum in \eqref{incgammaeq1}, we obtain \eqref{Eq:2014Apr21eq1}.

\chapter{Appendix for Chapter~4}

\section{Derivation of the Fokker-Planck equation (discrete-time version) }
\label{sec:FP_disc}

We derive the Fokker-Planck equation which describes the time evolution of the probability density $P(x;t)$. Here we give a derivation for the discrete-time version. As for the continuous-time version, see e.g.,~\cite{Damgaard:1987rr}.

We start with the descretized Langevin equation,\footnote{
Sometimes, the stochastic variable is normalized as $\tilde{\eta}=\sqrt{\epsilon}\eta$. With this normalization, the Langevin equation is 
\begin{align}
x(t+\epsilon) = x(t)  - \bibun{S}{x} \eps  +  \sqrt{\eps}\, \eta(t), 
\end{align}
and the noise average is $\nsave{\tileta(t_1) \tileta(t_2)} = 2 \delta(t_1-t_2)$. The choice of the normalization is just a matter of taste, as long as the normalization is set consistently including the noise average. }
\begin{align}
x(t+\epsilon) = x(t)  + \biggl(- \bibun{S}{x}  + \eta(t)\biggr) \epsilon.
\end{align}
Let $\delta x(t)\equiv x(t+\epsilon)-x(t)$. The expectation value of $\calO$ changes as
\begin{align}
\nsave{\calO(x(t+\epsilon))} -  \nsave{\calO(x(t))} &= \nsave{ \delx \calO(x(t)) \, \delta x(t) +  \frac{1}{2} \delx^2 \calO(x(t)) \, (\delta x(t))^2}, \nn \\
                         &= \biggl\langle \delx \calO(x(t)) \, ( - \delx S + \eta(t)  ) \epsilon + \frac{1}{2} \delx^2 \calO(x(t)) \,(- \delx S + \eta(t))^2 \epsilon^2 \biggr\rangle.
\label{Eq:2015Aug07eq2}
\end{align}
The first line is the expansion with respect to $\delta x$ up to the second order, and in the second line we replaces $\delta(x)$ with $\biggl(- \delx S + \eta(t)\biggr) \epsilon$ by using the Lanvevin equation. \eqref{Eq:2015Aug07eq2} contains $\eta(t)^1$ and $\eta(t)^2$. From \eqref{eq:2016Jan09eq1}, the integral of $\eta^2$ is 
\begin{align}
\langle \eta(t)^2 \rangle_\eta & = \frac{\int_{-\infty}^\infty d \eta(t)\, \eta(t)^2 e^{- \frac{\eps}{4} \eta(t)^2}}{
\int_{-\infty}^\infty d \eta(t)\, e^{- \frac{\eps}{4} \eta(t)^2}} \nn \\
 & = \frac{2}{\eps}. 
\end{align}
Odd powers of $\eta$ average to zero: $\langle \eta(t)^n \rangle_\eta = 0, \{n = 2m-1,\, m\in \mathbb{Z}\}$. Therefore, \eqref{Eq:2015Aug07eq2} can be rewritten as~\footnote{
Note that, although $x(t)$ depends on $\eta(\tau)$ at $\tau<t$,
it does not depend on $\eta(t)$. Therefore, $\frac{\partial \calO(x(t))}{\partial x} \eta(t)$ is linear in $\eta(t)$, and hence, 
\begin{align}
\left\langle \frac{\partial \calO(x(t))}{\partial x} \eta(t)\right \rangle =0. 
\end{align}
}
\begin{align}
\nsave{\calO(x(t+\epsilon))} - \nsave{\calO(x(t))}&= \biggl\langle \delx \calO(x(t)) \, ( - \delx S) \, \epsilon + \frac{1}{2} \delx^2 \calO(x(t)) \biggl( (\delx S)^2 \epsilon^2 + \tileta^2 \epsilon^2 \biggr) \biggr\rangle_\eta, \nn \\
& = \epsilon \biggl\langle - \bibun{\calO(x(t))}{x}\bibun{S}{x} + \frac{ \partial^2 \calO(x(t))}{\partial x^2} \biggr\rangle_\eta + \calO(\epsilon^2).
\label{Eq:2015Aug10eq1}
\end{align}

We rewrite the integral over $\eta$ in \eqref{Eq:2015Aug10eq1} to the integral over $x$, by using the probability distribution $P(x;t)$. Then the first term in \eqref{Eq:2015Aug10eq1} becomes 
\begin{align}
\nsave{(\delx \calO) \, \delx S} & = \int dx \, (\delx \calO)  \, (\delx S ) \,  P, \nn \\
 & = \int dx \, \biggl(  \delx ( \calO (\delx S) P) - \calO \delx ( (\delx S)P ) \biggr) , \nn \\
 & = \biggl[ \calO (\delx S) P \biggr]_{x\in d\Omega} + \int dx\, \calO \biggl( - \delx (P\delx S ) \biggr).
\end{align}
Here $d\Omega$ stands for the surface term, which can be neglected when $P(x;t)$ falls sufficiently fast. The second term in \eqref{Eq:2015Aug10eq1} becomes
\begin{align}
\nsave{ \delx^2 \calO} & = \int dx \, (\delx^2 \calO) P, \nn \\
 & = \int dx \, \biggl[\delx  \biggl(  (\delx \calO) P \biggr) - (\delx \calO) ( \delx P) \biggr], \nn \\
 & = \bigl[ (\delx \calO) P \bigr]_{x \in d\Omega} - \int dx\, (\delx \calO) ( \delx P) , \nn \\
 & =  - \int dx \,  \biggl[ \delx ( \calO (\delx P)) - \calO (\delx^2 P) \biggr], \nn \\
 & = - \biggl[ \calO (\delx P) \biggr]_{x\in d\Omega} + \int dx \, \calO (\delx^2 P), \nn \\
 & = \int dx \, \calO (\delx^2 P).  
\end{align}
Here we assume that both $P$ and its derivatives $\delx P$, $\delx^2 P$ vanish sufficiently fast. In this way, we obtain the Fokker-Planck equation
\begin{align}
\bibun{P}{t} = \bibun{}{x}\biggl( \bibun{}{x} + \bibun{S}{x}\biggr) P.
\end{align}

\section{Fokker-Planck Hamiltonian and equilibrium solution}
\label{sec:stationarysolution}

We construct the Fokker-Planck Hamiltonian $\HFP$ and prove that it is Hermitian.

Let $\psi(x;t)$ be 
\begin{align}
\psi(x;t) = P(x;t) e^{\frac12 S(x)}. 
\end{align}
The time derivative of this function is 
\begin{align}
 \partial_t \psi &= (\partial_t P) e^{\frac12 S}, \nn \\
      &= \bigl[\delx ( S' + \delx) P  \bigr] e^{\frac12 S}.
\end{align}
For any differentiable function $A$, we have 
\begin{align}
(\delx A) e^{\half S} & = \left(\delx - \frac{1}{2} S' \right) \left(A e^{\frac{1}{2} S}\right),
\end{align}
where $\partial_x A = A'$.
By using this relation, we can derive 
\begin{align}
- \partial_t \psi & =  - \biggl( \delx - \frac{1}{2}S' \biggr) \biggl(\delx + \frac12 S' \biggr) \psi, \nn \\
   & = \HFP \psi. 
\label{eq:2017Apr15eq1}
\end{align}
This $\HFP$ is called the Fokker-Planck Hamiltonian.

The FP Hamiltonian is Hermitian. Formally, it can be seen as 
\begin{align}
\HFP^\dagger& =   - \left( \biggl( \delx - \frac{1}{2}S' \biggr) \biggl(\delx + \frac12 S' \biggr) \right)^\dagger \nn \\
            & =   - \biggl(\delx + \frac12 S' \biggr) ^\dagger \biggl( \delx - \frac{1}{2}S' \biggr)^\dagger  \nn \\
            & =   - \biggl( - \delx + \frac12 S' \biggr) \biggl( - \delx - \frac{1}{2}S' \biggr) \nn \\
            & = \HFP.
\end{align}

Let us show the Hermiticity of the FP Hamiltonian more explicitly.
For that, we consider $\int d x ( \HFP \phi)^* \phi$, and prove  
\begin{align}
\int d x \,  (\HFP \phi)^* \phi = \int dx \, \phi^* \HFP  \phi. 
\end{align}
By writing $\HFP$ explicitly, we obtain 
\begin{align}
\int dx \, (\HFP \phi^*) \phi = -  \int dx \, \biggl[\biggl( \delx^2 + \frac{1}{2} \delx S' - \frac{1}{2} S' \delx - \frac{1}{4} S'^2 \biggr) \phi \biggr]^* \phi.
\end{align}
Because $S\in\mathbb{R}$, 
\begin{align}
\int dx \, (\HFP \phi^*) \phi = -  \int dx \, \biggl[\biggl( \delx^2 + \frac{1}{2} \delx S' - \frac{1}{2} S' \delx - \frac{1}{4} S'^2 \biggr) \phi^* \biggr] \phi.
\label{eq:2016Jan09eq2}
\end{align}
We integrate each term by part, as 
\begin{align}
\int dx (\delx^2 \phi^*) \phi &= \int dx \, \phi^* \delx^2 \phi + (\abst), \nn \\
\int dx [\delx (S' \phi^*) ] \phi &=  - \int dx \,  \phi^* (S' \delx ) \phi + (\abst), \\
\int dx \, S' ( \delx \phi^* ) \phi  & =  - \int dx \, \phi^* (\delx S' \phi ) + (\abst), \\
\int dx (S'^2 \phi^*) \phi & = \int dx \, \phi^* S'^2 \phi + (\abst).
\end{align}
Here $(\abst)$ represents the surface term $x\in d\Omega$.
By substituting these relations to \eqref{eq:2016Jan09eq2}, we obtain
\begin{align}
\int dx \, (\HFP \phi^* ) \phi &= 
- \int dx \, \phi^* \biggl( \delx^2 - \frac{1}{2} S' \delx + \frac{1}{2} \delx S' - \frac{1}{4} S'^2\biggr) \phi, \nn \\
& = \int dx \, \phi^* \HFP \phi.
\end{align}
Therefore, $\HFP^\dagger = \HFP$ holds. 

\section{Example of the stochastic quantization}
\label{sec:SQ_2ndqunt}
In order to understand the properties of the FP Hamiltonian, we consider the stochastic quantization of the system described by a simple action
\begin{align}
S = \frac{1}{2} k x^2.   
\end{align}
The Fokker-Planck Hamiltonian $\HFP$ is 
\begin{align}
\HFP &= - \frac{1}{2} \delx^2 + \frac{1}{8} k^2 x^2  - \frac{1}{4} k. 
\end{align}
We introduce the creation and annihilation operators $a^\dagger$ and $a$, 
\begin{align}
a &= \sqrt{\frac{1}{2\omega}} (\delx + \omega x ), \\
a^\dagger &= \sqrt{\frac{1}{2\omega}} ( -\delx + \omega x ).
\end{align}
where $\omega = k/2$. 
Equivalently, 
\begin{align}
\omega x & = \sqrt{\frac{\omega}{2}} ( a + a^\dagger), \\
  \delx  & = \sqrt{\frac{\omega}{2}} ( a - a^\dagger).
\end{align}
By using them, the FP Hamiltonian is expressed as 
\begin{align}
H_{\rm FP} &= \frac{\omega}{2} ( aa^\dagger + a^\dagger a) - \frac{1}{4}k , \nn \\
           &= \frac{\omega}{2} ( aa^\dagger + a^\dagger a) - \frac{1}{2}\omega , \nn \\
           &= \omega a^\dagger a.   
\end{align}
We can see that the zero-point energy is removed.

The $n$-particle state $|n\ket$ is given by 
\begin{align}
|n\ket = \frac{1}{\sqrt{n!}} (a^\dagger)^n |0\ket
\end{align}
It is the eigenstate of the FP Hamiltonian with the energy $\omega n$:
\begin{align}
H_{\rm FP} |n\ket = \omega a^\dagger a |n\ket = \omega n | n\ket
\end{align}
Let the solution of the FP equation be 
\begin{align}
\psi(x,t) = \sum_{n=0}^\infty a_n | n \ket. 
\end{align}
The left and right-hand sides of the FP Hamiltonian are
\begin{align}
(l.h.s) & =  -\frac{\partial}{\partial \tau} \psi  = - \sum_{n=0}^\infty \dot{a}_n | n\ket, \\
(r.h.s) & = H_{\rm FP} \psi = \sum_{n=0}^\infty \omega n a_n| n \ket. 
\end{align}
By noticing that $|n\ket$'s are orthogonal to each other, we conclude 
\begin{align}
\dot{a}_n = - \omega n a_n. 
\end{align}
Therefore, 
\begin{align}
 a_n = a_n^{(0)} e^{- \omega n \tau}, 
\end{align}
and
\begin{align}
\psi = \sum_{n=0}^{\infty} a_n^{(0)}e^{- \omega n \tau} | n \ket.
\end{align}
As $\tau\to \infty$, all modes but the zero mode vanish. 
Hence, the probability distribution $P$ behaves as 
\begin{align}
P(x,t) & = e^{-\frac{1}{2} S(x)} \psi(x,t), \nn \\
       & \to e^{-\frac{1}{2} S(x)}  a_0^{(0)} | 0 \ket.
\end{align}

\section{Langevin equation for lattice QCD}
\label{sec:QCDLangevinequation}

We derive the Langevin equation for lattice QCD. Via the path integral, the partition function is given by 
\begin{align}
Z = \int \calD U e^{-S(U) }, 
\end{align}
where $\calD U$ is the Haar measure. The link variable $U_{x\mu}$ can be expressed by using real variables $\omega_{ax\mu}$ as 
\begin{align}
U_{x\mu}=e^{ i \lambda_a \omega_{ax\mu}}. 
\label{eq:2016feb18eq1}
\end{align}
In terms of $\omega$, the Langevin equation can be defined by. 
\begin{align}
\dot{\omega}_{ax\mu} = - \frac{ \partial S}{\del \omega_{ax\mu}} + \eta_{ax\mu}.  
\end{align}
$\eta_{ax\mu}$ is the Gaussian noise. The distribution of the Gaussian noise is $\exp(- \eps \eta^2/4)$.

The discretized Langevin equation is 
\begin{align}
\omega_{ax\mu}(t+\eps)  = \omega_{ax\mu}(t) + \biggl( - \frac{ \partial S}{\del \omega_{ax\mu}} + \eta_{ax\mu}\biggr) \eps. 
\end{align}
By using the Gell-man matrices (generators of SU($N_c=3$)) $\lambda_a$,
\begin{align}
\sum_a \lambda_a \omega_{ax\mu}(t+\eps)  = \sum_a \lambda_a \biggl[\omega_{ax\mu}(t) + \biggl( - \frac{ \partial S}{\del \omega_{ax\mu}} + \eta_{ax\mu}\biggr) \eps \biggr].
\label{eq:2016feb18eq2}
\end{align}

We change the normalization of the noise such that it can be generated via the Box-Muller method. In the Box-Muller method, the Gaussian random number $x$ is generated with the weight $\exp(-x^2/2)$.
Hence we introduce $\eta'$ such that $\eps \eta^2/4 = \eta'^2/2$. By using the same symbol $\eta$ for this $\eta'$ to simplify the notation, we obtain 
\begin{align}
\sum_a \lambda_a \omega_{ax\mu}(t+\eps)  = \sum_a \lambda_a \left[ \omega_{ax\mu}(t) + \biggl( - \frac{ \partial S}{\del \omega_{ax\mu}} \biggr) \eps + \sqrt{2 \eps} \, \eta_{ax\mu} \right]. 
\end{align}

From \eqref{eq:2016feb18eq1} we obtain $ \sum_a \omega_a \lambda_a = -i  \ln U $, hence \eqref{eq:2016feb18eq2} is 
\begin{align}
\ln U_{x\mu}(t+\eps) & = \ln U_{x\mu}(t) + i X, \\
X &= \sum_a \lambda_a \biggl[ \biggl( - \frac{ \partial S}{\del \omega_{ax\mu}} \biggr) \eps  + \sqrt{2 \eps} \, \eta_{ax\mu}\biggr]. 
\end{align}
For infinitesimal $\eps$, we obtain 
\begin{align}
U_{x\mu}(t+\eps) = e^{iX} U_{x\mu}(t).  
\end{align}
This equation describes the Langevin-time evolution of the link variables.

In terms of the link variable, the derivative is given by 
\begin{align}
D_{ax\mu} f( \link{x}{\mu}) = \lim_{\delta\to 0} \frac{ f( e^{ i \delta \lambda_a} \link{x}{\mu}) - f( \link{x}{\mu})}{\delta}. 
\end{align}

\underline{Gauge drift term}

We write down the explicit form of $X$ for the plaquette action. 
Let the plaquette which contains a specific link variable $U$ be 
\begin{align}
S = c\,  \tr \left [ U A + A^{-1} U^{-1} \right], \, (c=-\frac{\beta}{6}).
\end{align}
Here we used $U^{-1}$ rather than $U^\dagger$, having the generalization to the complex Langevin method in mind.

To determine the derivative, we consider an infinitesimal variation 
\begin{align}
S(U) \to S(e^{i \lambda_a \delta } U) = c \, \tr \left[ e^{i \lambda_a \delta } U A + A^{-1} U^{-1} (e^{ i \lambda_a \delta })^{-1} \right]. 
\end{align}
Because $e^{i \lambda_a \delta}$ is unitary and $\lambda_a$ is Hermitian, $(e^{i \lambda_a \delta })^{-1} = (e^{ i \lambda_a \delta })^\dagger = e^{- i \lambda_a\delta }$.
By keeping up to the first-order terms, we obtain 
\begin{align}
S(e^{i \lambda_a \delta } U) = S(U) +   (i c \delta) \, \tr [ \lambda_a  U A -  A^{-1} U^{-1} \lambda_a ]. 
\end{align}
Therefore, the derivative is given by 
\begin{align}
D_{a} S = i \, c \, \tr \lambda_a (U A - A^{-1} U^{-1}). 
\end{align}

By substituting it to \eqref{eq:2016feb25eq2}, we obtain 
\begin{align}
X & = \sum_{a=1}^{N^2-1} \left( - D_{a x \mu} S + \eta_{ax\mu} \right) \lambda_a, \nn \\
  & = \sum_{a=1}^{N^2-1} \lambda_a \left( - i \, c \, \tr \left[\lambda_a (U A - A^{-1} U^{-1})\right]  + \eta_{ax\mu} \right), \nn \\
  & = \sum_{a=1}^{N^2-1} \lambda_a \left( \tr [\lambda_a B] + \eta_{a x \mu} \right) , 
\end{align}
where $B = -i c (UA - A^{-1} U^{-1}) = i ( \beta/6) (UA - A^{-1} U^{-1})$.

We can use the properties of the Gell-Mann matrices 
\begin{align}
\sum_{a=1}^{N_c^2-1} (\lambda^a)_{\alpha \beta} (\lambda^a)_{\gamma \delta} = 2 \left( \delta_{\alpha \delta} \delta_{\beta \gamma} 
- \frac{1}{N_c} \delta_{\alpha \beta} \delta_{\gamma \delta} \right) 
\label{eq:2016feb28eq1}
\end{align}
to simplify the expression as 
\begin{align}
X & =  2 \left(B - \frac{1}{N_c} (\tr B) \1   \right) + \sum_{a=1}^{N^2-1} \lambda_a \eta_{a x \mu}. 
\end{align}

\underline{Fermion drift term}

The fermion part of the action is given by 
\begin{align}
S_f = - \frac{N_f}{4} \ln \det M. 
\end{align}
The derivative is
\begin{align}
D_a S_f = - \frac{N_f}{4} \tr \left[ M^{-1} D_a M \right].
\end{align}
Therefore, the fermion part of $X$ is 
\begin{align}
X_f  & = - \sum_{a=1}^{N_c^2-1} (D_a S_f ) \lambda_a \epsilon, \nn \\
     & = \frac{N_f}{4} \sum_{a=1}^{N_c^2-1} \tr \left[ M^{-1} D_a M \right] \lambda_a \epsilon. \nn 
\end{align}

\bibliographystyle{utphys}
\bibliography{ref_list}

\end{document}